\documentclass{tADP2e}

\usepackage{natbib}
\usepackage{graphics,graphicx}
\usepackage{epsfig}
\usepackage{amssymb}
\usepackage[intlimits,centertags]{amsmath}
\usepackage{eepic}
\usepackage{array,longtable}
\usepackage{latexsym}

\renewcommand{\propto}{\sim}

\newcommand{\textcolor}[1]{}
\newcommand{\red}{}

\newcommand{\blue}{}
\newcommand{\textcite}[1]{Ref.\ \cite{#1}}
\newcommand{\void}{}

\begin{document}
 
\title{Electronic structure calculations\\
 using dynamical mean field theory}

\author{{K.\ HELD}$^*$\thanks{$^*$ E-mail: k.held@fkf.mpg.de}\\
Max-Planck-Institut f\"ur
Festk\"orperforschung, D-70569 Stuttgart, Germany}

\markboth{K.\ Held}{Electronic structure calculations using dynamical mean field theory}
\maketitle             

\begin{abstract}

The calculation of electronic properties of materials
is an important task of solid state theory, albeit
particularly difficult if electronic correlations
are strong, for example in transition metals, their oxides and in $f$-electron systems.
The standard approach to material calculations, the density functional theory in 
its local density approximation (LDA), incorporates  electronic correlations
only very rudimentarily and fails if the correlations are strong.
Encouraged by the success of 
dynamical mean field theory (DMFT) 
in dealing with strongly correlated model Hamiltonians,
physicists from the bandstructure and  the many-body community 
have joined  forces and developed a combined LDA+DMFT method recently.
Depending on the strength of electronic
correlations, this new approach
yields a weakly correlated metal as in LDA, a strongly correlated metal,
or a Mott insulator. By now, this approach is widely regarded as  a breakthrough
for electronic structure calculations of strongly correlated materials.

The author will review this LDA+DMFT method and also discuss
alternative approaches to employ DMFT in
electronic structure calculations, for example, by replacing the 
LDA part by the so-called GW approximation.
Different methods to solve the DMFT equations are introduced with a focus on
those that are suitable for realistic calculations with many orbitals.
An overview of the successful application of LDA+DMFT to a wide variety of materials, ranging
from Pu and Ce, to Fe and Ni, to numerous transition metal oxides, is given.

\hspace{2em}

\centerline{\bfseries Contents} 
 
\hspace{1em}

\noindent \ref{intro} {Introduction}\hfill \pageref{intro}\\
\ref{ESC} {Conventional electronic structure calculations}\hfill \pageref{ESC}\\
\phantom{1.} \ref{Sec:abinitioHam} {Ab initio Hamiltonian}\hfill \pageref{Sec:abinitioHam}\\
\phantom{1.} \ref{dft} {Density functional theory (DFT)}\hfill \pageref{dft}\\
\phantom{1.} \ref{lda}  {Local density approximation (LDA)}\hfill \pageref{lda}\\
\ref{dmft}  {Dynamical mean field theory (DMFT)}\hfill \pageref{dmft}\\
\phantom{1.} \ref{DMFTintro} {Derivation of the DMFT equations}\hfill \pageref{DMFTintro}\\
\phantom{1.} \ref{ClusterDMFT} {Extensions of DMFT}\hfill \pageref{ClusterDMFT}\\
\ref{AbinitioDMFT} {Merging conventional bandstructure approaches with DMFT}\hfill \pageref{AbinitioDMFT}\\
\phantom{1.} \ref{LDADMFT} {LDA+DMFT}\hfill \pageref{LDADMFT}\\
\phantom{1.1.1.} \ref{HamLDADMFT} {Hamiltonian formulation}\hfill \pageref{HamLDADMFT}\\
\phantom{1.1.1.} \ref{selfconsist} {Self-consistent LDA+DMFT calculations}\hfill \pageref{selfconsist}\\
\phantom{1.1.1.} \ref{constrainedLDA} {Constrained LDA calculations of the interaction parameters}\hfill \pageref{constrainedLDA}\\
\phantom{1.1.1.} \ref{SDFT} {Spectral density functional theory formulation}\hfill \pageref{SDFT}\\
\phantom{1.1.1.} \ref{SimpTMO} {Simplifications for transition metal oxides}\hfill \pageref{SimpTMO}\\
\phantom{1.} \ref{HDMFT} {Hartree+DMFT and Hartree-Fock+DMFT}\hfill \pageref{HDMFT}\\
\phantom{1.} \ref{GWDMFT} {GW+DMFT}\hfill \pageref{GWDMFT}\\
\ref{DMFTsolvers} {DMFT solvers suitable for material calculations}\hfill \pageref{DMFTsolvers}\\
\phantom{1.} \ref{HartreeFock} {Polarised Hartree-Fock (HF) approximation (LDA+U)}\hfill \pageref{HartreeFock}\\
\phantom{1.} \ref{HubbardI} {Hubbard-I, Hubbard-III and alloy-analogy approximation}\hfill \pageref{HubbardI}\\
\phantom{1.} \ref{IPT} {Iterated perturbation theory (IPT) and extensions}\hfill \pageref{IPT}\\
\phantom{1.} \ref{NCA} {Non-crossing approximation (NCA)}\hfill \pageref{NCA}\\
\phantom{1.} \ref{QMC} {Quantum Monte Carlo (QMC) simulations}\hfill \pageref{QMC}\\
\phantom{1.1.1.} \ref{HFQMC} {Hirsch-Fye algorithm}\hfill \pageref{HFQMC}\\
\phantom{1.1.1.} \ref{PQMC} {Projective quantum Monte Carlo  simulations for $T\!=\!0$}\hfill \pageref{PQMC}\\
\phantom{1.1.1.} \ref{Fourier} {Fourier transformation from $\tau$ to $i\omega_{\nu}$}\hfill \pageref{Fourier}\\
\phantom{1.1.1.} \ref{SecMEM} {Maximum entropy method}\hfill \pageref{SecMEM}\\
\phantom{1.} \ref{LaTiO3} {Comparing different DMFT solvers for La$_{1-x}$Sr$_{x}$TiO$_{3}$}\hfill \pageref{LaTiO3} \\
\ref{results} {Realistic material calculations with DMFT}\hfill \pageref{results}\\
\phantom{1.} \ref{felectrons} {$f$ electron systems}\hfill \pageref{felectrons}\\
\phantom{1.1.1.} \ref{Sec:Ce} {Volume collapse transition in cerium}\hfill \pageref{Sec:Ce} \\
\phantom{1.} \ref{TM} {Transition metals}\hfill \pageref{TM}\\
\phantom{1.1.1.} \ref{Sec:FeNi} {Ferromagnetism in Fe and Ni}\hfill \pageref{Sec:FeNi}\\
\phantom{1.} \ref{TMO} {Transition metal oxides}\hfill \pageref{TMO}\\
\phantom{1.1.1.} \ref{Sec:LaTiO3} {Ferro-orbital order in LaTiO$_3$}\hfill \pageref{Sec:LaTiO3}\\
\phantom{1.1.1.} \ref{V2O3} {Mott-Hubbard transition in  V$_2$O$_3$}\hfill \pageref{V2O3}\\
\phantom{1.1.1.} \ref{Sec:VO2} {Peierls transition in VO$_2$}\hfill \pageref{Sec:VO2}\\
\phantom{1.1.1.} \ref{Sec:ruthenates} {Orbital selective Mott-Hubbard transition in  Ca$_{2-x}$Sr$_x$RuO$_4$}\hfill \pageref{Sec:ruthenates}\\
\phantom{1.1.1.} \ref{SrVO3} {`Kinks' in SrVO$_3$} \hfill \pageref{SrVO3}\\
\phantom{1.1.1.} \ref{Sec:cobaltates} {$e_{g}'$ hole pockets in  Na$_x$CO$_2$}\hfill \pageref{Sec:cobaltates}\\
\phantom{1.1.1.} \ref{Sec:LiVO} {Heavy-Fermion behaviour in LiV$_2$O$_4$}\hfill \pageref{Sec:LiVO}\\
\phantom{1.1.1.} \ref{Sec:CMR} {Colossal magnetoresistance in manganites}\hfill \pageref{Sec:CMR}\\
\phantom{1.} \ref{OM} {Other Materials}\hfill \pageref{OM}\\
\phantom{1.1.1.} \ref{Sec:HMFM} {Half-metallic ferromagnetism in Heussler alloys}\hfill \pageref{Sec:HMFM}\\
\phantom{1.1.1.} \ref{Sec:fullerenes} {Superconductivity in A$_x$C$_{60}$}\hfill \pageref{Sec:fullerenes}\\
\phantom{1.1.1.} \ref{Sec:zeolites} {Mott-insulating zeolites}\hfill \pageref{Sec:zeolites}\\
\ref{summary} {Summary and outlook}\hfill \pageref{summary}\\
\phantom{1} {References}\hfill \pageref{references}\\

\end{abstract}


\section{Introduction}
\label{intro}
\label{introduction}

One of the most important challenges of theoretical physics is the development of reliable methods for the
quantitative calculation of material properties. In solid state theory,
we know the Hamiltonian to do these material calculations.
For example, if we  neglect 
relativistic corrections and employ the Born-Oppenheimer \cite{Born27a} approximation,
this Hamiltonian  consists
of three terms: the kinetic energy, the lattice potential and the Coulomb interaction
between the electrons:
\begin{equation}
 {H} = \sum_i
     \left[ \;\textcolor{blue}{ -\frac{\displaystyle  \hbar^2 \Delta_i }{\displaystyle 2 m_e}}\;+\; \textcolor{green}{\sum_l
\frac{\displaystyle - e^2}{\displaystyle 4\pi\epsilon_0} \; \frac{\displaystyle Z_l}{\displaystyle |{\bf r}_i-{\bf R}_l|}}  \right]\;\;\; +\;\;\;
    \frac{1}{2} \; \sum_{i\neq j} \; \textcolor{red}{\frac{\displaystyle  e^2}{\displaystyle 4\pi\epsilon_0}\; \frac{\displaystyle 1}{\displaystyle |{\bf r}_i -{\bf r}_j|}}. \label{AbInitioHam}
\end{equation}
Here,  ${\bf r}_i$ and ${\bf R}_l$ denote the position of electron $i$ and ion $l$
with charge $-e$ and $Z_l e$, respectively; $\Delta_i$ is the Laplace operator 
for the kinetic energy of electrons with mass $m_e$;
$\epsilon_0$ and $\hbar$ are the vacuum dielectric and  Planck constant.

\begin{figure}[tb]
\vspace{-1.2cm}

\phantom{a} \hspace{-.99cm} 
\hspace{.6cm} \epsfig{file=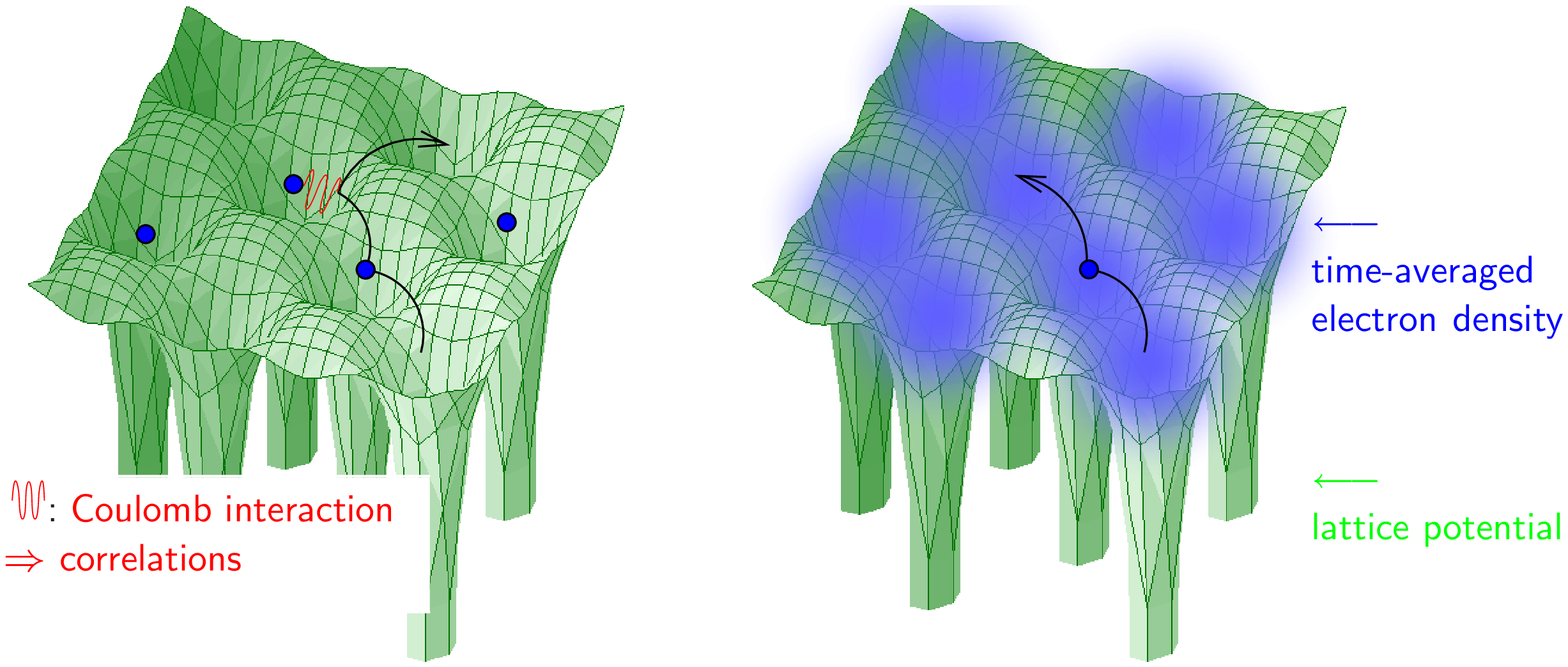,width=10.4cm,angle=90}
\vspace{-4.83cm}

{\caption{\label{FigHam}}}

\begin{tabular}{p{.45\linewidth} p{0.01\linewidth}  p{.44\linewidth}}
{\bf Solid State Hamiltonian}:&&  {\bf LDA approximation}:\\ 
 To calculate the physical properties
of a given material, one has to take into account three terms (Hamiltonian [\ref{AbInitioHam}]):
The  kinetic energy due to which the electrons move
(arrow),  the lattice potential of the ions and the
Coulomb interaction between the electrons. 
Due to the latter, the moving electron  is repelled by the
electron already located at this site. It is energetically favourable if the depicted electron 
hops somewhere else, as indicated by the arrow.
Hence, the movement
of every electron is correlated with that of every other. &&
LDA is an approximation which allows for calculating material properties, but
dramatically simplifies the electronic correlations:  The LDA bandstructure 
corresponds to the simplification that every electron moves
independently, i.e. uncorrelated, within 
a time-averaged local density of the other electrons
(visualised as static clouds in the figure).
 \end{tabular}
\end{figure}

The three terms of Hamiltonian [\ref{AbInitioHam}] describe the
movement of the electrons, the attractive  lattice potential of the (fixed) ions and the mutual
Coulomb repulsion of the electrons, see Fig.\ \ref{FigHam} for an illustration.
But while we know the Hamiltonian, we cannot solve it, even not numerically, if
more than a very few  [${\cal O}(10)$] electrons are involved.
This problem is due to the last term, the Coulomb interaction, which
correlates the movement of every electron
$i$  with every other electron $j$.  
The numerical effort to solve  Eq.\ [\ref{AbInitioHam}]
quantum-mechanically grows exponentially with the number of correlated electrons. 
Such an 
exponential problem cannot be solved for significantly 
many electrons, even
if computer power continued to  grow rapidly.

In this situation, we have two possibilities: 
Either we dramatically simplify the Hamiltonian
[\ref{AbInitioHam}], hoping that the simplified model
allows for a qualitative understanding including electronic
correlation. 
Or we employ 
equally dramatic approximations to deal with  
Eq.\ [\ref{AbInitioHam}] directly.
These two strategies have been followed by the two large communities of
solid state theory: the many-body model Hamiltonian and the
density functional community.

Within  the density functional theory (DFT) introduced by Hohenberg and Kohn \cite{Hohenberg64a},
the local density approximation (LDA)  turned out to be unexpectedly successful,
and established itself as {\em the} method for realistic solid state calculations
in the last century; for reviews see e.g.\ Refs.\ \cite{Jones89a,Harrison89,Dreizler90,Martin04}. 
 This is surprising because LDA represents
a substantial
approximation to the  Coulomb interaction. Basically, an
electron  at $r_i$ sees a  time-averaged {\em local} density $\rho({\mathbf r}_i)$
of the other electrons,  with a corresponding
{\em local} LDA potential $V_{\rm LDA}(\rho({\mathbf r}_i))$,
as visualised in Fig.\ \ref{FigHam}. This reduces the many-body 
problem to a single electron calculation in a {\em local} potential.

The success of LDA shows that this treatment is actually sufficient
for many materials, both for calculating ground state energies and
bandstructures,  implying that electronic correlations are rather weak in these materials. But,
there are important  classes of materials where LDA fails, such as transition metal oxides
or heavy Fermion systems. For example, LDA
predicts
La$_{2}$CuO$_{4}$, Cr-doped V$_2$O$_3$ and NiO  to be 
 metals above the antiferromagnetic ordering temperature
 whereas, in reality, these materials are 
 insulators \cite{Leung88a,Zaanen88a,Pickett89a,Mattheiss72a,Mattheiss72b}.   
Similarly, the LDA  bandstructure is in strong disagreement with
experiment for $f$-electron systems. In these two classes of materials, the valence electrons partially occupy the
$d$ or $f$ orbitals.
If there are two electrons in these narrow orbitals on the same lattice site,
the distance $|{\bf r}_i -{\bf r}_j|$ is particularly short, and  electronic correlations are thus
particularly strong.
Hence, the approximate LDA treatment of  Hamiltonian [\ref{AbInitioHam}]  fails.

\begin{figure}[tb]
\vspace{-.525cm}

\epsfig{file=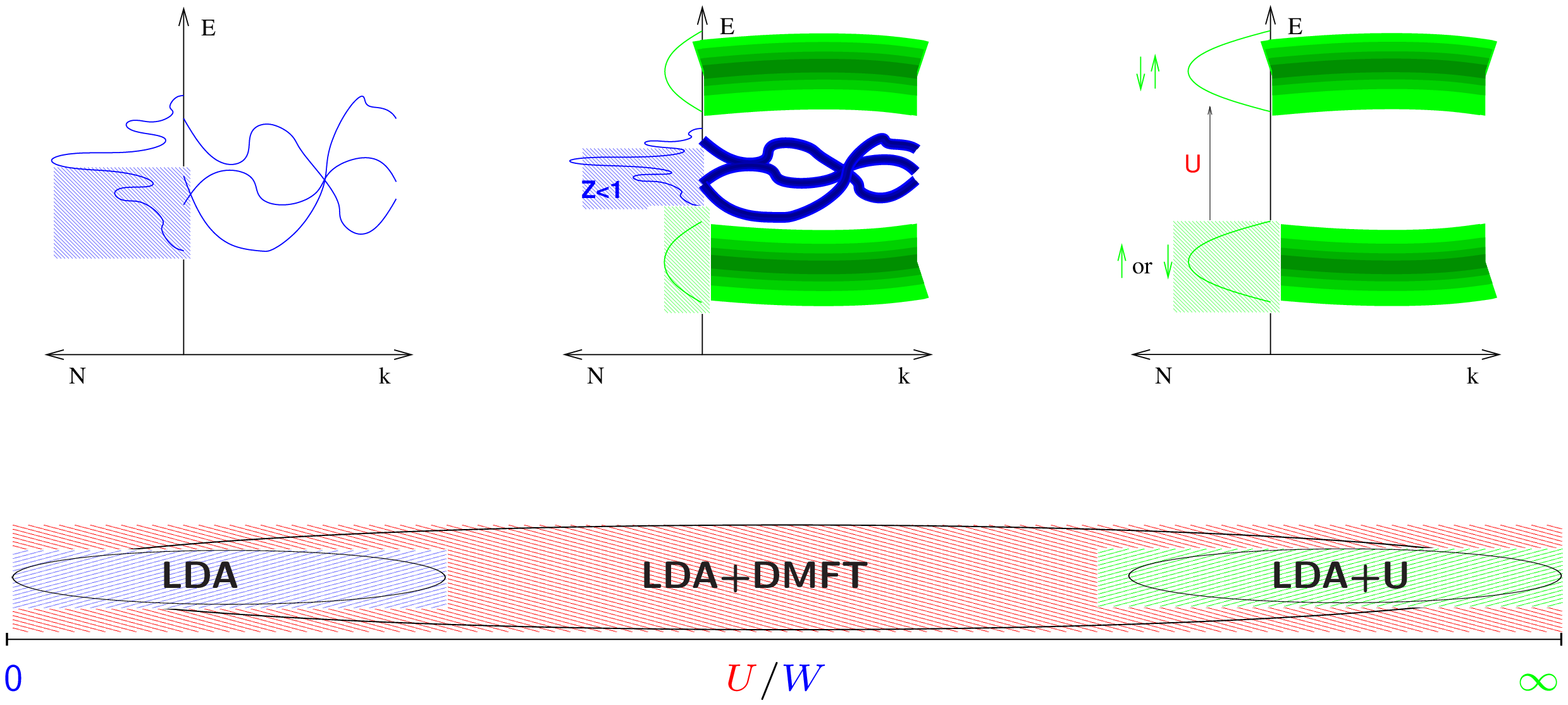,width=10.4cm,angle=90}
\vspace{-4.6cm}

\caption{\label{LDALDAULDADMFT}
\label{figLDADMFT}}
\vspace{.2cm}

\begin{tabular}{p{.26\linewidth}
 @{\hspace{0.05\linewidth}}
 p{.35\linewidth}
 @{\hspace{0.05\linewidth}}
 p{.27\linewidth}}

 {\textcolor{blue}{\bf Weakly }} &
{\textcolor{red}{\bf Strongly}} &
 {\textcolor{green}{\bf Mott insulator}}:\\
 {\textcolor{blue}{\bf correlated metal}}: &
{\textcolor{red}{\bf  correlated metal}}: &
 {\textcolor{green}{\bf}}\\
 The on-site Coulomb interaction is weak compared to the
 LDA bandwidth: $U\! \ll \! W$; and LDA gives the correct answer, typically
 a (weakly correlated) metal for which we schematically draw a density of
 states $N$ at  energies $E$
and the bandstructure, i.e. $E$ vs.\ wave vector ${\mathbf k}$.
&
 In this intermediate regime,
 one has already Hubbard bands, like for $U/W\!\gg\! 1$ (right hand side),
 but at the same time a remainder of the weakly correlated LDA metal (left hand side),
 in  form of a quasiparticle peak: The ($U\!=\!0$) LDA bandstructure is reproduced,
 albeit with its width and weight reduced by a factor $Z$
 and life time effects which result in a Lorentzian broadening of the quasiparticle levels.
 &
 If the Coulomb interaction $U$ becomes large ($U\!\gg\! W$),
 the LDA band splits into two Hubbard bands, and we have a Mott insulator with only the lower
 band occupied (at integer fillings). Such a splitting can be described  by
 the so-called LDA+U method with the drawbacks discussed in the text.
\end{tabular}
\vspace{.1cm}

\centerline{The entire parameter regime is described by LDA+DMFT.}
\end{figure}

To describe these $d$- and $f$-electron systems correctly,
genuine many-body effects have to be taken into account.
The  transition metal oxides mentioned above are, for example,
Mott  insulators \cite{Mott49a,Mott68a,Gebhard97a}.
Since the $d$ orbitals are only partially filled the LDA bandstructure
predicts metallic behaviour, see the left panel of Fig.\ \ref{LDALDAULDADMFT}.
But instead,
the on-(lattice-)site Coulomb repulsion $U$ splits  the metallic LDA bands 
into two sets of Hubbard \cite{Hubbard63a} bands, 
as in the right panel of Fig.\ \ref{LDALDAULDADMFT}. Let us for a moment assume
that the average number of $d$ or $f$ electrons is 
one per lattice site (`integer filling'). Then, one can envisage the lower Hubbard band as
consisting of all  states with  exactly one electron on every lattice site
and the upper Hubbard band as those states where two electrons are in 
 $d$ or $f$ orbitals on the
same lattice site. Since it costs  an energy $U$ to have
two electrons on the same lattice sites, the latter states are
completely empty and the former completely filled with a gap of size $U$ in between. 
Such kind of Mott-Hubbard physics and the associated
energy gain is completely missing in the LDA. 

This shortcoming can be overcome by the self interaction correction (SIC) to LDA
by Perdew and Zunger \cite{Perdew81,Svane90,Szotek93}
or
by  the  LDA+U method of
Anisimov {\em et al.} \cite{Anisimov91}  which we will discuss in
Section \ref{LDAU}.
In the presence of orbital or magnetic ordering  these approaches
 yield insulating spectra, similar to that of the paramagnetic
Mott insulator displayed
in the right panel of Fig.\ \ref{LDALDAULDADMFT}.
Important differences are that the LDA+U spectra
are completely coherent and that
 the LDA+U solution has not the high entropy
of the finite temperature paramagnetic solution, because of the
 orbital or magnetic ordering.
For ordered systems on the other hand,
polarons \cite{Brinkman70a,Bulaevskii68,Szczepanski90,Martinez91,Capelluti02} result in additional peaks in
the  Hubbard bands. As was noted
by Sangiovanni {\em et al.} \cite{Sangiovanni05} 
these polaron peaks   are present
in dynamical mean field theory \cite{Strack92a}
but not in the static  LDA+U
mean field theory.
Another drawback is that  SIC-LDA and LDA+U
almost automatically yield localised electrons
with split (Hubbard) bands,
even if this is not correct. 

Missing in both LDA and LDA+U is  also the  strongly correlated metallic phase
found in many transition metal oxides
and heavy Fermion systems either at intermediate
values of $U$ or at
a non-integer number of $d$ or $f$ electrons per site
so that the Mott insulator is doped and becomes metallic.
This strongly correlated metal already has Hubbard bands
as the Mott insulator, but the low-energy behaviour is
dominated by {\em quasiparticle} 
physics, see the central panel of Fig.\ \ref{LDALDAULDADMFT}. 
We can think of these quasiparticles as  dressed electrons
which move independently from each other, although
with  a larger  effective  mass $m$ since the quasiparticle
consists of the initial electron plus 
the electron-electron interaction with its environment.
The mass enhancement $m/m_e$ becomes more pronounced 
with increasing $U$, and the weight of the 
quasiparticle peak  $Z= (m/m_e)^{-1}$ is reduced correspondingly
until it vanishes completely 
at  the Mott-Hubbard transition where the system becomes  insulating.

A modern, non-perturbative technique
to describe these  many-body effects is the dynamical mean field theory (DMFT)
developed primarily by  Metzner and Vollhardt \cite{Metzner89a}, 
M\"uller-Hartmann 
\cite{MuellerHartmann89a}, Georges and Kotliar
\cite{Georges92a},  and Jarrell
\cite{Jarrell92a}; for more details see Section \ref{DMFT}.
One of the strong points of  DMFT
is that it can describe such strongly correlated metals and Mott insulators
in a single framework.

Recently, physicists from the many-body and DFT community 
have developed
a new approach to correlated materials which merges
two of the most successful approaches of the two communities:
LDA+DMFT. 
 The initial work by Anisimov {\em et al.} \cite{Anisimov97a} and 
Lichtenstein and Katsnelson \cite{Lichtenstein98a}
was followed by a rapid development which is the subject of this Review.
Shorter LDA+DMFT summaries have already been published in
the form of  conference proceedings, lecture notes and a  Psi-k Highlight,
see  Refs.\
\cite{LDADMFT2,LDADMFT3,Held03,LichtensteinCP,KotliarCP01,Oudovenko04a,Georges04}; a full review was submitted by
Kotliar {\em et al.} \cite{Kotliar06} during the completion of this work.
Instead of LDA, one can also use the so-called GW
 approximation of Hedin \cite{Hedin},
resulting in the related GW+DMFT approach formulated by 
Biermann {\em et al.} \cite{Biermann03} and presented in more detail in \textcite{BiermannCP04} and \textcite{AryasetiawanCP04}. 

The LDA+DMFT approach does not 
 only contain the correct quasiparticle physics and energy of the strongly correlated metallic phase, but also
reproduces the correct results in the limit of small and large $U$, 
see Fig.\ \ref{LDALDAULDADMFT}.
Hence, LDA+DMFT correctly accounts for the {main contributions} of electronic
correlations (there are corrections
due to  the $\mathbf k$-dependence of the self energy as discussed in Section \ref{DCA}).
In contrast,
LDA yields an uncorrelated metal even for strongly-correlated metals or 
Mott insulators. LDA+U on the other hand
typically predicts an insulator for the {\em ab-initio} calculated $U$ values of $3d$  transition metal oxides, even 
for materials which are metallic.
Considering this,  it might not be so astonishing that
LDA+DMFT
turned out to be a breakthrough for the calculation
of materials with strong electronic correlations.

In the following, we will review these
recent developments at the former borderline between
DFT and many-body theory. Thereby we
take a Hamiltonian point of view which allows 
us to understand  physically what LDA+DMFT describes,
and what it does not.
Complementarily,  the review of Kotliar {\em et al.} \cite{Kotliar06}
focuses on the spectral density functional theory 
for which LDA+DMFT is  the standard approximation
just as LDA is for DFT.

We start, in Section \ref{ESC},
with a brief recapitulation of conventional DFT/LDA band structure calculation. For more LDA and DFT details,
we advise the reader to  consult  the excellent review by Jones and 
Gunnarsson \cite{Jones89a} or the books Refs.\ \cite{Jones89a,Harrison89,Dreizler90,Martin04}. 
In the present Review, we will not elaborate on how to construct an
optimal set of wave functions within LDA and how to choose maximally localised orbitals
for the further many-body DMFT calculation.
To this end, the  linearised
muffin tin orbitals (LMTO) of Andersen  \cite{LMTO1}, for example, in the third generation \cite{Andersen99,AndersenPsik},
or the Wannier \cite{Wannier37} function projection by Marzari and Vanderbilt  \cite{Marzari97} can be employed; also see Refs.
\cite{Souza02,Ku02,Schnell02a,Schnell03a,Anisimov05,Lechermann06a}.
We refer the reader to
these original and review publications for the full specification of the orbital basis sets
whose construction is a  science on its own.

In Section \ref{DMFT}, we give a compendious introduction to DMFT, without
discussing all  derivations
of the DMFT equations and the physics of the Mott transition  since this has been reviewed by Georges {\em et al.}
\cite{Georges96a}.  The central part of this review is
Section \ref{AbinitioDMFT} in which
the recently proposed methods 
employing DMFT for realistic material calculations
are introduced. 
In particular, we will discuss the LDA+DMFT method
in Section  \ref{LDADMFT}, a combination of the Hartree approximation and
DMFT  in Section \ref{HDMFT},
 and the GW+DMFT approach in Section \ref{GWDMFT}.
For more details on the GW approximation itself, we recommend the review by
Aryasetiawan and Gunnarsson \cite{Aryasetiawan98}. 
Section \ref{DMFTsolvers} is devoted to
several methods, numerically exact and approximate ones, 
for solving the DMFT equations.  The respective advantages and disadvantages
are discussed. 

In Section \ref{results}, we review the successful 
electronic structure calculations with DMFT accomplished
for $f$-electron systems, transition metals and their oxides.
Finally in Section \ref{summary} we give a
 summary and outlook.


\section{Conventional electronic structure calculations}
\label{ESC}

\subsection{Ab initio Hamiltonian}
\label{Sec:abinitioHam}
Let us start by rewriting the {\em ab initio} Hamiltonian [\ref{AbInitioHam}]
in second, instead of first, quantised form:
\begin{eqnarray}
\hat{H} &=&\sum_{\sigma }\int \!d^{3}r\;\hat{\Psi}^{+}({\bf
r},\sigma )\Big[
{-\frac{\hbar ^{2}}{2m_{e}}\Delta + \underbrace{\textcolor{green}{\sum_l
\frac{\displaystyle - e^2}{\displaystyle 4\pi\epsilon_0} \; \frac{\displaystyle Z_l}{\displaystyle |{\bf r}-{\bf R}_l|}}}_{\equiv V_{{\rm ion}}({\bf r})}}\Big] \hat{\Psi}(%
{\bf r},\sigma )  \nonumber \\
&&\!+\frac{1}{2}\sum_{\sigma \sigma ^{\prime }}\int \!d^{3}r \, d^{3}r^{\prime }\;
\hat{\Psi}^{+}({\bf r},\sigma )\hat{\Psi}^{+}({\bf r^{\prime
}},\sigma
^{\prime })\;\underbrace{ \textcolor{red}{\frac{\displaystyle  e^2}{\displaystyle 4\pi\epsilon_0}\; \frac{\displaystyle 1}{\displaystyle |{\bf r} -{\bf r^{\prime }}|}}}_{\equiv V_{{\rm ee}}({\bf r}\!-\!{\bf r^{\prime }})}\;\hat{\Psi}({\bf %
r^{\prime }},\sigma ^{\prime })\hat{\Psi}({\bf r},\sigma ),
\label{abinitioham}
\end{eqnarray}%
where now $ \hat{\Psi}^{+}({\bf r},\sigma )$ and $\hat{\Psi}({\bf
r},\sigma )$ are field operators that create and annihilate an
electron at position ${\bf r}$ with spin $\sigma$;
$V_{{\rm ion}}({\bf r})$ and $V_{\rm ee}({\bf r})$ denote the lattice potential
and the electron-electron interaction, respectively; for the  other
symbols see Eq.\  [\ref{AbInitioHam}].

Not included in Eq.\ [\ref{abinitioham}] are the kinetic energy of the lattice ions (phonons)
and relativistic corrections, in particular, the spin-orbit coupling.
Both are known to be important in some materials. For example,
the electron-phonon coupling plays  an important role in 
 Jahn-Teller distorted manganites according to Millis {\em et al.} \cite{Millis95a}.
The spin-orbit coupling on the other hand is important for $f$-electron
systems.

\subsection{Density functional theory (DFT)}
\label{dft}

\begin{figure}[t]
\centerline{ \epsfig{file=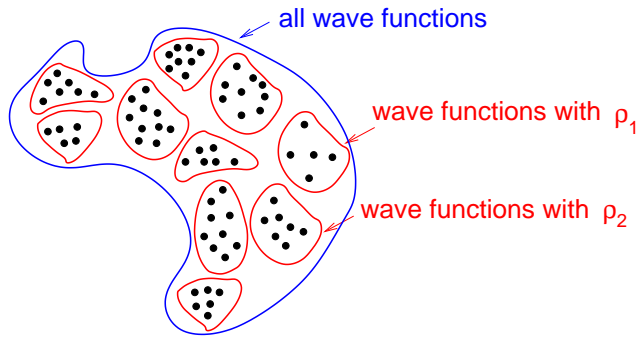,width=8.4cm}}
{\caption{\void Construction of the DFT functional according to  Levy
\cite{Levy}.
The DFT energy  functional for a given electron density $\rho_1({\mathbf r})$
is the minimal expectation value
of all wave functions yielding $\rho_1({\mathbf r})$.
\label{FigDFTpicture}}}
\end{figure}

The basic idea of density functional theory (DFT) is to
work with a simple quantity, i.e.\ the electron density $\rho({\bf r})$,
instead of trying to solve the {\em ab initio} Hamiltonian 
[\ref{AbInitioHam},\ref{abinitioham}] through complicated many-body
wave functions. This is possible, at least for the ground state energy and its
derivatives, thanks to the   Hohenberg-Kohn \cite{Hohenberg64a} theorem,
which states that the ground state energy is a
functional of the electron density $E[\rho({\bf r})]$ which is minimised 
at the ground state density $\rho({\bf r})=\rho_0({\bf r})$. 
 Following Levy
\cite{Levy},  this theorem can be easily proven and the functional
even be constructed by a two-step minimisation process,
see Fig.\ \ref{FigDFTpicture}.
In a first step, we select all many-body 
wave functions  $\varphi
({\bf r}_{1}\sigma _{1},...\,{\bf r}_{N}\sigma _{N})$
for a fixed number of electrons $N$ which yield
a certain  electron density $\rho ({\bf
r})$, and minimise the energy expectation value for these wave functions 
\begin{equation}
E[\rho ]=\min \Big\{\langle \varphi |\hat{H}|\varphi \rangle \;\;\Big|%
\;\;\langle \varphi |\sum_{i=1}^{N}\delta ({\bf r}-{\bf
r_{i}})|\varphi \rangle =\rho ({\bf r})\Big\}.
\label{ELDA}
\end{equation}
This is  exactly the DFT functional for the ground state energy:
A second minimisation yields the ground state
energy
$E_0=\min_{\rho} E[\rho ]$ at the ground state density $\rho_0$,
since for the ground state density $E[\rho_0]$ includes
the energy expectation value w.r.t\ the ground state wave function,
see Fig.\ \ref{FigDFTpicture} for an illustration.

While this construction proves the Hohenberg-Kohn theorem,
we did actually not gain anything:
For obtaining $E[\rho ]$ we  have
to calculate the expectation value $\langle \varphi |\hat{H}|\varphi \rangle$
for complicated many-body wave functions $\varphi$.
Only  the
 ionic potential 
$E_{{\rm ion}}[\rho ]=\int d^{3}r\;V_{{\rm ion}}({\bf r})\;\rho
({\bf r)}$
and  the Hartree term
  $E_{{\rm Hartree}}[\rho ]=\frac{1}{2}%
\int d^{3}r^{\prime }\,d^{3}r\;V_{{\rm ee}}({\bf r}\!-\!{\bf r^{\prime }}%
)\;\rho ({\bf r^{\prime })\rho (r)}$
can be expressed easily through the electron
density. Denoting by $E_{{\rm kin}}[\rho ]$ the kinetic energy functional,
we can hence write
\begin{equation}
E[\rho ]=E_{{\rm kin}}[\rho ]+E_{{\rm ion}}[\rho ]+E_{{\rm
Hartree}}[\rho ]+E_{{\rm xc}}[\rho ],  \label{Erho}
\end{equation}
where all the difficulty is now hidden in the 
 exchange and correlation term $E_{\rm xc}[\rho ]$.
This term is unknown.
An important aspect of DFT is  however
that the functional $E[\rho ]-E_{{\rm ion}}[\rho ]$
does not depend on the material investigated; 
for a prove  see e.g.\  \textcite{Jones89a}.
Hence, if we knew the DFT functional for one material, we could calculate all 
materials by simply adding $E_{{\rm ion}}[\rho ]$.

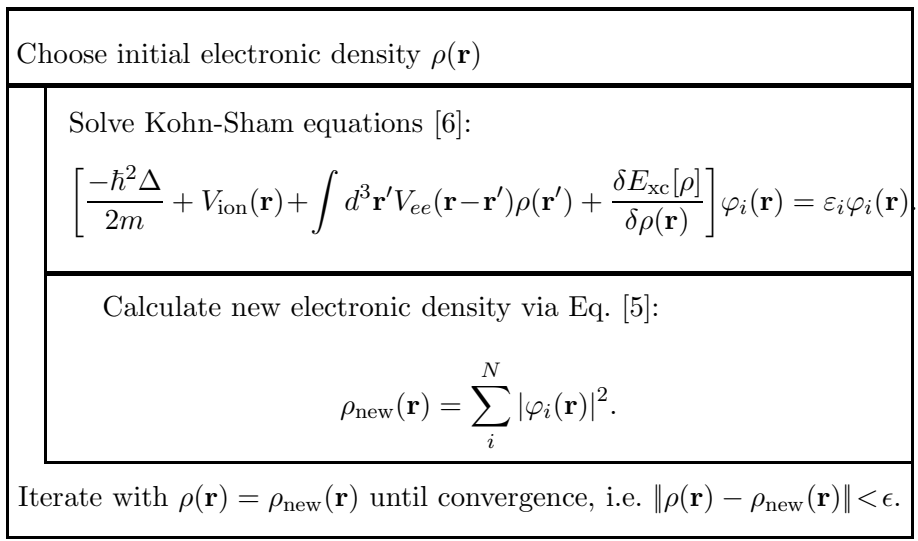
\begin{figure}[tb]
\begin{center}
\unitlength1mm
\begin{picture}(130,72)
\thicklines

\put(5.00,60.00){\framebox(120.00,10.00)}
\put(6.00,63.00){\parbox{80mm}{Choose initial electronic density 
                 $ \rho ( {\bf r} ) $}}

\put(5.00,0.00){\framebox(120.00,60.00)[cc]{}}

\put(10.00, 35.00){\framebox(115.00,25.00)[cc]
{\parbox{109mm}{
\vspace{.3cm}

 Solve Kohn-Sham equations [\ref{KohnSham}]:
 \vspace{-.3cm}

      \!     \!   \!  \[
                 \left[ \frac{- \hbar^2 \Delta }{ 2 m }
                         +  V_{\rm ion} ({\bf r}) 
                        \!+ \!\int d^3 {\bf r}'
                            V_{ee} ({\bf r}\! -\! {\bf r}') \rho ({\bf r}')
                       + \frac{ \delta E_{\rm xc} [ \rho ] }
                              { \delta \rho ({\bf r}) }
                 \right] \!\varphi_i ({\bf r})
                    =  \varepsilon_i \varphi_i ({\bf r}).
                 \]
               }}}

\put(10.00, 10.00){\framebox(115.00,25.00)[cc]
{\parbox{100mm}{ \vspace{.6cm}

Calculate new electronic density via Eq.\ [\ref{rhophi}]:
                 \[
                 \rho_{\rm new} ({\bf r})   =   \sum_i^N | \varphi_i ({\bf r}) |^2 .
                 \]
               }}}

\put(5.00,4.00){
\parbox{118mm}{ Iterate with  $\rho ({\bf r})= \rho_{\rm new} ({\bf r})$ until
convergence, i.e. $|\!|\rho ({\bf r}) -\rho_{\rm new} ({\bf r}) |\!|\! <\! \epsilon$. }}

\end{picture}
\end{center}
\caption{\void Flow diagram of the DFT/LDA calculations.} 
\label{flowLDA} 
\end{figure}

For calculating the ground state energy and density, we 
have to minimise  $\delta
\{E[\rho ]-\mu (\int d^{3}r \rho({\mathbf r})-N)\}/\partial \rho({\mathbf r})=0$
where the Lagrange parameter $\mu$ fixes the number of electrons to $N$.
To avoid the difficulty of expressing the kinetic energy  $E_{{\rm kin}}[\rho ]$
through $\rho({\mathbf r})$,
Kohn and Sham \cite{Kohn65} introduced an (at this
point auxiliary) set of  one-particle wave functions $\varphi _{i}$
yielding the density
\begin{equation}
\rho ({\bf r})=\sum_{i=1}^{N}|\varphi _{i}({\bf r})|^{2},
\label{rhophi}
\end{equation}
and minimised w.r.t.\ the $\varphi _{i}$'s instead of $\rho ({\bf r})$.
That is, we 
minimise $\delta
\{E[\rho ]-\varepsilon _{i}[\int d^{3}r|\varphi _{i}({\bf
r})|^{2}]-1\}/\delta \varphi _{i}({\bf r})=0$
where the  Lagrange parameters $%
\varepsilon _{i}$ guarantee the normalisation of the
 $\varphi _{i}$'s.
This minimisation leads us to the Kohn-Sham  \cite{Kohn65} equations
\begin{equation}
\left[ -{\frac{\hbar ^{2}}{2m_{e}}\Delta +V_{{\rm ion}}({\bf r})}+
\int d^{3}{r^{\prime }}\;
{V_{{\rm ee}}({\bf r}\!-\!{\bf r^{\prime }})} {\rho ({\bf r^{\prime }})}
+{{\frac{\delta {E_{{\rm xc}}[\rho ]}}{\delta \rho ({\bf r)}}}}%
\right] \varphi _{i}({\bf r})=\varepsilon _{i}\;\varphi _{i}({\bf
r}). \label{KohnSham}
\end{equation}
These  Kohn-Sham equations are Schr\"odinger equations, describing 
single electrons moving in 
a time-averaged potential
\begin{equation}
V_{\rm eff}({\bf r})=V_{{\rm ion}}({\bf r})+\int d^{3}{r^{\prime }}\;
{V_{{\rm ee}}({\bf r}\!-\!{\bf r^{\prime }})}{\rho ({\bf r^{\prime }})}
 +{{\frac{\delta {E_{{\rm xc}%
}[\rho ]}}{\delta \rho ({\bf r)}}}} \label{VDFT}
\end{equation}
of all electrons. The
Kohn-Sham  equations and
the electron density have to be calculated self-consistently,
  see the flow diagram Fig.~\ref{flowLDA}.

Let us note that the 
 one-particle Kohn-Sham equations
 serve, in principle, only  the purpose of minimising the DFT energy,
and have no physical meaning.
If we knew the exact $E_{\rm xc}$, which is non-local in $\rho({\mathbf r})$, 
we would obtain the exact ground state
energy and density. However, in practise, one has to make
approximations to $E_{\rm xc}$ such as the local density approximation
discussed in the next Section. We can then think of these
approximations as describing single electrons
moving in an approximated potential ${\delta {{E_{\rm xc}
}[\rho ]}}/{\delta \rho ({\bf r)}}$, 
as illustrated in the
Introduction, Fig.\ \ref{FigHam}.

Let us also note that the  kinetic energy in Eq.\ [\ref{VDFT}],
$E_{{\rm kin}}[\rho _{{\rm min%
}}]=-\sum_{i=1}^{N}\langle \varphi _{i}|{\hbar ^{2}\Delta }/{(2m_{e})}%
|\varphi _{i}\rangle $,
is that of independent (uncorrelated) electrons.
The true kinetic energy functional for the many-body problem
is different. 
We hence have to add the difference between the  true kinetic energy functional for the many-body problem 
and the above uncorrelated kinetic energy to
${E_{{\rm xc}}}$, so that all many-body difficulties are buried in ${E_{{\rm xc}}}$.

\subsection{Local density approximation (LDA)}
\label{lda}\label{LDA}

Since we do not know ${E_{{\rm xc}}}$ exactly, we have to make
approximations, and the most widely employed approximation
 is the local density approximation
(LDA). In LDA, the complicated non-local functional  $E_{\rm xc}[\rho ]$
is replaced by a local (LDA) exchange energy density\footnote{Note that in the literature, e.g. in \cite{Jones89a}, 
often the exchange energy density {\em per particle}  $\epsilon_{\rm xc}^{{\rm LDA}}(\rho ({\bf r}))=E_{\rm xc}^{{\rm LDA}}(\rho ({\bf r}))/\rho ({\bf r})$ is used.}
which is a function of the local density only:
\begin{equation}
E_{\rm xc}[\rho ]\;\stackrel{\rm LDA}{\approx}\;\int d^{3}r\;E_{\rm xc}^{{\rm LDA}}(\rho ({\bf r}
)).
\end{equation}
In practise, $E_{\rm xc}^{{\rm LDA}}(\rho (%
{\bf r}))$ is calculated from the  perturbative solution
 \cite{jellium1a,jellium1b}
 or the numerical
simulation \cite{jellium2} of the jellium model which is defined 
by $V_{{\rm ion}}({\bf r})={\rm const}$.
Due to translational symmetry, the jellium model has a constant 
electron density $\rho({\mathbf r})=\rho_0$.
Hence, 
with the correct jellium $E_{\rm xc}^{{\rm LDA}}$, we could
 calculate the energy of any material with a constant
electron density exactly.
However, for real materials   $\rho({\mathbf r})$ is
varying, less so for  $s$ and $p$ valence electrons but
strongly for  $d$ and $f$ electrons.
While  LDA
turned out to be unexpectedly successful for many materials,
it fails for materials with strong
electronic correlations between the $d$ or $f$ electrons.
In this paper, we will 
review recent advances to improve on LDA
for such correlated materials
by using DMFT.

To understand these advances physically, 
let us now turn to the LDA calculation of
band structures, one of the main applications 
of LDA. If we calculate a LDA band structures we will leave
the firm ground of DFT, which strictly speaking
only allows for calculating the ground state energy and its 
derivatives. Instead, for the LDA bandstructure,
we interprete the Lagrange parameters
$\varepsilon _{i}$ as the physical
(one-particle) energies of the system. This corresponds exactly to the
picture Fig.\ \ref{FigHam} of Section \ref{introduction}.
Physically, the one-particle LDA bandstructure
corresponds to approximating
the {\em ab initio} many-body Hamiltonian
by
\begin{eqnarray}
\!\hat{H}_{{\rm LDA}}\! &\!=\! \sum\limits_{\sigma} \!\int \!d^{3}r\;\hat{\Psi}^{+}({\bf r}%
,\sigma )&\left[ -\frac{\hbar ^{2}}{2m_{e}}\Delta + V_{{\rm ion}}({\bf r}%
)+  \right. \nonumber \\ &&
\left.\int\! d^3{r^{\prime }}\,{\rho ({\bf r^{\prime }})}{V_{{\rm ee}}({\bf r}\!-\!%
{\bf r^{\prime }})}
+{{\frac{\partial {%
E_{\rm xc}^{{\rm LDA}}}(\rho({\mathbf r}))}{\partial \rho ({\bf r)}}}}\right] \hat{\Psi}({\bf r},\sigma).  \label{HLDA0}
\end{eqnarray}

In practise, one solves the three-dimensional
one-electron Kohn-Sham equations given by
Hamiltonian [\ref{HLDA0}] through expanding
the wave functions in sophisticatedly chosen
basis sets.
These basis sets  use either the
atom and so called muffin tin potentials, or
free electrons in, e.g. ultrasoft \cite{Vanderbilt90a}, pseudopotentials
as a starting point; for a detailed discussion see \cite{Harrison89}.
An important aspect for merging LDA with DMFT is that
 we  need
to identify the correlated $d$ or $f$ orbitals.
For the  linearised muffin-tin
orbital (LMTO)  basis of Andersen \cite{LMTO1}, and its $N$th order extension
(NMTO) \cite{Andersen99,AndersenPsik},
these orbitals are inherently defined.
For other basis sets, e.g. plane waves,
one has first to project onto  Wannier orbitals  \cite{Marzari97}
before doing the DMFT calculation.

Let us now transform our ${\mathbf r}$-basis
to such a basis with localised $d$ or $f$ orbitals:
\begin{eqnarray}
\hat{\Psi}^+({\bf r},\sigma) &=& \sum_{i l} \hat{c}_{i l}^{\sigma \dagger} \varphi^*_{i l}({\bf r}). \label{basistrafo}
\end{eqnarray}
Here, $l$ denotes the orbital index
(which, for problems with more than one atom in the unit cell,
 will also index  orbitals on different sites in this unit cell.)
Transforming the  Hamiltonian [\ref{HLDA0}] to this orbital
basis, we obtain
\begin{eqnarray}
 \hat{H}_{\rm LDA}  &=& \sum_{il\,{\rm }jm \,\sigma } {t_{il\,jm}}\;{\hat{c}}_{il}^{\sigma \dagger }{\hat{c}}%
_{jm}^{\sigma }, \label{HLDA}
\end{eqnarray}
where  ${t_{il\,jm}}$ denotes the tight-binding hopping matrix elements
\begin{eqnarray}
t_{il\,jm}=\int d^3 r \,
\varphi_{il}^*({\bf r})&& \Big[-\frac{\hbar
^{2}\Delta }{2m_{e}}+V_{{\rm ion}}({\bf r})+ \nonumber \\ &&
\int d^3{r^{\prime }}{\rho ({\bf %
r^{\prime }})}{V_{{\rm ee}}({\bf r}\!-\!{\bf r^{\prime }})}+{{\frac{\partial {%
E_{\rm xc}^{{\rm LDA}}}(\rho({\mathbf r}))}{\partial \rho ({\bf r)}}}}\Big]\varphi
_{jm}({\bf r}). \label{tInt}
\end{eqnarray}
According to the Bloch theorem, we can diagonalise
 Hamiltonian [\ref{HLDA}]
by Fourier transforming to $\hat{c}_{{\bf k} l}=\frac{1}{\sqrt{L}}\sum_i \hat{c}_{i l} e^{-i {\bf k} {\bf R}_i}$ ($L$ denotes the number of lattice sites ${\mathbf R}_i$): 
\begin{eqnarray}
 \hat{H}_{\rm LDA}  &=& \sum_{{\bf k} \sigma \, l  \,m} \epsilon^{\rm LDA}_{
l \,m} ({\bf k})\;   \hat{c}_{{\bf k} l}^{\sigma \dagger } \hat{c}_{{\bf k}m}^{\sigma}.\label{HLDAk}
\end{eqnarray}
Here,
\begin{equation}
\epsilon^{\rm LDA}_{l \,m} ({\bf k})= \frac{1}{L}\sum_{ij} t_{il\,jm} 
e^{i {\bf k}({\bf R}_i -{\bf R}_j)} \label{tFTeps}
\end{equation}
is still a matrix in our orbital indices which can be fully diagonaliced
by a simple orbital rotation. This orbital rotation yields
the Kohn-Sham eigenvalues, i.e. the LDA bandstructure.
Let us emphasise that this rotation is
different one for every ${\bf k}$ point.

As already mentioned, LDA is  highly successful
for calculating both, static physical properties and
bandstructures.
But LDA  is not reliable when applied to correlated
materials and can even be completely wrong because it
  treats electronic {\em correlations} only very rudimentarily:
The exchange correlation functional has been approximated by a functional
which depends only on the {\em local} density. It has been calculated from the jellium problem,
a weakly correlated model with extended orbitals.
Basically, LDA is a one-electron approach 
where the effect of the other electrons is through a static
(time independent) mean field.
In the following, we will discuss how to 
treat electronic correlations in a better way 
with dynamical mean field theory (DMFT).


\section{Dynamical mean field theory (DMFT)}
\label{dmft}\label{DMFT}

The conventional approach to strongly correlated electron systems 
is perturbation theory, either
starting from the limit of zero Coulomb interaction ($U=0$;
weak coupling) or from the opposite limit of zero 
 kinetic  energy ($W=0$;
strong coupling). Such a treatment is, however, not appropriate 
for the important `in-between' regime where
  we have a strongly correlated 
metal and possibly a Mott-Hubbard transition
between this correlated metal and a  Mott insulator (see Fig.\ \ref{LDALDAULDADMFT}).

Metzner and Vollhardt
\cite{Metzner89a} introduced a new limit to correlated  electron systems, the limit of infinite 
dimensions $d\rightarrow \infty$ or equivalently an infinite number 
of neighbouring lattice sites.
In this limit,
the competition between kinetic energy $W$ and Coulomb  
interaction $U$ is maintained, albeit resulting in a simplified,
momentum-independent self energy: $\Sigma_{\bf k}(\omega) \stackrel{d\rightarrow\infty}{\longrightarrow} \Sigma(\omega)$.
The work by Metzner and Vollhardt \cite{Metzner89a} initiated a rapid development:
M\"uller-Hartmann 
\cite{MuellerHartmann89a,MuellerHartmann89b,MuellerHartmann89c} showed
 only the local
Coulomb interaction yields  dynamic ($\omega$-dependent) correlations 
whereas the non-local density-density interactions are
 reduced to the Hartree contribution which is $\omega$-independent.
Some models like the Falikov-Kimball model  \cite{Falicov69a}
and the Kondo lattice model for 
classical spins
 could even be solved 
exactly by Brandt and Mielsch 
\cite{Brandt89a,Brandt90a,Brandt91a,vanDongen90a,vanDongen92a,Freericks03} and Furukawa  \cite{Furukawa94a,Furukawa99a},
respectively.
The next important step was to  put the initial ideas on the footings
of a mean field theory, coined {\em dynamical}  mean field theory (DMFT) because of
the frequency dependence of  $\Sigma(\omega)$.  For the Falicov-Kimball model,
Brandt and Mielsch
 \cite{Brandt89a,Brandt90a,Brandt91a} already succeeded 
in this respect, and  Jani\v{s} and Vollhardt \cite{Janis91a,Janis92a} 
generalised the coherent potential approximation (CPA) to
include the dynamics of correlated electron models
in the  $d\rightarrow \infty$ limit.
But the breakthrough represents the work
by Georges and Kotliar \cite{Georges92a} (also see \textcite{Jarrell92a} and \textcite{Ohkawa91a,Ohkawa91c})
who showed that  a many-body model like
the Hubbard model is mapped onto the self-consistent solution of an auxiliary Anderson impurity model for $d\rightarrow \infty$
(for the physics of this Anderson impurity model see e.g. \cite{Anderson84a,Hewson}). This mapping was crucial for the 
following development since physicists were thenceforth  able to employ well-known
solvers for the Anderson impurity model to
deal now with lattice many-body models in the DMFT framework:
Approximate solvers like
iterated perturbation theory \cite{Georges92a},
self-consistent   perturbation theory \cite{MuellerHartmann89c,Schweitzer91a},
and the non-crossing approximation \cite{Jarrell93c,Jarrell93d,Pruschke93a,Pruschke93b};
and numerically exact solvers like 
quantum Monte Carlo simulations \cite{Jarrell92a,Rozenberg92a,Georges92c}, exact diagonalisation \cite{Caffarel94a} and the
numerical renormalisation group \cite{Sakai94a,Bulla98a}.

The DMFT results turned out to be a big step forward for our
understanding of, for example, the Hubbard model and the periodic Anderson model,
and improved our insight, particularly into the Mott-Hubbard transition;
see Georges {\em et al.} \cite{Georges96a} for a review and
Kotliar and Vollhardt \cite{Kotliar04a} for a first reading.

 Since DMFT
is central to this review, which addresses readers from different communities,
we feel a self-contained derivation
of DMFT is helpful.
In  Section \ref{DMFTintro}, we present
such a  derivation based on the  local nature of 
the weak-coupling 
perturbation theory, as first noticed by Metzner and Vollhardt \cite{Metzner89a} and 
M\"uller-Hartmann
\cite{MuellerHartmann89a}. 
 Alternative derivations of the DMFT equations have used the
cavity method \cite{Georges96a}, the expansion around the atomic
limit \cite{Metzner91a} and a generalisation of the
 coherent potential approximation
 \cite{Janis91a,Janis92a}. The reader
who is not interested in this derivation might
directly turn to the DMFT self-consistency scheme
summarised on p.\   \pageref{DMFTSCscheme} ff.
Shortcomings and extensions of DMFT by non-local correlations 
are discussed 
in Section \ref{ClusterDMFT}.

\subsection{Derivation of the DMFT equations}
\label{DMFTintro}
\label{DMFTtutorial}

Let us in the following
consider a generalised many-body Hamiltonian
\begin{eqnarray}
 \hat{H}  \!&\!=\!&\! \sum_{i l\,j m\,\sigma}
 {t_{il\,j m}}\;{\hat{c}}_{il}^{\sigma \dagger }{\hat{c}}_{jm}^{\sigma } 
+ \sum_{i\, l m n o\,\sigma\sigma^{\prime}}
U_{l m n  o}
\,
{\hat{c}}_{il}^{\sigma \, \dagger }
{\hat{c}}_{im}^{\sigma^{\prime}  \, \dagger}
{\hat{c}}_{in}^{\sigma^{\prime}} 
{\hat{c}}_{io}^{\sigma},\label{DMFTH}
\end{eqnarray}
where ${\hat{c}}_{il}^{\sigma \dagger }$ (${\hat{c}}_{il}^{\sigma}$) 
creates (annihilates) an electron with spin $\sigma$ and
 orbital index  $l$ at lattice site $i$; $t_{i l\,j m}$ 
is a hopping amplitude between
lattice sites $i$ and $j$ and orbitals $l$ and $m$
[of the same form as in Eq.\  [\ref{HLDA}]); finally, $U_{l m n o}$ 
denotes a general local Coulomb interaction.\\

{\em Proper scaling in the limit $d \rightarrow \infty$--}
Let us consider an extension of the lattice at hand
so that we have a  
large number ${{\cal Z}_{|\!|i\!-\!j|\!|}}$ of
equivalent (neighbouring) sites, with the same 
`distance' $|\!|i\!-\!j|\!|$ to site $i$.
Later, we will consider the limit  ${{\cal Z}_{|\!|i\!-\!j|\!|}}\rightarrow \infty$.
The reader might envisage, for example,   the generalisation
of the cubic lattice to a $d$-dimensional hypercubic lattice.
This hypercubic lattice has ${{\cal Z}_{1}}=2d$ (6 for $d=3$)
nearest neighbours and, in general,  ${{\cal Z}_{|\!|i\!-\!j|\!|}} \propto d^{|\!|i\!-\!j|\!|}$
equivalent sites at  distance $|\!|i\!-\!j|\!|$  (to leading order in $1/d$). For the cubic lattice,  $|\!|\ldots |\!|$ is the so-called Manhattan norm.
The advantage of the notation ${{\cal Z}_{|\!|i\!-\!j|\!|}}$ is that
all classes of equivalent sites and different lattice
 topologies can be treated on the same footings. 
The reader will however realise  that
these  ${{\cal Z}_{|\!|i\!-\!j|\!|}}$'s are in general not independent.

How do the two terms of Hamiltonian [\ref{DMFTH}] scale
with increasing ${{\cal Z}_{|\!|i\!-\!j|\!|}}$ in
the limit ${{\cal Z}_{|\!|i\!-\!j|\!|}} \rightarrow \infty$?
Obviously, the purely local interaction $U_{lmno}$ and the associated potential
energy per site  scales like
\begin{equation}
\left\langle \sum_{l m n o\, \sigma \sigma^{\prime} }
U_{l m n o }
\, {\hat{c}}_{il}^{\sigma \, \dagger }
{\hat{c}}_{im}^{\sigma^{\prime}  \, \dagger}
{\hat{c}}_{in}^{\sigma^{\prime}} 
{\hat{c}}_{io}^{\sigma}  \right \rangle  \stackrel{{{\cal Z}_{|\!|i\!-\!j|\!|}\rightarrow \infty}}{\longrightarrow} {\rm const.},
\end{equation}
i.e. stays at a finite constant, neither going to zero nor infinity. Here and in the following,
\begin{equation}
\langle \hat{\cal O} \rangle= \frac{{\rm Tr}\, \hat{\cal O} e^{-\beta \hat{H}}}{{\rm Tr}\, e^{-\beta \hat{H}}}
\end{equation}
denotes the thermal expectation value at inverse temperature $\beta=1/T$; ${\rm Tr}$ is the trace.

For the first term of Hamiltonian [\ref{DMFTH}], the kinetic energy,
we have ${{\cal Z}_{|\!|i\!-\!j|\!|}}$ equivalent $j$ terms in the sum.
Hence, the kinetic energy per site $i$  diverges unless we
rescale 
 ${t_{il\,j m}}$ in the limit ${{\cal Z}_{|\!|i\!-\!j|\!|}}\rightarrow \infty$.
To avoid this divergence, let us consider the following ansatz for the
scaling of these hopping amplitudes:
\begin{equation}
 {t_{il\,j m}} = \frac{{t^*_{il\,j m}}}{\sqrt{{\cal Z}_{|\!|i\!-\!j|\!|}}} \label{tscaling}
 \end{equation}
with  ${t^*_{il\,j m}}$ staying constant upon increasing
${{{\cal Z}_{|\!|i\!-\!j|\!|}}}$.
Since the non-interacting ($U=0$) Green function\footnote{
Depending on the problem at hand, we will work with Green functions
for real time/frequencies or
imaginary time/Matsubara frequencies [$\omega_{\nu}=(2\nu+1)\pi/\beta$]; see \textcite{Abrikosov63} for an introduction.
Both representations are
connected via the analytical continuation in the complex plane.
Let us define for completeness the real time  Green function
\begin{equation*}
G_{il\,jm}(t-t')= -i \Theta(t-t') \langle T c_{il}^{\phantom{\dagger}}(t) c^{\dagger}_{jm}(t')\rangle
\end{equation*}
and the imaginary time  (thermal) Green function
\begin{equation*}
G_{il\,jm}(\tau-\tau')= - \langle T c_{il}^{\phantom{\dagger}}(-i\tau) c^{\dagger}_{jm}(-i\tau')\rangle.
\end{equation*}
Here, $T$ is 
the Wick time-ordering operator w.r.t.\ $t$ or $\tau$ and
$c_{il}(t)=\exp(i\hat{H}t)c_{il}\exp(-i\hat{H}t)$ in both cases. 
The corresponding frequency-dependent Green functions follow by
Fourier transformation.
}
${G}_{il \, jm}^0(\omega)$ is directly connected to $t_{il\, jm}$ it scales exactly in the same way\footnote{Expressing the matrix elements of the
inversion in Eq.\ [\ref{G0scaling}] in terms of minors one directly sees 
that the off-diagonal elements are a factor $t_{il\, jm}/(\omega-t_{il\, im})$ smaller than the diagonal elements.}:
\begin{equation}
{G}_{il \, jm}^0(\omega)=\left[(\omega \underline{\underline{\mathbf   1}}-\underline{\underline{\mathbf   t}})^{-1}\right]_{il\, jm}\propto\frac{1}{\sqrt{{\cal Z}_{|\!|i\!-\!j|\!|}}}.
\label{G0scaling}
\end{equation}
Here, bold symbols with
double underlines denote matrices w.r.t.\
orbital and site indices ($\underline{\underline{\mathbf   1}}$ is the unit and $\underline{\underline{\mathbf   t}}$
the hopping matrix).
Let us assume for a moment that the very same scaling also holds for the
interacting Green function
\begin{equation}
{G}_{il \, jm}(\omega)\propto\frac{1}{\sqrt{{\cal Z}_{|\!|i\!-\!j|\!|}}}.
\label{Gscaling}
\end{equation}
Later, we will see that indeed all Feynman diagrams for ${\mathbf G}(\omega)$ scale
like Eq.\ [\ref{Gscaling}] or fall off even faster.

With the scaling  Eq.\ [\ref{Gscaling}], the kinetic energy per site
scales properly, i.e. stays finite:
\begin{equation}
 \sum_{ l\,j m\,\sigma}
 {t_{il\,j m}}\;
\left\langle{\hat{c}}_{il}^{\sigma \dagger }{\hat{c}}_{jm}^{\sigma }
\right\rangle \stackrel{{{\cal Z}_{|\!|i\!-\!j|\!|}}\rightarrow \infty}{\longrightarrow} {\rm const}.
\end{equation}
To see this, note that there are 
${{\cal Z}_{|\!|i\!-\!j|\!|}}$ terms in the sum
for every class of equivalent sites $j$.
This factor is however canceled by
a factor  $\frac{1}{\sqrt{{\cal Z}_{|\!|i\!-\!j|\!|}}}$
for ${t_{il\,j m}}$ and another factor
 $\frac{1}{\sqrt{{\cal Z}_{|\!|i\!-\!j|\!|}}}$
for $\langle{\hat{c}}_{il}^{\sigma \dagger }{\hat{c}}_{jm}^{\sigma }
\rangle$ which is directly connected to
the Green function 
\begin{equation}
G_{il\, jm}(\tau=0+)=
-\delta_{ij}\delta_{lm}+\langle {\hat{c}}_{il}^{\sigma \dagger }{\hat{c}}_{jm}^{\sigma }  \rangle.
\end{equation}
Only  the scaling of Eq.\ [\ref{tscaling}] provides a non-trivial limit
in which the interplay between kinetic and Coulomb interaction energy
remains intact.
\\

\begin{figure}[tb]
\begin{center}
\epsfig{file=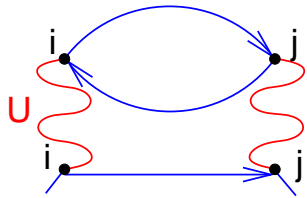,width=4.cm}
\end{center}
\caption{A second-order diagram for the Green function (or the
self energy if the legs are amputated).
A simple power counting of the three Green functions connecting $i$ and $j$ 
shows that the contribution of this diagram vanishes
 in the $d\rightarrow \infty$ limit.
Only the local contribution with $i=j$ survives. \label{FeynIrrelevant}}
\end{figure}

{\it Locality of diagrams--}
Now, let us consider the Feynman diagram
for the Green function depicted in Fig.\ \ref{FeynIrrelevant}. 
The contribution of this diagram, but generally every  diagram where
 two  sites $i \neq j$ are connected
by three independent lines (${\mathbf G}^0_{ij}$'s)\footnote{Here and in the following, bold symbols denote matrices in the orbital index.}
scales like  $ 1/\sqrt{{\cal Z}_{|\!|i\!-\!j|\!|}}^3$
 for individual $i$'s and $j$'s.
This is a factor $1/{{\cal Z}_{|\!|i\!-\!j|\!|}}$ smaller than the
direct contribution Eq.\ [\ref{G0scaling}], i.e.,
a Green function line without interaction.
Hence, such non-local diagrams
for the Green function (or self energy) become {\em irrelevant} in the limit ${\cal Z}_{|\!|i\!-\!j|\!|}\rightarrow \infty$.

\begin{figure}[tb]
\begin{center}
\epsfig{file=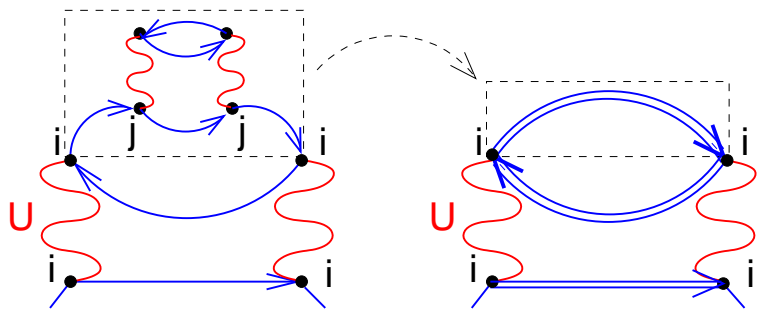,width=10cm}
\end{center}
\caption{\void Left hand side: A  diagram for the Green function or
self energy which contributes
in the $d\rightarrow \infty$ limit
since $i$ and $j$ are connected only by two Green function lines.
However, this diagram  is contained in the {\em local} diagram
on the right hand side in terms of the full (interacting)
Green function (double line), since it includes   interaction or 
self energy parts as the one on the left side.
For $d\rightarrow \infty$, all  ({\em skeleton}) diagrams
in terms of the interacting Green function
 are {\em purely local}, i.e. only involve one site $i$.\label{FeynRelevant}}
\end{figure}

There are also Feynman diagrams as in  Fig.\ \ref{FeynRelevant},
where $i \neq j$ are connected by only two lines.
Taking into account a factor ${{\cal Z}_{|\!|i\!-\!j|\!|}}$ for the sum over different
$j$, we see that
the overall  contribution of this diagram scale like ${{\cal Z}_{|\!|i\!-\!j|\!|}}^0$.
However, if
we write the Feynman diagrams in terms of the 
 {interacting}  (full)
Green function ${\mathbf G_{ij}}$ instead of the  non-interacting  ${\mathbf G^0_{ij}}$ this
diagram is already contained in the {\em local} diagram
on the right hand side of Fig.\ \ref{FeynRelevant}.
This is because ${\mathbf G_{ij}}$ includes 
self-energy inclusions as on the left hand side
 of Fig.\ \ref{FeynRelevant}.

Generally, the relation between  ${\mathbf G}_{ij}$ (double lines) and  ${\mathbf G^0_{ij}}$ (single lines)
is given by the Dyson equation which reads
\begin{equation}
 { \mathbf G}_{i j}(\tau) = {\mathbf G^0}_{i j}(\tau) +  \sum_{i^{\prime} j^{\prime}} \int_{0}^{\beta}{\rm d} \tau^{\prime}\int_{0}^{\beta} {\rm d} \tau^{\prime\prime}\,
{\mathbf G}^0_{i i^{\prime}}(\tau^{\prime}) {\mathbf \Sigma}_{i^{\prime} j^{\prime}}(\tau^{\prime\prime}-\tau^{\prime}) {\mathbf G}_{j^{\prime} j} (\tau-\tau^{\prime\prime}) \label{Dyson1}
\end{equation}
or graphically \\
\epsfig{file=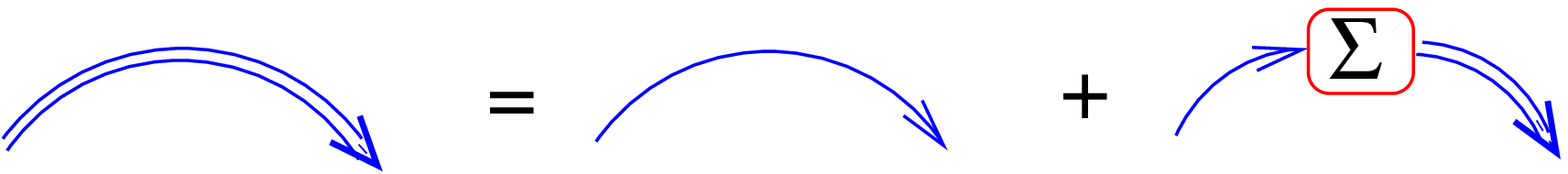,width=\textwidth,height=1.35cm}\\
Fourier transformed, Eq.\ [\ref{Dyson1}] simplifies to
\begin{equation}
  {\mathbf G}_{\bf k}(i \omega_{\nu})^{-1} = {\mathbf G}^0_{\bf k}(i\omega_{\nu})^{-1} - {\mathbf \Sigma}_{\bf k} (i\omega_{\nu}).
\label{Dyson2}
\end{equation}

If we now write our diagrams in terms of 
  ${\mathbf G}_{ij}$, we have to restrict 
ourselves to a subset of
all Feynman diagrams,
the so-called {\em skeleton} diagrams which
 do not contain any part 
connected to the rest of the diagram by only two  ${\mathbf G}_{ij}$ lines because such
diagrams are generated if ${\mathbf G}_{ij}$ is expanded 
in terms of  ${\mathbf G}^0_{ij}$. If these were included in the
{\em skeleton} diagrams they would we counted twice.

How do the {\em skeleton} diagrams
scale with  ${{\cal Z}_{|\!|i\!-\!j|\!|}}$?
If there are two sites $i \neq j$, both sites have to be connected by three Green function lines (directly or indirectly via additional sites), which follows directly from the
definition of the  {\em skeleton} diagrams.
Therefore,
the scaling for individual sites  $j \neq i$
goes at least like
 $1/\sqrt{{\cal Z}_{|\!|i\!-\!j|\!|}}^3$ as in Fig.\ \ref{FeynIrrelevant}.
These contributions
become irrelevant for ${{\cal Z}_{|\!|i\!-\!j|\!|}}\rightarrow \infty$.
Consequently,
all  {\em skeleton} diagrams are {purely local}.
An electron may still leave site $i$, interact on other sites,
and return to site $i$ retardedly. But these processes
are entirely described by the local interacting Green function
${\mathbf G}_{ii}$, containing
self energy inclusions as in  Fig.\ \ref{FeynRelevant}.

As a result, the self energy
itself is  {\em purely local}:
\begin{equation}
{\mathbf \Sigma}_{i j}(\omega) \stackrel{{{\cal Z}_{|\!|i\!-\!j|\!|}}\rightarrow \infty}{\longrightarrow} \delta_{ij} {\mathbf \Sigma}(\omega)
\label{LocalSelfE}
\end{equation}
or Fourier-transformed ${\mathbf k}$-independent: ${\mathbf \Sigma}_{\mathbf{k}}(\omega)={\mathbf \Sigma}(\omega)$. From this and the
Dyson equation [\ref{Dyson1}], it follows that
${G}_{il \, jm}(\omega)\propto{1}/{\sqrt{{\cal Z}_{|\!|i\!-\!j|\!|}}}$,
 {\em a posteriori} confirming our assumption Eq.\ [\ref{Gscaling}].

\begin{figure}[tb]
\begin{center}
\epsfig{file=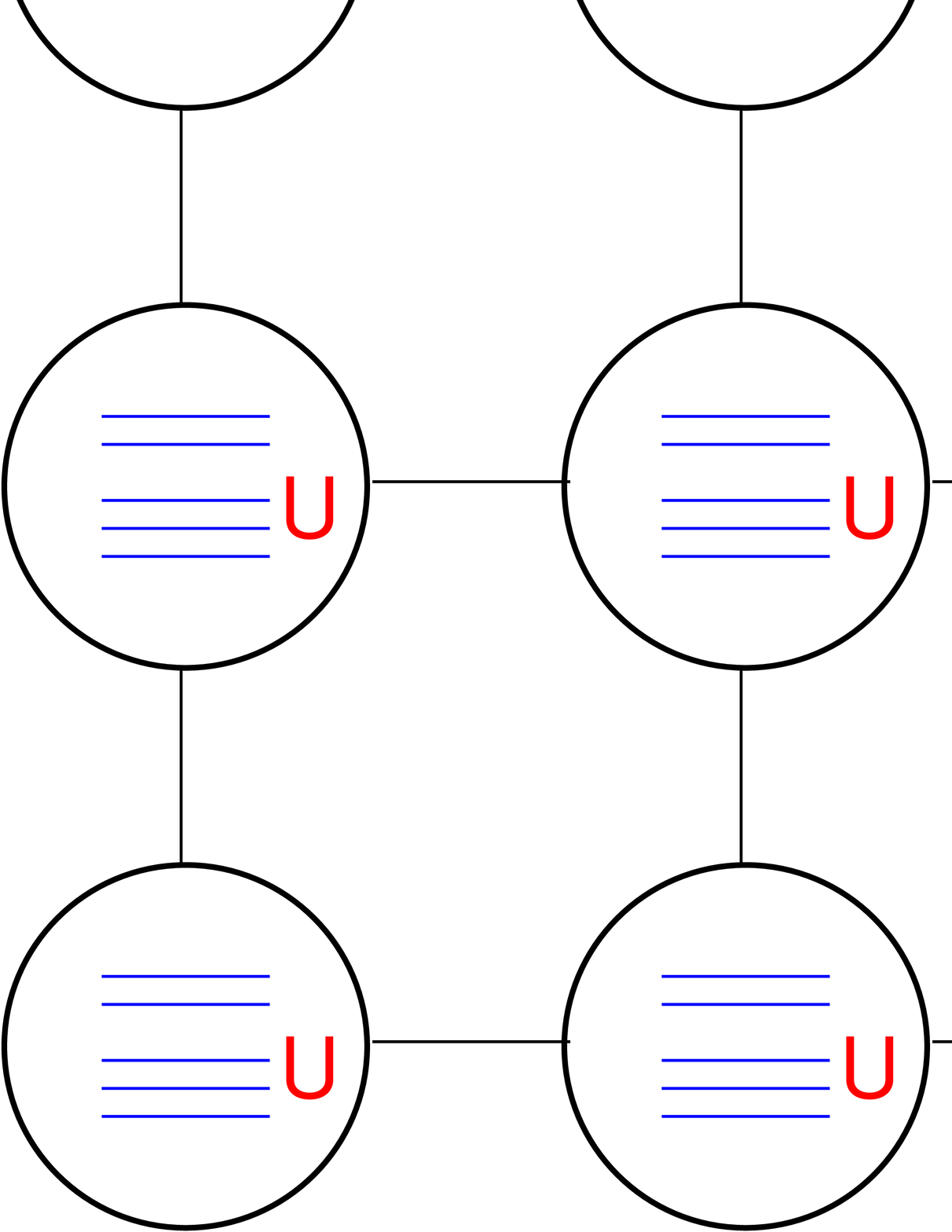,width=11.cm}
\end{center}
\caption{\void DMFT maps the lattice many-body problem with interactions $U$
on every site (left side) onto  a single
site problem where the interaction has been replaced by
the self energy $\Sigma$ except for a single site
(right side). All  irreducible Feynman diagrams are hence local:
electrons, leaving and returning to this single interacting site, are dressed
by  $\Sigma$ (double line in Fig.\ \ref{FeynRelevant}).
The DMFT mapping becomes exact for $d\rightarrow \infty$ and is an approximation in finite dimensions.
\label{figDMFT}}
\end{figure}

To conclude  this scaling analysis, we can say that
DMFT represents the local contribution
of all  {\em skeleton} Feynman diagrams, thereby replacing
the lattice many-body problem by a local (single site) problem,
see Fig. \ref{figDMFT}. This simplification becomes
exact for $d\rightarrow \infty$ and is an approximation
if applied to a finite dimensional problem.\\

{\it Mapping onto the Anderson impurity model--}
Diagrammatically, DMFT corresponds to  the local contribution of all 
topologically distinct Feynman diagrams.
Exactly the same diagrams can be obtained
via an Anderson impurity model if its
on-site interaction has the same form as the original Hamiltonian [\ref{DMFTH}]:
\begin{eqnarray}
\!\! \!\hat{H}_{\rm AIM}  \!&\!=\!&\! \sum_{{\bf k} l\sigma}\! \epsilon_{l}^{\phantom{\dagger}}\! ({\bf k})\,
{\hat{a}}_{{\bf k} l}^{\sigma \dagger }{\hat{a}}_{{\bf k}l}^{\sigma }
+ \!\! \sum_{{\bf k} l\sigma}\! \big[ V_{ lm}({\bf k})\, {\hat{a}}_{{\bf k}l}^{\sigma \dagger} {\hat{c}}_{m}^{\sigma }
\!+\! {\rm h.c.} \big]\!
+ \!\!\!\!\!\!\sum_{ l m n o\, \sigma \sigma^{\prime}}\!\!\!\!\!
U_{l m n o}
\,
{\hat{c}}_{l}^{\sigma  \dagger }
{\hat{c}}_{m}^{\sigma^{\prime}   \dagger}
{\hat{c}}_{n}^{\sigma^{\prime}} \!
{\hat{c}}_{o}^{\sigma}.
\label{AIMH}
\end{eqnarray}
Here, ${\hat{a}}_{{\bf k} l}^{\sigma \dagger }$ (${\hat{a}}_{{\bf k}l}^{\sigma }$) are creation
and annihilation operators for non-interacting conduction electrons at wave vector ${\mathbf k}$ which
have a dispersion $\epsilon_{l}$({\bf k})  and hybridise with the localised
interacting  electrons ${\hat{c}}_{m}^{\sigma \, \dagger }$  via
$V_{l m}({\bf k})$.
 
Let us now switch  to the language of functional integrals in terms of
Grassmann variables $\psi^{\phantom\ast }$ and $\psi ^{\ast }$, see e.g.  \textcite{Negele87a}.
By a simple Gaussian integration we can get rid of the
conduction electrons since these only enter  quadratically in Hamiltonian
[\ref{AIMH}] and, hence, in the functional integral.
Integrating out these conduction electrons, we arrive at an
effective problem for the interacting electrons  ${\hat{c}}_{m}^{\sigma \, \dagger }$
whose Green function reads 
\begin{equation}
G_{lm}^{\sigma }(i \omega_{\nu})=-\frac{1}{{\cal Z}}\int {\cal D}[\psi ]{\cal
D}[\psi
^{\ast }]\psi _{\nu l}^{\sigma \phantom \ast }\psi _{\nu m}^{\sigma \ast
}e^{%
{\cal A}[\psi ,\psi ^{\ast },(\mbox{\small \boldmath ${\cal G}$}^0)^{-1}]}.  \label{siam}
\end{equation}%
Here, $\nu$  denotes  imaginary Matsubara frequencies
  $\omega_{\nu}=\pi(2\nu+1)/\beta$;
\begin{equation}
{\cal Z}=\int {\cal D}[\psi ]{\cal D}[\psi
^{\ast }]e^{{\cal A}[\psi ,\psi ^{\ast },(\mbox{\small\boldmath ${\cal G}$}^0)^{-1}]}
\end{equation}
is the partition function and
the single-site action ${\cal A}$ has the form 
\begin{eqnarray}
{\cal A}[\psi ,\psi ^{\ast },(\mbox{\boldmath ${\cal G}$}^0)^{-1}]&=&\sum_{\nu
\sigma \,l m}\psi _{\nu
m}^{\sigma \ast }[{\cal G}^{\sigma 0}_{ m n}(i \omega_{\nu})]^{-1}\psi _{\nu n}^{\sigma
{\phantom\ast }}\nonumber\\&&
-\!\!\!\!\sum_{l m n o\,\sigma\sigma^{\prime}}
\!\!\!\!U_{l m n o}
    \int\limits_{0}^{\beta }d\tau \,\psi _{l}^{\sigma \ast }(\tau )\psi
_{n}^{\sigma^{\prime } \phantom\ast }(\tau )\psi _{m}^{\sigma
^{\prime}\ast }(\tau )\psi _{o}^{\sigma \phantom%
\ast }(\tau). \label{effAction}
\end{eqnarray}
In Eq.\ [\ref{effAction}], the interaction part
of ${\cal A}$ is in terms of  $\tau$,  the
Fourier transform of the Matsubara frequencies $\omega_{\nu}$.
The non-interacting Green function of the effective problem is given by
\begin{equation}
[{\cal G}^{\sigma 0}_{mn}(i \omega_{\nu})]^{-1} =  i \omega_{\nu} + t_{im \, in}+ \mu - 
\sum_{{\bf k}l} \frac{V_{lm}({\bf k})^*V_{ln}({\bf k})}
{ i \omega_{\nu}+\mu  - \epsilon_{l}({\bf k})}, \label{G0AIM}
\end{equation}
with the same local term $t_{im \, in}$ as the
original many body Hamiltonian [\ref{DMFTH}].

The topology of the {\em irreducible}
diagrams of this effective Anderson impurity model
is exactly the same as the DMFT single site problem: simply the local contribution of all Feynman diagrams.
If also the lines of the diagrams are the same
we will hence obtain the same self energy.
For the local {\em irreducible} diagrams,
this is achieved if the interacting  Green function ${\mathbf G}(\omega)$
of the Anderson  impurity model also equals the local 
DMFT Green function ${\mathbf G}_{i i}(\omega)$.

Which Anderson impurity model has now this ${\mathbf G}(\omega)$ as its interacting
Green function?\footnote{Since Green function  ${\mathbf G}(\omega)$
and self energy ${\mathbf \Sigma}(\omega)$ are the same for DMFT and 
the Anderson impurity model we employ the same symbols.}
 Given its interacting Green function ${\mathbf G}(\omega)$ and self energy 
${\mathbf \Sigma}(\omega)$,
the impurity model's
{\em non-interacting} Green function $\mbox{\boldmath ${\cal G}$}^0$ has to fulfil
\begin{equation}
 [{\mbox{\boldmath ${\cal G}$}}^0(\omega)]^{-1}= [{\mathbf G}(\omega)]^{-1} + {\mathbf \Sigma} (\omega).
\label{AIMDyson}
\end{equation}
This is the analogy to the Dyson equation [\ref{Dyson1}], but now for the auxiliary Anderson impurity model.
Note, that if we expand the ${\mathbf G}(\omega)$ of the {\em irreducible} diagrams
in terms of $\mbox{\boldmath ${\cal G}$}^0(\omega)$, we will get {\em reducible} inclusions
and all  local Feynman diagrams.
While the reducible diagrams for the Anderson impurity model
are different from the (non-local) reducible DMFT diagrams,
the irreducible diagrams are exactly the same. Hence, we can
calculate the interacting DMFT Green function or self energy by solving
an Anderson impurity model.

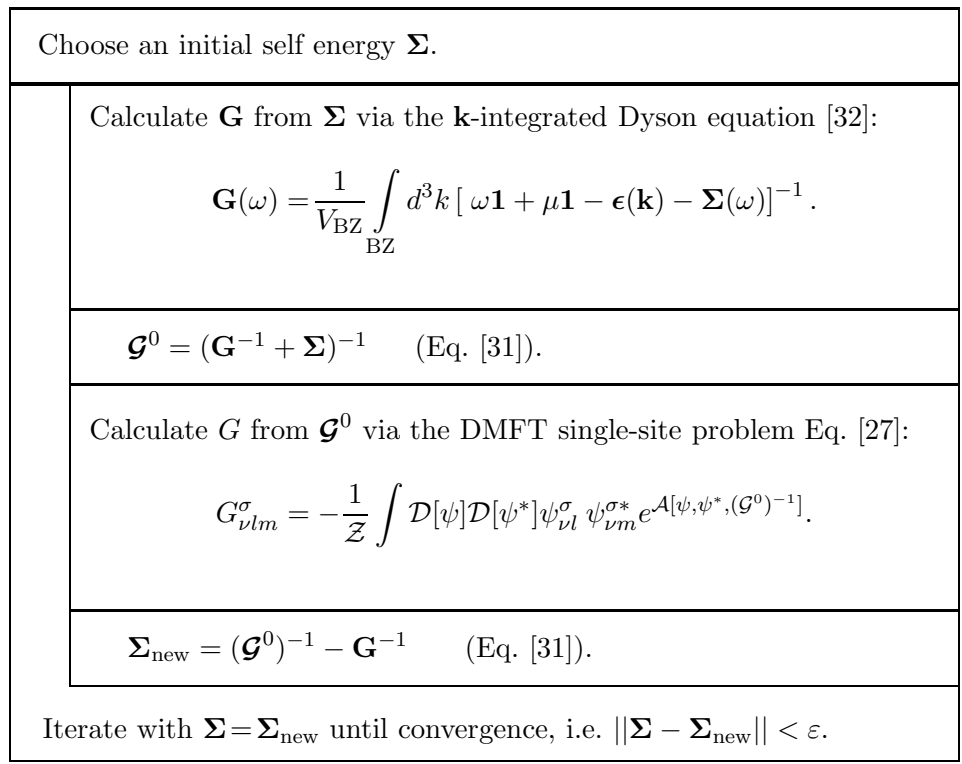
\begin{figure}[tb]
\unitlength=1mm
\linethickness{0.4pt}
\begin{picture}(130.00,105.00)
\put(4.00,5.00){\framebox(126.00,90.00)[cc]{}}
\put(4.00,95.00){\framebox(126.00,10.00)[cc]{
\parbox{12cm}{Choose an initial self energy \protect${\mathbf \Sigma}$.}}}

\put(12.00,15.00){\framebox(118.00,80.00)[cc]{}}

\put(7.00,8.00){
\parbox{10.5cm}{Iterate with  $ {\mathbf \Sigma} \!=\!
  {\mathbf \Sigma}_{\rm new}$ until convergence, i.e.\ $ ||
   {\mathbf \Sigma} - {\mathbf \Sigma_{\rm new}}|| < \varepsilon $. }}

\put(12.00,65.00){\framebox(118.00,30.00)[cc]
 {\parbox{11.3cm}{

Calculate ${\mathbf G}$ from ${\mathbf \Sigma}$ via the ${\mathbf k}$-integrated Dyson equation [\ref{Dyson}]:
\[{\mathbf G}(\omega )=\!\frac{1}{V_{\rm BZ}}\int_{\rm BZ}
{{
d^{3}}{k}}\left[ \;\omega {\mathbf 1}+\mu  {\mathbf 1}-  \mbox{\boldmath $\epsilon$}({\bf k})
-{\mathbf \Sigma}(\omega )\right]^{-1}. \]
}}}

\put(12.00,55.00){\framebox(118.00,10.00)[cc]
{\parbox{10.3cm}{\(\mbox{\boldmath ${\cal G}$}^0 = ( {\mathbf G}^{-1}+{\mathbf \Sigma})^{-1}\)\hspace{.5cm} {(Eq.\ [\ref{AIMDyson}]).}}}}

\put(12.00,25.00){\framebox(118.00,30.00)[cc]
{\parbox{11.3cm}{

    Calculate $G$ from $\mbox{\boldmath ${\cal G}$}^0$ via the DMFT single-site problem Eq.\ [\ref{siam}]:
\[
G_{\nu lm}^{\sigma }=-\frac{1}{{\cal Z}}\int {\cal D}[\psi ]{\cal
D}[\psi
^{\ast }]\psi _{\nu l}^{\sigma \phantom\ast }\psi _{\nu m}^{\sigma \ast
}e^{%
{\cal A}[\psi ,\psi ^{\ast },({\cal G}^0)^{-1}]}.
\]
 }}}

\put(12.00,15.00){\framebox(118.00,10.00)[cc]
{\parbox{10.3cm}{
 ${{\mathbf \Sigma}_{\rm new}}= (\mbox{\boldmath ${\cal G}$}^0)^{-1} - {\mathbf G}^{-1}$ \hspace{.5cm} {(Eq.\ [\ref{AIMDyson}]).}}}}
\end{picture}
\caption{\void Flow diagram of the DMFT self-consistency cycle, also see
 \textcite{Held03}.
\label{DMFTflow}\label{flowDMFT}}
\end{figure}

{\it Self consistency scheme--}
\nopagebreak
\label{DMFTSCscheme}
Building upon these considerations, we can now formulate the self-consistent
DMFT scheme, which is
summarised in the DMFT flow diagram Fig.\ \ref{DMFTflow}.
We start with a trial self energy ${\mathbf \Sigma}(\omega)$, which might just be zero. Then, we calculate the local 
Green function ${\mathbf G}(\omega)$ by {\bf k}-integration of the
Dyson equation [\ref{Dyson2}]:
\begin{equation}
  {\bf G}(\omega) = \frac{1}{V_{\rm BZ}}\int_{\rm BZ} d^3k \; [\omega {\mathbf 1}+\mu{\mathbf 1}- \mbox{\boldmath $\epsilon$}({\bf k}) - {\bf \Sigma} (\omega)]^{-1}.
\label{Dyson}
\end{equation}
Here, we have made use of ${{\mathbf {G}}^0}({\mathbf k})^{-1}= \omega{\mathbf 1}+\mu{\mathbf 1}-\mbox{\boldmath $\epsilon$}({\bf k})$
where $\mbox{\boldmath $\epsilon$}({\bf k})$ is the Fourier transform of $t_{il \, jm}$ like
in Eq.\ [\ref{tFTeps}],
${\mathbf 1}$ the orbital unit matrix, and ${V_{\rm BZ}}$ the volume of the  Brillouin zone (BZ).

Given ${\mathbf G}(\omega)$ and  ${\mathbf \Sigma}(\omega)$, we define
the effective non-interacting Green function of the Anderson model
$\mbox{\boldmath ${\cal G}$}^0(\omega)$ via Eq.\ [\ref{AIMDyson}].
Now comes the difficult part: We have to solve the Anderson impurity model
and calculate its Green function  ${\mathbf G}(\omega)$.
Different methods to this end are presented in Section \ref{DMFTsolvers}.
Using
 Eq.\ [\ref{AIMDyson}] a second time, 
we obtain a new self energy. (Sometimes, in particular in perturbation theory,
it is more convenient to calculate the self energy of the Anderson
impurity model directly, like in our diagrammatic presentation in Section \ref{DMFTintro}.)
 With this new self energy  ${\mathbf \Sigma}(\omega)$,
the algorithm is iterated
until a convergence criterion $\varepsilon$ in some norm $|\!| ... |\!|$,
e.g. the maximum or sum norm, is met.

\subsection{Extensions of DMFT}
\label{ClusterDMFT}
\label{DCA}

DMFT accounts for a major, local part of electronic correlations.
This part is responsible for the 
strong quasiparticle renormalisation and the Mott-Hubbard
transition in materials
with  strong electronic correlations such as transition metal oxides
and heavy Fermion systems, see the discussion in Section \ref{introduction}.
But also many other physical phenomena  have been described by DMFT:
the crossover from Slater \cite{Slater51a} to Heisenberg \cite{Heisenberg28a}
antiferromagnetism
\cite{Jarrell92a} and associated metamagnetism \cite{Held96a,Held97a},
the  influence of the lattice  \cite{Ulmke98a,Wahle98a} and
Hund's exchange on ferromagnetism   \cite{Held98a},
scattering at impurities \cite{Janis93c,Ulmke95b,Dobrosavljevic93a,Dobrosavljevic94a} or spins  \cite{Furukawa94a,Furukawa99a},
the effect of  phonons
\cite{Freericks93d,Freericks94a,Freericks94b}, particularly on the
 `colossal magnetoresistance' in manganites \cite{Millis96a,Millis96b,Michaelis03a}
and superconductivity \cite{Capone02a,Han03a}.
This list is far from complete, with only a bare minimum of references.
Nonetheless, it shows that DMFT yields  a wide variety of
physical phenomena, posing the question:
What does DMFT not describe?

What DMFT neglects completely are non-local correlations.
Such non-local correlations are responsible for
a variety of physical phenomena,
ranging from valence bonds,
pseudo gaps and (possibly) $d$-wave superconductivity
 to (para-)magnons, quantum critical behaviour, and generally
the critical behaviour in the vicinity of phase transitions.
All these effects are not included in DMFT.
Some of these effects stem from rather short range correlations, e.g.
the valence bond formation of
pairs of spins into singlets; others require long-range correlations, e.g  
quantum criticality.

There have been various attempts to include non-local
correlations beyond DMFT. 
Most successfully and widely employed
  are  cluster extensions of DMFT.
Since these have been reviewed recently by
Maier {\em et al.}
\cite{Maier04},
we will only briefly discuss the basic ideas
and different cluster schemes,  without  giving referenced to
all applications  to model systems.

\begin{figure}[tb]
\centerline{ \epsfig{file=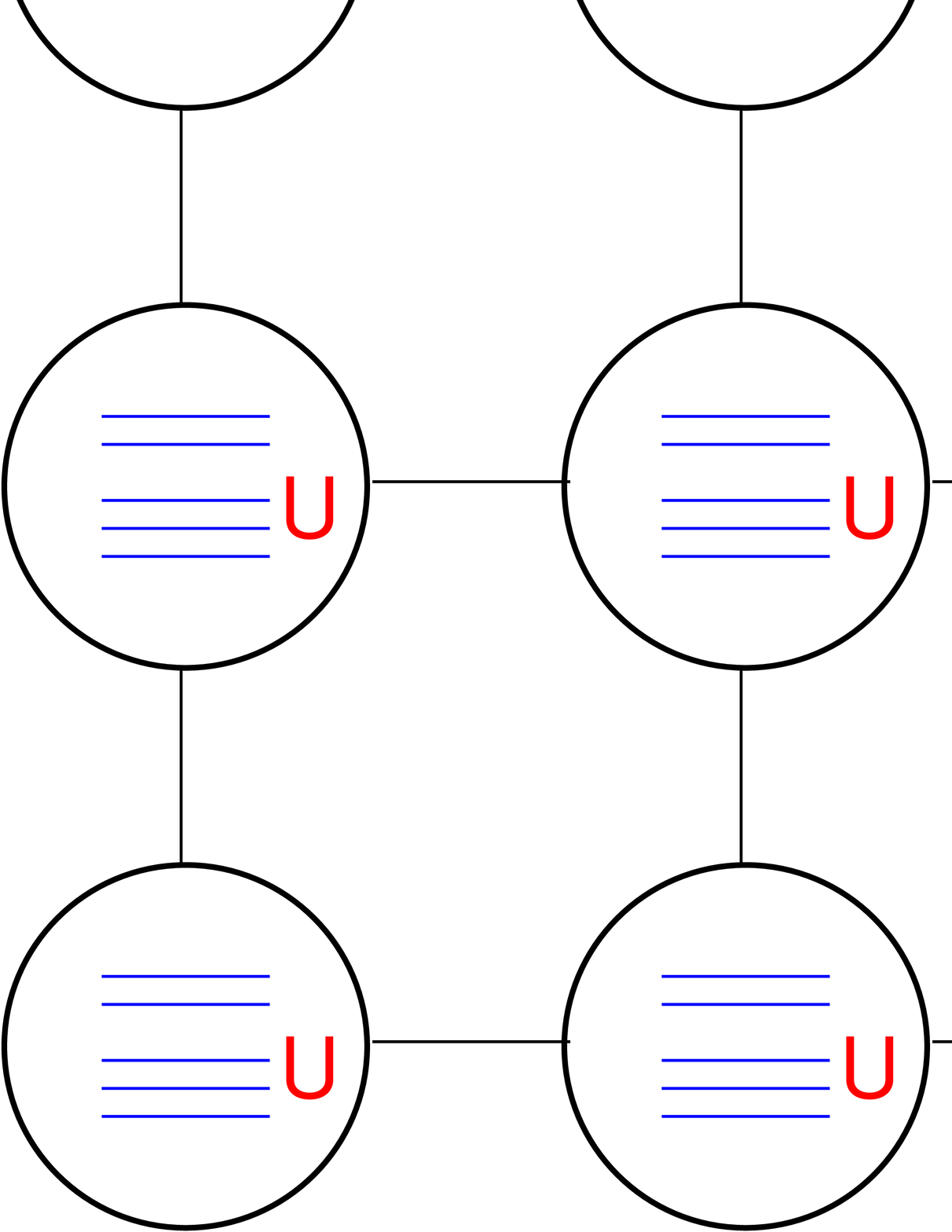,width=12.cm}}
\caption{\void In cluster  DMFT 
the lattice many-body problem 
(left side) is mapped onto a 
 DMFT cluster (right hand side) which consists of
several (here two) interacting sites.
Nonlocal correlations between the cluster sites
are taken into account. The DMFT field couples to the border of the cluster.
\label{clusterDMFT}}
\end{figure}

The idea of these cluster extensions is to treat, 
instead of the single DMFT site,
a whole cluster of sites, see  Fig.\ \ref{clusterDMFT}. 
A natural choice for the cluster is a super cell
in real space, or the LDA unit cell if it
contains more than one site  with interacting
$d$ or $f$ orbitals. Then the DMFT basis block is not
a single site but this super cell.
Non-local correlations  within the super cell
are taken into account, whereas such correlations
between different super cells are neglected.
Beyond the local DMFT self energy,  there are now also 
off-diagonal elements $\Sigma_{ij}(\omega)$ 
for two sites $i\neq j$ within the cluster.
This  path is followed in the so-called cluster
DMFT approaches \cite{Georges96a,clusterDMFT,LichtensteinDCA,Potthoff}. 
Cluster DMFT turned out to yield surprisingly 
precise results for the one-dimensional Hubbard model \cite{Capone04a}.
With the number of cluster-sites, the numerical effort 
grows considerably, in particular if realistic calculations require
the inclusion of several orbitals.
Such realistic LDA+cluster DMFT calculation have
been employed so-far by Poteryaev {\em et al.} \cite{Poteryaev04a}
 for studying  Ti$_2$O$_3$, by Mazurenko {\em et al.} \cite{Mazurenko02a}
for NaV$_2$O$_5$, and
 by Biermann {\em et al.}  \cite{Biermann05a},
who found VO$_2$ to be insulating because
spins form non-local singlets so that
a Peierls gap opens. In both cases the LDA unit cell
provided for a natural choice of a two-site cluster.

Another scheme, named dynamical cluster approximation (DCA), has been proposed by Hettler {\em et al.}
\cite{DCA}.  While cluster DMFT can be
best understood in real space, 
the idea of DCA is to patch the 
self energy in ${\mathbf k}$-space. 
To this end, the  first Brillouin zone is divided
 into $N_c$ patches around  ${\mathbf k}$-vectors ${\mathbf K}$.
In contrast to the constant DMFT self energy,
the DCA self energy $\Sigma_{\mathbf K}$ is only 
constant within the patch, but varying from patch to patch. 
This way, ${\mathbf k}$-dependencies of the self energy are taken into
account.
The basic differences between
DCA and cluster DMFT in real space are:
The DCA cluster has periodic  boundary conditions, whereas
that of cluster DMFT has open boundary conditions. 
Moreover,
the dynamical mean field couples to every site of the cluster in DCA
and to the boundary sites in cluster
DMFT. In the limit of infinite cluster size ($N_c \rightarrow \infty$),
both approaches become exact.
While cluster DMFT might be more suitable for dealing with
particular short-range correlations in real space, such as the formation 
of a spin singlet on neighbouring sites,
DCA more naturally preserves translational symmetry.
Let us also mention three further alternatives: 
the cluster perturbation theory by Gros and Valenti \cite{Gros93a},
the variational cluster perturbation theory of Potthoff {\em et al.} \cite{Potthoff},
and the (very general) self energy functional approach of Potthoff \cite{Potthoff03a,Potthoff06a}.
In the context of classical spin models
cluster extensions such as the 
embedded cluster method \cite{Kikuchi51}
have been used for a long time, leading to early proposals
\cite{Georges96a} to employ similar approaches
for DMFT.

In the DMFT context,
cluster extensions
have  been applied extensively by-now for studying superconductivity.
Indeed, $d$-wave superconductivity in the two-dimensional Hubbard model
was found for a $2\times2$ cluster \cite{DCASC,LichtensteinDCA,Capone06a}, and a pseudogap phase at elevated
temperatures 
\cite{Civelli05a}.
Whether the  $d$-wave superconductivity persists for
larger cluster sizes and in the
$N_c\rightarrow \infty$ limit is however less obvious
\cite{Maier04,Maier05a,Arita06a}. 
Let us also note that
Jarrell and coworkers 
\cite{Aryanpour03a,Hague03} employed a hybrid cluster solve for treating 
larger cluster sizes, where the
the fluctuation exchange approximation (FLEX) by Bickers {\em et al.}
\cite{Bickers89b} is employed
for  longer-range correlations.

A different route to extend DMFT by non-local correlations
has been   proposed
by  Schiller and Ingersent \cite{Schiller,Zarand00}. 
The authors extended DMFT diagrammatically through
all diagrams to  order $1/d$.
This leads to a theory with a single-site and a two-site cluster whose
Green functions have to be subtracted.
Necessarily, this $1/d$ approach is hence restricted
to nearest neighbour correlations.

All extensions of DMFT mentioned so-far, are restricted
to relatively short-range correlations.
Often, however, long-range correlations are of vital importance,
e.g. for
magnons, screening and quantum criticality.
If we want to account for such long-range correlations, 
other extensions of DMFT are necessary.
Very recently, Toschi {\em et al.} \cite{Toschi06a}
introduced a way to do so (also see \textcite{Kusunose06a,Slezak06a}): 
the dynamical vertex approximation (D$\Gamma$A).
The basic idea is to extend DMFT diagrammatically:
 Toschi {\em et al.} \cite{Toschi06a}  take 
 the local, fully irreducible two-particle vertex
as a starting point and construct from this vertex all possible 
(local and non-local) self energy diagrams. A restriction of this approach
to the particle-hole channel yields (para-)magnons,
a restriction to the particle-particle channels
yields the cooperon diagrams. Hence, ladder diagrams
describing long-range correlations for weakly 
correlated systems are recovered, but with the
vertex instead of the bare Coulomb interaction
so that strong correlations are accounted for.
Independently, Kusunose  \cite{Kusunose06a} 
extended DMFT 
by the  transverse particle-hole channel 
using iterated perturbation theory,
and  Slezak {\em et al.} \cite{Slezak06a}
proposed to use similar ladder diagrams 
for solving larger DCA clusters.
Previous attempts  supplement the
DMFT self energy by spin-fluctuations from
 the spin-fermion model \cite{Kuchinskii05a,Sadovskii05a,Kuchinskii06a,Kuchinskii06b}
 or  from the self-consistent renormalisation theory
\cite{Saso99a}. Bolech {\em et al.} \cite{Bolech03a}
explored the possibility of 
a renormalisation group extension of
cluster DMFT and Rubtsov {\em et al.}
extended DMFT by including non-local correlations through
the dual Fermion approach \cite{Rubtsov06,Rubtsov07}.


\section{Merging conventional bandstructure approaches with DMFT}
\label{AbinitioDMFT}
The standard 20th century approach 
to calculate materials realistically, i.e. density functional theory (DFT) in its 
local density approximation (LDA),
 was introduced in Section \ref{ESC}.
As pointed out in Section \ref{intro},
this conventional approximation does not work for materials with $d$ or $f$ electrons
because electronic correlations are too strong.
Of prime importance for these correlations are
the {\em local} Coulomb interactions
between the  $d$ or $f$ electrons.
This is because the local Coulomb interaction is the largest one and
also because other density-density interactions
are to leading order in $1/Z_{|\!|i\!-\!j|\!|}$ (one over the number
of sites with the same distance, see  Section \ref{dmft})  only given by the Hartree
term as was proved by M\"uller-Hartmann \cite{MuellerHartmann89a}.
This Hartree term is already contained in the LDA.

In this Section, we will review recent approaches which
start from the  {\em ab initio} Hamiltonian  [\ref{abinitioham}]
and merge conventional bandstructure approaches (Section \ref{ESC})
with DMFT (Section \ref{DMFT}) in order to
reliably account for the {\em local} correlations induced by the
{local} Coulomb interaction.
In  Section \ref{LDADMFT}, we 
will discuss how conventional LDA calculations  are merged with DMFT,
presenting two points of view: 
(i) a physically-motivated
Hamiltonian formulation
in Section \ref{HamLDADMFT}, which
can be done self-consistently as pointed out in Section \ref{selfconsist}
and whose parameters can be calculated by constrained
LDA as discussed in Section \ref{cLDA};
and (ii) in  Section \ref{SDFTLDADMFT} a
formulation in terms of a spectral density functional theory
 which appreciates the Hohenberg-Kohn
\cite{Hohenberg64a} extremum principle for the energy.
Section \ref{SimpTMO} is devoted to simplified LDA+DMFT implementations
for transition metal oxides.
In Section  \ref{HDMFT},
the Hartree+DMFT and the Hartree-Fock+DMFT approach
are introduced. The former  is very natural from the DMFT point of view
concerning non-local interactions.
Section \ref{HDMFT} also paves the way for the more profound alternative to LDA+DMFT
which is a combination of the
GW approximation  and DMFT, finally addressed in Section \ref{GWDMFT}.

\subsection{LDA+DMFT}
\label{LDADMFT}
\subsubsection{Hamiltonian formulation}
\label{HamLDADMFT}

In Section \ref{LDA}, we deduced the
 LDA Hamiltonian [\ref{HLDA}] from the {\em ab initio} Hamiltonian
 [\ref{abinitioham}]. 
To account for  the important {\em local} correlations, 
we will now supplement this LDA Hamiltonian  [\ref{HLDA}]
by the  {\em local} Coulomb interaction.
The resulting multi-band many-body problem will then be solved
by DMFT. A predecessor of this LDA+DMFT approach is the so-called LDA+U
method of Anisimov {\em et al.} \cite{Anisimov91}, which constructs the same many-body problem but
then solves it by (symmetry-broken) Hartree-Fock, i.e. without 
taking into account electronic correlations if these  are defined as being 
beyond  Hartree-Fock. As an approximate solution
of the DMFT self-consistency equations we will rediscover this
LDA+U scheme in Section \ref{HartreeFock}.

In principle, one could use the most general local Coulomb interaction
as in Hamiltonian [\ref{DMFTH}] of Section \ref{dmft} and
calculate the Coulomb matrix elements by
\begin{equation}
U_{l m n  o} ={\frac{\displaystyle  e^2}{\displaystyle 4\pi\epsilon_0}
\int d^3 r\int d^3 r^{\prime} \,
\varphi_{il}({\bf r})^*\varphi_{im}({\bf r^{\prime}})^*
{ \textcolor{red} \frac{\displaystyle 1}{\displaystyle |{\bf r} -{\bf r^{\prime }}|}}}
\varphi_{in}({\bf r^{\prime}})\varphi_{io}({\bf r}). \label{CoulombME}
\end{equation}
\label{Uimpractical}
However, this approach is not practical. First of all, it is often 
sufficient to take into account only the most important matrix
elements of the Coulomb interaction, including only
two distinct orbital indices and restricting the interaction
to the localised $d$ or $f$ orbitals only.
Another aspect which makes Eq.\ [\ref{CoulombME}] impractical is screening:
If we add an extra electron to the $d$ or $f$ orbitals at
our local site $i$, the $sp$  electrons will redistribute and,
therefore, reduce the energy required to add this extra electron.
Since  non-local Coulomb interactions as well as local interactions between $d$ 
and $sp$ electrons
are responsible for this  screening process,
a Hamiltonian with only a local Coulomb interaction like Eq.\ 
 [\ref{DMFTH}] does not describe these processes anymore.
Therefore, the effective local Coulomb interaction has to be smaller (screened)
than the overlap integral Eq.\ [\ref{CoulombME}].
Presently, two methods are employed  for the calculation of
 these screened local Coulomb interactions:
the constrained LDA method
in the context of LDA+DMFT and the random phase approximation (RPA)
within the GW+DMFT approach.
This will be discussed in more detail in Section \ref{cLDA}
and in Section \ref{GWDMFT}, respectively.\\

{\it Set-up of the LDA+DMFT Hamiltonian--}
Based on these considerations,  the starting point of the LDA+DMFT approach is 
usually the following Hamiltonian, where $\hat{H}_{\rm LDA}$ denotes
the LDA Hamiltonian [\ref{HLDA}]:
\begin{eqnarray} \hat{H} &=&
\!\!\hat{H}_{\rm LDA}+
 U {\sum^{\prime }_{\stackrel{\scriptstyle i \sigma}{\scriptstyle l \in L_U}}}
\hat{n}_{il}^{\sigma}\hat{n}_{il}^{\bar{\sigma}}
+ \!\!\!  \!\!\! {{\sum^{\prime }_{\stackrel{\scriptstyle i \sigma\sigma^{\prime}}{\scriptstyle l \in L_U \neq m \in L_U}}}}\!\!\!  \!\!\! 
(V-\delta_{\sigma \sigma^{\prime}}J)\,
\hat{n}_{il}^{\sigma}\hat{n}_{im}^{{\sigma^{\prime}}}
\nonumber \\&&
\!-\frac{J}{2}  \!\!\!  \!\!\! {{\sum^{\prime }_{\stackrel{\scriptstyle i \sigma}{\scriptstyle l \in L_U \neq m \in L_U}}}}\!\!\!  \!\!\!
\hat{c}^\dagger_{il\sigma }
\hat{c}^{\phantom{\dagger}}_{il\bar{\sigma}}
\hat{c}^\dagger_{im\bar{\sigma}} 
\hat{c}^{\phantom{\dagger}}_{im\sigma} 
-\frac{\tilde{J}}{2}  \!\!\!  \!\!\! {{\sum^{\prime }_{\stackrel{\scriptstyle i \sigma}{\scriptstyle l \in L_U \neq m \in L_U}}}}\!\!\!  \!\!\!
\hat{c}^\dagger_{il\sigma }
\hat{c}^\dagger_{il\bar{\sigma}} 
\hat{c}^{\phantom{\dagger}}_{im\sigma} 
\hat{c}^{\phantom{\dagger}}_{im\bar{\sigma}} \label{Hint}
-{\sum_{\stackrel{\scriptstyle i \sigma}{\scriptstyle l \in L_U}}} \Delta\epsilon \,\hat{n}_{il\sigma}
\label{LDADMFTH}\label{hint}.
\end{eqnarray}
 Here, the prime on the sum indicates that
every term of the sum is counted only once;
$\bar{\sigma}= \downarrow\!(\uparrow)$ for $\sigma = \uparrow\!(\downarrow)$; 
$\hat{n}_{il}^{\sigma}=\hat{c}_{il}^{\sigma \dagger}\hat{c}_{il}^{\sigma \phantom{\dagger}}$;
$\delta_{\sigma\sigma^{\prime}}$ denotes the Kronecker symbol.
Moreover,
$l \in L_U$ indicates that the interaction is only taken into
account on the subset $L_U$ of the `U-interacting' orbitals in the
unit cell $i$. For example,  $L_U$ might denote the set of
$d$ (or the $t_{2g}$) orbitals  of a transition metal  or the $4f$ orbitals of rare earth elements.

\begin{figure}
\begin{center}
\centerline{\includegraphics[clip=true,width=4cm]{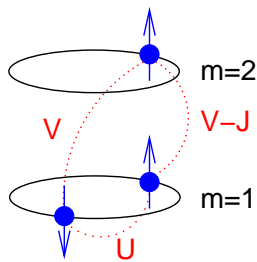}}
\end{center}
\caption{\void Pictogram of the intra- and inter-orbital 
repulsion $U$ and $V$ and the Hund exchange $J$ for two orbitals in Hamiltonian [\ref{LDADMFTH}].\label{hund}}
\end{figure}

The Coulomb interaction terms consist of the intra-orbital Coulomb repulsion
$U$, the inter-orbital Coulomb repulsion $V$, the 
Hund exchange term $J$ and a pair-hopping term $\tilde{J}$, see Fig. \ref{hund}.
The last term of Eq.\  [\ref{LDADMFTH}] describes the contribution
of $U$, $V$ and $J$ already included in $\hat{H}_{\rm LDA}$, which
has to be
subtracted to avoid a double counting.
This last term led to some criticism. It arises, however,
very naturally from the
constrained LDA calculation, as we will discuss below.
If the Coulomb interaction is the same for all orbitals,
the double counting term corresponds to a simple shift of the
chemical potential and has no effect.
This is the case if, for example, only $d$ orbitals
are taken into account, see Section \ref{SimpTMO}.

In Eq.\ [\ref{LDADMFTH}], the Hund exchange has been separated into
its $z$-$z$ and the spin-flip components
which altogether yield 
a SU(2) symmetric contribution of the form 
\begin{equation}
\hat{H}_{J} = - 2 J \sum_{im}\left[\hat{s}_{im}  \hat{s}_{il}+\frac{1}{4}\hat{n}_{il}\hat{n}_{im} \right]
\end{equation}
for the coupling of the spins  $\hat{s}_{im}$ between different orbitals $l$,$m$
at site $i$.
For real-valued wave functions
 $\varphi_{im}({\bf r})$,
it immediately follows from Eq.\ [\ref{CoulombME}] that the pair-hopping amplitude 
equals that of the exchange interaction: $\tilde{J}=J$.
Note, that purely real wave functions can be chosen
if  time reversal symmetry is not broken, for example by an external magnetic field.
 The pair-hopping term has not yet been
included in LDA+DMFT calculations because one commonly assumes 
that configurations  are rare in which one orbital is 
doubly occupied while another is empty if the 
Hund exchange and the hopping terms are included.
For quantum Monte-Carlo (QMC) simulations, the spin-flip term of the Hund exchange poses a (sign) problem
\cite{Held99phd} and has, hence, not been included in LDA+DMFT calculations
with QMC as an impurity solver so far.
Different ways to overcome this sign-problem 
are discussed in Section \ref{QMC}. These improved
methods, such as the continuous time QMC, may hence 
allow to do such LDA+DMFT calculations in the future.

For degenerate orbitals, e.g. the $t_{2g}$ orbitals in a cubic symmetry,
Hamiltonian  [\ref{LDADMFTH}] has to be invariant under
orbital rotations, necessitating $V=U-2J$. 
Then, with $M$  interacting orbitals, the average Coulomb interaction is
\begin{equation}
\bar U=\frac{U+(M-1)(U-2J)+(M-1)(U-3J)}{2M-1}.
\end{equation}
Even  if degeneracy is only fulfilled approximately one often employs
 $V=U-2J$ as an approximation, which is often quite
good and allows us to set up  Hamiltonian  [\ref{LDADMFTH}] with
only three parameters: the average Coulomb repulsion $\bar{U}$,
the exchange interaction $J$ and the double counting correction
$\Delta \epsilon$. All these parameters can be calculated by constrained LDA,
see Section \ref{constrainedLDA}.

\begin{figure}[tb]
\begin{center}
\unitlength1mm
\begin{picture}(130,130)
\thicklines

\put(0.00,108.00){\framebox(128.00,10.00)[cc]
{\parbox{124mm}{Do a LDA calculation as in flow diagram Fig.\ \ref{flowLDA}, Section \ref{LDA}:
	$\Rightarrow \rho({\bf r})$.	}}}

\put(0.00,0.00){\framebox(128.00,108.00)[cc]{}}

\put(10.00, 68.00){\framebox(118.00,40.00)[cc]
{\parbox{110mm}{

 Calculate  the  $t_{il\,jm}$'s (Eq.\ [\ref{tInt}]) of the LDA Hamiltonian [\ref{HLDA}]:

\vspace{-.8cm}

               \,  \begin{eqnarray*}
                 t_{il\,jm}\!=\!\!\int\! d^3 r 
\varphi_{il}({\bf r}) && \Big[\frac{-\hbar
^{2}\Delta }{2m_{e}}+V_{{\rm ion}}({\bf r})+\\&&
 \!\int\! d^3{r^{\prime }}{\rho ({\bf %
r^{\prime }})}{V_{{\rm ee}}({\bf r}\!-\!{\bf r^{\prime }})}+{{\frac{\delta {%
E_{\rm xc}^{{\rm LDA}}}[\rho ]}{\delta \rho ({\bf r)}}}}\Big]\varphi
_{jm}({\bf r}),
                 \end{eqnarray*}

\vspace{-.2cm}
	       and the Fourier transform of   $t_{il\,jm}$, i.e. the LDA band structure $\mbox{\boldmath $\epsilon$}^{\rm LDA}{({\bf k})}$ (Eq.\ [\ref{tFTeps}]).
               }}}
\put(10.00, 48.00){\framebox(118.00,20.00)[cc]
{\parbox{110mm}{ \vspace{.2cm}

Do a constrained LDA calculation, see Section \ref{cLDA}:\\
 $\rho({\bf r}) \stackrel{\rm Eq. [\ref{cLDAU})}{\longrightarrow} U$;
  $\;\;\rho ({\bf r})\stackrel{\rm Eq. [\ref{LDADMFTHJ})}{\longrightarrow} J$;
  $\;\;\rho ({\bf r})\stackrel{\rm Eq. [\ref{cLDADe})}{\longrightarrow}\Delta \epsilon$.
               }}}
\put(10.00, 30.00){\framebox(118.00,18.00)[cc]
{\parbox{110mm}{ \vspace{.1cm}

Solve the so-defined many-body problem Eq.\ [\ref{LDADMFTH}] by DMFT as in flow diagram Fig.\ \ref{flowDMFT}
 with the parameters \({\mbox{\boldmath ${\epsilon}$}}^{\rm LDA}{({\bf k})}\), $U$, $J$ and $\Delta \epsilon$.
               }}}
\put(10.00, 10.00){\framebox(118.00,20.00)[cc]
{\parbox{110mm}{ \vspace{.6cm}

Calculate a new density $\rho_{\rm new}$ via
   \[ {\mathbf \Sigma}(\omega) \stackrel{\rm Eq. [\ref{StoG}]}{\longrightarrow} {\mathbf G}_{il\, jm}  \stackrel{\rm Eqs. [\ref{Gtocc}],[\ref{cctorho}]}{\longrightarrow} \rho_{\rm new}.
 \] 
               }}}
\put(0.00,4.00){
\parbox{118mm}{ Iterate with  $\rho ({\bf r})= \rho_{\rm new} ({\bf r})$ until
convergence, i.e. $|\!|\rho ({\bf r}) -\rho_{\rm new} ({\bf r}) |\!|\! <\! \epsilon$ }}.

\end{picture}
\end{center}
\caption{\void Flow diagram of the LDA+DMFT algorithm.} 
\label{flowLDADMFT} 
\end{figure}
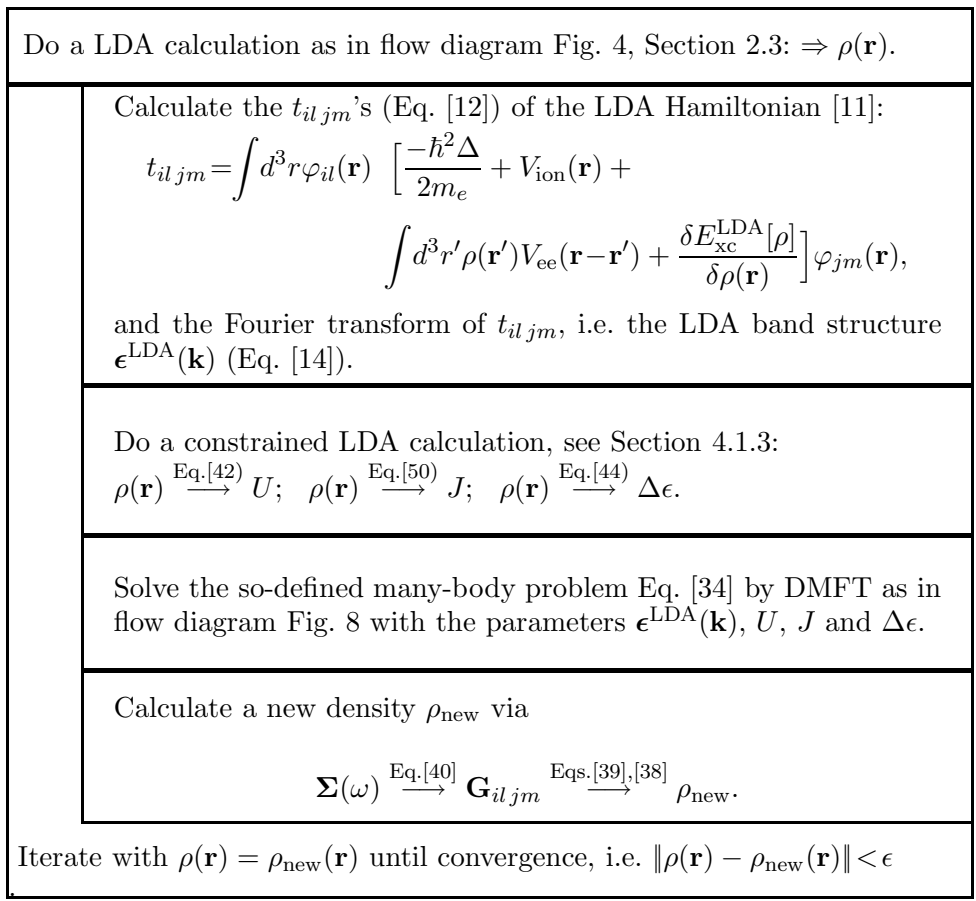

After  setting up the multi-band many-body Hamiltonian [\ref{LDADMFTH}],
we have to solve it. This can be done, without genuine electronic correlations,
in the LDA+U approach of Anisimov {\em et al.} \cite{Anisimov91} or more sophisticatedly by DMFT,
as in the pioneering work of 
Anisimov {\em et al.} \cite{Anisimov97a} and Lichtenstein and Katsnelson 
\cite{Lichtenstein98a}.
Within DMFT, Hamiltonian [\ref{LDADMFTH}] is mapped
onto an auxiliary Anderson impurity problem with the same Coulomb
interaction and an effective
 non-interacting Green function $(\mbox{\boldmath ${\cal G}$}^0)^{-1} = ( {\mathbf G}^{-1}+{\mathbf \Sigma})$,
 see Section \ref{DMFT}.
This Anderson impurity model has to be solved self-consistency together with the ${\bf k}$-integrated
Dyson equation [\ref{Dyson}], relating the local Green function ${\mathbf G}$ and self energy ${\mathbf \Sigma}$. The
LDA part $\hat{H}_{\rm LDA}$ of Hamiltonian [\ref{LDADMFTH}] 
is only entering in this Dyson equation [\ref{Dyson}]. Most conveniently,  $\hat{H}_{\rm LDA}$ is formulated
in ${\bf k}$-space with matrix elements as in Eq.\ [\ref{HLDAk}], and  the double counting correction
$\Delta \epsilon$ is included in this term:
\begin{eqnarray}
 \hat{H}_{\rm LDA}^{\rm dc}= \hat{H}_{\rm LDA} -{\sum_{i \sigma \, l \in L_U}} \Delta\epsilon \,\hat{n}_{il\sigma}
  &=& \sum_{{\bf k} \sigma \, l  \,m} \tilde{\epsilon}^{\rm LDA}_{
l \,m} ({\bf k})\;  c_{{\bf k} l}^{\sigma \dagger } c_{{\bf k}m}^{\sigma} \label{HLDAdc}
\end{eqnarray}
with 
$\tilde{\epsilon}^{\rm LDA}_{l \,m} ({\bf k})={\epsilon}^{\rm LDA}_{l \,m} ({\bf k}) - \delta_{lm}  \Delta\epsilon \;\forall\, l\!\in\! L_U$ (for all $U$-interacting orbitals) and
$\tilde{\epsilon}^{\rm LDA}_{l \,m} ({\bf k})=\tilde{\epsilon}^{\rm LDA}_{l \,m} ({\bf k}) \;
\forall\, l\! \not\in\! L_U$ (for all non-interacting orbitals).
Employing $\mbox{\boldmath $\tilde{\epsilon}$}^{\rm LDA}({\mathbf k})$ of Eq.\ [\ref{HLDAdc}] for the ${\bf k}$-integrated
Dyson equation  [\ref{Dyson}], the 
LDA+DMFT approach consists of the following steps,
see flow diagram Fig.\ \ref{flowLDADMFT}:
\begin{enumerate}
\item We start with a conventional LDA calculation, as in the flow diagram Fig.\ \ref{flowLDA}.
\item With the given density we can calculate the hopping matrix elements $t_{il \; jm}$
  and from these, via a Fourier transformation, the  bandstructure  ${\epsilon}^{\rm LDA}_{l \,m}$. 
\item In a constrained LDA calculation we determine $\bar U$, $J$ and $\Delta \epsilon$.
\item  This defines the Hamiltonian [\ref{LDADMFTH}]  which we solve by DMFT,  see  flow diagram Fig.\  \ref{flowDMFT}. Thereby,
       $\tilde{\epsilon}^{\rm LDA}_{l \,m}$ enters in the Dyson equation; $\bar U$ and $J$ 
        as interaction parameters in the Anderson impurity model.
\item  From the DMFT self energy ${\mathbf \Sigma}(\omega)$, we can calculate a new electron density. 
       With this,  we can go back to step 2. to calculate  a new  
       LDA Hamiltonian, if we do a self-consistent LDA+DMFT calculation as will be discussed now.
\end{enumerate}
  
\subsubsection{Self-consistent LDA+DMFT calculations}
\label{selfconsist}

In the first four steps of the LDA+DMFT scheme above, 
the LDA bandstructure calculation
 and the inclusion of electronic correlations by DMFT are
performed sequentially.
In general, the DMFT solution will result in  changes of  the
electron density $\rho ({\bf r})$, see Eq.\ [\ref{cctorho}] below.
The new  $\rho ({\bf r})$ then leads to a new
LDA-Hamiltonian $\hat{H}_{\rm LDA}$ [\ref{HLDA}] since both the Hartree term
and the exchange correlation term of $\hat{H}_{\rm LDA}$
depend on $\rho ({\bf r})$. Also, the Coulomb
interaction $U$ changes and needs to be determined by a new
constrained LDA calculation, resulting
in a new many-body problem.

This gives an iteration scheme as in the flow diagram Fig.\ \ref{flowLDADMFT},
which without Coulomb interaction  reduces to the
self-consistent solution of the Kohn-Sham \cite{Kohn65} equations since ${\mathbf \Sigma}(\omega)=0$.  A similar self-consistency
scheme was employed by Savrasov and Kotliar \cite{SAVRASOV2} 
in their calculation of Pu, albeit without self-consistency for $\bar U$, and
without $J$ and $\Delta\epsilon$.
The quantitative difference between non-self-consistent
and self-consistent LDA+DMFT calculations depends on the change of the number of electrons
in the different bands after the DMFT calculation. This change, of course, depends on the
problem at hand. For example for the Ce calculation presented in Section \ref{Ce}, the change of the orbital occupation (from LDA to LDA+DMFT)
was very minor in the vicinity of $\alpha$-$\gamma$ transition but more
significant at lower volumes. Hence, a self-consistent LDA+DMFT calculation does not
appear to be necessary in the vicinity of the Ce $\alpha$-$\gamma$ transition.
Indeed, most LDA+DMFT calculations reported so far are not
self-consistent. A notable exception is  the work of Savrasov and Kotliar \cite{SAVRASOV2} 
on Pu already mentioned.

We still have to describe how to
calculate the new  electron density $\rho ({\bf r})$ after the DMFT calculation.
We can relate $\rho ({\bf r})$ to the DMFT self energy  ${\mathbf \Sigma}(\omega)$
as follows:\\
Expressing $\rho({\mathbf r})$ via an expectation value
of the  field operator $\hat{\Psi}^+({\mathbf r,\sigma})$
and doing the basis transformation to the 
orbital basis set $\varphi_{il}({\mathbf r})$ (Eq.\ [\ref{basistrafo}]),
we obtain 
\begin{equation}
\rho ({\bf r}) = \sum_{\sigma} \langle\hat{\Psi}^+({\mathbf r,\sigma}) \hat{\Psi}({\mathbf r,\sigma}) \rangle 
	   = \sum_{\sigma \, il\, jm} \varphi^*_{il}({\mathbf r})  \varphi^{\phantom{*}}_{jm}({\mathbf r}) \langle\hat{c}^{\sigma \dagger}_{il} \hat{c}^{\sigma}_{jm} \rangle. \label{cctorho}
\end{equation}
Here, the  expectation value can be calculated from the equal-time Green function
\begin{equation}
G^{\sigma}_{il\,jm}(\tau=0^+)= - \delta_{ij} \delta_{lm} + \langle\hat{c}^{\sigma \dagger}_{il} \hat{c}^{\sigma}_{jm} \rangle.
\label{Gtocc}
\end{equation}
This Green function is, in turn, the Fourier transformation with respect to space and (imaginary) time of
\begin{equation}
{\mathbf G}^{\sigma}_{\mathbf k}(i \omega_{\nu})=  \big[i\omega_{\nu} {\mathbf 1} + \mu  {\mathbf 1} - 
{\mbox{\boldmath $\tilde{\epsilon}$}}^{\rm LDA}({\mathbf k})- {\mathbf \Sigma}(i\omega_{\nu}) \big]^{-1}; \label{StoG}
\end{equation}
(bold symbols denote matrices in
the orbital index).
In principle, the frequency dependent DMFT self energy ${\mathbf \Sigma}(i \omega_{\nu})$ and 
the ${\mathbf k}$-dependent  LDA  ${\mbox{\boldmath $\tilde{\epsilon}$}}({\mathbf k})$ can be spin-dependent.
In the present presentation however, such a possible spin-dependence is not taken into account.
We also have not discussed  the peculiarities of the LDA basis set; rather we
assume that the LDA Hamiltonian has been calculated in an appropriate basis set with sufficiently localised orbitals.
For the self-consistent LDA+DMFT calculation, this basis set should also be 
optimised again with the new LDA+DMFT density $\rho({\mathbf r})$.

A remark: Since the solution of the DMFT Anderson impurity model is often the 
computationally most expensive task, it might be a good idea to do only 
one DMFT iteration for every iteration loop in the flow diagram Fig.\ \ref{flowLDADMFT}.
When converged, this will produce the same result as a scheme in which, for {\em every} new $\rho({\bf r})$,
the DMFT equations are iterated  until convergence. One might also omit the initial LDA loop.
However, we would expect that  the overall numerical effort grows.

\subsubsection{Constrained LDA calculations of the interaction parameters}
\label{constrainedLDA}
\label{cLDA}

The task of calculating $\bar{U}$,
$J$ and $\Delta \epsilon$ in Hamiltonian  [\ref{LDADMFTH}]
is not at all trivial as pointed out on p.\ \pageref{Uimpractical}
 and requires additional approximations. Such an approximation,
which works in practice and gives very reasonable values for
the screened Coulomb interaction,
is the constrained LDA method by Dederichs {\em et al.} \cite{Dederichs84}, McMahan {\em et al.} \cite{McMahan88} and Gunnarsson {\em et al.} \cite{Gunnarsson89}.
Hence, constrained LDA allowed us to perform parameter-free, i.e. truly
 {\em ab initio},
LDA+DMFT calculations for transition metal oxides and $f$ electron
systems, yielding results in good agreement with experiment
\cite{Held01b,Sekiyama03}. 

\begin{figure}
\begin{center}\mbox{}
\centerline{\includegraphics[clip=true,width=11.5cm]{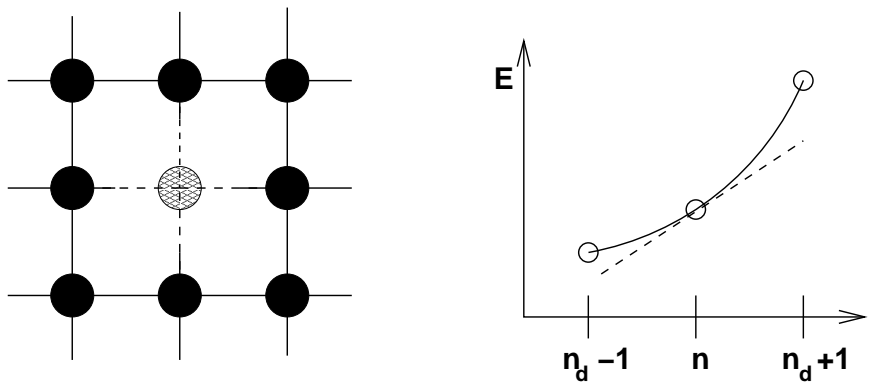}}
\end{center}
\caption{\void Left: In the constrained LDA calculation the interacting $d$ or $f$ electrons
on one lattice site are kinetically decoupled from the rest of the system, i.e.
they cannot hop to other lattice sites. By contrast, the {\em non}-interacting electrons can
still hop to other sites (indicated by dashed lines)
and screen the  $d$ or $f$ electrons.
 Right: This allows us to change
the number $n_d$ of interacting  $d$ or $f$ electrons on the decoupled site and to calculate the
corresponding LDA energies $E(n)$ [circles]; the {\em interacting}
Hamiltonian  Eq.~[\ref{LDADMFTH}] reproduces these constrained LDA energies if $\bar U$ and $\Delta \epsilon$ 
are set properly;
the dashed line sketches the behaviour of
$\hat{H}_{\rm LDA}$  defined in Eq.~[\ref{HLDA}].\label{cLDAfig}}
\end{figure}

The basic idea of constrained LDA is to do LDA calculations
for a slightly modified problem, i.e. a problem where the
interacting $d$ or $f$ electrons of one site are kinetically decoupled from the rest of the system.
This is achieved by setting the hopping matrix elements  between these localised orbitals $l$ 
on site $i$  and all other orbitals to zero: ${t_{il\,jm}}=0 \; \forall j,m$, see Fig.~\ref{cLDAfig}.
This allows us to change the number of interacting electrons on this decoupled site;
they cannot hop away.
At the same time, screening effects of the other electrons  are taken 
into account, since they redistribute if 
the number of $d$ or $f$ electrons is changed on the decoupled site.

Usually, we know
how many  $d$ or $f$ electrons $n_d$ to expect, for example,
from the formal oxidation, and typically $n_d$ is close to an integer number.
Then, we can do constrained LDA calculations for $n_d-1$, $n_d$, and $n_d+1$ electrons
on the decoupled site within the interacting orbitals ($l\in L_U$), 
This leads to three corresponding total energies, see Fig.~\ref{cLDAfig}.
The LDA Hamiltonian $\hat{H}_{\rm LDA}$ which was calculated at a fixed
density $\rho({\mathbf r})$, see Section \ref{LDA},  would predict a linear change 
of the LDA energy with the number of interacting electrons, i.e. $E(n_d)=E_0+\epsilon_d^{\rm LDA}\; n_d$. This linear behaviour
does not take into account
the Coulomb interaction $\bar U$ which requires a higher energy cost to add the ($n+1$)th electron
than to add the $n$th electron. This effect leads to the curvature in  Fig.~\ref{cLDAfig}
and is taken into account in the Hamiltonian  [\ref{LDADMFTH}] which yields for $J=0$:
\begin{equation}
 E(n_d)=E_0+ \frac{1}{2}\,  \bar{U} n_d(n_d-1)+ (\epsilon^{\rm LDA}_d+\Delta\epsilon) n_d.
\label{ECLDA}
\end{equation}
Here, $\epsilon^{\rm LDA}_d$   is the energy contribution
of $\hat{H}_{\rm LDA}$ in Eq.\  [\ref{LDADMFTH}] for $d$ or $f$ orbitals
on  the decoupled site.
If $m$ denotes the (degenerate)  $d$  or $f$ orbital this
contribution is given by the diagonal hopping matrix element
(Eq.\ [\ref{HLDA}]):
$\epsilon^{\rm LDA}_d\equiv t_{i m\, i m}$ (if so, orbitally averaged). Note that  all hopping elements $t_{i m\, j l}$ to other orbitals 
 and sites are set zero in the truncated problem
of the constrained LDA.
Since the $n_d$ electrons are kinetically decoupled from the rest of the
system, we can calculate the interaction energy in Eq.\ [\ref{ECLDA}] directly from the 
number of $n_d$ electrons.\footnote{Operators $\sum_m \hat{n}_m$ 
stay constant all the time and can be replaced by this constant, i.e. $n_d$.}
Note that part of $\Delta\epsilon$ just arises to cancel the Coulomb contribution
$ (1/2)\, d[\bar{U} n_d(n_d-1)]/d n_d=\bar{U}(n_d-1/2)$. 

Knowing how the energy $E(n)$ changes when going from $n_d-1$ to $n_d$ and $n_d+1$
electrons allows us to determine $\bar{U}$ {\em and}
$\Delta\epsilon$ since there are two parameters and two energy differences.
In the actual calculation, we obtain these energy differences from
the constrained LDA eigenlevels of the decoupled $d$ or $f$ orbitals,
using the relation
$\epsilon_d(n_d)=\frac{dE({n_d})}{d n_d}$  \cite{Slater74,Janak78}.
From this we directly determine
\begin{eqnarray}
\bar{U} &=& E(n_d+1)+E(n_d-1)-2E(n_d) \nonumber\\
&\approx& \epsilon_d(n_d+\frac12) - \epsilon_d(n_d-\frac12).\label{constrainedLDAU}\label{cLDAU}
\end{eqnarray}
 $\Delta\epsilon$ is obtained from the energy difference
to add a single electron.  Eq.\ [\ref{ECLDA}] yields
\begin{eqnarray}
E(n_d)-E(n_d-1) &=&   \bar{U} (n_d-1)+ (\epsilon^{\rm LDA}_d+\Delta\epsilon)
\label{epsUn}
\end{eqnarray}
so that 
\begin{eqnarray}
\epsilon^{\rm LDA}_d+\Delta\epsilon &=& E(n_d)-E(n_d-1) -  \bar{U} (n_d-1) \nonumber\\
	&\approx& \epsilon_d(n_d-\frac12) -  \bar{U} (n_d-1). \label{cLDADe}
\end{eqnarray}
This shift of the $d$ or $f$ eigenlevel can be taken into account
by replacing $\epsilon^{\rm LDA}_d=t_{i m\, i m}$ in $\hat{H}_{\rm LDA}$ [\ref{LDADMFTH}]
by the term $\epsilon_d(n_d-\frac12) -  \bar{U} (n_d-1)$ from the constrained
LDA calculation. This replacement thus includes $\Delta \epsilon$ in $\hat{H}_{\rm LDA}$.

With this procedure, we guarantee that 
 Hamiltonian  [\ref{Hint}] correctly reproduces
the constrained  LDA energies for
the truncated problem where hopping from and to the
localised $d$ or $f$ orbitals onto one site $i$ 
is forbidden. It is slightly different from the approach by
Anisimov {\em et al.} \cite{Anisimov91} who directly related the
double counting correction to $\bar U$:
\begin{equation}
\Delta \epsilon = {\bar U}(n_d-\frac12). 
\end{equation}
A third path has been proposed by Lichtenstein {\em et al.} \cite{FeNi,LichtensteinCP},
identifying an orbital-dependent term
\begin{equation}
\Delta \epsilon_m =  \frac{1}{2}\sum_{\sigma} \Sigma_{mm}^{\sigma}(\omega=0). 
\end{equation}

A cautionary remark:  While the total
LDA spectrum is rather insensitive to the choice of the basis, the constrained LDA
calculations of $\bar U$ depend more strongly on the shape of the orbitals
which are considered to be interacting, e.g. the $d$ or $f$ orbitals.
For example, in the case of  LaTiO$_3$ 
at a Wigner-Seitz radius of 2.37 a.u.\ (atomic units) for Ti,
a LMTO-ASA calculation by Nekrasov {\em et al.} \cite{Nekrasov00} using the TB-LMTO-ASA code \cite{LMTO2nd} 
yielded $\bar{U}=4.2$ eV while
a ASA-LMTO calculation  within the
orthogonal representation by Solovyev {\em et al.} \cite{Solovyev} gave $\bar{U}=3.2$ eV.
Thus, an appropriate basis
 is mandatory, and, even then, a significant uncertainty in $\bar U$
remains. For conventional LDA bandstructure calculations,
the choice of basis is not so crucial since with a large enough basis set the same 
bandstructure $\mbox{\boldmath $\epsilon$}^{\rm LDA}({\mathbf k})$ will be obtained.
In this respect, LDA+DMFT is more challenging since it now becomes important for
which orbitals the Coulomb interaction is taken into account.

Similarly as $\bar U$, the Hund exchange $J$
can be obtained from constrained LDA calculations.
Instead of the number of electrons, however,
the spin-polarisation has to be changed.
Since the electrons within the interacting orbitals 
on our decoupled lattice site cannot hop,
we can change the number of $n_{d\uparrow}$ and  $n_{d\downarrow}$
electrons, while keeping $n_d=n_{d\uparrow}+ n_{d\downarrow}$ fixed.
Then the Coulomb repulsion $\bar U$ will not be affected,
but nonetheless $E(n_{d\uparrow},n_{d\downarrow}=n_d-n_{d\uparrow})$
will change because of the Hund rule exchange.
Instead of the change of the total energy with $n_{d\uparrow}$, we can, as in Eq.\ [\ref{constrainedLDAU}],
use the now spin-dependent energy levels $\epsilon_{d\sigma}({n_{d\sigma}},n_{d{\bar{\sigma}}})=\frac{dE({n_{d\sigma}},n_{d{\bar{\sigma}}})}{d n_{d \sigma}}.$
From this Anisimov {\em et al.} \cite{Anisimov91} determine $J$ as
\begin{equation}
J_{\rm cLDA}=\epsilon_{d\uparrow}(n_{d\uparrow}=\frac{n_d}{2}+\frac12,n_{d{\downarrow}}=\frac{n_d}{2}-\frac12)
-\epsilon_{d\downarrow}(n_{d\uparrow}=\frac{n_d}{2}+\frac12,n_{d{\downarrow}}=\frac{n_d}{2}-\frac12).
\label{cLDAJ}
\end{equation}
Taking such a value of $J$,  Anisimov {\em et al.} \cite{Anisimov91} assume that the Hund exchange enters via a term 
\begin{equation}
\hat{H}_{J}= -{\red J} \sum_{ l\in L_U\neq m\in L_U\, \sigma}  \hat{n}_{l \sigma} \hat{n}_{m \sigma}. \label{HundLDAU}
\end{equation}
Nature and the LDA+DMFT Hamiltonian [\ref{LDADMFTH}] are, however, different.
As depicted in Fig.\ \ref{hund}, the Hund exchange also
influences the difference between inter-orbital and intra-orbital 
Coulomb interaction, given by $V=U-2J$ in  Hamiltonian [\ref{LDADMFTH}]. This 
goes beyond Eq.\ [\ref{HundLDAU}] which only gives a reduction of the 
 inter-band interaction for two spin aligned electrons.

For example, let us consider the case of a V$^{3+}$ ion in a cubic crystal field. Then we have two
 $d$ electrons ($n_d=2$) in three degenerate $t_{2g}$ orbitals. 
These two electrons can be spin aligned or not.
For Hamiltonian   [\ref{HundLDAU}], the energy difference between these
two configurations is simply
\begin{equation}
E({n_{d \uparrow}=2,n_{d\downarrow}=0})-E({n_{d \uparrow}=n_{d\downarrow}=1})={\blue } {\red J}.
\end{equation}
In contrast, the difference also depends on the orbital configuration
 for the LDA+DMFT Hamiltonian (Eq.\ [\ref{LDADMFTH}]).
For the configuration $(n_{d \uparrow}=2,n_{d\downarrow}=0)$, the two electrons have to be in 
different orbitals and their energy is $V-J=U-3J$. For the configuration
$(n_{d \uparrow}=1,n_{d\downarrow}=1)$ and without the spin-flip term,
we have the possibility that the two electrons are in the same orbital with 
$E(n_{d \uparrow}=n_{d\downarrow}=1)=U$ or in different orbitals with
$E(n_{d \uparrow}=n_{d\downarrow}=1)=V=U-2J$.
Averaging over the (unknown) orbital configurations, 
this yields an average value $E(n_{d \uparrow}=n_{d\downarrow}=1)=U-4/3J$. Hence, we have
\begin{equation}
E({n_{d \uparrow}=2,n_{d\downarrow}=0})-E({n_{d \uparrow}=n_{d\downarrow}=1})={\blue \frac{5}{3} {\red J}}
\label{LDADMFTHJ}
\end{equation}
for Hamiltonian [\ref{LDADMFTH}] if the spin-flip and pair-hopping term
are not taken into account.

For example, the constrained LDA calculation by Solovyev {\em et al.} \cite{Solovyev}
yields a value of $J_{\rm cLDA}=0.93\,$eV for V$^{3+}$. To reproduce this result
with Hamiltonian Eq.\ [\ref{LDADMFTH}], the left hand side of Eq.\ [\ref{LDADMFTHJ}]
has to be  $0.93\,$eV and, therefore, the parameter $J$ in  Eq.\ [\ref{LDADMFTHJ}]
has to be $J={3}/{5}\,0.93\,$eV$=0.56\,$eV. Such a reduced value of the Hund exchange coupling
is also more reasonable when we compare it
with the atomic value of $J=0.689\,$eV which is precisely measurable from the atomic spectrum\footnote{
How to extract the Racah parameters from
the atomic spectra is discussed e.g.\ by Griffith \cite{Griffith71}. He arrives at the following fit formulas for the
Racah parameters $B$ and $C$:
\begin{eqnarray}
B&=& \frac{3}{50} \big[E(^3P)-E(^3F)\big]+\frac{1}{70}\big[E(^1G)-E(^1D)\big] \\
C &=& \frac{1}{350} \big[21 E(^3F)-84 E(^3P)+40E(^1G)+37E(^1D)+28E(^1S)\big]
\end{eqnarray} 
where $E(X)$ denotes the energy of the atomic state 
$X$.
Looking up these atomic energies for the vanadium 3$d^2$ configurations at the
NIST Atomic Spectra data base ({\tt http://physics.nist.gov/cgi-bin/AtData/main\_asd},
which stem in turn from the overview article by Sugar and Corliss \cite{Suga85a}),
we obtain $B=0.1085\,$eV and $C=0.4181\,$eV so that
the Hund exchange coupling is $J=\frac{5}{2}B+C=0.689\,$eV
for the free atom.}, see Table \ref{tabJ}.
In particular, note that it is implausible that the Hund exchange is larger
than the atomic value since we expect that
 screening effects   reduce $J$ in the crystal.

\begin{table}[tb]
\begin{center}
\begin{tabular}{p{4.5cm}|p{3.2cm}|p{3.2cm}}
constrained LDA $J_{\rm cLDA}$ &  $J=\frac{3}{5}J_{\rm cLDA}$ & atomic value\\
\hline
$0.9\,$eV & $0.6\,$eV & $0.7\,$eV
\end{tabular}
\end{center}
\caption{\void Hund exchange for a V$^{3+}$ ion as calculated by constrained LDA, corrected value for  Eq.\ [\ref{LDADMFTH}]
and the atomic value from Suga and Corliss \cite{Suga85a}.\label{tabJ}}
\end{table}

Similarly, the constrained LDA $J_{\rm cLDA}$ has to be corrected by
$J=\frac{3}{10}J_{\rm cLDA}$ for $n_d=3$ before it is included in Hamiltonian Eq.\ [\ref{LDADMFTH}].
These two factors determine all correction factors for the exchange coupling $J$ between
$t_{2g}$ orbitals because
of  the unimportance of $J$ for
$n_d=1$ and because particle hole symmetry gives the same correction
factors for $n_d=4$ and 5 as for $n_d=2$ and 1, respectively.

\subsubsection{Spectral density functional theory formulation}
\label{SDFT}
\label{SDFTLDADMFT}

Our previous derivation of the LDA+DMFT method was physically motivated.
We
started from the assumption that the Kohn-Sham equations, i.e.
the LDA part, yield
the correct  results for the  weakly correlated $s$ or $p$ bands,
while the DMFT-part  takes into account the local Coulomb interactions
of the strongly correlated $d$ or $f$ bands.
Let us now consider an alternative way to formulate the
LDA+DMFT method as a {\em spectral density functional theory},
which goes back to  Chitra and Kotliar \cite{Chitra00} and
Savrasov {\em et al.} \cite{SAVRASOV,SAVRASOV2}, for more information also see \textcite{KotliarCP01} and \textcite{LichtensteinCP} and particularly the recent
review by Kotliar {\em et al.} \cite{Kotliar06}.
The basic idea is to replace the Hohenberg-Kohn \cite{Hohenberg64a} DFT functional
$E[\rho]$ by an energy functional $E[\rho,{\mathbf G}]$
which depends
on  the electron density $\rho({\mathbf r})$
and the local Green function $G_{il\,im}(\omega)$ of the interacting orbitals
($m,l\in L_U$). Since the local Green function is related to the ${\mathbf k}$-integrated spectrum
Savrasov {\em et al.}  \cite{SAVRASOV,SAVRASOV2} coined the name
{\em spectral density functional theory}.

As a starting point 
let us take the standard Luttinger-Ward \cite{Luttinger60a} many-body functional,
 which in compact form
(suppressing site $i$, orbital $l$, spin $\sigma$ and Matsubara frequency $\omega_{\nu}$ indices) reads
\begin{eqnarray}
\Omega[\Sigma,G] &=& \Phi[G]-{\rm Tr}\Sigma G - {\rm Tr} \ln ( (G^0)^{-1}-\Sigma).
\label{LWfunctional}
\end{eqnarray}
Here ${\rm Tr}$ 
denotes the trace ($\frac{1}{\beta}\sum_{\nu il\sigma}$) and $\Phi[G]$ the full set of two-particle irreducible diagrams
with lines $G$ and without external legs.
From the stationary conditions
\begin{eqnarray}
\frac{\delta \Omega[\Sigma,G]}{\delta G} & = & 0
\end{eqnarray}
we obtain the diagrammatic representation of the 
self energy:
\begin{eqnarray}
\Sigma = \frac{\delta \Phi[G]}{\delta G}.
\label{SigmaLW}
\end{eqnarray}
The second stationary condition
\begin{eqnarray}
\frac{\delta \Omega[\Sigma,G]}{\delta \Sigma} & = & 0
\end{eqnarray}
yields the usual Dyson equation, i.e. Eq.\ [\ref{Dyson}] without ${\mathbf k}$ integration:
\begin{equation}
G=  [(G^0)^{-1}-\Sigma]^{-1}.
\end{equation}
Such a construction is then a `conserving approximation' in the sense of Baym and Kadanoff \cite{Baym61a,Baym62a}.
On the basis of this functional theory, albeit
with a purely local Green function and self energy,
Jani\v{s} \cite{Janis86a,Janis89a} formulated 
a generalised  coherent potential approximation (CPA),
and Jani\v{s} and Vollhardt \cite{Janis92a} the DMFT self-consistency approach.

On the other hand, DFT was also formulated 
as an
 effective action  by Fukuda {\em et al.} \cite{Fukuda,Fukuda96}.
Now,  the electron density $\rho({\mathbf r})$ plays the role of the
Green function $G$ in Eq.\ [\ref{LWfunctional}],
and the Kohn-Sham potential $V_{\rm KS}({\mathbf r})$
replaces the self energy.
In terms of these variables, the DFT functional reads
\begin{eqnarray}
\Omega_{\rm DFT}[V_{\rm KS},\rho] &=& 
 \int d^3 r V_{\rm ion}({\bf r}) \rho({\mathbf r})+
\frac{1}{2} \int d^3 r d^3 r^{\prime} \frac{ \rho({\mathbf r}) \rho({\mathbf r^{\prime}})}
{|{\bf r}-{\bf r^{\prime}}|} + E_{\rm xc}[\rho]  \nonumber \\
&& - \int d^3 r \, V_{\rm KS}({\bf r}) \rho({\mathbf r}) - {\rm Tr}\,{\ln [i\omega_{\nu} +{{\hbar ^{2}}/{2m_{e}}\Delta - V_{\rm KS}({\mathbf r})]}}. \label{DFTF}
\end{eqnarray}
Here, the first line corresponds to  the many-body functional
$\phi[G]$  which is now given by the ionic potential, the Hartree and the
exchange correlation term;
${\rm Tr}$ denotes again the trace (now w.r.t.\ the operator following and
 any basis, e.g. the  $|{\mathbf r}\rangle$ basis)
as well as the sum over Matsubara frequencies as before.

Minimisation of the functional w.r.t.\ $\rho$, i.e.
 $\delta \Omega_{\rm DFT}[\rho,V_{\rm KS}]/\delta \rho=0$,
reproduces the correct form the Kohn-Sham potential
in the Kohn-Sham equation  [\ref{KohnSham}],
\begin{equation}
V_{\rm KS}({\bf r}) = V_{{\rm ion}}({\bf r})+
\int d^{3}{r^{\prime }}\;
{V_{{\rm ee}}({\bf r}\!-\!{\bf r^{\prime }})} {\rho ({\bf r^{\prime }})}
+{{\frac{\delta {E_{{\rm xc}}[\rho ]}}{\delta \rho ({\bf r)}}}},
\label{VKSF}
\end{equation}
in analogy to Eq.\ [\ref{SigmaLW}] which gives the correct form of the
self energy.
The second minimisation,  $\delta \Omega_{\rm DFT}[\rho,V_{\rm KS}]/\delta V_{\rm KS} =0$,
gives the Kohn-Sham equation  [\ref{KohnSham}] itself if
 $V_{\rm KS}({\bf r})$ is replaced by the right hand side of Eq.\ [\ref{VKSF}],
which of course is guaranteed if self-consistency is obtained.
More precisely, it yields $\rho ({\mathbf r})$ directly, expressed via the
Kohn-Sham equation:
\begin{equation}
\rho({\mathbf r})= \frac{1}{\beta}\sum_{\nu} \left\langle {\mathbf r}
\left|\left[
i \omega_{\nu} +{\hbar^{2}}/{2m_{e}}\Delta - V_{\rm KS}\right]^{-1}
 \right|  {\mathbf r} \right\rangle.
\label{rhoKS}
\end{equation}

In the case of LDA+DMFT, we have to deal with a mixed representation
in terms of the density $\rho({\mathbf r})$ (LDA) and the local Green function
of the interacting orbitals ${\mathbf G}(\omega)$ (DMFT).
Savrasov {\em et al.}  \textcite{SAVRASOV,SAVRASOV2} constructed such a functional:
\begin{eqnarray}
\Omega[{\mathbf \Sigma},{\mathbf G},V_{\rm KS},\rho] &=& 
 \int d^3r V_{\rm ion}({\bf r}) \rho({\mathbf r}) + \frac{1}{2} \int d^3 r d^3 r^{\prime} \frac{ \rho({\mathbf r}) \rho({\mathbf r^{\prime}})}
{|{\bf r}-{\bf r^{\prime}}|} + E_{\rm xc}[\rho] \nonumber \\
&& -\Phi[{\mathbf G}]- \Phi_{\rm dc}[{\mathbf G}] \nonumber\\
&& - \int d^3 r \, V_{\rm KS}({\bf r}) \rho({\mathbf r}) 
- {\rm Tr}\, {\mathbf \Sigma}(i \omega_{\nu}) {\mathbf G}(i \omega_{\nu})
 \nonumber\\&&
- {\rm Tr}\,{\ln [i\omega_{\nu} \!+\!{{\hbar ^{2}}/{2m_{e}}\Delta\! -\! V_{\rm KS}\! -\!\Sigma(i\omega_{\nu})_{lm}\varphi_{il}^*({\mathbf r^{\prime}})\varphi_{im}({\mathbf r}) ]}}.
\label{LDADMFTf}
\end{eqnarray}
Here,
$\Sigma(i\omega_{\nu})_{lm}\varphi_{il}^*({\mathbf r^{\prime}})\varphi_{im}({\mathbf r})$
is the transformation of the local orbital self energy to the real space
formulation in terms of ${\mathbf r}$ and ${\mathbf r^{\prime}}$,
and  $\Phi_{\rm dc}[{\mathbf G}]$ is an additional double counting correction term.

In the spirit of Hohenberg and Kohn \cite{Hohenberg64a}, we are looking for the
density $\rho({\mathbf r})$ and the local Green function ${\mathbf G}(\omega)$ 
at which  the functional [\ref{LDADMFTf}] takes its minimum.
(The minimisation w.r.t.\ ${\mathbf \Sigma}$ and $V_{\rm KS}$ gives two additional equations, 
defining  ${\mathbf \Sigma}$ and $V_{\rm KS}$.)
The minimisation procedure
yields the exact ground state energy, density and  local Green function
if  $E_{\rm xc}[\rho]$ and $\Phi[{\mathbf G}(\omega)]$ are themselves the exact ones.
However, we do not know these exact 
  $E_{\rm xc}[\rho]$ and $\Phi[{\mathbf G}(\omega)]$  and have to use approximations.
LDA+DMFT is one such approximation where we replace  $E_{\rm xc}[\rho]$
by the LDA exchange-correlation energy  $E_{\rm xc}^{\rm LDA}[\rho]$ and
$\Phi[{\mathbf G}(\omega)]$ by all local two-particle irreducible diagrams,
i.e. the DMFT diagrams $\Phi_{\rm DMFT}[{\mathbf G}(\omega)]$. The double counting term
  $\Phi_{\rm dc}[{\mathbf G}]$ will have one of the 
forms discussed in Section \ref{cLDA}.

With this LDA+DMFT approximation of the spectral density functional [\ref{LDADMFTf}],
we recover the LDA+DMFT equations formulated earlier in this Section:
\begin{itemize}
\item
The DMFT self energy ${\mathbf \Sigma}(\omega)$ minus the double counting correction term
is obtained from
\begin{equation}
\frac{\delta \Omega[{\mathbf \Sigma},{\mathbf G},V_{\rm KS},\rho]}{\delta{\mathbf G}}=
\frac{\delta \Phi_{\rm DMFT}[{\mathbf G}]}{\delta{\mathbf G}} -{\mathbf \Sigma} -\frac{\delta \Phi_{\rm dc}[{\mathbf G}]}{\delta{\mathbf G}} = 0.
\end{equation}
\item
From
\begin{equation}
\frac{\delta \Omega[{\mathbf \Sigma},{\mathbf G},V_{\rm KS},\rho]}{\delta{\mathbf \Sigma}}=0
\end{equation}
and expressing ${-{\hbar ^{2}}/{2m_{e}}\Delta + V_{\rm KS}}$ in the basis $\varphi_{il}({\mathbf r})$,
i.e. replacing it by the orbital matrix (the LDA bandstructure)
$\mbox{\boldmath ${\epsilon}$}^{\rm LDA}({\mathbf k})$ defined via Eq.\ [\ref{tFTeps}],
we obtain the  Dyson equation:
\begin{equation}
{\mathbf G}({\mathbf k},i\omega_{\nu}) = [i\omega_{\nu} -\mbox{\boldmath ${\epsilon}$}^{\rm LDA}
({\mathbf k}) -{\mathbf \Sigma}(i\omega_{\nu}) ]^{-1}
\end{equation}
\item
Similarly, 
\begin{equation}
\frac{\delta \Omega[{\mathbf \Sigma},{\mathbf G},V_{\rm KS},\rho]}{\delta V_{\rm KS}}=0
\end{equation}
yields the  Kohn-Sham equations, or more precisely,
 $\rho({\mathbf r})$ expressed through the 
Kohn-Sham equations as in Eq.\ [\ref{rhoKS}], as in Eq.\ [\ref{rhoKS}] but now with
the self energy included:
\begin{equation}
\rho({\mathbf r})= \frac{1}{\beta}\sum_{\nu} \left\langle {\mathbf r}
\left|\left[
i \omega_{\nu} +{\hbar^{2}}/{2m_{e}}\Delta - V_{\rm KS} -\Sigma(i\omega_{\nu})_{lm}\varphi_{il}^*({\mathbf r^{\prime}})\varphi_{im}({\mathbf r})
\right]^{-1}
 \right|  {\mathbf r} \right\rangle.
\end{equation}
This equation can also be formulated in the orbital basis $\varphi_{il}({\mathbf r})$ instead of the
real space basis $|{\mathbf r}\rangle$.
\item
Finally, the correct form of the Kohn-Sham potential follows from
\begin{equation}
\frac{\delta \Omega[{\mathbf \Sigma},{\mathbf G},V_{\rm KS},\rho]}{\delta \rho}= 0
\end{equation}
as before.
\end{itemize}

Let us not forget to mention a conceptional deficiency 
of spectral density functional theory (or LDA+DMFT)
compared with standard DFT: 
What we define as local and interacting in DMFT is basis dependent.
Different basis sets will hence produce different results,
unless more and more orbitals are considered as interacting in DMFT
and more and more non-local correlations are taken into account.
This poses the so-far open question: What is the optimal basis set
for LDA+DMFT?

\subsubsection{Simplifications for transition metal oxides with well separated $e_{g}$ and $t_{2g}$ bands}
\label{SimpTMO}

Many transition metal oxides are cubic perovskites, often
with a slight distortion of the ideal cubic crystal structure. In these
systems, the cubic crystal field of the
oxygen ions splits  the $d$-orbitals
into three degenerate $t_{2g}$ and two degenerate 
$e_{g}$ orbitals. This splitting is often so strong that the $t_{2g}$ or 
$e_{g}$ bands at the Fermi energy are rather well separated from all
other bands.
For the low energy physics, it is hence sufficient 
to take only those bands into account which cross the Fermi energy,
e.g. the three  $t_{2g}$ bands. Note that these effective bands 
at the Fermi energy are not
purely of $d$ character. These are the effective bands corresponding
to the LDA eigenenergies. Hence,
they also contain contributions
of the other orbitals, in particular the oxygen $p$ orbitals due to hybridisation effects.

From a conventional LDA bandstructure calculation, these effective low-energy bands can be
accurately determined via the NMTO downfolding approach of
Andersen {\em et al.} \cite{Andersen99,AndersenPsik} or the projection  to Wannier orbitals
 \cite{Marzari97}.

If the transition metal oxide is cubic or only slightly distorted,
a further simplification can be employed for the
(almost) degenerate 
bands at the Fermi energy: Instead of the full LDA Hamiltonian 
$\hat{H}_{\rm LDA}$ in Eq.\ [\ref{LDADMFTH}]
only the total DOS of the low-lying bands needs to be 
taken into account within DMFT: Without symmetry breaking, the
Green function and the self energy of these bands 
remain degenerate, i.e. $%
G_{l\,m}(\omega)=G(\omega)\delta
_{l\,m}$ and $\Sigma
_{l\,m}(\omega)=\Sigma (\omega)\delta
_{l\,m}$ for $l,m\in L_{U} $
 (i.e. the 
interacting orbitals at the Fermi energy). Downfolding to a basis with
these
degenerate $L_{U}$ bands results in an effective Hamiltonian $H_{%
{\rm LDA}}^{{\rm eff}}$ or a corresponding
$\mbox{\boldmath $\tilde{\epsilon}$}^{\rm LDA}({\mathbf k})$. 
 From this reduced Hamiltonian, the diagonal element of the Green function
 is calculated via
 \begin{eqnarray} G(\omega)  {\mathbf 1}
&=&\frac{1}{
{V_{B}}} \int {\rm d}^3 k \;
[\omega {\mathbf 1}+\mu  {\mathbf 1}-\mbox{\boldmath $\tilde{\epsilon}$}({\mathbf k})-  \Sigma(\omega) {\mathbf 1} ]^{-1}.
\end{eqnarray}
 Due to the diagonal structure of the self energy,
the degenerate interacting Green function can be expressed via the
 non-interacting Green function $G^{0}(\omega)$:
\begin{eqnarray}
G(\omega)&\!=\!&G^{0}(\omega-\Sigma (\omega))=\int d\epsilon\,
\frac{N^{0}(\epsilon )}{\omega-\Sigma (\omega)-\epsilon}.
\label{intg}
\end{eqnarray}
Thus, it is possible to use the Hilbert transformation of the
unperturbed
LDA-calculated density of states (DOS) $N^{0}(\epsilon )$, i.e. Eq.\ [\ref%
{intg}], instead of the Dyson equation (Eq. [\ref{Dyson}]) with the full Hamilton matrix $\mbox{\boldmath $\tilde{\epsilon}$}^{\rm LDA}({\mathbf k})$. 
This simplifies the
calculations considerably. With Eq.\ [\ref{intg}] also some
conceptual simplifications arise: (i) the subtraction of the
double counting correction $\Delta\epsilon$ only results in an
(unimportant) shift of the chemical potential in Eq.\ [\ref{intg}] and, thus, the exact
form of $\Delta\epsilon$ is irrelevant; (ii) the theorem of 
M\"uller-Hartmann
\cite{MuellerHartmann89c}
of a fixed spectral function at the Fermi energy holds within a Fermi liquid; (iii) as the number of electrons
within the different bands is fixed, a self-consistent LDA+DMFT 
calculation (step 5 in Section \ref{HamLDADMFT})
is dispensable.
If the distortion from the cubic symmetry is only weak or if the off-diagonal
elements of $\mbox{\boldmath $\epsilon$}^{\rm LDA}{({\bf k})}$ are small,
a slight modification of Eq.\ [\ref{intg}] can be used:
\begin{eqnarray}
G_{m m}(\omega)&\!=\!&G^{0}(\omega-\Sigma (\omega))=\int d\epsilon
\frac{N^{0}_m(\epsilon )}{\omega-\Sigma_{mm} (\omega)-\epsilon}
\end{eqnarray}
which now depends on the orbitals, albeit
neglecting the orbital admixture.

It should also be noted that the approximation Eq.\
[\ref{intg}] is justified only if the overlap between the $t_{2g}$
orbitals and the other orbitals is rather weak.

\subsection{Hartree+DMFT and Hartree-Fock+DMFT}
\label{HDMFT}
Within the concept of DMFT, a combined
Hartree+DMFT approach is a very natural approximation, since the non-local (density-density) Coulomb interaction $U_{ij}$
enters only  via the Hartree \cite{Hartree28a} term in the $d\rightarrow \infty$ limit \cite{MuellerHartmann89a}:
There are  ${{\cal Z}_{|\!|i\!-\!j|\!|}}$  equivalent $U_{ij}$ terms so that
 $U_{ij}$ has to scale like  $1/{{\cal Z}_{|\!|i\!-\!j|\!|}}$ in order that the Hartree energy contribution stays finite. This is exactly the analogy to the Weiss \cite{Weiss07a} mean field theory
for spin models, in which the summed contribution of the neighbouring sites gives
an effective Weiss field $h^{\rm Weiss}_{i}=\sum_{j\neq i} J_{ij} \langle \hat{S}_j \rangle$
so that the proper scaling is $J_{ij} \propto 1/{{\cal Z}_{|\!|i\!-\!j|\!|}}$.

\begin{figure}
\begin{center}
\includegraphics[clip=true,width=6cm]{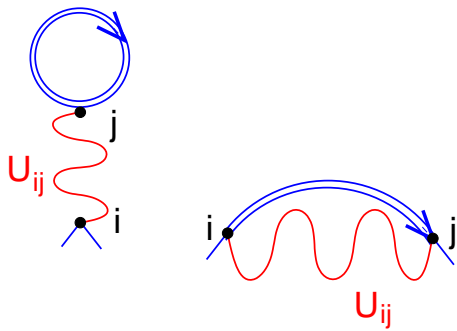}
\end{center}
\caption{\void Left: Hartree diagram for the non-local Coulomb interaction. Right: Exchange diagram
(Fock term). Both are  taken into account within Hartree-Fock. \label{FigHF}}
\end{figure}

Consequently, other contributions 
scale at least like  $1/\sqrt{{\cal Z}_{|\!|i\!-\!j|\!|}}$.
 For the Fock term, i.e.  Fig.\ \ref{FigHF} (right), this scaling is obvious
since there is one Green function  ${\mathbf G}_{i j} \propto 1/\sqrt{{\cal Z}_{|\!|i\!-\!j|\!|}}$
connecting $i$ and $j$.\footnote{The general way of such scaling  arguments
was elaborated
in Section \ref{DMFTtutorial}.}
Note that Si and Smith \cite{Smith96,Si96} and  Chitra and Kotliar \cite{Chitra00b,Chitra01}
employed a different scaling  $U_{ij}\propto 1/\sqrt{{\cal Z}_{|\!|i\!-\!j|\!|}}$
in the so-called extended DMFT, also see \textcite{Parcollet99}. This scaling results in additional diagrams,
but requires to treat the Hartree term separately since
it would otherwise diverge.

The reader might think of the  Hartree+DMFT approach as a special case
of LDA+DMFT, in which the LDA exchange and correlation potential $E_{\rm xc}[\rho ]=0$.
In our orbital basis $\varphi_{il}({\bf r})$, the Hartree term  contributes as
\begin{equation}
\Sigma^{\rm Hartree}_{il\, jm}\equiv t^{\rm Hartree}_{il\,jm}=\int d^3 r \,
\varphi_{il}^*({\bf r}) \int d^3{r^{\prime }}{\rho ({\bf %
r^{\prime }})}{V_{{\rm ee}}({\bf r}\!-\!{\bf r^{\prime }})}\varphi
_{jm}({\bf r}), \label{SigmaHartree}
\end{equation}
which is the left diagram  of Fig.\ \ref{FigHF} and gives, Fourier-transformed,
a ${\mathbf k}$-dependent self energy ${\mathbf \Sigma}^{\rm Hartree}_{\mathbf k}$.
An advantage is that Hartree+DMFT is a diagrammatic approach,
consisting of the Hartree diagram of
Fig.\ \ref{FigHF} and the local DMFT contributions of all 
irreducible diagrams.

Because we know the Hartree contribution diagrammatically,\footnote{It also follows from the variational principle
as was stressed by Slater \cite{Slater30a}.}
we are in full control of the double counting term of the many-body
Hamiltonian Eq.\ [\ref{LDADMFTH}]. Diagram Fig.\ \ref{FigHF} yields in terms
of the on-site equal-time Green function $G^{\sigma}_{il\,im}(\tau=0^+)$
the following local correction for the self energy of the interacting
orbitals $l\in L_U$:
\begin{equation}
\Sigma^{\rm Hartree}_{\rm dc}= \frac{1}{\sum\limits_{\sigma l}}\Big[ \sum_{\sigma l}
 U  G_{ll}^{{\sigma}}(\tau\!=\!0^+\!)
+\!\! \sum_{\stackrel{\scriptstyle m\in L_U}{\scriptstyle m\neq l}}\!\!\! V G_{im\,im}^{\bar{\sigma}}(\tau\!=\!0^+\!)
+  (V\!-\!J)  G_{im\,im}^{{\sigma}}(\tau\!=\!0^+\!)\Big].
 \label{HartreeDC}
\end{equation}
Altogether, the Hartree+DMFT self energy consists of the $\omega$-independent, but ${\mathbf k}$-dependent,
Hartree contribution and the $\omega$-dependent, but ${\mathbf k}$-independent, DMFT contribution
with the double counting term subtracted:
\begin{equation}
\Sigma^{\rm Hartree+DMFT}_{ij\, lm}(\omega)=
 \Sigma^{\rm Hartree}_{il\, jm}+ \delta_{ij} \Sigma^{DMFT}_{lm}(\omega)- \delta_{ij} \delta_{lm}\Sigma^{\rm Hartree}_{\rm  dc}.
\label{SigmaHDMFT}
\end{equation}
With this self energy, one  needs to calculate a new electron density $\rho({\mathbf r})$
which in turn gives a new $\Sigma^{\rm Hartree+DMFT}_{ij\, lm}$ and has to
be iterated until convergence.

The advantages of Hartree+DMFT are that it follows the $d\rightarrow \infty$ concept and that it
is a very clear and diagrammatically controlled scheme. The disadvantage, however,
is that by neglecting the LDA exchange and correlation term all exchange and
correlation contributions need to stem from DMFT which only includes
the local exchange and correlations. With the success of LDA and its
superiority over a simple Hartree calculation, this seems not to be 
sufficient. This  explains the reluctance, in particular, of the physicists of
the LDA community, to implement such a scheme. Indeed, Hartree+DMFT calculations
have not been done in the context of realistic calculations; for model Hamiltonians see e.g.
\textcite{Wahle98a}. A more thorough investigation of how well such a relatively simple scheme
works for realistic calculations is mandatory. Also note that
with cluster DMFT calculations, which were briefly discussed in  Section \ref{DCA},
more and more non-local exchange and correlation contributions can be taken into account
so that the treatment of the remaining Coulomb interaction terms
by the Hartree approximation is a less severe approximation.
The calculation of the screened Coulomb interaction parameters, however,
remains challenging.

If one combined the Hartree-Fock \cite{Fock30a} approximation and DMFT, at least the non-local exchange
(given by the right diagram of Fig.\ \ref{FigHF}) would be additionally included.
The problem with the exchange term is that the
 Kohn-Sham equations become much more complicated due to
the non-local nature of the exchange term:
In terms of the first quantised wave functions  $\varphi _{i}({\bf r})$, 
the Hartree-Fock equations read
\begin{eqnarray}
\left[ -{\frac{\hbar ^{2}}{2m_{e}}\Delta +V_{{\rm ion}}({\bf r})}+
\int d^{3}{r^{\prime }}\;
{ \textcolor{red}{\frac{\displaystyle  e^2}{\displaystyle 4\pi\epsilon_0}\; \frac{\displaystyle 1}{\displaystyle |{\bf r} -{\bf r^{\prime }}|}}} {\rho ({\bf r^{\prime }})}\right] \varphi _{i}({\bf r})&& \nonumber \\
- \sum_j \int d^{3}{r^{\prime }} 
{ \textcolor{red}{\frac{\displaystyle  e^2}{\displaystyle 4\pi\epsilon_0}\; \frac{\varphi^*_{j}({\bf r^{\prime}})
\varphi_{i}({\bf r^{\prime}})}{\displaystyle |{\bf r} -{\bf r^{\prime }}|}}}
\varphi _{j}({\bf r})&=&\varepsilon _{i}\;\varphi_{i}({\bf
r}), \label{HFEq}
\end{eqnarray}
instead of the Kohn-Sham equation [\ref{KohnSham}]. But because it is not local in ${\mathbf r}$, Eq.\ [\ref{HFEq}] 
is mathematically much more complicated to solve.
Nonetheless, with increasing computational resources  exact exchange calculations 
have become possible and were performed by Stadele {\em et al.} \cite{Stadele97}.

Closely related are also the  Hartree-Fock calculations by Schnell {\em et al.} \cite{Schnell02a,Schnell03a}.
The authors start with a conventional bandstructure calculation
which is, however, only used to construct maximally localised Wannier
orbitals by the Marzari-Vanderbilt \cite{Marzari97} projection approach.
With these orbitals Schnell {\em et al.} \cite{Schnell02a,Schnell03a} calculate the 
non-interacting hopping matrix elements, i.e. the
overlap integral of these Wannier wave functions w.r.t.\ $-{\frac{\hbar ^{2}}{2m_{e}}\Delta +V_{{\rm ion}}({\bf r})}$,
and the Coulomb interaction by the overlap integral Eq.\ [\ref{CoulombME}]. 
This defines a multi-band many-body Hamiltonian for
 which, because of the localised
orbitals, the hopping and Coulomb interaction between far apart orbitals is neglected.
 Schnell {\em et al.} \cite{Schnell02a,Schnell03a} solve this Hamiltonian within a reduced set of Wannier 
orbitals by Hartree-Fock. But their approach
also offers the advantage that more sophisticated many-body
approaches like DMFT can be employed  in the future.
The calculation of the screened Coulomb interaction remains problematic.
 Schnell {\em et al.} \cite{Schnell02a,Schnell03a} propose to employ
the Thomas-Fermi theory of screening to this end.

\subsection{GW+DMFT}
\label{GWDMFT}

\begin{figure}
\begin{center}
\includegraphics[clip=true,width=11.5cm]{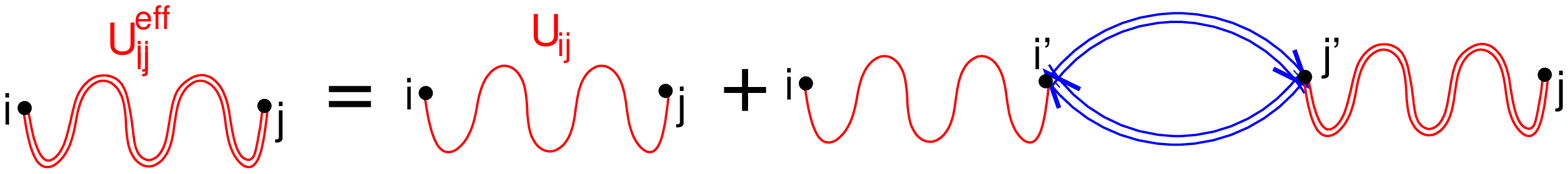}
\end{center}
\vspace{.6cm}

\caption{\void The bare Coulomb interaction (wiggled line) is, within RPA, screened by
particle-hole excitations (Green function bubbles; double lines), resulting in
a screened Coulomb interaction (wiggled double line) given
by an infinite series of bubble diagrams which can be calculated
self-consistently as indicated.\label{figRPA}}
\end{figure}

A serious disadvantage of both,
Hartree+DMFT and Hartree-Fock+DMFT, is that the
screening of the  Coulomb interaction is not yet included.
Most important for this screening are the random phase approximation (RPA) bubble diagrams
depicted in Fig.\ \ref{figRPA}.
Particle-hole excitations, even at relatively high energy,
 dramatically reduce the effective  Coulomb interaction.
This is taken into account by  the  infinite series of RPA diagrams, replacing 
the bare Coulomb interaction (wiggled line) by the screened Coulomb interaction (wiggled double line).

For a many-body physicist, a diagrammatically controlled approach for the challenging task
of realistic  material calculations is desirable. The minimal set of diagrams to this
end are (i) the Hartree and Fock terms of Fig.\ \ref{FigHF} which make up a major contribution
of the Coulomb interaction, (ii) the RPA diagrams of Fig.\ \ref{figRPA} for calculating the screened Coulomb interaction,
and
(iii) the local DMFT  contribution of all irreducible diagrams  for $d$ or $f$ materials with strong electronic correlations.
The former two terms are contained in the so-called GW approach, and Biermann {\em et al.} \cite{Biermann03}
recently proposed to include the local dynamics of DMFT in a GW+DMFT method,
also presenting first results for ferromagnetic Ni.

Let us start by briefly recapitulating the GW equations introduced by Hedin \cite{Hedin},
for a review see \textcite{Aryasetiawan98}.
The GW self energy consists of the Hartree part, Eq.\ [\ref{SigmaHartree}], and the
exchange contribution, the right diagram of  Fig.\ \ref{FigHF}. The latter is given by
\begin{equation}
\Sigma^{\rm GW}({\mathbf r}, {\mathbf r^{\prime}};\omega) = {i} \int \frac{d \omega^{\prime}}{{2 \pi}}
 G({\mathbf r}, {\mathbf r^{\prime}};\omega+\omega^{\prime}) W({\mathbf r}, {\mathbf r^{\prime}};\omega^{\prime}).
\label{SigmaGW}
\end{equation}\label{GWname}
This form of the self energy, Green function $G$ times screened interaction
$W$, coined the name of the `GW' approximation.  The imaginary unit $i$ in front of
$G$ stems from the standard definition of the real time or real frequency Green function and the 
rules for evaluating the diagram  Fig.\ \ref{FigHF}, see \textcite{Abrikosov63}.

To take the RPA screening into account, Eq.\ [\ref{SigmaGW}] employs the screened interaction $W$, i.e. the
double line in Fig.\ \ref{figRPA}. This screened $W$ is given by a geometric series, defined iteratively in
Fig.\ \ref{figRPA}, which yields
\begin{equation}
 W({\mathbf r}, {\mathbf r^{\prime}}; \omega) = \int d^3 {r^{\prime\prime}}
V_{\rm ee}({\mathbf r},  {\mathbf r}^{\prime\prime}) \epsilon^{-1}( {\mathbf r^{\prime\prime}}, {\mathbf r^{\prime}};\omega).
\label{WGW}
\end{equation}
Here, $V_{\rm ee}({\mathbf r},  {\mathbf r}^{\prime\prime})$ is the bare Coulomb interaction
(Eq.\ [\ref{abinitioham}] in Section \ref{ESC}] which does not depend on $\omega$, and  
$\epsilon^{-1}( {\mathbf r}, {\mathbf r^{\prime}}; \omega)$ denotes
the inverse of the dielectric function
\begin{equation}
\epsilon( {\mathbf r}, {\mathbf r^{\prime}}; \omega) = \delta_{ {\mathbf r}{\mathbf r^{\prime}}}
 - V_{\rm ee}({\mathbf r},  {\mathbf r^{\prime}}) P^{\rm GW}( {\mathbf r}, {\mathbf r^{\prime}};\omega).
\label{epsGW}
\end{equation}
Together, Eqs.\ [\ref{WGW}] and [\ref{epsGW}] have the 
usual pole-structure of the geometric series in which the GW polarisation 
\begin{equation}
P^{\rm GW} ( {\mathbf r}, {\mathbf r^{\prime}}; \omega) = - 2 i  \int \frac{d \omega^{\prime}}{{2 \pi}}
 G({\mathbf r}, {\mathbf r^{\prime}};\omega+\omega^{\prime}) G({\mathbf r}, {\mathbf r^{\prime}};\omega^{\prime})
\label{GWP}
\end{equation}
is the factor for the  bubble consisting of two Green functions.
In Eq.\ [\ref{GWP}], the prefactor $2$  stems from the spin summation, assuming here
that $G$ and $\Sigma$ are spin-independent in a paramagnetic phase.
Because of the screening the effective interaction $W$ becomes frequency dependent.

Since the  computational burden of the more complicated GW  approximation is
much higher than that of the simpler LDA, plane-wave basis calculations  become very costly
because of the  big basis set. Therefore,  the LMTO basis of Andersen \cite{LMTO1} or a Gaussian basis is 
 preferable.

To merge GW with DMFT, first the
local contribution in the GW equations has to be subtracted since
this part will later be included within DMFT:
\begin{equation}
{\mathbf P}^{{\rm GW+DMFT}} ({\mathbf k}, \omega) ={\mathbf P}^{\rm GW} ({\mathbf k}, \omega)
- \underbrace{\frac{1}{V_{\rm BZ}}\int_{\rm BZ} d^3 k\, {\mathbf P}^{\rm GW}({\mathbf k}, \omega)}_{\rm local\, contribution}+
{\mathbf P}^{\rm DMFT} (\omega),
\label{GWDMFTP}
\end{equation}
where we 
Fourier-transformed  $P^{\rm GW} ( {\mathbf r}, {\mathbf r^{\prime}}; \omega)$ of Eq.\ [\ref{GWP}]
from real to ${\mathbf k}$-space,
and also employed an orbital basis, indicated by the bold symbols.
To switch between  the representation in terms of spatial coordinates 
${\mathbf r}$ and the representation in terms of orbitals $l$,
one has to calculate the  overlap integral w.r.t.\ the orbital wave functions
$\varphi_{il}({\mathbf r})$  or to multiply by  $\varphi_{il}({\mathbf r})$,
respectively.
In practice,  Aryasetiawan {\em et al.} \cite{AryasetiawanCP04} propose to calculate the two-particle polarisations and interactions
within another basis, the optimal product basis, see \textcite{Aryasetiawan98}.

Similarly, we have
\begin{eqnarray}
{\mathbf \Sigma}^{{\rm GW+DMFT}} ({\mathbf k}, \omega) &=&{\mathbf \Sigma}^{\rm GW} ({\mathbf k}, \omega)
- \frac{1}{V_{\rm BZ}}\int_{\rm BZ} d^3 k\,{\mathbf \Sigma}^{\rm GW} ({\mathbf k}, \omega) \nonumber\\ &&+ {\mathbf \Sigma}^{\rm Hartree}  ({\mathbf k}, \omega)
- {\mathbf \Sigma}^{\rm Hartree}_{\rm dc}+{\mathbf \Sigma}^{{\rm DMFT}} ({\mathbf k}, \omega).
\label{totalS}
\end{eqnarray}
Here, also the Hartree self energy and its 
local part,
Eqs.\ [\ref{SigmaHartree}] and [\ref{HartreeDC}], respectively,  have been included.
The local DMFT self energy contribution  is that of the auxiliary Anderson model
or, equivalently, that of the local contribution of all irreducible diagrams. Note, that the combined contribution
${\mathbf \Sigma}^{{\rm DMFT}} ({\mathbf k}, \omega)
- {\mathbf \Sigma}^{\rm Hartree}_{\rm dc}$
is actually the exchange-correlation part of the DMFT self energy.

To calculate the DMFT polarisation, let us start with the local impurity interaction
of the DMFT Hamiltonian Eq.\ [\ref{DMFTH}], considering only
 density-density type of interaction between two orbitals, i.e.
 $U_{l m n o}= \delta_{mn}\delta_{lo} U_{lm}$ in Eq.\ [\ref{DMFTH}].
Within DMFT, this local interaction is screened, yielding a local screened interaction
\begin{equation}
{W}_{lm}(\omega)=U_{lm}(\omega)-\sum_{m^{\prime} l^{\prime}} U_{ll^{\prime}}(\omega) \chi_{l l^{\prime} m^{\prime} m}(\omega) U_{m^{\prime} m}(\omega),
\label{localW}
\end{equation}
which 
can be calculated from the
 local susceptibility
\begin{equation}
\chi_{lmno}(\tau)=\langle T c^{\dagger}_l(\tau) c^{\phantom{\dagger}}_m(\tau) c^{\dagger}_n(0) n^{\phantom{\dagger}}_o(0) \rangle.
\label{DMFTsusz}
\end{equation}
Here, $T$ is the time-ordering operator.

From the difference between the inverse  matrices  in the orbital index
${\mathbf W}$ and ${\mathbf U}$, the 
DMFT local polarisation can be calculated by
an  inversion w.r.t.\ the orbital indices:
\begin{equation}
{\mathbf  P}^{\rm DMFT}(\omega) = {\mathbf U}^{-1}(\omega)- {\mathbf W}^{-1}(\omega). 
\label{UDMFT}
\end{equation}
This relation follows directly from the definition of the local polarisation analogous
to Eq.\ [\ref{WGW}], after substituting $\epsilon(\omega)$ of Eq.\ [\ref{epsGW}]
and solving for ${\mathbf P}(\omega)$, with 
 ${\mathbf U}(\omega)$ playing the role of $V_{\rm ee}$.

Now, however, the interaction  ${\mathbf U}(\omega)$ is frequency dependent or (in imaginary time)
$\tau$-dependent. This has to be taken into account in the effective action
of the Anderson impurity model, where we have now to employ the following
${\cal A}[\psi ,\psi ^{\ast },({\cal G}^0)^{-1},{\mathbf U}]$ in
Eq. [\ref{siam}]:
\begin{eqnarray}
{\cal A}[\psi ,\psi ^{\ast },({\cal G}^0)^{-1},{\mathbf U}]&=&\sum_{\nu
\sigma \,l m}\psi _{\nu
m}^{\sigma \ast }({\cal G}^{\sigma 0}_{\nu m n})^{-1}\psi _{\nu n}^{\sigma
{\phantom\ast }}\nonumber\\&&
\!\!\!\!\!+\!\!\!\sum_{l m\,\sigma\sigma^{\prime}}
\!
    \int\limits_{0}^{\beta }\!\!d\tau \psi _{l}^{\sigma \ast }(\tau )
\psi_{l}^{\sigma}(\tau )
\,U_{l m}(\tau\!-\!\tau^{\prime})\,
\psi _{m}^{\sigma^{\prime} \ast }(\tau^{\prime} ) \psi_{m}^{\sigma^{\prime}}(\tau^{\prime} ). 
\label{effActionGWDMFT}
\end{eqnarray}

\begin{figure}[htb]
\begin{center}
\unitlength1mm
\small
\hspace{-.15cm}\begin{picture}(130,184)
\thicklines

\put(3.00,175.00){\framebox(127.00,7.00)[cc]
{\parbox{120mm}{Do LDA calculation, Fig.\ \ref{flowLDA}, yielding 
	${\mathbf G}_{\mathbf k}(\omega)\!=\! [\omega{\mathbf 1}\!+\!\mu{\mathbf 1}\!-\! {\mathbf \epsilon}^{\rm LDA}{({\bf k})}]^{-1}$ (Eq.\ [\ref{GLDA}]).
        }}}

\put(3.00,32.00){\framebox(127.00,143.00)[cc]{}}

\put(10.00,167.00){\framebox(120.00,8.00)[cc]
{\parbox{117mm}{ \vspace{.05cm}

Calculate GW  polarisation 
${\mathbf P}^{\rm GW} (\omega) \!= \! - 2 i  \! \int \! \frac{d \omega^{\prime}}{{2 \pi}}
 {\mathbf G}(\omega+\omega^{\prime}) {\mathbf G}(\omega^{\prime})\; \mbox{(Eq.\ [\ref{GWP}])}.
$
}}}

\put(10.00,151.00){\framebox(120.00,16.00)[cc]
{\parbox{117mm}{\vspace{.2cm}

If  DMFT polarisation ${\mathbf P}^{\rm DMFT}$  is known (after the 1. iteration), include it
\vspace{-.6cm}

\[
\! \!{\mathbf P}^{{\rm GW+DMFT}} ({\mathbf k}, \omega)\! =\!{\mathbf P}^{\rm GW} ({\mathbf k}, \omega)
-\frac{1}{V_{\rm BZ}}\!{\int d^3 k\, {\mathbf P}^{\rm GW}({\mathbf k}, \omega)}+
{\mathbf P}^{\rm DMFT} (\omega) \; \mbox{(Eq. [\ref{GWDMFTP}])}.
\]
}}}

\put(10.00, 139.00){\framebox(120.00,12.00)[cc]
{\parbox{117mm}{\vspace{.15cm}

With this polarisation, calculate the screened interaction (Eqs.\ [\ref{WGW}], [\ref{epsGW}]):
\vspace{-.2cm}

\[
 {\mathbf W}({\mathbf k}; \omega) =
{\mathbf V}_{\rm ee}({\mathbf k})
 [{\mathbf 1}
 - {\mathbf V}_{\rm ee}({\mathbf k}) {\mathbf P}({\mathbf k};\omega)]^{-1}.
\]}}}

\put(10.00, 132.00){\framebox(120.00,7.00)[cc]
{\parbox{117mm}{ Calculate  ${\mathbf \Sigma}^{\rm \scriptscriptstyle  Hartree}_{\mathbf k}\!$ (Eq. [\ref{SigmaHartree}]) and
 double counting correction $\Sigma^{\rm\scriptscriptstyle Hartree}_{\rm\scriptscriptstyle dc}$ (Eq.\ [\ref{HartreeDC}]).
               }}}

\put(10.00,125.00){\framebox(120.00,7.00)[cc]
{\parbox{117mm}{
 Calculate
$
 \Sigma^{\rm\scriptscriptstyle GW}({\mathbf r}, {\mathbf r^{\prime}};\omega) = {i} \int \frac{d \omega^{\prime}}{{2 \pi}}
 G({\mathbf r}, {\mathbf r^{\prime}};\omega+\omega^{\prime}) W({\mathbf r}, {\mathbf r^{\prime}};\omega^{\prime})
\;  \mbox{(Eq.\ [\ref{SigmaGW})]}.$
             }}}

\put(10.00,118.00){\framebox(120.00,7.00)[cc]
{\parbox{117mm}{ \vspace{.1cm}

Calculate the DMFT self energy ${\mathbf \Sigma}^{\rm DMFT}$ and polarisation ${\mathbf P}^{\rm DMFT}$ as follows:}}}

\put(15.00,106.00){\framebox(115.00,12.00)[cc]
{\parbox{110mm}{ \vspace{.3cm}

From the local  Green function
 ${\mathbf G}$ and old self energy ${\mathbf \Sigma}^{\rm DMFT}$ 
calculate
\vspace{-.6cm}

   \[ 
(\mbox{\boldmath ${\cal G}$}^0)^{-1}(\omega)\!=\! {\mathbf G}^{-1}(\omega)\! + \! {\mathbf  \Sigma}^{\rm DMFT}(\omega)\;\;\mbox{(Eq.\ [\ref{AIMDyson}])}; {\mathbf \Sigma}^{\rm DMFT}\!\!=\!0 \,\, \mbox{in 1. iteration}.
\]   
}}}

\put(15.00, 94.00){\framebox(115.00,12.00)[cc]
{\parbox{110mm}{ \vspace{.2cm}

Extract the local screening contributions from ${\mathbf W}$  (Eq.\ [\ref{UDMFT}]):
\vspace{-.2cm}

   \[ {\mathbf U}(\omega)= [{\mathbf W}^{-1}(\omega) -  {\mathbf  P}^{\rm DMFT}(\omega)]^{-1}.
\]  

 }}}

\put(15.00, 71.00){\framebox(115.00,23.00)[cc]
{\parbox{112mm}{ \vspace{-.02cm}

With  ${\mathbf U}$  and $\mbox{\boldmath ${\cal G}$}^0$,
solve  impurity problem with effective action  (Eq.\ [\ref{effActionGWDMFT}])
\vspace{-.72cm}

\[
\!\! {\cal A}\!=\!\!\!\sum_{\nu
\sigma \,l m}\!\psi _{\nu
m}^{\sigma \ast }({\cal G}^{\sigma 0}_{\nu m n})^{-1}\psi _{\nu n}^{\sigma
{\phantom\ast }}
\!+\!\!\!\sum_{l m\sigma\sigma^{\prime}}
\!
    \int\limits_{0}^{\beta }\!d\tau\psi _{l}^{\sigma \ast }(\tau )
\psi_{l}^{\sigma}(\tau)
\,U_{l m}(\tau\!-\!\tau^{\prime})
\psi _{m}^{\sigma^{\prime} \ast }(\tau^{\prime} ) \psi_{m}^{\sigma^{\prime}}(\tau^{\prime} ),
\]\vspace{-.42cm}

resulting in  ${\mathbf G}$  and susceptibility $\chi$  (Eq.\ [\ref{DMFTsusz}]).
              }}}

\put(15.00,59.00){\framebox(115.00,12.00)[cc]
{\parbox{110mm}{ \vspace{-.05cm}

From  ${\mathbf G}$ and  $\chi$,
calculate 
   $  {\mathbf  \Sigma}^{\rm DMFT}(\omega)=  (\mbox{\boldmath ${\cal G}$}^0)^{-1}(\omega)-{\mathbf G}^{-1}(\omega)
\mbox{ (Eq.\ [\ref{AIMDyson}])}$,
 \vspace{.05cm}

$ {\mathbf  P}^{\rm DMFT}(\omega) = {\mathbf U}^{-1}(\omega)-
[{\mathbf U}- {\mathbf U} {\mathbf \chi}{\mathbf U}]^{-1}(\omega)
\mbox{  (Eqs.\ [\ref{localW}], [\ref{UDMFT}])}.
$
 }}}
\put(10.00,45.00){\framebox(120.00,80.00)[cc]{}}

\put(10.00, 45.00){\framebox(120.00,14.00)[cc]
{\parbox{115mm}{\vspace{.25cm}

Combine this to the total GW self energy (Eq.\ [\ref{totalS}]):
\vspace{-.7cm}

\[
{\mathbf \Sigma}^{{\rm\scriptscriptstyle GW \!+\!DMFT}} \! ({\mathbf k}, \! \omega)\! =\!{\mathbf \Sigma}^{\rm\scriptscriptstyle GW}\! ({\mathbf k}, \omega)
\!-  \! \! \!\int \! \!  \!d^3 k\, {\mathbf \Sigma}^{\rm\scriptscriptstyle GW} ({\mathbf k}, \! \omega) 
+ {\mathbf \Sigma}^{\rm \scriptscriptstyle Hartree}  \! ({\mathbf k}) \! \!
- {\mathbf \Sigma}^{\rm\scriptscriptstyle \small Hartree\, dc}\!+{\mathbf \Sigma}^{{\rm\scriptscriptstyle DMFT}} \! (\omega).
\]
}}}

\put(10.00, 39.00){\framebox(120.00,6.00)[cc]
{\parbox{115mm}{\vspace{.2cm}
From  this and   ${\mathbf G}^0$, calculate $ 
 {\mathbf G}^{\rm new}_{\bf k}(\omega)^{-1} = 
{\mathbf G}^0_{\bf k}(\omega)^{-1} - {\mathbf \Sigma}_{\bf k} (\omega) .
$
\vspace{.15cm}

}}}

\put(5.00,34.00){
\parbox{110mm}{ Iterate with  $ {\mathbf G}_{\bf k}={\mathbf G}^{\rm new}_{\bf k}$ until
convergence, i.e. $|\!|{\mathbf G}_{\bf k}-{\mathbf G}^{\rm new}_{\bf k}|\!|\! <\! \epsilon$.
}}
\end{picture}
\end{center}
\vspace{-3.4cm}

\caption{\void Flow diagram of the GW+DMFT algorithm.} 
\label{flowGWDMFT} 
\end{figure}
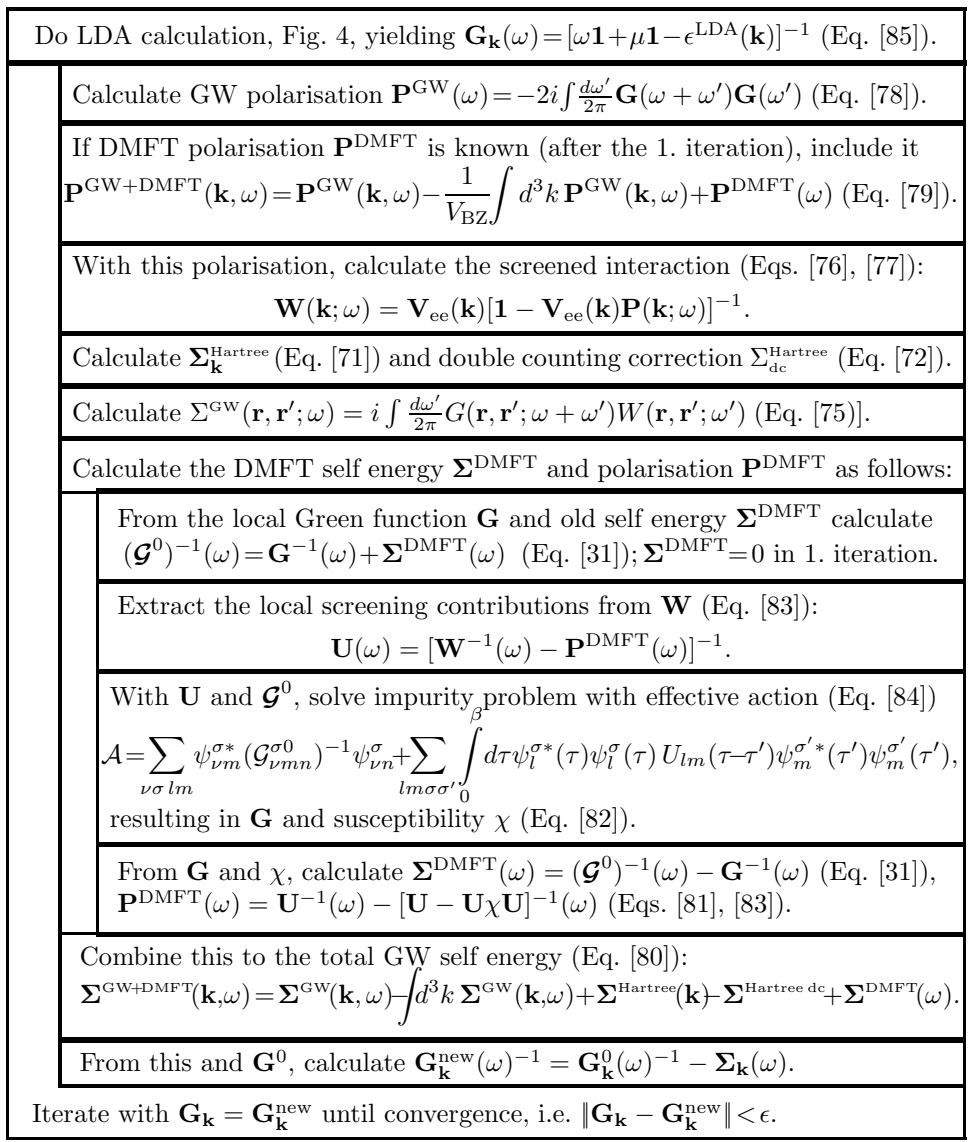

Altogether, we are now in the position of formulating the
 GW+DMFT  
scheme, introduced by Biermann {\em et al.} \cite{Biermann03,BiermannCP04, AryasetiawanCP04}, 
see flow diagram Fig.\ \ref{flowGWDMFT}:
Starting point is a conventional LDA calculation, yielding an electron density $\rho({\mathbf r})$
and also a LDA Green function
\begin{equation}
{\mathbf G}^{\rm LDA}_{{\mathbf k}}(\omega)= [\omega{\mathbf 1}+\mu{\mathbf 1}- \mbox{\boldmath $\epsilon$}^{\rm LDA}{({\bf k})}]^{-1}.
\label{GLDA}
\end{equation}
This Green function is inserted into  Eq.\ [\ref{GWP}] to calculate  the GW  polarisation
${\mathbf P}^{\rm GW}({\mathbf k},\omega)$ and,  via Eq.\ [{\ref{epsGW}}], the screened interaction 
${\mathbf W}({\mathbf k}, \omega)$, Eq.\ [\ref{WGW}].

Now three self energies have to be calculated:
(i) the GW  self energy ${\mathbf \Sigma}^{\mathrm GW}({\mathbf k}, \omega)$, i.e. the
Fourier transform of Eq.\ [\ref{SigmaGW}];
(ii) the  Hartree self energy 
and double counting correction, Eqs.\   [\ref{SigmaHartree}] and [\ref{HartreeDC}], respectively;
and (iii) the DMFT self energy. For the latter, we first have to
define  the non-interacting Green function of
the auxiliary DMFT impurity problem, i.e. ${\mathbf {\cal G}}^0$.
Moreover, we have to unscreen the local Coulomb interaction
since the DMFT diagrams include the local screening bubble, the contribution of Fig. \ref{figRPA}
with $i^{\prime}=j^{\prime}$.
This contribution would be doubly counted if we
started DMFT with the fully screened ${\mathbf W}$.

The result of the calculation of the impurity problem  will be the impurity  Green function,
which is identical to the  local GW+DMFT Green function, and the local
susceptibility.
From these two quantities we can determine the local DMFT self energy and polarisation.
Altogether, this then allows us to calculate the total GW+DMFT self energy
and GW+DMFT polarisation. The former in turn allows us to calculate the GW+DMFT
Green function by the Dyson equation [\ref{Dyson2}] in which the non-interacting
Green function is simply the solution of the {\em ab initio} Hamiltonian
(Eq.\ [\ref{abinitioham}]) without electron-electron interaction.
With this new Green function ${\mathbf G}$, the DMFT self energy ${\mathbf \Sigma}^{\rm DMFT}$,
and the GW+DMFT  polarisation ${\mathbf P}^{\rm GW+DMFT}$, we can restart
our iteration loop until convergence.

Certainly, such a fully self-consistent  GW+DMFT scheme is a formidable task, and
Biermann {\em et al.}  \cite{Biermann03} employed a simplified implementation for their calculation of Ni:
For the DMFT impurity problem, only the local  Coulomb interaction between $d$ orbitals was included
and its frequency dependence was neglected ${\mathbf W}(\omega)={\mathbf W}(0)$.
Moreover, only one iteration step has been done, calculating the
inter-site part of the self energy by GW with the LDA Green function as an input
and the intra-site part of the self energy by DMFT (with the usual DMFT
self-consistency loop, see flow diagram Fig.\ \ref{flowDMFT}).
The GW polarisation ${\mathbf P}^{\rm GW}$ was calculated from the LDA instead of the GW Green function.
This is, actually,  common practice even for 
 conventional GW calculations.

Notwithstanding, Aryasetiawan {\em et al.} \cite{Aryasetiawan04} emphasised the importance of the frequency dependence
of the Coulomb interaction, even for the effective low-energy physics.
For a model Hamiltonian, the Hubbard model with nearest-neighbour interaction,
  Sun and Kotliar \cite{Sun02,Sun04}  already performed such self-consistent 
 GW+DMFT calculation and  included further local correlations
through the extended (E-)DMFT \cite{Si96,Smith96,Kajueter96,Chitra01}.
Also note, that one can embed GW+DMFT (similar to LDA+DMFT) as a 
Luttinger-Ward
\cite{Luttinger60a} 
free energy functional  which is, here, a functional of the Green function ${\mathbf G}$ and the 
screened Coulomb interaction ${\mathbf W}$; for more details see \textcite{BiermannCP04}.


\section{DMFT solvers suitable for material calculations}
\label{DMFTsolvers}

The equivalence of the DMFT  single-site problem
and the Anderson impurity problem, noted by Georges and Kotliar \cite{Georges92a}, also see Jarrell \cite{Jarrell92a},
allows us to employ a variety of well established techniques
to solve the DMFT equations. Since the DMFT results depend
 on this solver, 
we should be careful to employ the respective solvers only 
in the parameter regime where they are
applicable, and e.g.\  not a perturbative
approximation in the non-perturbative regime.
Otherwise, we run the risk of reporting a good agreement
with experiment which stems from a cancellation of errors.
LDA+DMFT is an approximation and misses non-local
correlations which are certainly important
for some materials.
If we introduce a second error due to the DMFT impurity 
solver, good agreement with experiment might 
be by chance. Possibly, we then
overlook important non-local correlations.

For these reasons,
Nekrasov {\em et al.} 
\cite{Nekrasov00} coined the notation DMFT(X) and LDA+DMFT(X) where
X denotes the solver for the  auxiliary DMFT  impurity problem.

In the following we will review the most common
techniques X which are promising for material calculations with DMFT.
At the beginning, we will  discuss approximate techniques,
starting with the Hartree approximation in Section \ref{Hartree}
which makes the LDA+DMFT approach equivalent to  LDA+U.
This allows us to describe an insulating phase with
Hubbard bands, albeit only if there is spin or orbital 
order. In the case of alternating antiferromagnetic or orbital order, however, LDA+U fails to yield the 
correct shape of the Hubbard bands
since it misses the formation of polaron side bands
\cite{Sangiovanni05}.
Alternatives, in particular for the
 paramagnetic phase without symmetry breaking, are
the approximations by Hubbard \cite{Hubbard63a,Hubbard64b} and extensions which are
discussed in Section \ref{HubbardI}.
In the opposite limit of weak coupling (weak Coulomb interaction),
one can do iterated perturbation theory (IPT) which is second
order perturbation theory in the local Coulomb interaction of
the Anderson impurity model. IPT will be discussed in Section \ref{IPT}.

A good description
of the behaviour at strong coupling is possible by
$1/N$ type of approaches  such as
the non-crossing approximation (NCA), which is a
resolvent perturbation in the hybridisation of the Anderson impurity model.
These approaches become exact for a large number of degenerate
orbitals $N$ and are good for the insulator
but also for the metal if the temperature
is higher than the Kondo temperature of the impurity model.
In Section \ref{NCA}, we will introduce the NCA.

In Section \ref{QMC},  the concept of numerically exact
quantum Monte Carlo (QMC) simulations is introduced,
which directly solve the Anderson impurity model,
albeit on the imaginary axis. 
Therefore, the maximum entropy method \cite{MEM} is needed
for an analytical continuation to real frequencies
if, for example,  the spectral function is calculated.
The numerical effort
of QMC grows cubically with decreasing temperature
so that it is possible to do room temperature, or say
100$\,$K, calculations nowadays. Since the effort only
grows cubically (not exponentially) with temperature
calculations at lower temperatures will be possible
in the future, although not at really low temperatures.
To this end, Feldbacher {\em et al.} \cite{Feldbacher04} recently developed
a projective QMC algorithm for `T=0' calculations which
we will also discuss.

We will not present details of
the exact diagonalisation (ED) 
and the numerical renormalisation group (NRG) method of Wilson \cite{Wilson75},
which  have been very successfully employed for the
one-band Hubbard model: ED  by Caffarel and Krauth \cite{Caffarel94a}, also see \textcite{Georges96a,Capone05a},
 and NRG  by Bulla {\em et al.}  \cite{Bulla99,Bulla01,Bulla07}.
Hence, let us briefly state the idea and the limitations
for multi-orbital calculations here.
Both of these methods diagonalise the Anderson impurity model.
In the case of ED 
the Anderson impurity model is diagonaliced directly
for a limited number of non-interaction `bath' sites.
Astonishingly, Potthoff \cite{Potthoff01a} demonstrated
that even a single `bath' site gives a very good estimate
of the critical $U$ value of the Mott transition.
In the case of NRG, one diagonalises within a restricted energy window
which is renormalised to lower and lower energies. Since the 
number of quantum mechanical states grows exponentially with
the number of orbitals in the Anderson impurity model,
these methods are severely restricted concerning the number of  orbitals:
one orbital is manageable; a sound calculation with two
orbitals is already almost exceeding the computational limits, but
can with great numerical effort still be done; a 
reliable treatment of three orbitals will be impossible even if 
computer resources grow considerably. Hence the usefulness of ED and NRG
for realistic material calculations is very limited and
we will not discuss these methods here, referring the reader  to
the references above for more information.

Most recently, also the related (dynamical) dynamical matrix renormalisation group (DMRG) approach  \cite{White92a,Jeckelmann02a,Schollwoeck05a}
has been employed by different groups \cite{Gebhard03a,Raas04a,Nishimoto04a,Garcia04a,Raas05a,Karski05a}.
DMRG is a powerful alternative to ED and 
NRG and might become a standard impurity solver for DMFT in the future. For realistic multi-orbital calculations, it however
also scales exponentially with the number of orbitals involved.

Another method which one might subsume under the $d\rightarrow \infty$ limit
is the Gutzwiller approximation.
As was shown by Metzner and Vollhardt
\cite{Metzner89a,Metzner89b,Metzner91b} and Gebhard \cite{Gebhard90a},
this approximate treatment of the Gutzwiller
\cite{Gutzwiller63a,Gutzwiller64a,Gutzwiller65a} wave function 
becomes exact for $d\rightarrow \infty$.
Gebhard
\cite{Gebhard90a,Gebhard91c} also showed that the $d\rightarrow\infty$
Gutzwiller wave function is equivalent to the Kotliar-Ruckenstein \cite{Kotliar86a} slave Boson approximation
at zero temperature (also note recent slave-rotor \cite{Florens02a} and slave-spin \cite{Medici05a} variants).
The Gutzwiller wave function starts with the Fermi sea onto which
the so-called Gutzwiller correlator $g^{\hat{d}}$ is applied, where
${\hat{d}}$ is the operator of local double occupations, i.e.
${\hat{d}}=\sum_i\hat{n}_{i\uparrow}\hat{n}_{i\downarrow}$
for the one-band Hubbard model. With $g$ as a variational parameter,
this wave function emulates a central effect of the local Coulomb interaction,
the reduction of the number of doubly occupied sites.
An important step for our understanding of the Mott-Hubbard transition
was the work by Brinkman and Rice \cite{Brinkman70a} who recognised that the Gutzwiller approximation
describes a Mott-Hubbard metal-insulator transition, indicated
by the disappearance of the quasiparticle peak. This approximation,
stemming from the metallic side, however misses a correct description of the Hubbard bands.
Gebhard \cite{Gebhard91c} generalised the   Gutzwiller approximation to the multi-orbital situation,
allowing subsequently  Strack and Vollhardt \cite{Strack91a,Strack91b} and 
Gul\'asci {\em et al.}
\cite{Gulacsi93a}
to apply it to model Hamiltonians. More
recently,
it was employed for realistic calculations of ferromagnetic transition metals
by B\"unemann {\em et al.} \cite{Buenemann97a,Buenemann97b,Buenemann98a}, for a summary of
these results see \textcite{Weber01}. Closely related to this
are also material calculations with the local ansatz,
which has been employed for material calculations
long before DMFT, see, e.g. Refs.\ \cite{Stollhoff81a,Oles81a,Oles84a}.

\subsection{Polarised Hartree-Fock (HF) approximation (LDA+U)}
\label{HartreeFock}
\label{Hartree} \label{LDAU}

The simplest way to deal with the auxiliary Anderson impurity problem
of DMFT, Eq.\ [\ref{AIMH}], is to treat the interaction in the Hartree-Fock (HF)
approximation. 
The Hartree diagram has already been shown in  Fig.\ \ref{FigHF}
(left) and corresponds to the decoupling
\begin{equation}
\langle
{\hat{c}}_{l}^{\sigma \, \dagger }
{\hat{c}}_{l}^{\sigma}  
{\hat{c}}_{m}^{\sigma^{\prime}  \, \dagger}
{\hat{c}}_{m}^{\sigma^{\prime}} \rangle
\,\stackrel{\rm Hartree}{\longrightarrow}\,
 {\hat{c}}_{l}^{\sigma \, \dagger }
{\hat{c}}_{l}^{\sigma}  
\langle{\hat{c}}_{m}^{\sigma^{\prime}  \, \dagger}
{\hat{c}}_{m}^{\sigma^{\prime}} \rangle
+\langle{\hat{c}}_{l}^{\sigma \, \dagger }
{\hat{c}}_{l}^{\sigma}   \rangle
{\hat{c}}_{m}^{\sigma^{\prime}  \, \dagger}
{\hat{c}}_{m}^{\sigma^{\prime}} 
-\langle{\hat{c}}_{l}^{\sigma \, \dagger }
{\hat{c}}_{l}^{\sigma}   \rangle
\langle{\hat{c}}_{m}^{\sigma^{\prime}  \, \dagger}
{\hat{c}}_{m}^{\sigma^{\prime}}  \rangle.
\label{HFeq}
 \end{equation}
Hence,  correlations given by
\begin{equation}
\langle {\hat{c}}_{l}^{\sigma \, \dagger }
{\hat{c}}_{l}^{\sigma}  
{\hat{c}}_{m}^{\sigma^{\prime}  \, \dagger}
{\hat{c}}_{m}^{\sigma^{\prime}} \rangle-\langle{\hat{c}}_{l}^{\sigma \, \dagger }
{\hat{c}}_{l}^{\sigma}   \rangle
\langle{\hat{c}}_{m}^{\sigma^{\prime}  \, \dagger}
{\hat{c}}_{m}^{\sigma^{\prime}}  \rangle
 \end{equation}
are neglected within the Hartree approximation, i.e. within the approximate 
expectation value of the right hand side of Eq.\ [\ref{HFeq}]. 
In principle, we can treat
the Fock term on a similar footing with a decoupling into non-diagonal expectation
values of the form
$\langle{\hat{c}}_{l}^{\sigma  \, \dagger} {\hat{c}}_{m}^{\sigma^{\prime}}  \rangle$.
However, for the local Fock term to contribute we need
either a (non-diagonal) magnetisation in the $xy$ plane or a corresponding kind of orbital ordering.
At least for degenerate orbitals,
these Fock expectation values, or the off-diagonal Green function required
in the diagram Fig.\ \ref{FigHF} (right), are zero.
Therefore, the local Fock term is usually not considered, only
the Hartree term.

If the Coulomb interaction has the form of Eq.\ [\ref{LDADMFTH}]
with an intra-orbital Coulomb interaction $U$, an inter-orbital interaction
$V$ and Hund's exchange coupling $J$,
Eq.\ [\ref{HFeq}] or equivalently the Green function of Fig.\ \ref{FigHF} (left)
result in the Hartree self energy
\begin{equation}
\Sigma^{\rm Hartree}_{l} =  
 U  \langle \hat{n}_{l}^{\bar{\sigma}}\rangle
+ \sum_{m\in L_U|_{\neq l}} V \langle \hat{n}_{m}^{\bar{\sigma}}\rangle
+ (V-J) \langle \hat{n}_{m}^{{\sigma}}\rangle.
\label{HFS}
\end{equation}
This is (up to a constant) the same as  Eq.\ [\ref{HartreeDC}] if  orbitally averaged and expressed in terms
of Green functions instead of expectation values (Eq. [\ref{Gtocc}]).

Doing such a LDA+DMFT(HF) calculation is completely equivalent to 
the LDA+U approach \cite{Anisimov91}, also see \textcite{Anisimov97b}.
For a simple Hartree calculation, the DMFT formulation is not necessary and instead of decoupling the
interaction of the Anderson impurity model one can directly apply the Hartree
decoupling to the LDA-constructed many-body multi-orbital problem
Eq.\ [\ref{LDADMFTH}]. This is how  Anisimov {\em et al.} \cite{Anisimov91} arrived at
the LDA+U scheme.

Because of the purely local interaction, the Hartree self energy, Eq.\ [\ref{HFS}], is
${\mathbf k}$-independent. Moreover,  it is static, i.e. $\omega$-independent. 
What kind of physics can we then expect to be described?
For the paramagnetic phase with an equal orbital occupation,
$\Sigma^{\rm Hartree}_{l}$ is independent of $l$. Hence, the self energy
is  reduced to a constant shift like a chemical potential,
which moreover should be  canceled by the orbitally-averaged double-counting
correction given by Eq.\ [\ref{HartreeDC}]. There is no effect at all.

This changes if the spin or orbital degrees of freedom are ordered ({\em polarised}),
either in a homogenous way like the Stoner \cite{Stoner51a} ferromagnet
or in a more complicated pattern such as the checkerboard alternating 
N\'eel \cite{Neel32a} state, which gives rise to the  Slater \cite{Slater51a}
bands. In these polarised phases, the bands of different spin or orbital species split 
into subbands since
the self energy Eq.\ [\ref{HFS}] differs. 
For a large average Coulomb interaction $\bar U = [U+(M-1)V+(M-1)(V-J)]/(2M-1)$ 
($M$: number of orbitals)
this splitting results in two completely separated sets of bands
with a typical distance  $\bar U$ at large  $\bar U$. For an integer filling, the lower set of bands will be completely 
filled and the upper set completely empty,  resembling the Hubbard bands of a Mott insulator,
shown schematically in Fig.\ \ref{figLDADMFT} of Section \ref{intro}. 

Since the number of double occupations is minimal for this
large $\bar U$ phase, the symmetry-broken Hartree-Fock  solution also
correctly produces the total energy in the $\bar U\rightarrow\infty$ limit.
The corrections to this limit are of order $W^2/{\bar U}^2$  and
depend on the kind of ordering, leading to the LDA+U prediction for the  
symmetry breaking in a specific material.
Together with the insulating spectrum, this explains why
the LDA+U method has been successfully applied for the {\em ab initio}
calculation of insulators, including their low temperature ordering.

This good description of the insulator might be astonishing if one considers that
the starting point, the
Hartree self energy Eq.\ [\ref{HFS}], is first-order perturbation theory 
in the Coulomb interaction, which suggests that Hartree is good
at weak coupling (Coulomb interaction) only.
However, except for almost uncorrelated systems with $\bar U/W\ll 1$,
the Hartree approximation is rather bad for the metallic phase.
The reason is that it strongly overestimates the energy of the paramagnetic phase.
This paramagnetic energy grows $\propto \bar U$ 
since double occupations cannot be avoided without symmetry breaking
in the Hartree approximation.
Therefore, the tendency towards orbital or magnetic ordering is heavily
overestimated, and LDA+U almost automatically yields split-bands even
if this is not correct at small and intermediate strengths of the Coulomb interaction
where the system is paramagnetic or has a much smaller magnetisation than the Hartree approximation
predicts.

Nonetheless, a paramagnetic insulator 
with Hubbard bands with separation $\bar U$  can be
described by LDA+U if one considers the symmetry breaking
 as an artificial means to produce the correct spectrum
and energy. Of course, then the magnetisation is incorrect,
but also the free energy  since the entropy of paramagnetic
uncoupled spins is much higher than that of ordered spins.
These obstacles can be overcome  by the Hubbard-I approximation
which also has the advantage that the weights of the Hubbard bands
are better reproduced for non-integer fillings, see the following Section.

A cautious remark is also in place if one applies LDA+U
for studying antiferromagnetic or orbitally ordered phases:
While LDA+U then yields the correct {\em static}
properties, {\rm dynamic} properties  are 
completely wrong.
This was shown by Sangiovanni {\em et al.} 
\cite{Sangiovanni05}, who particularly
pointed out that the  DMFT Hubbard bands
are very different from the 
coherent, narrowed Hubbard bands of LDA+U:
The DMFT Hubbard bands are incoherent,
have spin-polaron side peaks and their 
widths correctly converges to
that of the non-interacting bands
for $\bar U\rightarrow \infty$ \cite{Sangiovanni05}.

\subsection{Hubbard-I, Hubbard-III and alloy-analogy approximation}
\label{HubbardI}

Starting point of the Hubbard \cite{Hubbard63a,Hubbard64b} approximations 
is the atomic limit ($W=0$), in which we can calculate the (purely local) Green function exactly,
including the full multiplet structure with the 
spin-flip and pair-hopping term of the LDA+DMFT Hamiltonian (Eq.\ [\ref{LDADMFTH}]).
Due to the itinerancy of the system these atomic levels are, however, broadened.
This broadening is included in different ways in the
Hubbard-I \cite{Hubbard63a} and  Hubbard-III approximation \cite{Hubbard64b},
as well as in the alloy-analogy approximation.

The Hubbard approximations can be derived by decoupling the higher order Green functions
in the equations of motion. 
This procedure is not controlled and allows for  many different
decoupling schemes, i.e. approximations. 
For infinite dimensions,
the simplified momentum summations allow for the
decoupling of higher order Green functions \cite{Gros94a}.

For realistic material calculations, 
Lichtenstein and Katsnelson
\cite{Lichtenstein98a} formulated
the  LDA+Hubbard-I, or in our notation LDA+DMFT(Hubbard-I), approach as one of their
LDA++ approaches. For the $\gamma$-phase of cerium such
 LDA+DMFT(Hubbard-I) calculations  by McMahan {\em et al.} \cite{McMahan03} 
were quite successful.  These cerium results will be presented in Section \ref{Ce}.

Here we will motivate the Hubbard-I approximation physically without
attentiveness to the original equation of motion derivation by Hubbard \cite{Hubbard63a,Hubbard64b,Gebhard97a}
In the atomic limit, the hybridisation of our DMFT
Anderson impurity model (Eq.\ [\ref{AIMH}]) vanishes and the
impurity problem is simply the isolated atom for which
the Green function is known exactly. 
Expressed via the spectral representation, see \textcite{Abrikosov63} p.\ 163 ff., it reads
\begin{equation}
G_{lm}(\omega) = \sum_{\mu \nu} \frac{\langle\mu| c^{\phantom{\dagger}}_{l} |\nu\rangle \langle\nu | c^{\dagger}_{m} |\mu\rangle}{\omega + \mu -E_{\nu}+ E_{\mu}} e^{-\beta [E_{\nu}-\mu (N+1)]}
(1-e^{-\beta(E_{\mu}-E_{\nu}+\mu)}).
\end{equation}
Here, $|\mu\rangle$ ($|\nu\rangle$) are the exact many-body eigenvectors
 with eigenenergies $E_{\mu}$ ($E_{\nu}$)
for $N$ ($N+1$) electrons.
For the atomic limit we know these eigenenergies and vectors so that, 
without Hund's exchange ($J=0$, ${\bar U}=U=V$) and for degenerate levels with site energy $\varepsilon$,
the atomic Green function reads
\begin{equation}
G^{\rm at}(\omega) = \sum_{{ N}\!=\!0}^{2M-1}
\frac{w_{N}(\mu_{\rm at},T)}{\omega + \mu_{\rm at} - NU-\varepsilon}.
\label{HI-2}
\end{equation}
Thereby, the weights of the poles are given by
\begin{eqnarray}
w_{N} &\!=\!& \frac{(N+1) v_{N+1} + (2M-N)v_{N}}
{2M\sum_{N^{\prime}=0}^{2M} v_{N^{\prime}}}.
\label{HI-4}
\end{eqnarray}
These are determined in turn by the
 weight for having $N$ electrons on the atom with altogether $2M$ states
\begin{eqnarray}
v_N &\!=\!& \frac{2M!}{N!(2M-N)!}e^{-\beta[\frac12
N(N-1)U+N\varepsilon- N\mu_{\rm at}]},
\label{HI-5} 
\end{eqnarray}
consisting of the  Boltzmann weight and a combinatorial factor for the number
of orbital configurations with $N$ electrons in $M$ degenerate
orbitals. From  $G^{\rm at}$, we can
also calculate the average number of local electrons $n^{\rm at}$,
for example, via the following sum over Matsubara frequencies $\omega_{\nu}$:
\begin{equation}
n^{\rm at} = 2M\, T \sum_{\nu} G^{\rm at}(i\omega_{\nu})
e^{i\omega_{\nu} 0^+} +2M.
\label{HI-3}
\end{equation}
Since the non-interacting Green function of
the Anderson impurity model is known to be
${\cal G}^0(\omega)= 1/(\omega+\mu-\varepsilon)$ 
 the exact self energy in the atomic limit reads:
\begin{equation}
\Sigma^{\rm at}(\omega) = \omega + \mu-\varepsilon - 
[G^{\rm at}(\omega)]^{-1}.
\label{HI-1}
\end{equation}

Hubbard
\cite{Hubbard63a} now approximated the self energy
of the itinerant problem by this atomic self energy:
\begin{equation}
\Sigma(\omega)  = \Sigma^{\rm at}(\omega).
\end{equation}
This self energy allows us  to calculate the  ${\mathbf k}$-dependent
Green function via the Dyson equation [\ref{Dyson2}] or
the local Green function via Eq.\ [\ref{Dyson}]:
\begin{equation}
  {\mathbf G}(\omega) = \frac{1}{V_{\rm BZ}}\int_{\rm BZ} d^3k \; [\omega {\mathbf 1}+\mu {\mathbf 1}-
\mbox{\boldmath $\tilde{\epsilon}$}^{\rm LDA}({\bf k}) - {\Sigma} (\omega){\mathbf 1}]^{-1}.
\label{GHI}
\end{equation}
Here, $\mbox{\boldmath $\tilde{\epsilon}$}^{\rm LDA}$ denotes the LDA bandstructure minus double counting correction as
defined in Eq.\ [\ref{HLDAdc}].
For consistency, we require that the number of electrons in the auxiliary atomic problem
$n^{\rm at}$ equals the average number $n_d$ of electrons in the interacting $d$ or $f$ orbitals, which is
calculated from the corresponding orbitals of the Green function ${\mathbf G}(\omega)$ of Eq.\ [\ref{GHI}]. For this requirement to hold,
we have to adjust $\mu_{\rm at}$ correspondingly
\cite{Cyrot70a,Cyrot72a,Cyrot72b,Kawabata72a}.
Noting that the $w_j$'s
sum to one, we  also see that
$\Sigma^{\rm at}(\omega) \stackrel{\omega \rightarrow \infty}{\longrightarrow} ({2M-1})/({2M})\;  U n_d$
which is  the paramagnetic Hartree-Fock value and
correct in the high-temperature limit.

Let us now elucidate the features of the Hubbard-I approximation.
In contrast to the atomic limit, the Hubbard-I  Green function has `dynamic' Hubbard bands with a 
finite width. Usually  $T\ll U$, so that only two addends contribute to the atomic
Green function of Eq.\ [\ref{HI-2}]. Then, there will be two Hubbard
bands which are  centred around $-\mu$ and $U-\mu$ at large $U$.
If the average number of electrons in the interacting orbitals $n_d$
is integer, the lower Hubbard band  at $-\mu$ will be completely filled and
the upper Hubbard band at  $U-\mu$ completely empty.
Similar  as in the Hartree-Fock 
approximation of Section \ref{Hartree}, but now in the paramagnetic phase.
In contrast to the rigid bands of the Hartree approximation, these spectral weights shift, however,
when   $n_d$ is non-integer. This is very physical, since for non-integer $n_d$ one can add
an extra electron without paying more Coulomb interaction energy
than for the last electron added.

The major deficit of the Hubbard approximations is that the metal which is generally described 
at non-integer   $n_d$ is not a Fermi liquid. This is a consequence of the  construction with the
{\em atomic} self energy.
Hence, 
the Hubbard-I approximation is not an adequate method for  non-integer $n_d$,
except for such high temperatures at which the Fermi liquid behaviour has been lost.
Or if, as in the case of manganites, strong scattering destroys the Fermi liquid behavior, justifying the
application of the  Hubbard-I approximation for this system \cite{Edwards99a,Edwards02a,Meyer01a,Green02a} to some extend.

There have been several attempts to improve the Hubbard-I approximation,
starting with Hubbard's  own work.
The alloy-analogy approximation and the Hubbard-III approximation
 \cite{Hubbard64b} introduce an additional energy-dependent scattering correction $\Delta(\omega)$
to the self energy which then becomes
\begin{equation}
\Sigma(\omega)  = \Sigma^{\rm at}(\omega-\Delta(\omega)).
\end{equation}
In the alloy-analogy approximation, $\Delta(\omega)$ is fixed by 
requiring that the local Green function
calculated from $\Sigma(\omega)$ via Eq.\  [\ref{GHI}]
 still fulfils the atomic-limit Eq.\ [\ref{HI-1}]
\begin{equation}
G^{\rm at}(\omega) = [\omega+\mu_{at}-\varepsilon-\Delta(\omega)-\Sigma(\omega)]^{-1},
\label{Eq:HIII}
\end{equation}
but with the scattering correction $\Delta(\omega)$.
After substituting $\Sigma^{\rm at}$ from Eq.\ [\ref{HI-1}] into
 Eq.\ [\ref{Eq:HIII}],
we obtain 
an implicit equation for $\Delta(\omega)$.

The physical picture behind the alloy-analogy is that the electrons of one
orbital move in a background of static electrons of the other orbitals.
This reduces the Hubbard model to the simpler Falicov-Kimball model.
With this approximation,
the Coulomb interactions are  reduced to a static scattering potential which
is evaluated within the  coherent potential approximation (CPA).
Note that for disordered (non-interacting) systems
the CPA \cite{Soven67a,Taylor67a,Velicky68a,Elliot74a,Gyorffy72a,Kudronovsky87a}
becomes  exact for the scattering problem in $d\rightarrow \infty$ \cite{Schwartz72a}; for a CPA-inspired derivation of DMFT, see \cite{Janis92a,Janis93a,Janis93b}.
Another way of thinking is in terms of moments
of the interacting spectral function.
Then, the Hubbard-I approximation yields the first two moments correctly and further-going approximations
by Nolting and coworkers \cite{Nolting81a,Nolting89a,BeiderKellen90a,Potthoff96a,Beenen95a,Mehlig95a}
the first three moments.

The advantage of this alloy-analogy approximation and also the further going
Hubbard-III approximation are that the splitting of the bands now occurs at a finite $U$.
Moreover, the widths of the Hubbard-III bands at large $U$ equals the width of the non-interacting
system in agreement with general arguments. In contrast, this width is reduced in the Hubbard-I approximation
and even goes to zero for a Slater \cite{Slater51a} type of alternating 
ordering in the Hartree approximation.
In contrast, the DMFT Hubbard bands have the correct width $W$ 
also for staggered ordering \cite{Sangiovanni05}.

 The equation of motion decoupling scheme by Lacroix \cite{Lacroix81a,Costi86a}
for the Anderson impurity model is also
noteworthy here. It results in a three peak structure with a central resonance and two
split-off side bands. Within the DMFT self-consistency scheme, it
was employed by Jeschke and Kotliar \cite{Jeschke05a}.

Let us conclude here that  Mott insulating materials can be
described, at  strong coupling, by the Hubbard-I and -III approaches, the alloy-analogy or CPA,
and the LDA+U method. Important differences concern
the magnetic ordering tendencies and the shape of the Hubbard bands.  The development of an
{\em ab initio} approach
which can be more generally used  than LDA+U or LDA+DMFT(Hubbard-I) for insulating
transition metal oxides and other Mott insulators
seems to be manageable.

\subsection{Iterated perturbation theory (IPT) and extensions}
\label{IPT}

\begin{figure}
\begin{center}
\includegraphics[clip=true,width=11.5cm]{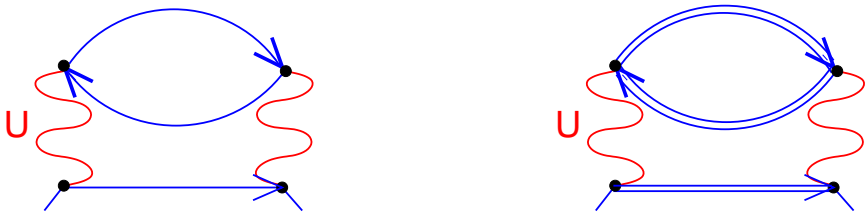}
\end{center}
\caption{\void Second order perturbation theory for the Anderson impurity model
with the bare Green function $\mbox{\boldmath ${\cal G}$}^0$ (left)
and the full Green function ${\mathbf  G}$ (right).\label{FigIPT}}
\end{figure}

The approaches discussed so far (Section \ref{Hartree} and \ref{HubbardI}) were only eligible
deep within the Mott insulating phase.  To describe the metallic phase, a natural
starting point is weak coupling perturbation theory in the Coulomb interaction $U$.
To first order in $U$, this is the Hartree-Fock  approximation which is, however,
better suited for a Mott insulator with magnetic or orbital ordering,
as was shown in  Section \ref{Hartree}.  To second order in $U$, we have to deal with the diagram shown
 in Fig.\ \ref{FigIPT}. As Green function we can either
 inset the bare non-interaction Green function
of the Anderson impurity problem $\mbox{\boldmath ${\cal G}$}^0$  or we can use the interacting Green function ${\mathbf  G}$
of the Anderson impurity model, generating an infinite
series of   $\mbox{\boldmath ${\cal G}$}^0$  diagrams which is  self-consistent 
in the sense of Baym and Kadanoff  \cite{Baym61a,Baym62a}.
 M\"uller-Hartmann \cite{MuellerHartmann89c} and Schweitzer and Czycholl \textcite{Schweitzer91a}  followed the latter path,
 which was extended  by Menge and  M\"uller-Hartmann \cite{Menge91a} to bubble and ladder summations.
However, while this self-consistent second order perturbation theory gives
to order $(U/W)^2$
the same result as the perturbation theory in the bare  $\mbox{\boldmath ${\cal G}$}^0$,
it lacks important physical aspects, in particular the formation of Hubbard bands.

These and even the correct qualitative features of the Mott-Hubbard transition
in the one-band Hubbard model are described by 
the iterated (second order) perturbation theory (IPT) of Georges and Kotliar \cite{Georges92a},
in which the  bare  $\mbox{\boldmath ${\cal G}$}^0$ of the auxiliary DMFT impurity model is used.
 Yosida and Yamada \cite{Yosida70a,Yosida75a} and Zlati\'{c} \cite{Zlatic85a} realised already that
this bare  perturbation theory for the Anderson impurity model is astonishingly good,
showing a three peak structure with  a qualitatively correct central quasiparticle resonance.
This motivated 
Georges and Kotliar \cite{Georges92a} to employ IPT as a DMFT solver. Later,
Zhang {\em et al.} \cite{Zhang93a} recognised that  IPT, albeit being a weak-coupling perturbation
theory, also becomes correct in the strong coupling limit ($U\rightarrow \infty$).
Hence, it might not be so astonishing that IPT describes the qualitative features of the Mott-Hubbard transition
correctly, see \textcite{Georges96a}.

Let us now write down the IPT equation. From the Feynman diagram
Fig.\ \ref{FigIPT}, we obtain in terms of imaginary time $\tau$ and Matsubara frequencies $\omega_n$:
\begin{equation}
{\mathbf \Sigma}^{\rm IPT} (i \omega_n) =  \frac{2M-1}{2M} n_d U \underbrace{-U^2\int_0^{\beta} d \tau e^{i\omega_n\tau}
\mbox{\boldmath ${\cal G}$}^0(\tau)\mbox{\boldmath ${\cal G}$}^0(\tau)\mbox{\boldmath ${\cal G}$}^0(-\tau)}_{\Sigma^{(2)}(i \omega_n)}.
\label{EqIPT}
\end{equation}
Here, the first part is the Hartree term, which yields the high-$\omega$ behaviour.
The second part stems from the second order diagram of Fig.\ \ref{FigIPT}.

Away from the, maybe coincidentally, good description for the
half-filled one-band Hubbard model, IPT is less good at intermediate coupling and fails to reproduce
the atomic limit.
Because of these shortcomings but
also because of the success of IPT for the half-filled one-band Hubbard model, several groups 
developed extrapolation schemes for the self energy which reproduce the correct self energy
to second order in $U/W$ and the atomic limit.
Edwards and Hertz \cite{Edwards90a,Edwards93a} extended the Hubbard approximations from Section \ref{HubbardI}
to yield the correct result to order $(U/W)^2$. This approach gives a Fermi liquid phase
at  small $U$, however, there is an unphysical paramagnetic, non-Fermi-liquid metallic phase at intermediate $U$ values before
the system becomes insulating. 

Kajueter and Kotliar \cite{Kajueter96a,Kajueter97a} made an ansatz for the self
energy of the form
\begin{equation}
\Sigma^{\rm IPT^{\prime}} (\omega) =  \frac{2M-1}{2M} n_d U + \frac{A\Sigma^{(2)}(\omega)}{1-B\Sigma^{(2)}(\omega)}
\end{equation}
where $\Sigma^{(2)}$  is the second order contribution from Eq.\ [\ref{EqIPT}],
analytically continuated to the real axis.
The parameters $B$ and $A$ are determined to give the correct self energy in the  strong coupling limit 
($U\rightarrow \infty$) and
the correct $1/\omega$ behaviour, respectively. The latter 
also guarantees that the first two spectral moments are correct.
However,  the quasiparticle peak comes out too small in the IPT approximation
by  Kajueter and Kotliar. Potthoff {\em et al.}
\cite{Potthoff97a,Potthoff98a} employed the same ansatz but adjusted the parameters 
to reproduce the correct third moment of the spectral function as well.
In essence, this scheme  is an interpolation scheme for the self energy between
IPT and  the strong coupling spectral density approach by Nolting and Borigie{\l} \cite{Nolting89a}.
The  ansatz by  Potthoff {\em et al.} \cite{Potthoff97a,Potthoff98a} also allows for the treatment of magnetic
phases.

Another  self energy interpolation scheme
was recently  proposed by Oudovenko, Savrasov {\em et al.} 
\cite{Oudovenko04,Savrasov05a}, using the Kotliar-Ruckenstein \cite{Kotliar86a} slave Boson approach
at weak and the Hubbard-I approximation at strong coupling.
Also the local moment approach of Logan {\em et al.} \cite{Logan96a,Logan97a,Logan00a}
fulfils this feature of reproducing the
correct  weak and strong coupling limit.

Anisimov {\em et al.}
\cite{Anisimov97a} used IPT with the Kajueter-Kotliar IPT \cite{Kajueter97a} interpolation 
as the impurity solver in the first LDA+DMFT calculations. As was demonstrated by Nekrasov {\em et al.} \cite{Nekrasov00}
in the context of realistic calculations for LaTiO$_3$, IPT
yields too small quasiparticle weights and violates
the Fermi energy pinning, which was proven  by M\"uller-Hartmann \cite{MuellerHartmann89c}
for the one-band Hubbard model repeating the  Luttinger-Ward \cite{Luttinger60a} arguments
and which generally holds for multi-band systems if
the self energy for {\em all} orbitals is degenerate. This is, of course,
generally not the case in realistic material calculations. It can be fulfilled
if the DMFT calculation works with a restricted set of orbitals, e.g. degenerate $t_{2g}$ orbitals for 
transition metal oxides.

An alternative weak-coupling perturbative approach is the
fluctuation exchange approximation (FLEX) of
Bickers and Scalapino  \cite{Bickers89a} which has been employed as a DMFT solver 
in realistic calculations by Lichtenstein and Katsnelson
 \cite{Lichtenstein98a}, Chioncel {\em et al.} \cite{Chioncel03} and 
Drchal {\em et al.} 
\cite{Drchal99,Drchal03}.
While this approach might be suitable for rather weakly correlated metals, i.e.
at weak coupling $U$, 
it does not capture the development of Hubbard bands. Therefore, it is not
appropriate at intermediate and strong coupling.

\subsection{Non-crossing approximation (NCA)}
\label{NCA}

The non-crossing approximation (NCA)
is a conserving approximation
corresponding to the resummation of a particular class of 
diagrams (those that do not cross). 
It can also be formulated as a 
resolvent perturbation theory in the hybridisation 
${\mbox{\boldmath ${\Delta}$}}(\omega )$  of the 
Anderson impurity problem \cite{Keiter70a,Bickers87b}. 
Hence, it is reliable 
at strong coupling where, for all orbital matrix elements, $\Delta_{lm}^{\sigma} (\omega )\ll U$.

Let us start by rewriting the non-interacting Green
function of the Anderson model (Eq.\ [\ref{siam}])
 in a form consisting of a local and a hybridisation part (bold symbols indicate orbital matrices):
\begin{equation}
 [{\mbox{\boldmath ${\cal G}$}}^0(\omega)]^{-1}= 
\omega+\mu-{\mbox{\boldmath ${\epsilon}$}}^{\rm at} -{\mbox{\boldmath ${\Delta}$}}(\omega),
\end{equation}
where we identify the local part by
\begin{equation}\label{h0loc}
{\mbox{\boldmath ${\epsilon}$}}^{\rm at}=\frac{1}{V_{\rm BZ}}\int d^3 k\, \mbox{\boldmath $\tilde{\epsilon}$}^{\rm LDA}({\bf k}).
\end{equation}
Then, the Dyson equation [\ref{AIMDyson}] of the Anderson impurity model
which connects the non-interacting Green function
 ${\mbox{\boldmath ${\cal G}$}}^0(\omega)$ of the Anderson model
and the local interacting Green 
function ${\mathbf G}(\omega)$
reads
\begin{equation}
[{\mathbf G}(\omega)]^{-1}=  \omega + \mu 
 -{\mbox{\boldmath ${\epsilon}$}}^{\rm at} -{\mbox{\boldmath ${\Delta}$}}(\omega) -  {\mathbf \Sigma} (\omega).
\end{equation}
Since NCA is a perturbation theory in terms of
the hybridisation function ${\mbox{\boldmath ${\Delta}$}}(\omega)$,
the first step is to diagonalise the atomic  problem (at)
as in the Hubbard approximations of Section \ref{HubbardI}.
In  Section \ref{HubbardI} we did so for the restriction
to degenerate orbitals 
and a single Coulomb parameter $U$.
Let us here diagonalise the atomic  problem by formally rewriting the local Hamiltonian
in terms of the local eigenvectors $|\alpha \rangle$ and eigenenergies $E_{\alpha}$:
\begin{eqnarray}		
 \hat{H}^{\rm at} &=&
 \sum_{lm\,\sigma}\hat{c}^\dagger_{l\sigma}
 {{\epsilon}}^{\rm at}_{lm}
\hat{c}^{\phantom{^\dagger}}_{m\sigma} 
+ \sum_{l m n o\,\sigma\sigma^{\prime}}
U_{l m n  o}
\,
{\hat{c}}_{l}^{\sigma \, \dagger }
{\hat{c}}_{m}^{\sigma^{\prime}  \, \dagger}
{\hat{c}}_{n}^{\sigma^{\prime}} 
{\hat{c}}_{o}^{\sigma}\\
&=&
 \sum_{\alpha} E_\alpha|\alpha\rangle\langle\alpha|.
 \end{eqnarray}

 In this
eigenbasis, the creation and annihilation operators read
\begin{eqnarray}
\hat{c}^\dagger_{l\sigma}  &=&
\sum\limits_{\alpha,\beta}
\left(D^{l\sigma}_{\beta\alpha}\right)^* |\alpha\rangle \langle \beta|,\\
\hat{c}^{\phantom{\dagger}}_{l\sigma}
&=&
\sum\limits_{\alpha,\beta}
D^{l\sigma}_{\beta\alpha} \; |\beta\rangle \langle \alpha|.
\end{eqnarray}
Here, the eigenstates denoted by $|\alpha\rangle$ have one electron more
than  the $|\beta\rangle$ states.

 A particular advantage of NCA is that this approach allows us to take into account the full 
 Coulomb matrix plus the spin-orbit coupling. Of course, one will
in practice reduce the matrix $U_{l m n  o}$ to two-orbital terms as
in Hamiltonian [\ref{HLDA0}] of Section \ref{AbinitioDMFT}, but the spin-flip contribution of 
the Hund exchange coupling
and the pair-hopping term can be included.

The key quantity for the resolvent perturbation theory is the resolvent
$\hat{R}(\omega)=(\omega-\hat{H})^{-1}$. Without hybridisation, this resolvent 
can be simply expressed in the eigenbasis of  $\hat{H}^{\rm at}$:
\begin{equation}
\hat{R}^{\rm at}(\omega)=\sum_{\alpha} \frac{1}{\omega-E_\alpha} |\alpha\rangle\langle\alpha|.
\end{equation}
If we now take into account the hybridisation 
 ${\mbox{\boldmath ${\Delta}$}}(\omega)$, we can define a self energy correction
due to this hybridisation. Denoting this self energy 
by
$\hat{S}(\omega)$,  the Dyson equation for the resolvent
reads:
\begin{equation}
\hat{R}(\omega)=\hat{R}^{\rm at}(\omega)
+\hat{R}^{\rm at}(\omega)\hat{S}(\omega)\hat{R}(\omega).
\end{equation}

Following  Keiter and Kimball \cite{Keiter70a} and Bickers {\em et al.} \cite{Bickers87b},
we now express   $\hat{S}(\omega)$ in a power series
in terms of the hybridisation and cut this power series after the lowest order term
(this is the non-crossing {\em approximation}):
\begin{equation}\label{Snca}
\begin{array}{c}\displaystyle
S_{\alpha\beta}(\omega)=\begin{array}[t]{l}
\displaystyle
\sum\limits_\sigma\sum\limits_{l m}
\sum\limits_{\alpha'\beta'}
\int\frac{d\varepsilon}{\pi}\,f(\varepsilon)\,
\left(D^{l \sigma}_{\alpha'\alpha}\right)^*
\Gamma_{l m}^\sigma(\varepsilon)
R_{\alpha'\beta'}(\omega+\varepsilon)
D^{m\sigma}_{\beta'\beta}\;\\[5mm]
\displaystyle
+\sum\limits_\sigma\sum\limits_{l m}
\sum\limits_{\alpha'\beta'}
\int\frac{d\varepsilon}{\pi}\,(1-f(\varepsilon))\,
D^{l \sigma}_{\alpha'\alpha}
\Gamma_{l m}^\sigma(\varepsilon)
R_{\alpha'\beta'}(\omega-\varepsilon)
\left(D^{m\sigma}_{\beta'\beta}\right)^*
\end{array}
\end{array}
\end{equation}
Here, $S_{\alpha\beta}(\omega)$  and $R_{\alpha\beta}(\omega)$ are the matrix elements of
$\hat{S}(\omega)$ and $\hat{R}(\omega)$, respectively, in the  atomic eigenbasis;
$f(\varepsilon)=1/[1+\exp(-\beta \varepsilon)]$ denotes the Fermi function,
and $\Gamma_{lm}^{\sigma}(\varepsilon)=-\Im m\left\{\Delta_{lm}^{\sigma}(\varepsilon+i0^+)\right\}$ is the
imaginary part of the hybridisation.
The resolvent $\hat{R}(\omega)$ in Eq.\ [\ref{Snca}] has to be determined self-consistently in order to yield a conserving approximation for the Anderson impurity model. If we expressed  Eq.\ [\ref{Snca}]
diagrammatically we would see that no conduction electron lines cross.

To employ NCA in the DMFT context, we have to determine
the key quantity of DMFT, i.e. the local Green function 
${\mathbf G}(\omega)$. This is achieved via
\begin{equation}\label{GfNCA}
G_{il\, im}^{\sigma}(\omega) = \frac{1}{{\cal Z}^{\rm at}}
\sum\limits_{\alpha,\alpha'}\sum\limits_{\beta,\beta'}
D^{l\sigma}_{\alpha\alpha'}\left(D^{m\sigma}_{\beta\beta'}\right)^*
\oint\frac{d \omega^{\prime} e^{-\beta \omega^{\prime}}}{2\pi i}
R_{\alpha\beta}(\omega^{\prime})R_{\alpha'\beta'}(\omega^{\prime}+\omega),
\end{equation}
with the atomic partition function
\begin{equation}
 {\cal Z}^{\rm at}=
\sum\limits_\alpha\oint\frac{d\omega \, e^{-\beta \omega}}{2\pi i} 
R_{\alpha\alpha}(\omega).
\end{equation}
With ${\mathbf G}(\omega)$ from Eq.\ [\ref{GfNCA}], we can 
continue the DMFT cycle, defining  a new auxiliary Anderson impurity
problem, solving it by NCA {\em etc.}

The particular advantages of NCA are that it is
a (computationally) relatively fast method---at least if not too
many orbitals are involved---, and 
a conserving approximation to the Anderson impurity model.
It is good at strong coupling and at temperatures
above the effective Kondo temperature of the Anderson impurity model.
The drawbacks are that NCA is known to violate Fermi liquid properties at 
low temperatures (below the Kondo temperature)
 and whenever charge excitations
become dominant \cite{MuellerHartmann84a,Pruschke93a}. Hence, in some parameter ranges it fails
in the most dramatic way and must therefore be applied with considerable
care \cite{Pruschke93a}. The NCA has been used intensively as a DMFT
solver, in particular, by Pruschke and coworkers. In the context of 
LDA+DMFT, Z\"olfl {\em et al.} \textcite{Zoelfl00,Zoelfl01} studied LaTiO$_3$ and cerium
by NCA.

The same shortcomings
are principally also true for the one-crossing approximation
\cite{Pruschke89a,Haule01} which goes beyond Eq.\ [\ref{Snca}] by taking into
account  additional  diagrams with a single line crossing, a vertex correction.
Haule {\em et al.} \cite{Haule04} used this extension recently to study the 
optical conductivity of cerium.
The conserving T matrix approximation (CTMA) includes further diagrams
beyond NCA. Kirchner {\em et al.} \cite{Kirchner04} extended this approach to calculate
dynamical properties which are necessary for
a DMFT solver. This approach fulfils the Fermi liquid properties
at low temperatures and might improve on most deficiencies of the NCA.
Hence,  this approximation might be suitable for DMFT, in particular, if
the deficiency of not yielding
the exact unitary limit,  
the exact height of the  spectral function at $\omega=0$,  is overcome \cite{Kirchner04}.
Most recently,
also 
a direct expansion of the Green function
in terms of the hybridisation has been proposed \cite{Dai05}.

\subsection{Quantum Monte Carlo (QMC) simulations}
\label{QMC}
In the previous Sections we have already 
introduced several methods  to calculate the 
Green function of the Anderson impurity model (Eq.\ [\ref{siam}]).
This is necessary for the DMFT self-consistently cycle
 (flow diagram Fig.~\ref{DMFTflow}),
and  quantum Monte Carlo (QMC) simulations
allow to do so in  a numerically exact way.
For the Anderson impurity model, the
QMC algorithm of  Hirsch and Fye \cite{Hirsch86a} is the
efficient, well established approach.
Hence, we will briefly review this QMC algorithm 
and a projective variant in the next two Sections.

Let us  mention however that there has been a rapid development 
in the field of QMC algorithms for the
Anderson impurity model most recently. In particular, continuous time QMC
algorithms have been developed \cite{Rubtsov04a,Rubtsov05a,Werner05a,Werner06a,Haule07a},  based  on a series expansions starting either from  the non-interacting
problem \cite{Rubtsov04a,Rubtsov05a} or the atomic limit \cite{Werner05a}.
Also  a combination of 
series expansion and  Hirsch-Fye algorithm is possible \cite{Sakai06a}.
One of these new  algorithms  might superseed the hirthto standard Hirsch-Fye
QMC approach in the future.  In the present state 
of flux, it is not clear 
 however which of the new algorithms or furthergoing ones
will prevail.

\subsubsection{Hirsch-Fye algorithm}
\label{HFQMC}

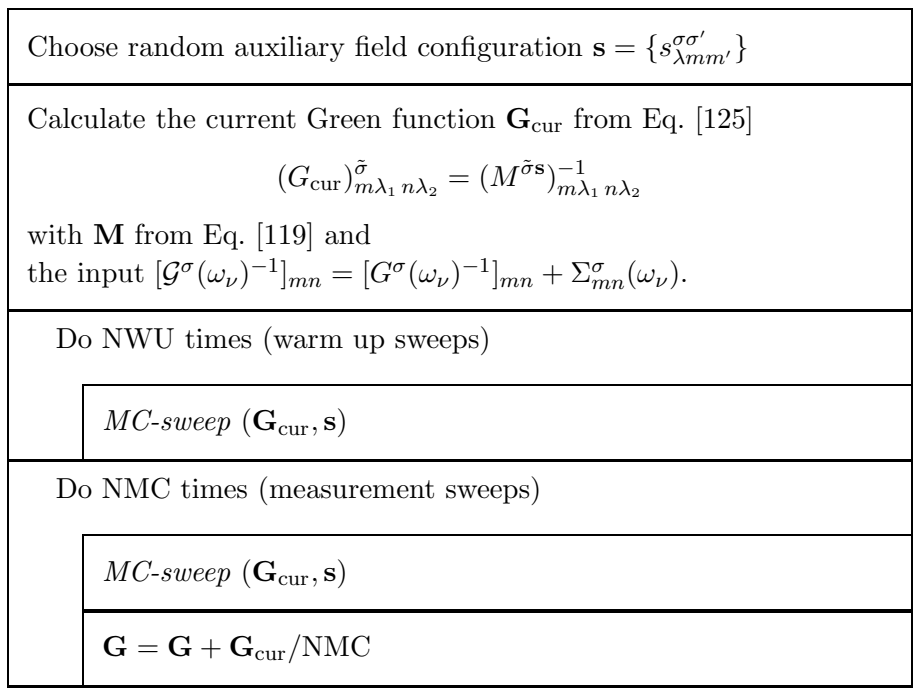
\begin{figure}[t]
\vspace{1cm}

\unitlength=1mm
\special{em:linewidth 0.4pt}
\linethickness{0.4pt}
\begin{picture}(130.00,85.00)


\put(0.00,80.00){\framebox(120.00,10.00)[cc]
{\parbox{11.5cm}{Choose random auxiliary field configuration $\mathbf{s}=\{s_{\lambda m m'}^{\sigma {\sigma'}}\}$}}}

\put(0.00,50.00){\framebox(120.00,30.00)[cc]
{\parbox{11.5cm}{
Calculate the current Green function ${\mathbf G}_{\rm cur}$ from Eq.\ [\ref{curG}]
\vspace{-.6cm}

 \[
({G_{\rm cur}})_{m \lambda_1\, n \lambda_2}^{\tilde{\sigma}} = 
 (M^{\tilde{\sigma}\mathbf{s}})^{-1}_{m \lambda_1 \, n \lambda_2}
\] 
\vspace{-.6cm}

with  ${\mathbf M}$ from Eq.\ [\ref{defM}] and\\
\vspace{-.4cm}

the input  $[{{\cal G}^{\sigma}}(\omega_{\nu})^{-1}]_{m n}=[{{G}^{\sigma}}(\omega_{\nu})^{-1}]_{m n}
+ \Sigma_{m n}^{\sigma}(\omega_{\nu})$.
}}}

\put(0.00,30.00){\framebox(120.00,20.00)[cc]{}}

\put(5.00,45.00){
\parbox{12.5cm}{Do NWU times (warm up sweeps)}}

\put(10.00,30.00){\framebox(110.00,10.00)[cc]
{\parbox{10.5cm}{ \em MC-sweep $({\mathbf G}_{\rm cur}, {\mathbf s})$}}}

\put(0.00,0.00){\framebox(120.00,30.00)[cc]{}}

\put(5.00,25.00){
\parbox{12.5cm}{Do NMC times (measurement sweeps)}}

\put(10.00,10.00){\framebox(110.00,10.00)[cc]
{\parbox{10.5cm}{ \em MC-sweep  $({\mathbf G}_{\rm cur}, {\mathbf s})$}}}

\put(10.00,0.00){\framebox(110.00,10.00)[cc]
{\parbox{10.5cm}{${\mathbf G} = {\mathbf G} + {\mathbf G}_{\rm cur}/ {\rm NMC}$}}}

\end{picture}
\vspace{.5cm}

\caption
{Flow diagram of the QMC algorithm to calculate the Green
function matrix ${\mathbf G}$ using the
procedure {\em MC-sweep} of Fig.~\ref{mcsweep} [closely following \cite{Held03}]. \label{QMCflow}}
\end{figure}

In essence, the QMC technique maps the interacting
 Anderson impurity problem (Eq.\ [\ref{siam}]) onto a sum of non-interacting
problems where the single particle moves in a fluctuating, 
time-dependent field. This sum is evaluated by Monte Carlo sampling,
see the flow diagram Fig.~\ref{QMCflow} for an overview.\\

\begin{figure}[t]
\vspace{.5cm}

\unitlength=1mm
\special{em:linewidth 0.4pt}
\linethickness{0.4pt}
\begin{picture}(130.00,70.00)

\put(5.00,0.00){\framebox(120.00,64.00)[cc]{}}

\put(5.00,55.00){
\parbox{11.5cm}{Choose $M (2M-1) \Lambda $ times an auxiliary field index
($\lambda \, m \, m' \, \sigma \, \sigma'$),\\  ${\mathbf s}_{\rm new}\equiv {\mathbf s}$ except 
for this index for which
$(s_{\rm new})_{\lambda m m'}^{\sigma {\sigma'}}=-s_{\lambda m m'}^{\sigma {\sigma'}}$.}}

\put(10.00,25.00){\framebox(115.00,25.00)[cc]
{\parbox{11cm}{
\vspace{.05cm}

Calculate flip probability
%
${\cal P}({\mathbf s}\rightarrow {\mathbf s}_{\rm new})  = {\rm min} \{1,P({\mathbf s}_{\rm new})/P({\mathbf s})\}$
\vspace{-.3cm}
\[
 \mbox{with}\; P({\mathbf s}_{\rm new})/P({\mathbf s})= \det \mathbf{M}_{m n}^{\sigma \mathbf{s}_{\rm new}}/ \det \mathbf{M}_{m n}^{\sigma \mathbf{s}}
\]

\vspace{-.4cm}

 and ${\mathbf M}$ from Eq.\ [\ref{defM}].}}}

\put(10,25){\line(5,-1){75}}
\put(125,25){\line(-8,-3){40}}
\put(85,10){\line(0,-1){10}}
\put(10,25){\line(0,-1){15}}

\put(50.00,21.00){
\parbox{6.2cm}  {\hspace{-.9cm}{\rm Random number} $\!\!\in\! {\mathrm (0,1)} \!<\!{\cal P}({\mathbf s}\!\rightarrow\! {\mathbf s}_{\rm new})$  ?}}
\put(20.00,12.00){
\parbox{3cm}{{\rm yes}}}
\put(110.00,12.00){
\parbox{2cm}{{\rm no}}}

\put(10.00,0.00){\framebox(115.00,10.00)[cc]{}}
\put(21.00,5.00){
\parbox{12cm}{\hspace{-.7cm}${\mathbf s}={\mathbf s}_{\rm new}$; recalculate  ${\mathbf G}_{\rm cur}$ via  Eq.\ [\ref{curG}] \hspace{1.399cm} Keep ${\mathbf s}$
}}

\end{picture}
\vspace{.5cm}

\caption
{
Procedure {\em MC-sweep} using the Metropolis\cite{Metropolis53a} 
rule to  change the sign of $s_{\lambda m m'}^{\sigma {\sigma'}}$. The recalculation of  ${\mathbf G}_{\rm cur}$, 
i.e. the matrix $\mathbf{M}$ of Eq.\ [\ref{defM}], simplifies
to ${\cal O}(\Lambda^2)$ operations if only one  
$s_{\lambda m m'}^{\sigma {\sigma'}}$ changes sign.
Then, Eqs.\ [\ref{newM1}] and  [\ref{newM}] can be employed
(closely following \cite{Held03}).
\label{mcsweep}}
\end{figure}
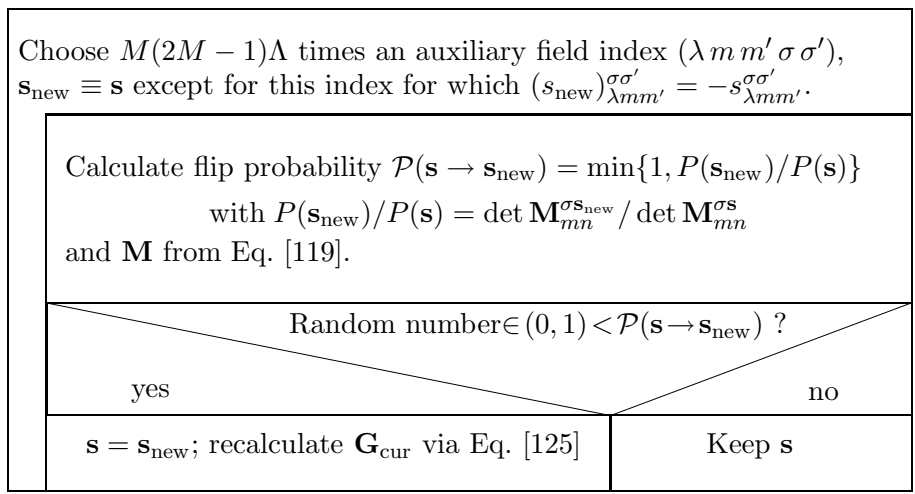

{\it Trotter discretisation}
\nopagebreak

Let us now discuss this approach in more detail.
In a first step, the imaginary time interval $[0,\beta]$ 
of the functional integral Eq.\ [\ref{siam}] is discretised into 
$\Lambda$ steps 
of size $\Delta \tau = \beta/\Lambda$, yielding support points
$\tau_\lambda=\lambda \Delta \tau$ with $\lambda=1 \dots \Lambda$. 
Using this Trotter discretisation, 
the integral $\int_0^\beta d\tau$ is
 transformed into the sum 
$\sum_{\lambda=1}^\Lambda \Delta \tau$ and
the exponential terms in Eq. [\ref{siam}] can be 
separated via the Trotter-Suzuki \cite{suzuki} formula for operators 
$\hat{A}$ and $\hat{B}$ 
\begin{equation}
e^{-\beta (\hat{A}+\hat{B})} = \prod_{\lambda=1}^\Lambda e^{-\Delta \tau \hat{A}} 
e^{-\Delta \tau \hat{B}} + {\cal O}(\Delta \tau),
\end{equation}
which is exact in the limit $\Delta \tau \rightarrow 0$. 
The single site action $\cal A$ of Eq.\ [\ref{effAction}] can now be written
 in the 
discrete,  imaginary time as
\begin{eqnarray}
{\cal A}[
     \psi,\psi^*,{\cal G}^{-1}] &=& \Delta \tau ^2 
     \sum_{ \sigma \,m n}\sum_{\lambda,\lambda'=
       0}^{\Lambda-1} {\psi_{ m \lambda}^{\sigma *}}  
       {{\cal G}_{ m n}^\sigma}^{-1} (\lambda\Delta
       \tau - \lambda'\Delta \tau) 
      \psi_{ n \lambda'}^\sigma
    \nonumber \\ && - \frac{1}{2} \Delta \tau {\sum_{m\sigma,m'\sigma'}^{\prime}}
    U_{m m'}^{\sigma \sigma'} \sum_{\lambda=0}^{\Lambda-1}
     {\psi_{m \lambda}^{\sigma *}}
     {\psi_{m \lambda}^\sigma}
     {\psi_{m' \lambda}^{\sigma' *}}
     {\psi_{m' \lambda}^{\sigma'}}.
\label{AQMC}
\end{eqnarray}
Here, the Coulomb interactions of the LDA+DMFT Hamiltonian [\ref{LDADMFTH}],
more specifically the inter-orbital and intra-orbital Coulomb repulsion
and the $Z$-component of the Hund exchange interaction,
were for convenience
put into a unified form with
\begin{equation}
    U_{m m'}^{\sigma \sigma'}  = \left\{
\begin{array}{ll}
0 & \mbox{ for }  m=m' \mbox{ and } \sigma = \sigma'\\
U & \mbox{ for }  m=m' \mbox{ and } \sigma \neq \sigma'\\
V-J & \mbox{ for }  m \neq m' \mbox{ and } \sigma = \sigma'\\
V & \mbox{ for }  m \neq m' \mbox{ and } \sigma  \neq \sigma'\\
\end{array}\right. .
\end{equation}

Also note, that the first term was Fourier-transformed from Matsubara frequencies
in  Eq.\ [\ref{effAction}]
to imaginary time.\\

{\it Hubbard-Stratonovich transformation}
\nopagebreak

In a second step, the $M(2M-1)$ interaction terms ($M$ denotes  the number 
of interacting orbitals)
of the single site action 
$\cal A$ (Eq.\ [\ref{AQMC}]) 
are decoupled by introducing a classical auxiliary field 
$s_{\lambda m m'}^{\sigma {\sigma'}}$:
\begin{eqnarray}
\lefteqn{ \exp \left\{
   \frac{\Delta \tau}{2} U_{m m'}^{\sigma \sigma'}
     ({\psi_{m \lambda}^\sigma}^*
     {\psi_{m \lambda}^\sigma}-
     {\psi_{m' \lambda}^{\sigma'}}^*
     {\psi_{m' \lambda}^{\sigma'}})^2 \right\} =} \qquad \qquad \nonumber \\ \!\!\!\!\!\!\!\!\!\!
&&  \!\!\!\!\! \frac{1}{2} \sum^{\prime}_{s_{\lambda m m'}^{\sigma {\sigma'}}\,  = \pm 1} \!\!
\exp \left\{ \Delta \tau {\cal J}_{\lambda m m'}^{\sigma {\sigma'}}
 s_{\lambda m m'}^{\sigma {\sigma'}} 
     ({\psi_{m \lambda}^\sigma}^*
     {\psi_{m \lambda}^\sigma}-
     {\psi_{m' \lambda}^{\sigma'}}^*
     {\psi_{m' \lambda}^{\sigma'}}) \right\},\label{Hirsch}
\end{eqnarray}
where  $\cosh ({\cal J}_{\lambda m m'}^{\sigma {\sigma'}}) = 
\exp(\Delta \tau U_{m m'}^{\sigma \sigma'}/2)$.
Since we need only one field for every
pair of orbitals [($m\sigma$) and ($m'\sigma'$)]
we can restrict the $s_{\lambda m m'}^{\sigma {\sigma'}}$ sum 
to one term per pair which is
 indicated by the prime.
This  so-called discrete 
Hirsch-Fye-Hubbard-Stratonovich transformation
can be applied to the Coulomb repulsion as well as to the $z$-component of 
Hund's rule coupling, all included in  $U_{m m'}^{\sigma \sigma'}$.
 One limitation of  QMC is that it is
very difficult to deal with terms which do not have this
density-density type of form.
In particular,  a Hubbard-Stratonovich
decoupling of the spin-flip term of  Hund's rule coupling 
leads to
a `minus-sign problem', see  \textcite{Held99phd}.
Therefore,  this spin-flip term and the
pair-hopping term, i.e. the second line of the LDA+DMFT Hamiltonian [\ref{LDADMFTH}],
are usually neglected.
In the particle-hole symmetric case, another decoupling
scheme which includes the spin-flip term is possible
without `minus-sign problem',
see \textcite{Motome97}.
More recently, also 
several new algorithms \cite{Sakai04a,Han04a,Rubtsov04a,Rubtsov05a,Werner05a,Werner06a,Haule07a} have been introduced 
which include the spin-flip and pair-hopping terms.
This is done by new kinds of Hubbard-Stratonovich
transformations \cite{Sakai04a,Han04a} or series expansions \cite{Rubtsov04a,Rubtsov05a,Werner05a,Haule07a,Sakai06a}.
Hence, LDA+DMFT(QMC) calculations with the correct
symmetry of the interaction should be possible in the future;
first steps have already been taken \cite{Haule07a}.

By means of Eq.\ [\ref{Hirsch}], we replace the interacting 
system by a sum of $\Lambda M (2M-1)$ auxiliary 
fields $s_{\lambda m m'}^{\sigma {\sigma'}}$.
This allows us to solve the functional integral  by a simple Gauss integration
since the Fermion operators only enter quadratically, i.e. for a
given configuration  ${\bf s}=\{ s_{\lambda m m'}^{\sigma {\sigma'}}\}$
of the auxiliary fields the system is non-interacting.
The quantum mechanical problem is then reduced to a matrix problem
\begin{eqnarray}
G_{{m} \lambda_1\, n \lambda_2}^{\tilde{\sigma}} = \frac{1}{\cal Z} \sum_\lambda
\sum^{\prime}_{m' \sigma', m'' \sigma''}
\sum^{\prime}_{s_{\lambda m'' m'}^{{\sigma''} {\sigma'}}\,  = \pm 1} 
(M^{\tilde{\sigma}\mathbf{s}})^{-1}_{m \lambda_1\, n \lambda_2} 
 \det \mathbf{M}_{\tilde{m} \tilde{n}}^{\sigma \mathbf{s}}.
\label{hidimsum}
\end{eqnarray}
Here, 
\begin{equation}
{\cal Z}= \sum_\lambda
{\sum^{\prime}_{m' \sigma', m'' \sigma''}}
\sum^{\prime}_{s_{\lambda m'' m'}^{{\sigma''} {\sigma'}}\,  = \pm 1} 
 \det \mathbf{M}_{m n}^{\sigma \mathbf{s}}
\end{equation}
 is  the partition function, the prime indicates that every distinct term is counted only once, the determinant includes  the orbital, imaginary time and  spin index,
 and $\mathbf{M}_{{m} {n}}^{\tilde{\sigma} \mathbf{s}}$ is the 
following matrix in the imaginary time indices:
\begin{eqnarray}
\mathbf{M}_{{m n} }^{\tilde{\sigma} \mathbf{s}} = \Delta \tau^2 
[({\mathbf{G}^{\sigma}}^{-1})_{m n} + \Sigma_{m n}^{\sigma}] 
e^{-\mbox{\boldmath ${\tilde{\cal J}}$}^{\sigma \mathbf{s}}_{n}} + \mathbf{1} - 
e^{-\mbox{\boldmath ${\tilde{\cal J}}$}^{\sigma \mathbf{s}}_{n}}\delta_{m n}.
\label{defM} 
\end{eqnarray}
The elements of the imaginary time matrix $\mbox{\boldmath ${\tilde{\cal J}}$}^{\sigma \mathbf{s}}_{m}$
are, in turn, given by
\begin{eqnarray}
\tilde{\cal J}^{\sigma \mathbf{s}}_{m \lambda \lambda'}= -\delta_{\lambda \lambda'} 
\sum_{m' \sigma'} {\cal J}_{m m'}^{\sigma \sigma'} 
\tilde{\sigma}_{m m'}^{\sigma \sigma'} s_{\lambda m m'}^{\sigma \sigma'},
\end{eqnarray}
where $\tilde{\sigma}_{m m'}^{\sigma \sigma'} = 2 \Theta(\sigma' - \sigma + 
\delta_{\sigma \sigma'}[m'-m]-1)$ changes sign if $(m \sigma)$ and 
$(m' \sigma')$ are exchanged. For more details and  a derivation why
the matrix  $\bf M$ enters in
 Eq.\ [\ref{defM}],
see Refs.\ \cite{Hirsch86a,Jarrell97,Georges96a}.\\

{\it Monte Carlo importance sampling}
\nopagebreak

Since the sum in Eq.\ [\ref{hidimsum}] consists of $2^{\Lambda M(2M-1)}$ 
addends, 
a complete summation for large $\Lambda$ is computationally impossible.
Therefore, the Monte Carlo method is employed, which is often an efficient
way  to calculate  
high-dimensional sums and integrals.  In this method, the integrand $F(x)$ is 
split into a normalised probability distribution $P$ and the remaining 
term $O$:
\begin{equation}
\int dx F(x) = \int dx \,O(x) \,P(x) \equiv \langle O\rangle_P
\end{equation}
with
\begin{equation}
\int dx P(x) = 1 \qquad {\rm and} \qquad P(x) \ge 0.
\end{equation}
In statistical physics, the Boltzmann distribution is often a good choice 
for the function $P$:
\begin{equation}
P(x) = \frac{1}{\cal Z} \exp[-\beta E(x)].
\end{equation}
For the sum of Eq.\ [\ref{hidimsum}], this probability distribution
 translates to
\begin{equation}
P(\mathbf{s}) = \frac{1}{\cal Z}  \det 
\mathbf{M}_{m n}^{\sigma \mathbf{s}}
\end{equation}
with the remaining term
\begin{equation}
O(\mathbf{s})_{{m} \lambda_1\, n \lambda_2}^{\tilde{\sigma}} = 
\left[ (M^{\tilde{\sigma}\mathbf{s}})^{-1} \right]_{m \lambda_1\, n \lambda_2}.
\label{curG}
\end{equation}

Instead of summing over all possible configurations, the Monte Carlo simulation
generates  configurations 
$x_i$ according to the probability distribution $P(x)$ and 
averages 
the observable $O(x)$ over these $x_i$. Therefore the relevant 
parts of the phase space with a large Boltzmann weight are taken into 
account to a greater extent than the ones with a small weight, coining the 
name {\it importance sampling}
 for this method. For $\cal N$ statistically independent 
addends $x_i$ drawn according to the probability $P(x)$ [this is indicated by
$x_i \in P(x)$ in the following equation], one gets by virtue of the central limit theorem
the following estimate
\begin{equation}
\langle O\rangle_P = \frac{1}{\cal N} \sum^{\cal N}_{\stackrel{\scriptstyle i=1} {\scriptstyle x_i \in P(x)}}
O(x_i) \pm \frac{1}{\sqrt{\cal N}} \sqrt{\langle O^2\rangle_P - \langle O\rangle_P^2}.
\label{MCsum}
\end{equation}
Here, the error and with it the number of needed addends $\cal N$ is nearly 
independent of the dimension of the integral. The computational effort for 
the Monte Carlo method is therefore only rising polynomially with the 
dimension of the integral and not exponentially as in a normal integration.
The so-called minus-sign problem occurs if the error [given by the variance in Eq.\ [\ref{MCsum}]) 
is large in comparison with the mean value. This happens particularly
if  $O(x)$ has contributions with positive and negative sign which almost cancel.

In order to pick configurations $x$ with the proper probability $P(x)$,
a Markov process is employed. Was $x$ realised, a new configuration $y$
is accepted with probability
 \begin{equation}
 {\cal P}(x \rightarrow y) = {\mathrm{min}} \left \{1,P(y)/P(x) \right\} 
 \label{uebwahr}
 \end{equation}
Since this transition probability  by Metropolis {\em et al.} \cite{Metropolis53a}
 fulfils the detailed balance
 \begin{equation}
   P(x) {\cal P}(x \rightarrow y) =
   P(y) {\cal P}(y \rightarrow x),
 \end{equation}
it guarantees that the series of configurations  $x$ 
obey the probability $P(x)$.

In our case, the probability ratio for deciding on the acceptance of a new auxiliary field configuration ${\mathbf s}_{\rm new}$ is given by
\begin{equation}
 P({\mathbf s}_{\rm new})/P({\mathbf s})= 
{\frac{\det  {\mathbf M}_{m n}^{\sigma {\mathbf  s_{\rm new}}}}{\det  {\mathbf M}_{m n}^{\sigma \mathbf s}}}.
\end{equation}

{\it Single spin-flip updates}
\nopagebreak

A very efficient algorithm is obtained by
considering a new configuration  of the auxiliary field 
(spins) ${\mathbf s}_{\rm new}$ which
differs from the old one by only one component ($\lambda\, m \, m'\,  \sigma {\sigma'}$):
  ${\mathbf s}_{\rm new}={\mathbf s}$ except for
$(s_{\rm new})_{\lambda m m'}^{\sigma {\sigma'}}=-s_{\lambda m m'}^{\sigma {\sigma'}}$.
This dramatically reduces the effort to calculate 
the probability ratio which then only depends on three numbers:
\begin{equation}
 P({\mathbf s}_{\rm new})/P({\mathbf s})= R^{\mathbf s_{\rm new} \mathbf  s}_{m m\sigma} R^{\mathbf s_{\rm new} \mathbf  s}_{m' m'\sigma'}-\delta_{\sigma \sigma'}R^{\mathbf s_{\rm new} \mathbf  s}_{m m' \sigma} 
\end{equation}
where
\begin{eqnarray} 
R^{\mathbf s_{\rm new}  \mathbf s}_{m m\sigma}&=& 1+
   \big(1-[(M^{\sigma\mathbf s})^{-1}]_{m \lambda\, m \lambda}\big)
\big( \underbrace{e^{2  {\cal J}_{\lambda m m'}^{\sigma \sigma'} \;
 \tilde{\sigma}_{m m'}^{ \sigma \sigma'} \; 
 {s}_{\lambda m m'}^{\sigma \sigma'} }}_{\equiv e^{\cal J}}-1\big),\\
R^{\mathbf s_{\rm new} \mathbf  s}_{m' m' \sigma'}&=&1+
   \big(1-[(M^{\sigma' \mathbf s})^{-1}]_{m'\lambda\, m' \lambda}\big)
( e^{-{\cal J}}-1), \label{Eq:R2}\\
R^{\mathbf s_{\rm new} \mathbf  s}_{m m' \sigma}&=&[(M^{\sigma\mathbf s})^{-1}]_{m \lambda\, m' \lambda}( e^{-{\cal J}}-1)
[(M^{\sigma\mathbf s})^{-1}]_{m' \lambda\, m \lambda}( e^{{\cal J}}-1). 
\end{eqnarray}
Similarly one can show with some matrix algebra
that the recalculation of  $ {\mathbf M}^{\sigma {\mathbf s_{\rm new}}}_{m n}$ requires  ${\cal O} (2 M^2 \Lambda^2)$
computational  
operations in two steps (without off-diagonal orbital-elements
this reduces to  ${\cal O}(2\times  \Lambda^2)$).
The first step is
 \begin{eqnarray}
\!  \!\!  \!  \! 
  [(M^{\sigma\mathbf s_{\rm new}}_{\rm 1.})^{-1}]_{\tilde{m}{\lambda}'\, \tilde{n} {\lambda}''} &=&  [(M^{\sigma\mathbf s})^{-1}]_{\tilde{m}{\lambda}'\, \tilde{n} {\lambda}''} + 
      \frac{ e^{{\cal J}}-1}
   {R^{\mathbf s_{\rm new}   s}_{m m\sigma}} \nonumber\\&&\!  \!  
 \times\big(  
   [({M}^{ \sigma \mathbf s})^{-1}]_{\tilde{m} {\lambda}'\, m {\lambda}}  -
   \delta_{{\lambda}' {\lambda}}  \delta_{\tilde{m}  m}\big) [(M^{\sigma\mathbf s})^{-1}]_{m {\lambda}\, \tilde{n} {\lambda}''}.  \label{newM1}
\end{eqnarray}
With the ${\mathbf M}^{\mathbf s_{\rm new}}_{\rm 1.}$ from the 1.\ step,
 $R^{\mathbf s_{\rm new} \mathbf  s}_{m' m' \sigma'}$
is recalculated according to Eq.\ [\ref{Eq:R2}]
and the second update incorporates the change because of $m'$: 
 \begin{eqnarray}
\! \! \! \!\! \! \!\!\! \!
   [(M^{\sigma'\mathbf s_{\rm new}})^{-1}]_{\tilde{m}{\lambda}'\, \tilde{n} {\lambda}''} &\!=\!&  [(M^{\sigma'\mathbf s_{\rm new}}_{\rm 1.})^{-1}]_{\tilde{m}{\lambda}'\, \tilde{n} {\lambda}''} + 
      \frac{ e^{- {\cal J}}-1}
   {R^{\mathbf s_{\rm new}   s}_{m' m'\sigma'}} \nonumber\\&&
\! \! \! \!  \! \!  \! \! \! \!  \! \! \! \!  \! \! \! \!
  \times\big(  
   [({M}^{ \sigma' \mathbf s_{\rm new}}_{\rm 1.})^{-1}]_{\tilde{m} {\lambda}'\, m' {\lambda}} \! -
   \delta_{{\lambda}' {\lambda}}  \delta_{\tilde{m}  m'}\big) [(M^{\sigma'\mathbf s_{\rm new}}_{\rm 1.})^{-1}]_{m' {\lambda}\, \tilde{n} {\lambda}''}.  \label{newM}
\end{eqnarray}
Note that these equations hold if $G(\tau>0)$ is negative, 
as in the standard definition for the Green function,
and if $G(0)=G(\tau=0-)$. Often in QMC,  the Green function
is defined differently, i.e. with the opposite (positive) sign.

Since roughly $M (2M-1)\Lambda$ single spin flips have to be tried before
arriving at a truly independent  $ {\mathbf s}_{\rm new}$,
the  overall cost of the algorithm is
\begin{equation}
2  M (2M-1) M^2 \Lambda^3 \times \text{number of MC-sweeps}
\end{equation}
in leading order of $\Lambda$.
It reduces considerably to 
\begin{equation}
2  M (2M-1) \Lambda^3 \times \text{number of MC-sweeps}
\end{equation}
if orbital off-diagonal elements are zero.

The advantage of the QMC method (for the algorithm see the flow diagrams Figs.~\ref{QMCflow} and \ref{mcsweep})
is that it is (numerically) exact. It allows one to calculate
the one-particle Green function as well as two-particle (or higher) Green 
functions. On present workstations the QMC approach
is able to deal with up to 
seven {\em interacting} orbitals and room temperature
or higher temperatures, for typical values of the Coulomb interaction $U$ and 
the LDA bandwidth $W$. Since the QMC approach calculates
$G(\tau)$ or $G(i \omega_n)$ with a statistical error, it also requires an
analytical continuation 
to obtain the Green function 
$G(\omega)$ at real (physical) frequencies $\omega$
or the physically relevant spectral
function
$A(\omega)=-\frac{1}{\pi} {\rm Im} G(\omega)$,
see Section \ref{SecMEM}.
Very
low temperatures are not accessible in QMC because the numerical effort
grows as $\Lambda^3\propto 1/T^3$. 
For these low temperatures or zero temperature,
a projective QMC method 
was developed recently.
We will discuss this variant in the next Section.

\subsubsection{Projective quantum Monte Carlo (PQMC) simulations for $T=0$}
\label{PQMC}

Often
interesting many-body physics occurs at low temperatures.
In this case, the Hirsch-Fye \cite{Hirsch86a} QMC algorithm,
which was introduced in the previous Section,
 is not applicable as discussed above.
For lattice QMC simulations an alternative projective
quantum Monte Carlo (PQMC) method was developed by White {\em et al.} \cite{White89b}.
This PQMC algorithm converges, according to
Assaad and Imada \cite{Assaad95a},  faster to the
groundstate than the finite temperature lattice QMC algorithm of
Blankenbecler {\em et al.}
\cite{Blankenbecler81a}.
The general idea of the PQMC is to 
start with a trial wave function $|\Psi_0\rangle$
and to project onto the ground state $|\Psi_{\rm GS}\rangle$ via
\begin{equation}
|\Psi_{\rm GS}\rangle = \lim_{\theta \rightarrow \infty}\frac{e^{-\theta/2\, \hat{H}}\,|\Psi_0\rangle}{\sqrt{
\langle\Psi_0|\, e^{-\theta \hat{H}}\, |\Psi_0\rangle}}.
\label{PQMCGS}
\end{equation}
If  $|\Psi_0\rangle$ has any overlap with the (unique) ground state only this state with the lowest energy
and, hence,  the largest contribution to $e^{-\theta \hat{H}}$
will prevail in the limit $\theta\rightarrow \infty$.
In the DMFT context, one often calculates
the Anderson impurity model in the thermodynamic limit with an 
infinite number of bath sites $N\rightarrow \infty$.
In this case, ground state and low-lying excited states have the same
energy to leading order in $1/N$ and yield the
same Green function. Hence, we can project
 onto such  low-lying excited states and obtain the
correct Green function and other expectation values,
so that
Anderson's orthogonality catastrophe \cite{Anderson67a,Anderson67b,Katsnelson06a}
is irrelevant for the PQMC algorithm, as was shown by 
Feldbacher {\em et al.} \cite{Feldbacher06}.

With the ground state given by Eq.\ [\ref{PQMCGS}], zero temperature observables can be calculated as follows:
\begin{equation}
\langle \hat{\cal O}\rangle =\langle\Psi_{\rm GS}|\, \hat{\cal O}\, |\Psi_{\rm GS}\rangle = \lim_{\theta \rightarrow \infty}\frac{\langle\Psi_|\,{e^{-\theta/2\, \hat{H}}\hat{\cal O} {e^{-\theta/2\, \hat{H}}\,|\Psi_0\rangle}}}{{
\langle\Psi_0|\, e^{-\theta \hat{H}}\, |\Psi_0\rangle}}.
\label{PQMCO}
\end{equation}
For a finite value of $\theta$, this expectation value can be 
calculated by (projective) QMC simulations which decouple the
projector ${e^{-\theta/2\, \hat{H}}}$ by Hubbard-Stratonovich transformations,
after discretising the imaginary time.

Let us now focus on the Anderson impurity model for which
Feldbacher {\em et al.} \cite{Feldbacher04} 
developed a new PQMC  method. This PQMC is related to the
one of White {\em et al.} \cite{White89b} for lattice many-body systems 
in a similar way as the QMC of Hirsch and Fye \cite{Hirsch86a} to that of Blankenbecler{\em et al.} \cite{Blankenbecler81a} for lattice QMC.
The biggest difference is that, for the Anderson impurity model, 
one is dealing with
matrices in the imaginary time index, 
instead of 
matrices in the lattice indices for the
 lattice QMC algorithm of Blankenbecler{\em et al.} \cite{Blankenbecler81a}:
The Hirsch-Fye algorithm is directly formulated in
terms of the non-interacting Green function of the Anderson impurity model
$\mbox{\boldmath ${\cal G}$}^0$. Within the DMFT iteration scheme,
we usually do not even define a lattice for the auxiliary Anderson
impurity model anymore.
This leads to the difficulty of how to define the trial wave function
which for a lattice problem is a more straightforward task.

Feldbacher {\em et al.}
 \cite{Feldbacher04} overcame this difficulty 
by considering
instead of Eq.\ [\ref{PQMCO}] an artificial finite temperature problem:
\begin{eqnarray}
\langle \hat{\cal O}\rangle_{\theta}
&=&\lim_{\beta \rightarrow \infty}
\frac {{\rm Tr} e^{-\beta/2\hat{H}_0}
e^{-\theta/2\, \hat{H}}\hat{\cal O} e^{-\theta/2\, \hat{H}_{\rm AIM}}e^{-\beta/2\hat{H}_0}}
{{\rm Tr} e^{-\beta\hat{H}_0}
e^{-\theta \hat{H}_{\rm AIM}}}.
\label{PQMCO2}
\end{eqnarray}
In the limit $\beta\rightarrow\infty$, this projects
onto the ground state of  the Hamilton operator 
$\hat{H}_0$. If this ground state is  $|\Psi_0\rangle$
and has energy  $E_0$ we have
\begin{equation}
\lim_{\beta \rightarrow \infty} {\rm Tr}  e^{-\beta\hat{H}_0} \hat{\cal O} \rightarrow 
e^{-\beta\hat{E}_0} \langle \Psi_0|\, \hat{\cal O} \,|\Psi_0\rangle.
\end{equation}
Hence, 
the
finite temperature problem of Eq.\ [\ref{PQMCO2}]
becomes equivalent to the projection
of Eq.\ [\ref{PQMCO}]. 

As a special case of  Eq.\ [\ref{PQMCO2}], the $\theta$-projected Green function is given by
\begin{eqnarray}
G_{lm}\left(  \tau,\tau^{\prime}\right)   &  =& - \langle T c_l(\tau) c_m^{\dagger}(\tau^{\prime})\rangle \\
&=&
\frac{\left\langle \Psi_{0}\right|
{T}e^{-\theta \hat{H}_{\rm AIM}}c_l(\tau)c_m^{\dagger}(\tau^{\prime})\left|  \Psi
_{0}\right\rangle }{\left\langle \Psi_{0}\right|  e^{-\theta \hat{H}_{\rm AIM}}\left|
\Psi_{0}\right\rangle }\label{projection2}\\
&  =& \lim_{\beta\rightarrow\infty}\frac{\operatorname*{Tr} {T}e^{-\beta \hat{H}_{0}%
}e^{-\theta \hat{H}_{\rm AIM}}c_l(\tau)c_m^{\dagger}(\tau^{\prime})}{\operatorname*{Tr}e^{-\beta \hat{H}_{0}%
}e^{-\theta \hat{H}_{\rm AIM}}}. \label{PQMCGF}
\end{eqnarray}
Here, $T$ is the Wick time ordering operator for the $\tau$'s
and $\hat{\cal O}(\tau)= e^{\tau\hat{H}} \hat{\cal O} e^{-\tau\hat{H}}$
as before.

A natural choice for the  trial wave function is
the ground state of a non-interacting Anderson impurity model
[$U=0$ in Hamiltonian [\ref{AIMH}]).
As was shown by Feldbacher {\em et al.} \cite{Feldbacher04},
one then arrives at the same algorithm as the Hirsch-Fye
 finite temperature algorithm.
The only difference is that in
 Eq.\ [\ref{defM}], instead of the finite temperature 
$(\mbox{\boldmath ${\cal G}$}^0)^{-1} = {\mathbf G}^{-1}+{\mathbf \Sigma}$,
a zero temperature $(\mbox{\boldmath ${\cal G}$}^0)^{-1}$
enters as a $\Lambda \times \Lambda$ matrix. While at half-filling the
finite temperature
$\mbox{\boldmath ${\cal G}$}^0_{mm} (\tau,0)$ decreases
(in terms of the absolute value)  from 
$\tau=0$ to $\tau=\beta/2$  and increases again
from $\tau=\beta/2$ to $\tau=\beta$,
the zero temperature $\mbox{\boldmath ${\cal G}$}^0_{mm} (\tau,0)$
continues to decrease from  $\tau=\theta/2$ to $\tau=\theta$.

With these rather small changes to the QMC code,
PQMC simulations are possible. 
In contrast to the finite temperature QMC, there are no thermal fluctuations.
This is the reason why these PQMC calculations converge much faster
to the ground state properties (as a function of $\theta$)
  than the Hirsch-Fye QMC (as a function of  $\beta$
which corresponds to the   same numerical effort). 
Let us mention that similar projective QMC versions
of the continuous time QMC algorithms  \cite{Rubtsov04a,Rubtsov05a,Werner05a} are possible \cite{Assaad07a}.

\subsubsection{Fourier transformation from $\tau$ to $i\omega_{\nu}$}
\label{Fourier}

In the previous two Sections, we have introduced 
the Hirsch-Fye QMC and projective QMC simulations for the
Anderson impurity model.
In the context of DMFT, this Anderson impurity model is determined self-consistently, see flow diagram Fig.\ \ref{flowDMFT}.
While the Dyson equation [\ref{Dyson}]
of this  self-consistency is formulated in terms
of Matsubara frequencies, 
the QMC calculations of the Anderson impurity model
use a discrete set of $\Lambda$ imaginary time
support points.
Hence, we have to overcome a final technical obstacle:
the Fourier transformation from $\tau$ to $i\omega_{\nu}$
and vice versa.
This is in particular problematic since we have 
only a limited number  $\Lambda$ of $\tau$ points,
which  only allow us to calculate the Green function
for an equal number of Matsubara frequencies by means of a 
discrete Fourier transformation.
On the other hand, the tail at high Matsubara frequencies is 
important since it is responsible for the jump of the Green
function at $\tau=0$: $G_{lm}(\tau=0^+)=-\delta_{lm}+G_{lm}(\tau=0^-)$.

Therefore, a discrete Fourier transformation does not work.
It would
yield a Green function which oscillates considerably
around the correct ${\cal G}(\tau)$. 
Different paths are used in the literature to overcome this obstacle:

Georges {\em et al.}
\cite{Georges96a} do a spline interpolation of ${\mathbf G}(\tau)$,
resulting in arbitrarily many support points and, hence, 
enough  Matsubara frequencies.
Jarrell 
\cite{Jarrell97} extend the number of
Matsubara frequencies by employing
 the iterated perturbation theory result at
high frequencies.

Ulmke {\em et al.}
\cite{Ulmke95b} use a smoothing
procedure which replaces ${\cal G}_{mm}^0(i \omega_{\nu})$ by
\begin{equation}
\tilde{\cal G}_{mm}^0(i \omega_{\nu}) \equiv \frac{\Delta\tau}
{1-\exp[-\Delta \tau/{\cal G}_{mm}(i \omega_{\nu})]} \, .
\label{smoothing}
\end{equation}
This $\mbox{\boldmath $\tilde{\cal G}$}^0$
is Fourier tranformed to imaginary time,
and this Fourier transform is in turn used as the
non-interacting Green function of the Anderson impurity
model.
After the QMC simulation
 yielded the output
$\tilde{G}_{mm}(\tau_\lambda)$, the process is reversed:
From the Fourier transform of $\tilde{G}_{mm}(\tau_\lambda)$, i.e.
$\tilde{G}_{mm}(i \omega_{\nu})$,
 the
inverse of Eq.\ [\ref{smoothing}] yields $G_{mm}(i\omega_{\nu})$.  The new self energy is then
$\Sigma_{mm}(i\omega_{\nu})= {\cal G}_{mm}(i\omega_{\nu})^{-1}-G_{mm}(i\omega_{\nu})^{-1}$.
This approach generates smooth Green functions $G(\tau_\lambda)$
and reproduces the correct $\Delta\tau\rightarrow 0$ limit.

In their implementation, McMahan {\em et al.}
\cite{McMahan03}  use  a constrained fit
to the output QMC impurity Green function $G(\tau_\lambda)$:
\begin{equation}
G(\tau) = \sum_i w_i f_i(\tau) .
\label{fit}
\end{equation}
The basis functions are $f_i(\tau)=
-e^{-\varepsilon_i\tau}/(e^{-\beta \varepsilon_i}\!+\!1)$ and  have
Fourier transforms $f_i(i\omega)=1/(i\omega\!-\!\varepsilon_i)$,
hence allowing for determining $G(i\omega_{\nu})$ at
$N_\omega > \Lambda$ Matsubara frequencies.
The difference to the  spline-fit  is that in Eq. [\ref{fit}]
every fit coefficient  is determined by the local behaviour in
a frequency interval, not by the local behaviour in
an imaginary time interval.

The fit of
Eq.\ [\ref{fit}] has some additional constraints:
 $w_i\ge 0$, $G(0^+)$
is precisely the QMC value, $G(0^+)\!+\!G(\beta^-)\!=\!-1$, and
$\frac{d}{d\tau}G(0^+) \!+\!\frac{d}{d\tau}G(\beta^-)\!=\!{\rm
g}_2$, where ${\rm g}_m$ is
the $(i\omega)^{-m}$ high-frequency moment of $G(i\omega)$.
For the last constraint, the second moment
${\rm g}_2$  is obtained from the
relation $G^{-1}(i\omega)={\cal G}^{-1}(i\omega)-\Sigma(i\omega)$
which implies ${\rm g}_2=g_2+{\rm s}_0$, where these are the
indicated moments of $G(i\omega)$, ${\cal G}(i\omega)$,
and $\Sigma(i\omega)$, respectively. Note that $g_2$ is
known since ${\cal G}$ is input to the QMC,
and for $s_0$ the analytical
high frequency behaviour (known from the
Hartree Fock) is taken:
$s_0 = \Sigma(\omega\!=\!\infty) = \Sigma^{\rm Hartree-Fock}$.
For M degenerate orbitals and without Hund's exchange coupling
this is:
$s_0 = 2M-1/2M [1\!+\!G(\tau\!=\!0^+)] U$.
Using these constraints and $\Lambda/4$  equally spaced
$\varepsilon_i$,  the agreement between Eq.\ [\ref{fit}]
with the QMC data for $G(\tau)$ is optimised.

For a faster convergence, McMahan {\em et al.} \cite{McMahan03}
also separated the self energy 
$\Sigma^{\rm
QMC}(i\omega) = \Delta\Sigma(i\omega) + \Sigma^{\rm Hartree-Fock}$
into a
constant Hartree-Fock contribution 
and a frequency-dependent rest $\Delta\Sigma(i\omega)$.
After every QMC calculation, the authors did computationally inexpensive
iterations during which  $\Delta\Sigma(i\omega)$ was kept fix,
but $\Sigma^{\rm Hartree-Fock}$ and the number of interacting electrons
were adjusted self-consistently until convergence.

For the PQMC simulations discussed in the previous Section,
the Fourier transformation is even more challenging since
the zero temperature Green functions extend from $0$ to $\infty$,
instead of $0$ to $\beta$ at finite temperatures $T=1/\beta$.
With the Green function known only at $\Lambda$ 
discrete $\tau$ points, one has not only to interpolate between the
$\tau$ points as for finite temperatures but also
to extrapolate to $\tau=\infty$.
To this end, Feldbacher {\em et al.} \cite{Feldbacher04}
employed the
maximum entropy method (see Section \ref{SecMEM})
which allows us to calculate from $\Lambda$ support points the
 zero temperature spectral function $A(\omega)$ at real
valued frequencies.
From this spectra function, the Green function at any Matsubara frequency
$\omega_{\nu}$ can be easily calculated as
\begin{equation}
G(  i\omega_{\nu})  =\int d\omega\frac{A(
\omega)  }{i\omega_{\nu}-\omega}.
\end{equation}

\subsubsection{Maximum entropy method}
\label{SecMEM}
Since, in QMC, the Green function is calculated on the imaginary (Matsubara) axis,
we have to do an analytical continuation to real frequencies for
getting the spectral function
$A(\omega)=-\frac{1}{\pi}{\rm Im} G(\omega)$
which is of direct physical interest since it can be measured, e.g.
by photoemission experiments.
Because of the statistical QMC error, the standard approach for
doing the analytical continuation is the maximum entropy method.
In the following the basic concept will be briefly discussed,
for a detailed review by Jarrell and Gubernatis see \cite{MEM}.

Starting point is the Fourier-transform of the spectral representation
of the Green function
 \begin{equation}
   G(\tau)=\int \limits_{-\infty}^{\infty} \! d \omega 
   \frac{e^{\tau (\mu -\omega)}}{1+e^{\beta(\mu-\omega)}} A(\omega).
   \label{GN}
 \end{equation}
This equation already shows that  the analytical continuation
is an ill-conditioned problem: The kernel of Eq.\ [\ref{GN}] 
is very small for large frequencies $\omega$ so that
large changes in $A(\omega)$ have only a small impact on $G(\tau)$.

In this problematic situation, the maximum entropy method is
used. It starts with the
entropy
 \begin{equation}
   S=-\int_{-\omega_0/2}^{\omega_0/2} d\omega \; A(\omega) 
   \ln \big[ A(\omega) \; \omega_0 \big]
 \end{equation}
as the {\em a priori} probability 
  $P(A) \propto \exp (T S)$ for a given spectrum
 $A(\omega)$, with adjustable parameter $T$.
Besides this constant {\em a priori} probability,
$A(\omega)$ has to yield the QMC-calculated $G(\tau)$.
How good this is achieved can be measured by the usual $\chi^2$ value
for the  quadratic difference between 
given (calculated)  $G(\tau)$ and the one obtained from
$A(\omega)$  via Eq.\ [\ref{GN}]. This  $\chi^2$ value
gives the conditional probability for $G(\tau)$ with
a given $A(\omega)$:
 \begin{equation}
   P(G|A) = e^{-\frac{1}{2} \chi^2},
   \label{BedW}
 \end{equation}
Following the Bayes theorem {\em a priori} and
conditional probability yield the {\em a posteriori} probability
for having a spectrum  $A(\omega)$ if $G(\tau)$ was calculated in QMC:
 \begin{equation}
   P(A|G) \propto P(G|A) P(A) \propto e^{T S-\frac{1}{2}\chi^2}.
\label{Eq:prob}
 \end{equation}
The result of the maximum entropy method is the most likely
spectrum $A(\omega)$, maximising $P(A|G)$.
This is a well defined statistical method for doing the
analytical continuation. Because of the statistical error
in the QMC and because Eq.\ [\ref{GN}] is ill conditioned for
large $|\omega|$ the maximum entropy can however not resolve
fine details at large frequencies, i.e. in the Hubbard band.
In contrast,
features at small frequencies, such as height and width of the central peak,
and the overall weight and position 
of the Hubbard bands are reliable.

\subsection{Comparing different DMFT solvers for La$_{1-x}$Sr$_{x}$TiO$_{3}$}
\label{LaTiO3} 

After the discussion of different methods X to solve the DMFT
self-consistency equations in the previous Sections,
let us now compare how this solver reflects in the
LDA+DMFT(X) results for a specific material:  La$_{1-x}$Sr$_{x}$TiO$_{3}$. 
Such a comparison has been carried out by Nekrasov {\em et al.} \cite{Nekrasov00},
and we will recapitulate their findings here.
The stoichiometric compound LaTiO$_{3}$
is a cubic perovskite with a small orthorhombic distortion
(the distorted angle is $\angle$ Ti-O-Ti $\approx
~155^{\circ }$; see \textcite{maclean79}), and is an antiferromagnetic insulator
 below $T_{N}=125$~K \cite{eitel86,gopel}. Above $T_{N}$, or at
low Sr-doping $x$, LaTiO$_{3}$ is a
strongly correlated, but---except for the distortion---simple paramagnet with  one 
3$d$
electron on the trivalent Ti sites. 
Since the aim of this Section is the comparison of DMFT solvers 
we neglect the small orthorhombic
distortion, i.e. consider a cubic structure with the same
volume.
\\

{\it LDA calculations for the cubic crystal structure}

\begin{figure}[tbp]
 \centering
\centerline{\includegraphics[clip=true,width=10.cm]
{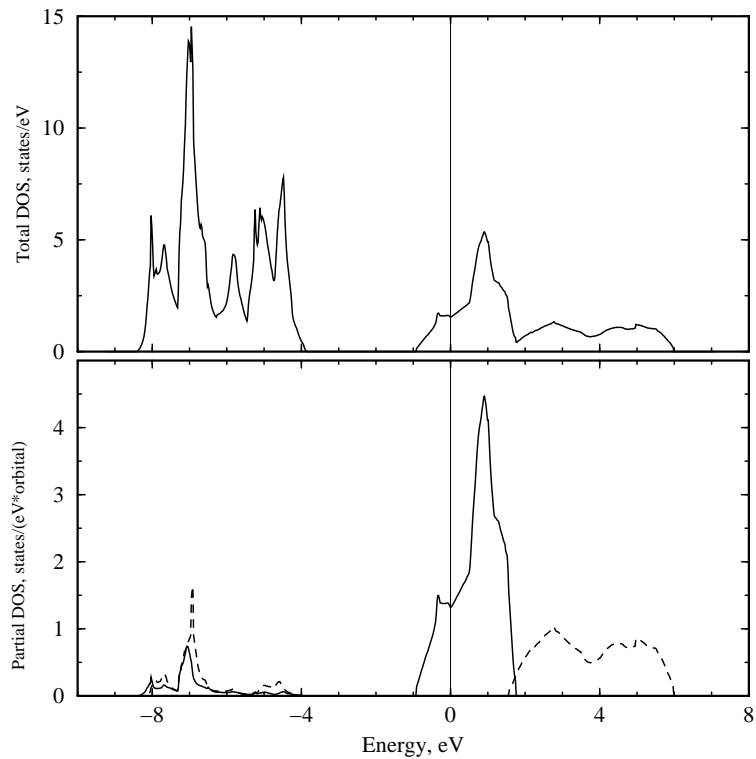}}

 \caption{\void Densities of states of LaTiO$_3$ calculated with LDA-LMTO.
Upper
figure: total DOS; lower figure: partial $t_{2g}$ (solid lines)
and $e_g$ (dashed lines) DOS (reproduced from \textcite{Nekrasov00}).} \label{ldados}
\end{figure}

Fig.\ \ref{ldados} shows the
LDA DOS 
for undoped   LaTiO$_{3}$. Thereby,  
Anisimov {\em et al.} 
\cite{Anisimov97a} and Nekrasov {\em et al.} \cite{Nekrasov00} 
approximated the crystal structure by a cubic one with the same volume.
LDA+DMFT calculations including the  orthorhombic distortion
have been recently carried out by  Pavarini {\em et al.} \cite{Pavarini03,Pavarini05} and 
Craco {\em et al.}
\cite{Craco03c}.
Here however, our main goal is the comparison of different
DMFT solvers in a realistic material calculation. 
For this intercomparison the  approximate cubic structure
is sufficient.

In  Fig.\ \ref{ldados},
the oxygen bands range from $-8.2$~eV
to $-4.0$~eV and are completely filled so that Ti is three-valent. 
The cubic crystal field splits the Ti 3$d$ bands into two
empty $e_{g}$-bands and three degenerate $t_{2g}$-bands.
well separated from the other bands.
For the low-energy behaviour it is
hence possible to restrict ourselves to these degenerate
$t_{2g}$ orbitals within the
 approximation introduced in Section \ref{SimpTMO}, i.e.
using the LDA DOS (Eq.\ [\ref{intg}])
instead of the full one-particle Hamiltonian $H_{\rm LDA}^0$ of
(Eq.\ [\ref{Dyson}]).
We take Sr-doping $x$ into account 
by adjusting the  LDA+DMFT chemical potential to  $n=1-x=0.94$ $t_{2g}$ electrons.

{\it Method matters}

\begin{figure}[tbp]
\centering
\centerline{\includegraphics[clip=true,width=9cm]{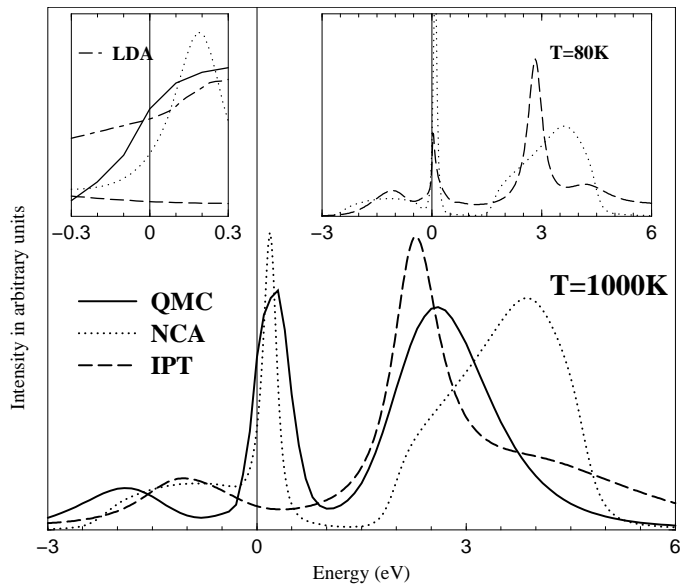}}

\caption{\void Spectrum  of La$_{0.94}$Sr$_{0.06}$TiO$_{3}$
 as calculated by LDA+DMFT(X) at $T=0.1$~eV ($\approx 1000$~K)
and $U=4$~eV employing the approximations X=IPT, NCA and
numerically exact QMC. Inset left: Behaviour at the Fermi level
including the LDA DOS. Inset right:  X=IPT and NCA spectra at
$T=80$~K
 (reproduced from \textcite{Nekrasov00}).}
\label{dmft_latio}
\end{figure}

In Fig.~\ref{dmft_latio}, 
we present the LDA+DMFT(X)  spectrum of
La$_{0.94}$Sr$_{0.06}$%
TiO$_{3}$, calculated  for the impurity solvers
 X=IPT, NCA and QMC.
Qualitatively, 
all three
methods X yield the characteristic three peak structure,
consisting of 
lower Hubbard band, quasiparticle peak and upper Hubbard band. 

Quantitatively, however,
we see pronounced differences:
The IPT quasiparticle peak is very narrow
and hence not visible at high temperatures (in the main panel),
also the shape of the IPT Hubbard bands is different.
NCA is much closer to the (numerically) exact QMC
result than IPT, but nonetheless
underestimates the width of the quasiparticle peak by a factor of
two. NCA
 also violates the Luttinger
pinning of the spectral function \cite{MuellerHartmann89c},
a known deficit  \cite{wernerH} seen clearly in the left inset,
and puts the lower Hubbard band to close to the Fermi energy.
The comparison of Fig.\ ~\ref{dmft_latio}
hence shows that, at least on  a quantitative level,
it {\em matters} which {\em method} is employed as
an impurity solver for the DMFT equations.


\section{Realistic material calculations with DMFT}
\label{results}

In this Section, we  will review  those electronic
structure calculations which have been accomplished
hitherto  using DMFT for actual material calculations.
The results are subdivided  into different
material classes:  $f$ electron systems
in Section \ref{felectrons},
transition metals  in Section \ref{TM}, their oxides in Section \ref{TMO} and other
materials in Section \ref{OM}.
Some highlights are discussed in more detail.
Not discussed are LDA+U calculations,
which would require a review on its own, 
and model DMFT calculations without
material-specific LDA or GW input.

\subsection{$f$ electron systems}
\label{felectrons}

The state-of-the-art
LDA+DMFT calculations for $f$-electron systems
take into account all $spdf$ valence orbitals and
all hybridizations between them, but
restrict the DMFT Coulomb interaction to
the $f$ orbitals in the LDA+DMFT Hamiltonian (Eq.\ [\ref{LDADMFTH}]).
This is a very reasonable starting point since the
$spd$ orbitals are much more extended and hence less
strongly interacting. Whether the interaction 
of the $d$ electrons leads to corrections,
which cannot be completely ignored,
remains however an open question since
presently a DMFT calculation with all 24 interacting
$d$ and $f$ orbitals would be too involved,  
at least when using a more rigorous impurity solver.

Two of the very early successes of 
LDA+DMFT were the calculations
of the Mott transition in plutonium by
Savrasov, Kotliar and Abrahams \cite{SAVRASOV}
and 
of the cerium volume collapse transition  
by Z\"olfl {\em et al.} \cite{Zoelfl01} and Held, McMahan and Scalettar \cite{Held01b}.
Savrasov {\em et al.} \cite{SAVRASOV,SAVRASOV2,Savrasov04a}
studied the $\delta$-phase of Pu which is not well described by LDA, as is evident from the 
underestimation of the Pu volume by 30\%. Within LDA+DMFT, electronic correlations 
drive the system towards a Mott transition. With these strong electronic correlations
and a Coulomb interaction of $U\approx4\,$eV,
the volume of $\delta$-Pu comes out correctly \cite{SAVRASOV}.
Subsequently, Dai {\em et al.} \cite{Dai03}
extended these calculations by including lattice dynamics,
allowing for the determination of
the first LDA+DMFT phonon spectrum. This theoretical prediction  well
agrees with the later experiments \cite{Wong03}.
Concerning the magnetic properties of  $\delta$ Pu, Pourovskii {\em et al.}
\cite{Pourovskii05b}
predicted the absence of dynamical and static magnetic
moments in agreement with experiment.
A detailed review of these activities, which are of possible
relevance for nuclear waste \cite{SAVRASOV}, can already 
be found in
\textcite{Kotliar06}. Hence, we will not discuss these calculations
in more detail here. But let us mention that,
more recently, also americium \cite{Savrasov06a},
 PuCoGa$_5$ \cite{Pourovskii06a}, different
actinide monochalcogenides \cite{Pourovskii05a},
 Pu and Am compounds \cite{Shick06a},
and the rare earth elements Nd and Pr \cite {McMahan05}
have been investigated.
In Ref.\  \cite {McMahan05}, also the effect of the spin-orbit coupling, which is certainly important for $f$ electron systems, was taken into account. 
Also, one of the very first 
LDA+DMFT(Hubbard-I)
calculations was for the mixed valence
4f material TmSe \cite{Lichtenstein98a}.

\subsubsection{Volume collapse transition in cerium}
\label{Sec:Ce} 
\label{Ce}
Let us now discuss in detail elemental cerium, 
the material hitherto
most intensively studied with LDA+DMFT
by various groups:
Z\"olfl {\em et al.} \cite{Zoelfl01},  McMahan {\em et al.}  \cite{Held01b,McMahan03,McMahan05},
Haule {\em et al.} \cite{Haule04},
and Amadon {\em et al.}  \cite{Amadon06a}.
Under pressure, Ce undergoes the volume collapse, or $\alpha$-$\gamma$ transition,
with a volume change of 15\% at room temperature. This transition is isostructural 
within the face-centred-cubic (fcc) lattice structure
and fades away for temperatures above  the critical point at $T\!=\!600\pm 50\,$K \cite{Ce}, for  reviews see \cite{Ce,Benedict93,Holzapfel95,JCAMD}.
The volume collapse reflects in
a dramatic transfer of spectral weight:
In the $\alpha$ phase there is 
a large peak at the Fermi energy; whereas this peak is very much reduced
in the $\gamma$ phase,
albeit there is still some spectral weight at the Fermi energy
as to be expected for a metal.
In accord with these findings,
the optical conductivity is higher in the $\alpha$ phase
where the frequency dependent scattering rate is characteristic
for a Fermi liquid behaviour with an effective mass of about
$20\,m_e$, see
\textcite{Eb01}. 
Concerning the magnetic susceptibility, $\alpha$ Ce 
behaves like a Pauli  paramagnet
at room temperature but has a  Curie-Weiss
form at higher temperatures and in the
$\gamma$ phase \cite{Ce}. 
Despite these dramatic differences in the spectrum, optical conductivity,
and susceptibility, the number
of $4f$ electrons does not change significantly and is close to
one across the $\alpha$-$\gamma$ phase transition,
as was revealed by Myon decay experiments
 \cite{Ce}.

Although some alternative theories have been proposed, see e.g. \cite{Eliashberg98a,Nikolaev99,Nikolaev00}, the general belief is that the $\alpha$-$\gamma$ transition has an electronic origin.
The first electronic theory,
the promotional model \cite{Pauling47a,Zachariasen49a},
which assumed the
 electronic configuration to change
from  $4f^0(spd)^4$ in $\alpha$ Ce to $4f^1(spd)^3$ for
the $\gamma$ phase, was however dismissed since it is at  odds with 
the above mentioned  Myon experiments
and also with LDA calculations 
\cite{Johansson74a}. Instead,
Johansson \cite{Johansson74a} proposed 
  a Mott transition (MT) model 
for the $4f$ electrons,
assuming the $4f$ electrons in the $\alpha$
phase to be  itinerant whereas they are localised (Mott insulating)
in the $\gamma$  phase which has a reduced LDA $4f$ bandwidth.
Of course, $\gamma$ Ce as a material remains metallic 
due to the other ($spd$) electrons.
In subsequent efforts to 
treat this scenario within LDA,
 Johansson \cite{JOHANSSON2} employed
 standard  LDA calculations for the $\alpha$ phase
and  treated  the $4f$
electrons in the $\gamma$ phase  as localised spins.
Qualitatively, this yields similar results as the more
sophisticated  self-interaction corrected
LDA calculation and the LDA+U method which were
later performed \cite{Svane97,Svane94a,Szotek94,Sandalov95,Shick01} along with orbitally polarised calculations
\cite{Eriksson90,Soderlind02,Svane97}.

At the time Johansson was working out the MT scenario,
another phenomenon of electronic correlations
was finally understood:
the Kondo effect.
Based on the  physics
of the Anderson impurity model,
Allen and Martin  \cite{Allen82a,KVC2}
proposed a  Kondo volume collapse
(KVC) scenario for the $\alpha$-$\gamma$ transition
which appeared to be in conflict with the MT scenario.
Both pictures agree that, at the experimental temperatures, the
large volume $\gamma$ phase has strongly correlated (localised)
$4f$ electrons with a 
$4f^1$ moment  and a Curie-Weiss magnetic
susceptibility.
But they differ considerably for $\alpha$ Ce:
The MT scenario of Johansson assumes a weakly correlated
(itinerant) $\alpha$ phase, with a `single'
peak at the Fermi energy in the $4f$ spectrum   as on the left hand side of
 Fig.\ \ref{figLDADMFT}, only with some additional structures.
As in the LDA, this itinerant phase has no
 $4f^1$ moment at all.
The  KVC scenario on the other hand
 assumes continued strong correlation
in the $\alpha$ phase with a three peak structure, including
a central Abrikosov-Suhl resonance and 
two side peaks at considerably higher and lower energies \cite{Liu92c},
as in the middle of Fig.\ \ref{figLDADMFT}. 
While the MT scenario envisages the coexistence of two phases and a 
first order transition between these, the
KVC assumes
the $f$-valence hybridisation to increase upon
pressure so that the Kondo temperature which depends exponentially on
this hybridisation changes dramatically.
This leads to a correlation contribution to the energy
(which is roughly proportional to the Kondo temperature)
with a negative curvature as a function of volume.
Therefore, one can lower the energy with a mixed
low and high volume phase by
 a Maxwell (tangent) construction \cite{Allen82a,KVC2},
indicating a first order transition
similar to the vapour-liquid transition.
Attempts to do realistic material calculations
on the basis of this KVC scenario started with  \textcite{Liu92c}
and were later continued \cite{Noam88,Laegsgaard99}.
 In these  calculations,
the $f$-valence hybridisation is determined from
LDA followed by a many-body treatment of the Anderson impurity
model with the seven $4f$ orbitals.

Studying the Hubbard model and the periodic Anderson model\footnote{While
the KVC
is based on the Anderson impurity model, a more realistic treatment
should start with its periodically extended version since
the $4f$ electrons on every Ce  site interact.}
which are the simplest models for the MT and KVC scenario,
respectively, Held {\em et al.} \cite{Held00a,Held00b} more
 recently stressed the similarity of these
two models within DMFT and, hence,  the MT and KVC scenarios.
Actually, one can integrate out the valence electrons of the periodic Anderson model,
resulting in an effective one-orbital model such as the Hubbard model
but with a frequency dependent kinetic energy term.
Hence, it might not be surprising that also the physics of the two
models is very much the same:
The spectrum shows a three peak structure
with a lower and an upper Hubbard band 
and a central quasiparticle resonance in between.
The local moment of the $\alpha$ phase is screened
at low energies for both the Hubbard and the periodic Anderson model.
This shows that the differences between the MT and the KVC scenario 
are due to the LDA treatment of the $\alpha$ phase:
In the $\alpha$ phase, the MT scenario
misses  electronic correlations
and, therefore, the three peak structure of the spectrum.
While many physical aspects of these models are similar,
there are also some noteable differences.
In particular, there is no  Mott-Hubbard transition in the  periodic Anderson model
at zero temperature, even though the finite temperature behavior is very similar
to that of the Hubbard model including a region of two coexisting
phases, i.e. metal and insulator
 \cite{deMedici05a}.

\begin{figure}[tb]
\centerline{ \includegraphics[width=6.0cm]{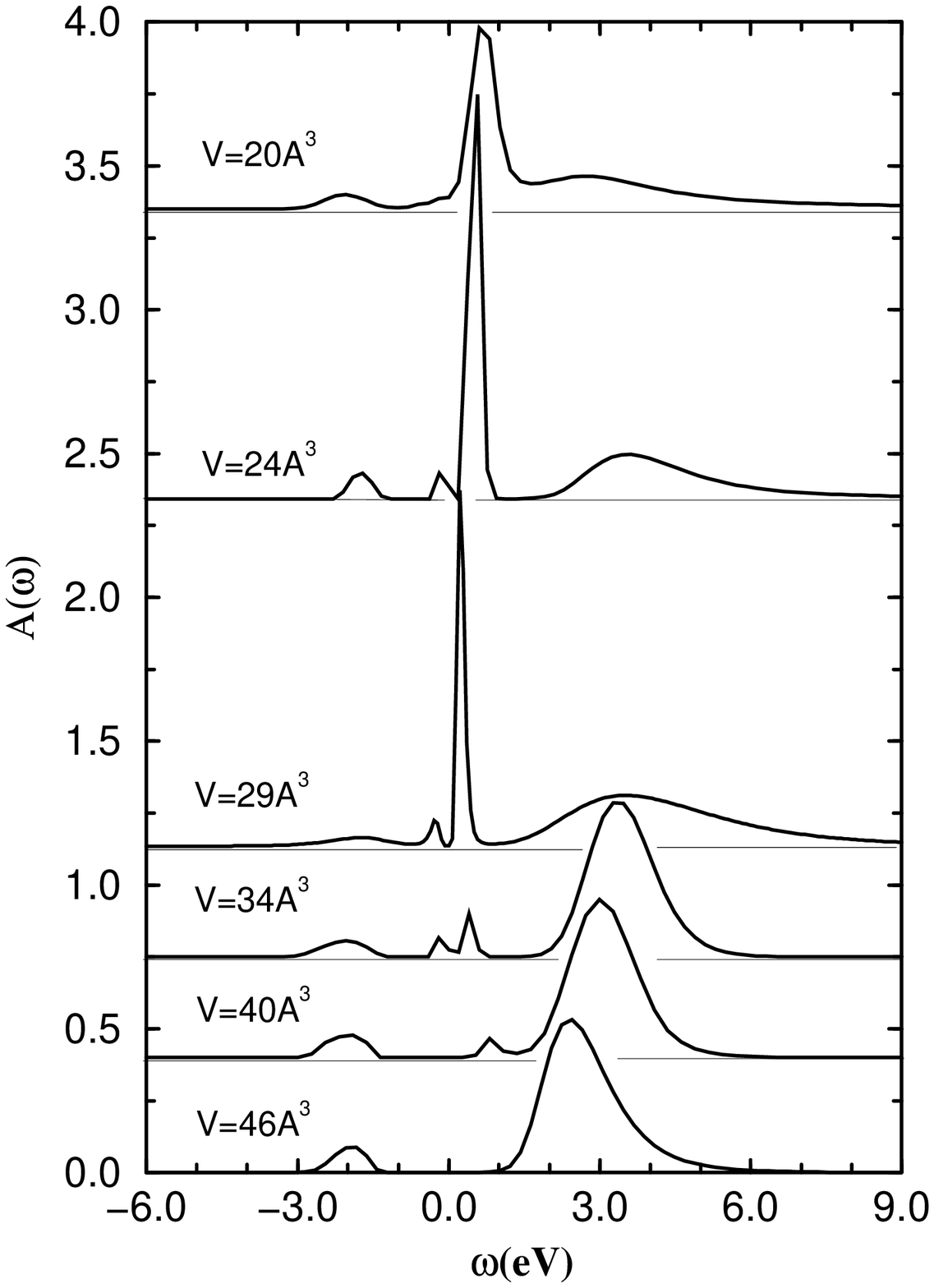}
\hspace{0.1cm} \includegraphics[width=6.1cm]{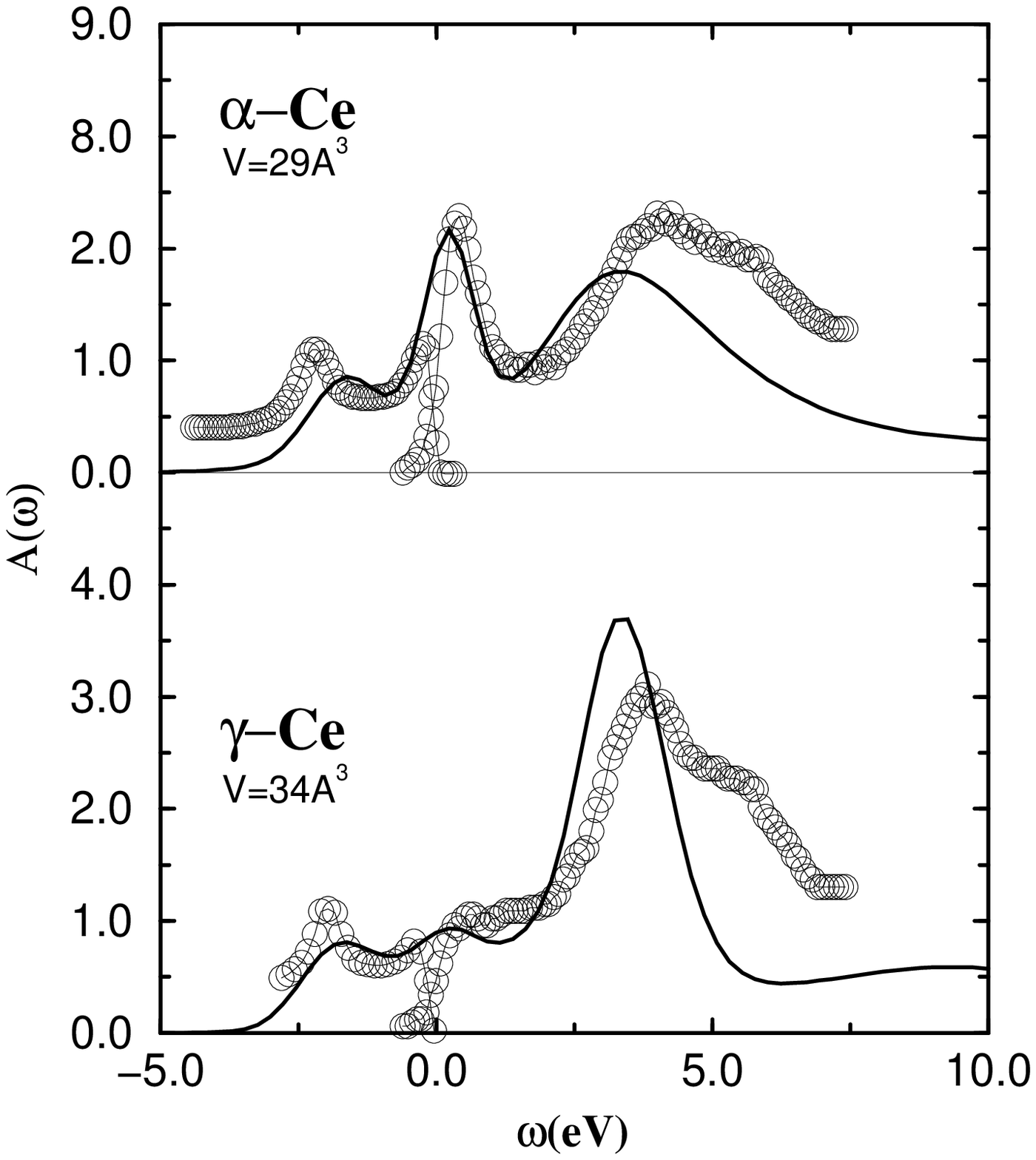}}
\vspace{.5cm}

\caption{\void Left: Evolution of the $4f$ spectral function $A(\omega)$
with volume at $T=632\,$K. The curves at different volumes
are  shifted as indicated by the base line. When
going from small to large volumes, the
weight of the central quasiparticle peak is dramatically
reduced  at volumes corresponding to
the experimental $\alpha$-$\gamma$ transition
from $V=29$  to  $34$\AA$^3$.
Right: Comparison of the parameter-free LDA+DMFT(QMC) spectrum
 with experiment (circles), as collected in Ref.\ \cite{Liu92c}
 (reproduced from \protect\cite{McMahan03}).
\label{fSpectrum}}
\end{figure}

Following these model studies, realistic LDA+DMFT calculations provided for an accurate
{\em ab-initio} description of Ce.
Fig.\ \ref{fSpectrum} (left panel) shows the evolution of
the ${\mathbf k}$-integrated
$4f$ spectral function 
with increasing volume.
At a very small volume, $V\!=\!20\,$\AA$^3$, most
of the spectral weight is in a central quasiparticle
peak or Abrikosov-Suhl resonance at the Fermi energy,
similar as in the LDA.
But, Hubbard side structures are already 
discernible, indicating that there are already electronic correlations,
albeit these are not yet extraordinarily strong.
With increasing volume, more and more spectral
weight is transferred to the Hubbard side bands;
electronic correlations increase.
Approaching the experimental volumes
of the  $\alpha$-$\gamma$ transition, which occurs 
between $28.2$ and $34.4\,$\AA$^3$ at room temperature,
the three peak structure becomes
much more pronounced: We see a sharp
quasiparticle  resonance
at $V=29\,$\AA$^3$.
From the experimental $\alpha$ to the experimental $\gamma$ phase
volume (from $V=29\,$\AA$^3$ to $34\,$\AA$^3$ in Fig. \ref{fSpectrum}), 
the weight of the quasiparticle peak shrinks dramatically
and fades away at even larger volumes.
This large volume phase with a two peak structure can also
be described by LDA+DMFT(Hubbard-I) and LDA+U calculations
(not shown).
Across the transition from  $\gamma$ to $\alpha$ Ce, also 
the  local magnetic moment 
$\langle m_Z^2\rangle$  is  reduced by 5\%
\cite{McMahan03} but not lost,
in contrast to the Mott transition scenario
of Johansson \cite{JOHANSSON2}.

Adding the $spd$ valence spectrum to the $4f$ spectrum and multiplying with the
Fermi (inverse Fermi) function, the total LDA+DMFT spectrum
is compared to the photoemission spectroscopy
(PES) spectrum  \cite{Wieliczka84} below the Fermi energy
and to  the Bremsstrahlung isochromatic
spectroscopy (BIS) \cite{Wuilloud83a} spectrum above the Fermi energy
in Fig.\ \ref{fSpectrum} (right panel).
 The agreement between
theory and experiment is very good. Note, that
there are no free parameters in the
LDA+DMFT(QMC) results of McMahan, Held and Scalettar 
\cite{Held01a,McMahan03} since the $f$-electron Coulomb interaction
and the double counting correction have been determined by
constrained LDA calculations and the experimental resolution has
been taken from \cite{Zoelfl01}.
 Particularly good is the
agreement of the quasiparticle peak  around the Fermi energy for both $\alpha$
{\it and} $\gamma$ Ce, but also the position of the upper and lower Hubbard
bands are approximately correct.  The biggest differences can be found in the
upper band which is broader in experiment and has some inner structure.
It was argued \cite{Zoelfl01,McMahan03}
that this is due to the Hund exchange
interaction which was neglected in both, LDA+DMFT(NCA)  \cite{Zoelfl01}
and
LDA+DMFT(QMC)\cite{McMahan03}, calculations.
This exchange interaction has only a minor effect for the
occupied states since these consist mainly of single occupied configurations.
But, it is important for the upper Hubbard band because the Hund exchange
splits these doubly occupied states into multiplets.

\begin{figure}[tb]
 \includegraphics[width=6.3cm]{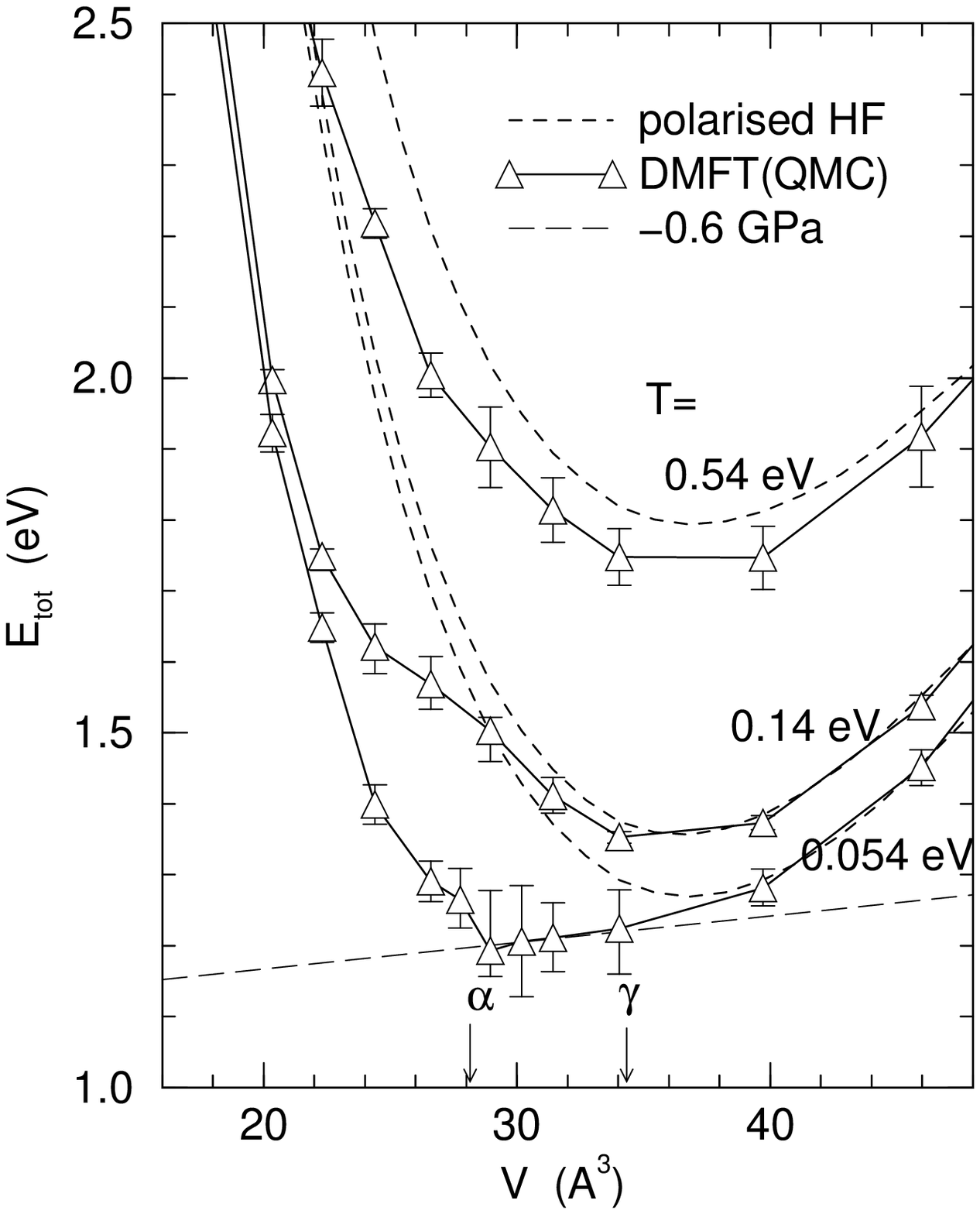}
 \includegraphics[width=6.2cm]{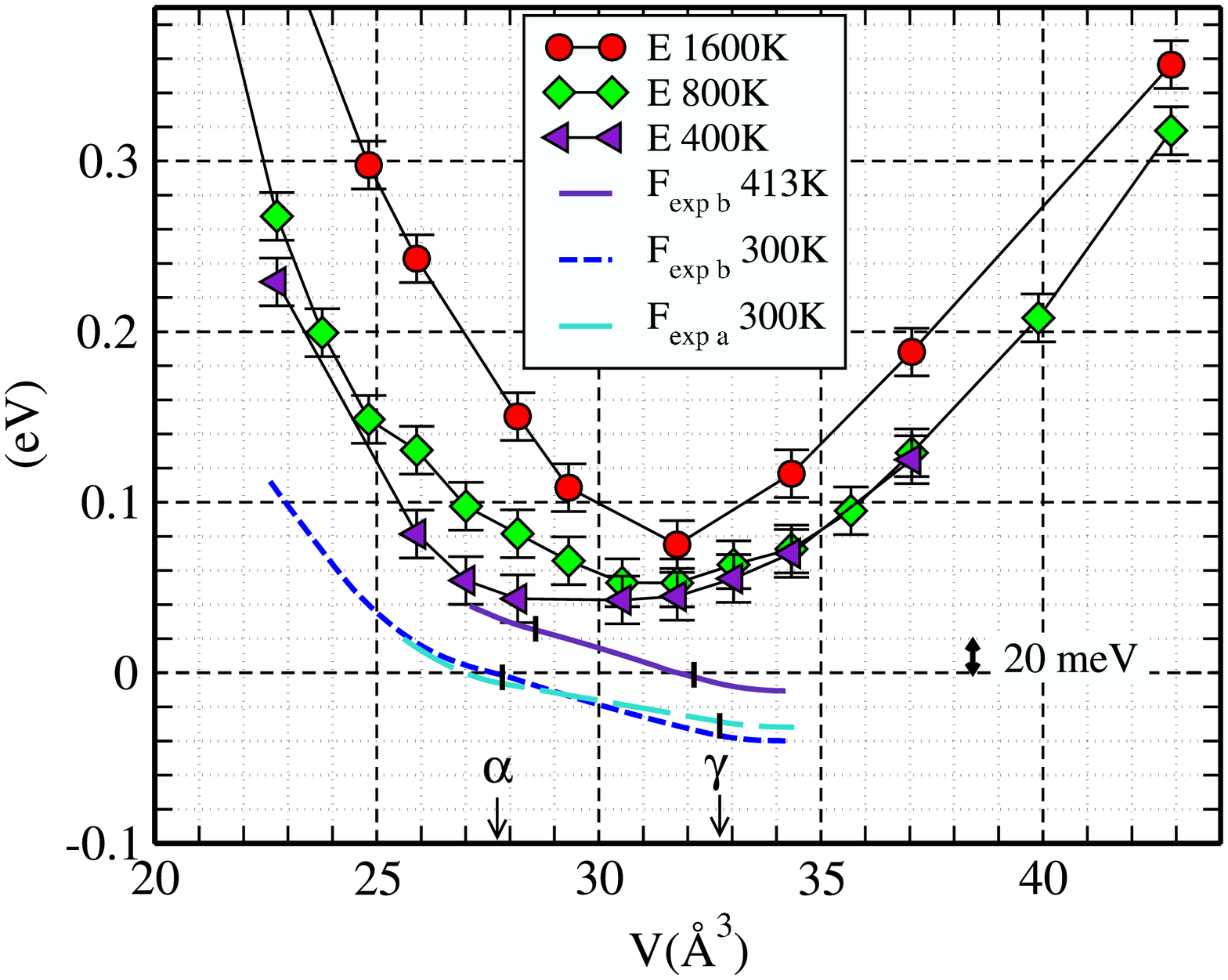}
\caption{\void Left:
Total LDA+DMFT(QMC) and LDA+DMFT(polarised HF)
 energy as a function
of volume at three temperatures. The
negative
curvature of the correlation energy  results in
the development of a side structure, visible as a deviation from the
 LDA+DMFT(polarised HF)
 energy.
 The long dashed line is the curve which
corresponds to the pressure of the $\alpha$-$\gamma$ transition:
$E=-P_{\rm exp}V$ (reproduced from \protect\cite{McMahan03}).
Right: Results for the total energy from \cite{Amadon06a}, including also  the experimental free energy. (reproduced  from \cite{Amadon06a}).
\label{Etotfig}}
\end{figure}

With decreasing volume, we saw the development of a
quasiparticle peak in the $4f$ electron spectrum.
This quasiparticle physics is associated with an energy
gain which cannot be captured by LDA 
and which gives rise to a correlation energy with a negative curvature
\cite{Held01a,McMahan03} at low temperatures (not shown).
Hence, it also reflects in the total energy shown in Fig.\ \ref{Etotfig}:
At high temperatures and for the  LDA+DMFT(polarised HF)
results, we see a simple $E_{\rm tot}(V)$ curve with a single
minimum, the equilibrium volume. 
But at lower temperatures, the negative curvature of the
correlation energy leads to  a side
structure. Since all energy contributions except for the
correlation energy have a positive curvature,
the  negative curvature
of the correlation energy is largely compensated, and we see  
a very shallow region
at $T=0.054\,$eV. Within the numerical error bars, it is difficult to
decide whether we already have a negative curvature or whether slightly lower 
temperatures are needed.
A negative curvature of the total energy
will give rise to a Maxwell construction and a first order
phase transition. The region in which the LDA+DMFT(QMC) results  show these
tendencies are in agreement with the experimental $\alpha$-$\gamma$ 
transition which is marked by the arrows in  Fig.\ \ref{Etotfig}.
 The slope of the shallow minimum is also
consistent with an experimental pressure of $-0.6$ GPa (long-dashed line)
given by the  $\alpha$-$\gamma$ transition pressure 
extrapolated to $T\!=\!0$ \cite{JOHANSSON2}.

\begin{figure}[tb]
\centerline{ \includegraphics[width=7cm]{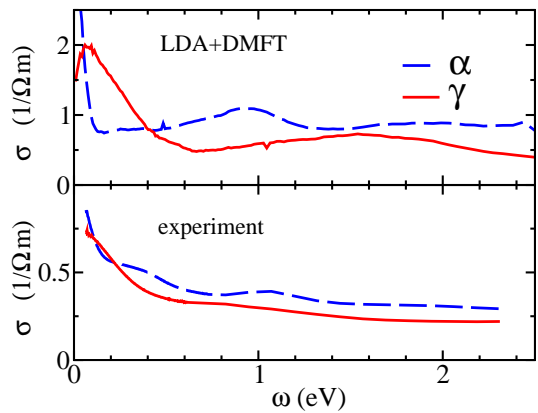}}
\caption{\void Comparison of  LDA+DMFT optical conductivity 
(top panel; \textcite{Haule04}) with experiment (lower panel; \textcite{Eb01}).
The LDA+DMFT spectra are for 580 K ($\alpha$-Ce)
and 1160 K ($\gamma$-Ce); the experimental temperature was 
5 K ($\alpha$-Ce) and 300 K ($\gamma$-Ce), respectively.
 (reproduced from \protect\textcite{Haule04}).
\label{Ce:OC1}}
\end{figure}

Since vertex corrections do not contribute to the optical conductivity within DMFT \cite{Pruschke93a,Pruschke93b}, 
the optical conductivity can be calculated directly from
the DMFT ${\mathbf G}({\mathbf k},\omega)$ and the dipole matrix elements.
Haule {\em et al.} \cite{Haule04} calculated these 
dipole transition matrix from the LDA wave functions and
 used  the one-crossing approximation (OCA)
for solving the
auxiliary DMFT impurity model.
The resulting optical conductivity
 in Fig.\ \ref{Ce:OC1} shows   shows a clear Drude peak for the $\alpha$ phase,
while for the $\gamma$-phase, 
the conductivity is much smaller for  $\omega \rightarrow 0$ and no  
Drude peak is discernible.
The basic features of the theoretical optical conductivity
 agree with experiment (lower panel of Fig.\ \ref{Ce:OC1}).
Hence, we can altogether conclude that LDA+DMFT correctly 
describes the thermodynamic and spectral
properties of the $\alpha$-$\gamma$ transition, as well as the
$4f$ occupation of $n_f\approx 1$ in the vicinity of the transition
\cite{McMahan03}.


\subsection{Transition metals}
\label{TM}

\subsubsection{Ferromagnetism in Fe and Ni}
\label{Sec:FeNi}

Electronic structure calculations for transition metals
and transition-metal--transition-metal alloys 
so-far concentrated mainly on the ferromagnets
Fe and Ni, starting with early 
LDA+DMFT calculations by Drchal {\em et al.}
\cite{Drchal99,Drchal03} and
Lichtenstein {\em et al.} \cite{FeNi}.
Since some aspects of ferromagnetism in Fe and Ni, 
in particular the ferromagnetic moment, are well 
described by conventional LDA calculations,
the question is: Are
transition metals  strongly correlated as the importance of 
the 
$3d$ orbitals suggest or not? In other words: Is an LDA+DMFT calculation  necessary
for iron and nickel or is LDA sufficient?

\begin{figure}[tb]
{\hspace{-0.5cm} \includegraphics[clip=true,width=7.099cm]{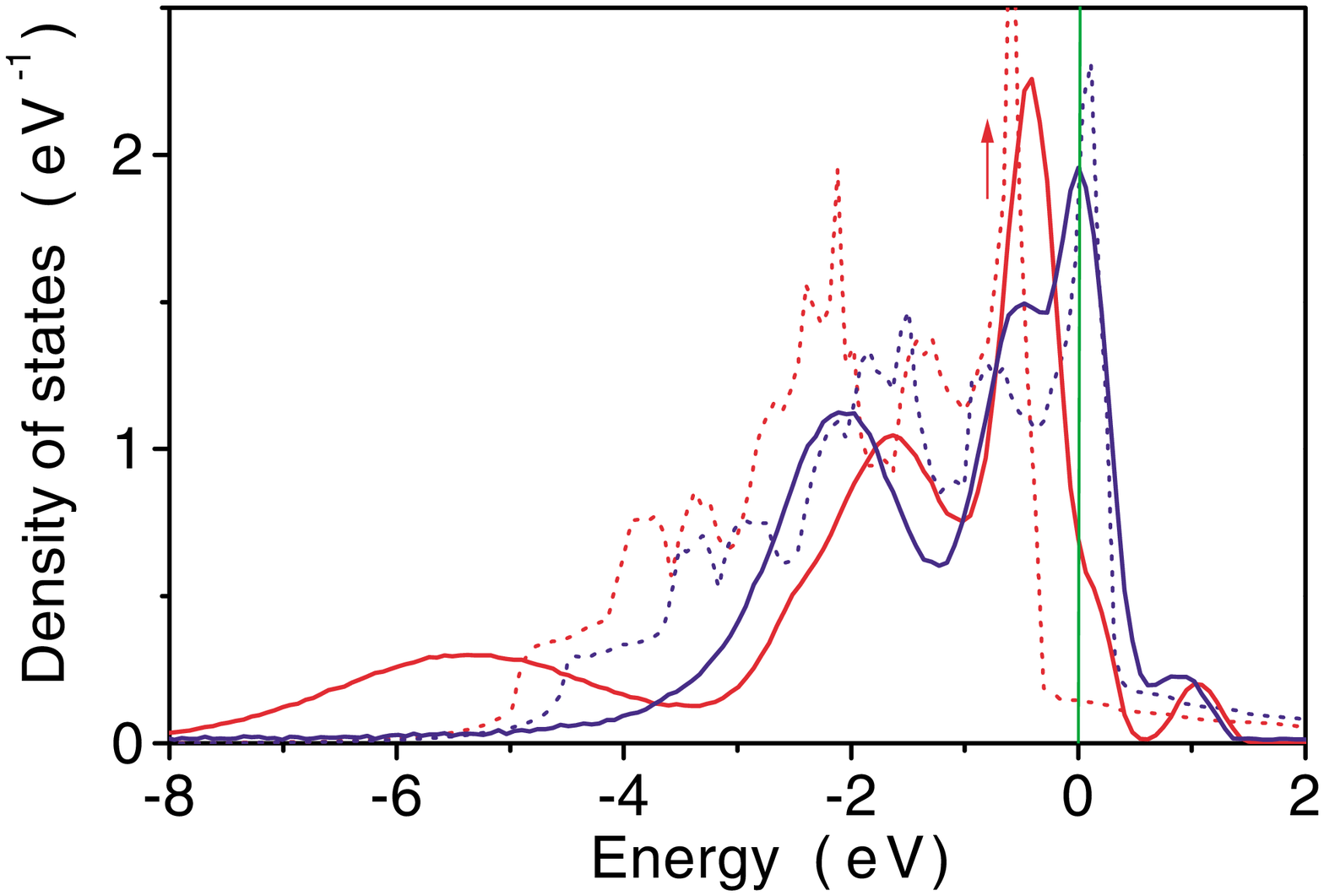}\\
\vspace{-5.3cm}

\hspace{6.11cm} \includegraphics[clip=true,width=7.8cm]{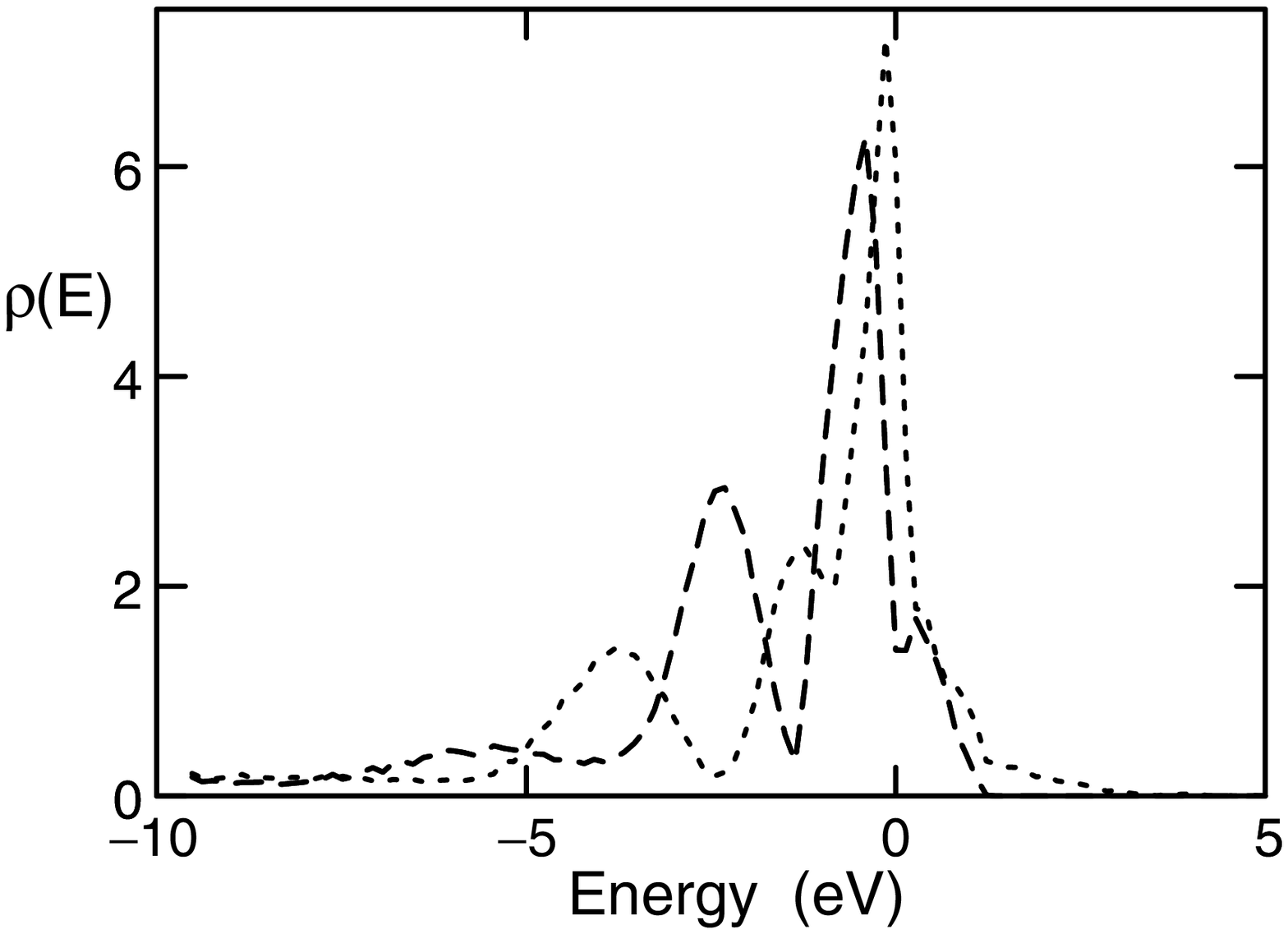}
}
\caption{\void \label{Fig:Ni}
Spectrum  ($\bf k$-integrated) of Ni [left: LDA+DMFT (solid lines), spinpolarised LDA (LSDA) (dotted lines); right: GW+DMFT].
The two lines represent the minority- and majority-spin spectrum respectively. 
At roughly -6eV, a satellite peak is clearly visible
in the majority-spin spectrum
(reproduced from \protect\cite{FeNi} and \protect\cite{Biermann03}, respectively).}
\end{figure}

Certainly not described by LDA  is the famous -6$\,$eV satellite in Ni. Using LDA+DMFT,
Lichtenstein, Katsnelson and Kotliar \cite{FeNi,LichtensteinCP} 
reinvestigated this element, and 
did indeed find
  a satellite at about -6$\,$eV,  see Fig.\ \ref{Fig:Ni} (left panel).
This spectral feature could hence be explained 
by LDA+DMFT as a Hubbard band in the majority-spin spectrum.
Later, this finding was also confirmed  by  GW+DMFT
calculations by Biermann,  Aryasetiawan  and Georges
\cite{Biermann03,BiermannCP04,AryasetiawanCP04}, reporting similar results as LDA+DMFT for Ni,
see Fig.\ \ref{Fig:Ni}.

\begin{figure}[tb]
\centerline{\includegraphics[clip=true,width=7.5cm]{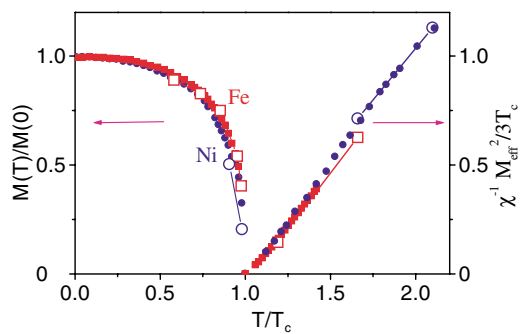}}
\caption{\void \label{Fig:FeNi}
Magnetisation below and magnetic susceptibility above
the Curie temperature for iron (squares) and Ni (circles). 
The LDA+DMFT results (open symbols) are compared to experiment
(full symbols) (reproduced from \textcite{FeNi};  experiments
from  \textcite{Vonsovsky74,Altmann00a}).
 }
\end{figure}

Along with the satellite peak,
 Lichtenstein {\em et al.} \cite{FeNi,LichtensteinCP} 
found for the paramagnetic phase 
a Curie susceptibility 
which indicates
the presence of local (unordered) magnetic moments
of size 3.09 and 1.50 $\mu_B$ for Fe and Ni, respectively.
This Curie susceptibility is in agreement with experiment,
see Fig.\ \ref{Fig:FeNi}. Along with the satellite, the local moment
is clearly a correlation
effect, since within L(S)DA the local magnetic moment
fades away with the magnetisation
Similarly, LDA+DMFT calculations for $\alpha$ and $\gamma$ Ce yield
 local magnetic moments in
the paramagnetic phase.
The absolute value for the Curie temperature
is somewhat overestimated, to a lesser extent in Ni (10\%) than in Fe (80\%).
This is (i) because the DMFT neglects non-local correlations such as spin waves and (ii) because the LDA+DMFT calculations 
employed a Z$_2$-symmetric Hund's exchange instead of a
SU(2)-symmetric so that  transverse spin fluctuations are suppressed.
Let us add that the general features of the 
susceptibility in   
Fig.\ \ref{Fig:FeNi} have
 been found  for the  ferromagnetic transition
of a simple model:  the single band Hubbard model on a fcc lattice
 \cite{Ulmke98a}.

More recently, Minar {\em et al.} \cite{Minar05a,Minar05b}
employed the Korringa-Kohn-Rostoker (KKR) method
for LDA+DMFT studies of Fe, Co, Ni and   the alloy
Fe$_x$Ni$_{1-x}$, describing -among others- the Fano effect 
in these ferromagnets,
and Grechnev {\em et al.} \cite{Grechnev06a}
studied  Fe, Ni and Co surface spectra.
Ferromagnetic Ni was also at the centre of the realistic Gutzwiller calculations
by B{\"u}nnemann {\em et al.} \cite{Buenemann97a,Buenemann97b,Buenemann98a}; for a summary of
these results see \textcite{Weber01} and, supplementarily,
\textcite{Buenemann02a} for the  angular resolved spectrum and
\textcite{Ohm02a} for the total energy.
LDA+DMFT calculations for magnetic multilayers
of transition metals
have been carried out recently by Chioncel {\em et al.} 
\cite{Chioncel03,Chioncel04}; also note Section \ref{Sec:HMFM}
discussing  half-metallic ferromagnetism in Heussler alloys.

\subsection{Transition metal oxides}
\label{TMO}

Transition metal oxides are an ideal laboratory for the study of electronic
correlations in solids, showing a rich spectrum of physical phenomena,
ranging from the Mott-Hubbard transition in  V$_2$O$_3$, 
to  high temperature superconductivity (among others in Sr$_2$RuO$_4$)
and the heavy Fermion behaviour in LiV$_2$O$_4$. 
The  materials mentioned and others have been studied meanwhile by LDA+DMFT.
 In these materials, the $3d$ bands are comparatively
narrow with width $W\approx~2-3\,$~eV so that electronic
correlations, induced by the local Coulomb interaction
$\bar{U}\approx~3-5\,$~eV, are strong.
Hence, in Fig.\ \ref{LDALDAULDADMFT}, transition metal oxides are neither 
in the  weakly correlated  region ($\bar{U}/W\ll~1$)
nor is the opposite limit ($\bar{U}/W\gg1$)  appropriate.
These materials are in the `in-between' regime,
$\bar{U}/W={\cal O}(1)$. If the $3d$ transition metal oxide is metallic
this phase is strongly correlated
 with a {\em quasiparticle} peak at the Fermi energy and Hubbard side bands.
But depending on the material, also Mott insulating 
behaviour and  the Mott-Hubbard metal-insulator transition can occur.

 In all LDA+DMFT calculations for transition metal oxides, 
the LDA+DMFT Hamiltonian (Eq.\ [\ref{LDADMFTH}]) was restricted to
the low-energy orbitals at the Fermi energy, typically the $t_{2g}$
or $e_g$ orbitals.
These Wannier orbitals represent mixtures between mainly 
the transition metal $d$ and the oxygen $p$
orbitals. So far, LDA+DMFT calculations have not succeeded
in taking into account a larger basis including
oxygen $p$ orbitals. The na\"ive inclusion of
non-interacting oxygen orbitals result
in strong deviation from an integer occupation
of the interacting $d$ orbitals.
Consequently, electronic correlations are too weak
\cite{Held03b}. This should be overcome if the
$p$-$d$ and the $p$-$p$ interactions are 
included. However, such calculations are computationally
very demanding  presently.
 Often, also the simplification for transition metal oxides
which we discussed in Section \ref{SimpTMO} and which
allows us to do the DMFT calculation with the LDA DOS only
was employed. For non-cubic systems this is an approximation,
which is however very reasonable as long as the crystal is not too strongly distorted
from the cubic symmetry. Typically,  every transition metal ion is still surrounded
by an octahedron of oxygen ions. However, this octahedron is then not
perfect anymore, but tilted and distorted.
Consequently, there are orbital off-diagonal elements between
the $e_g$ and $t_{2g}$ orbitals in the LDA.

In the following,  LDA+DMFT calculations for some
 transition metal oxides will be presented in detail.
 Besides these studies, LDA+DMFT has been also applied to
 Cr$_2$O$_3$ for which Craco, Laad and M\"uller-Hartmann  
\cite{Craco03b,Laad03a} 
analysed  orbital correlations and the orbital Kondo effect,
Sr$_2$(Ba$_2$)VO$_4$ under pressure - a potential $d^1$ superconductor \cite{Arita06d},
NiO  \cite{Lichtenstein98a,Savrasov03a,Ren06a,Wan06a,Kunes06a}, YTiO$_3$ \cite{Pavarini03},
TiOCl \cite{SahaDasgupta05a,Craco05a},
Tl$_2$Mn$_2$O$_7$ \cite{Craco06a},
MnO,  FeO and CoO \cite{Wan06a},
as well as the Verwey transition in Fe$_3$O$_4$
\cite{Craco05b}.

\subsubsection{Ferro-orbital order in LaTiO$_3$}
\label{Sec:LaTiO3}
LDA+DMFT calculations started with the investigation of
La$_{1-x}$Sr$_{x}$TiO$_{3}$ by Anisimov {\em et al.} \cite{Anisimov97a},
who used the IPT method as a DMFT solver. Subsequently,
Z\"olfl {\em et al.} \cite{Zoelfl00} repeated these calculations employing
LDA+DMFT(NCA), and Nekrasov {\em et al.} \cite{Nekrasov00} using 
LDA+DMFT(QMC). The latter authors also compared the application of 
different DMFT solvers, approximative and numerically exact ones,
for a realistic material calculation, i.e. doped LaTiO$_{3}$
which is a strongly correlated metal close to a Mott-Hubbard metal-insulator transition.
The results showed
that  the method for solving the DMFT equation matters,
 as has already been discussed in Section \ref{LaTiO3}
with Fig.\
\ref{ldados} showing the spectrum of lightly doped LaTiO$_3$.
Very recently, Craco {\em et al.} \cite{Craco03c} performed new LDA+DMFT(IPT) calculations for LaTiO$_3$,
including symmetry breaking. Pavarini {\em et al.} \cite{Pavarini03} and Craco {\em et al.} \cite{Craco03c,Craco06b} reported a ferro-orbital order. This is of particular interest since it rules out the orbital-liquid picture of Khaliullin {\em et al.} \cite{Khaliullin00,Khaliullin01}.
The LDA+DMFT results hence provided for an important piece
of information concerning the controversial debate on the physics of LaTiO$_3$.

Recently, also SrTiO$_3$/LaTiO$_3$ heterostructures have
 been studied
by LDA+DMFT and PES  \cite{Takizawa06a}.
Such  transition metal oxide heterostructures
promise to be a vivid area of research in the future,
 and LDA+DMFT can provide for the necessary 
theoretical support.

\subsubsection{Mott-Hubbard transition in  V$_2$O$_3$}
\label{V2O3}

\begin{figure}[tb]
\hspace{-3.7cm}\includegraphics[clip=true,width=11.5cm]{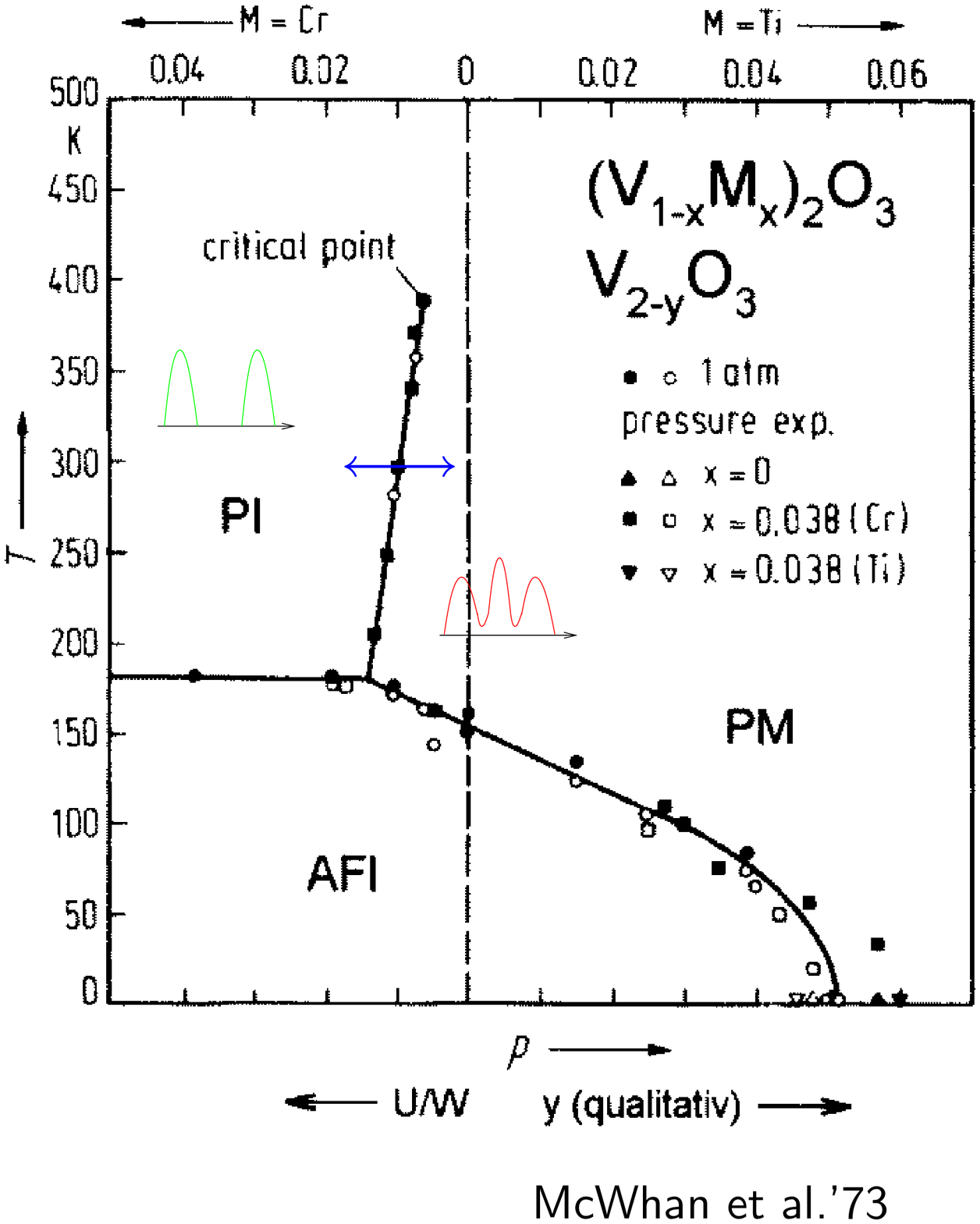}
\vspace{-5.2cm}

\hspace{6.4cm} \includegraphics[clip=true,width=6.6cm]{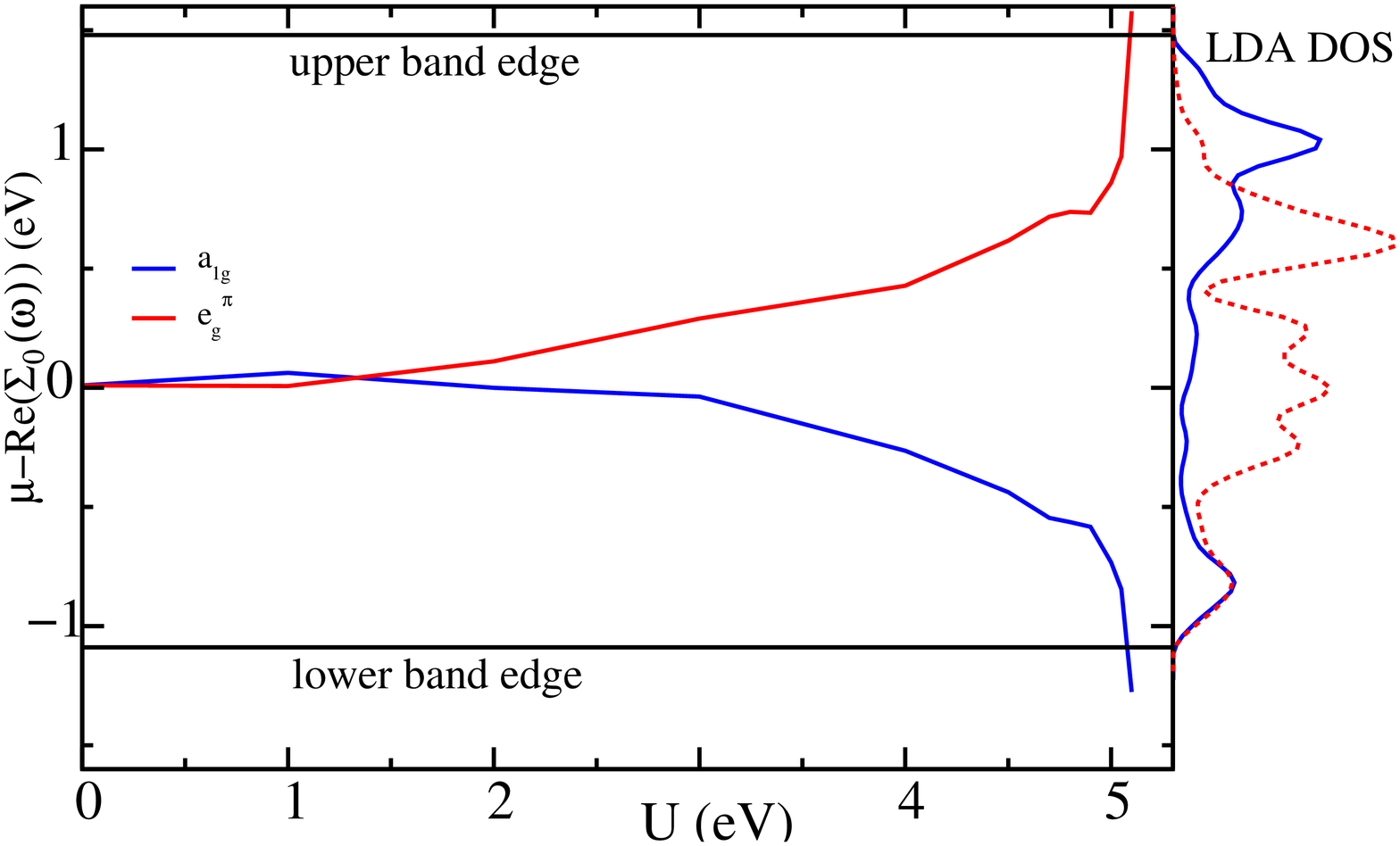}
\vspace{.3cm}

\caption{\void \label{Fig:V2O3}
Left: Phase diagram of V$_2$O$_3$ as a function of Cr
and Ti doping and/or pressure $P$. Within the 
paramagnetic phase a Mott-Hubbard transition 
between metallic (PM) and insulating phase (PI) occurs
upon increasing $U/W$ (reproduced from \cite{mcwhan73b}).
Right: LDA+DMFT result for the effective chemical potential
$\mu-{\rm Re} \Sigma (\omega_0)$ for the $a_{1g}$ and $e_g$
orbitals. (reproduced from \cite{Keller04a})}
\end{figure}

\begin{figure}[tb]
\centerline{\includegraphics[clip=true,width=6.5cm]{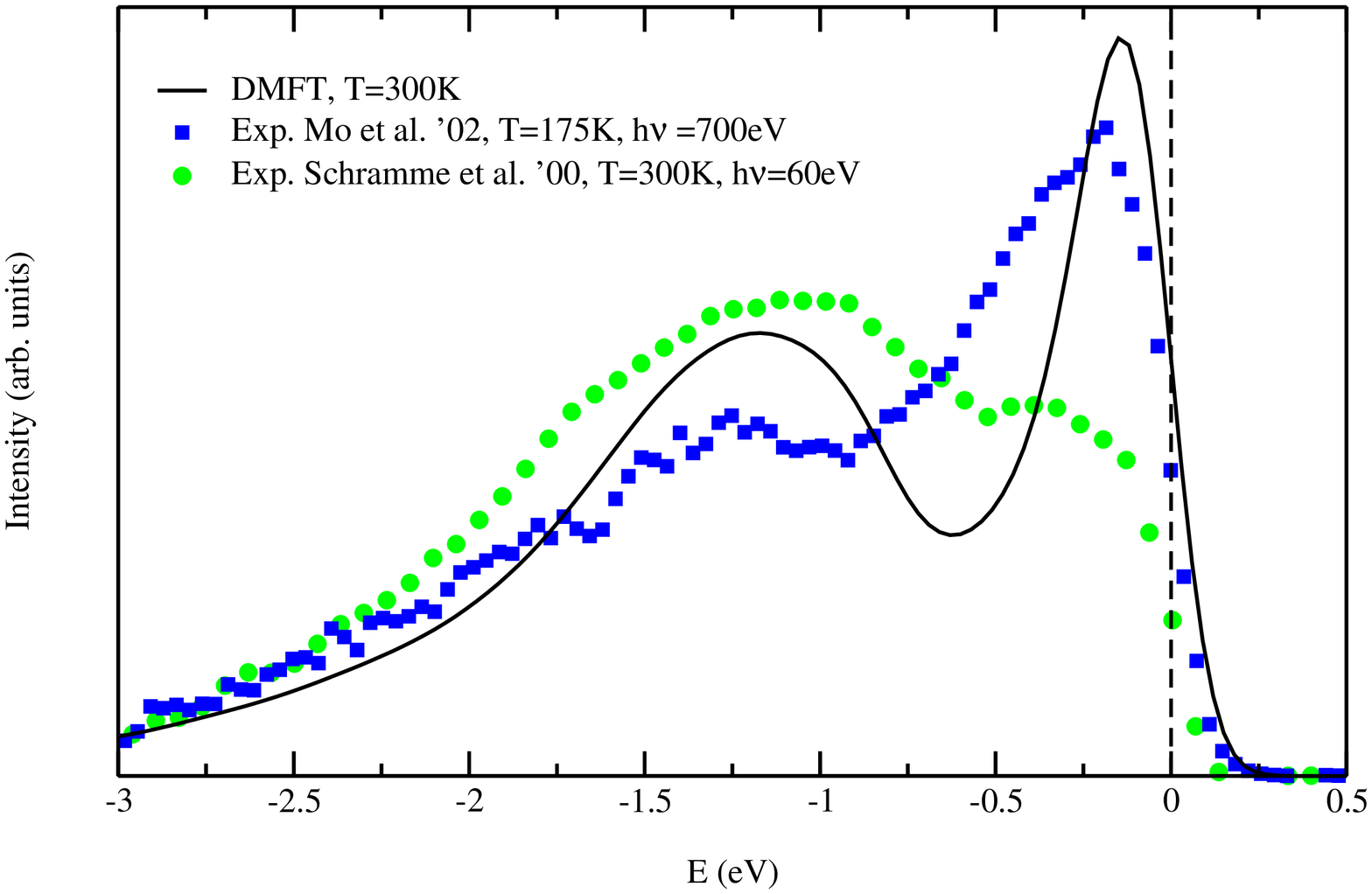}\includegraphics[clip=true,width=6.5cm]{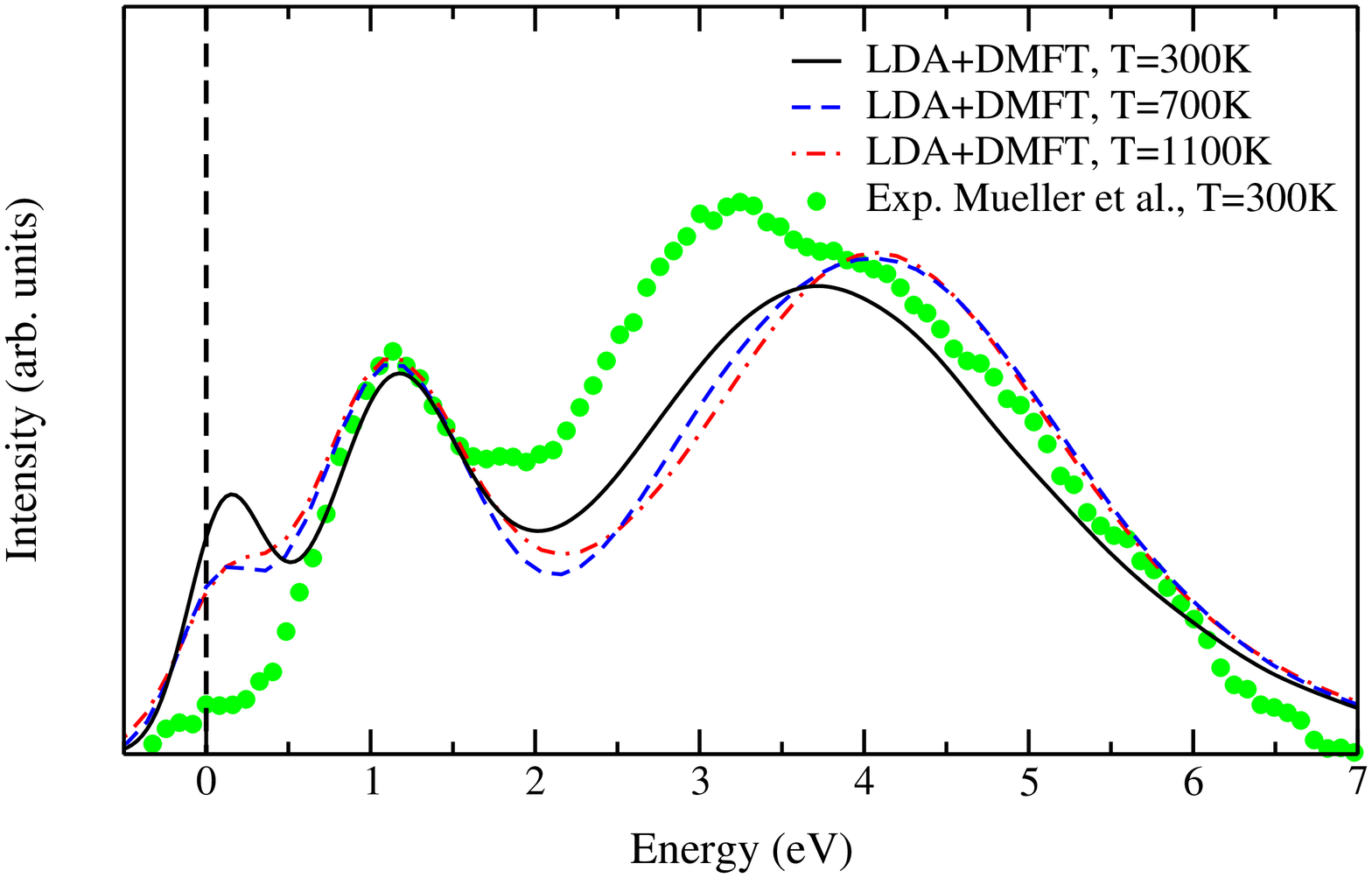}}
\caption{\void \label{Fig:V2O3spectrum}
Comparison of the LDA+DMFT(QMC) spectra with photoemission (left; \cite{Schramme00,Mo02})
and X-ray absorption (right; \cite{Mueller}). The photoemission experiments are for two different photon energies
 (reproduced from \cite{Keller04a}).}
\end{figure}

A particularly important system is V$_2$O$_3$ which undergoes the famous Mott-Hubbard 
metal-insulator transition \cite{mcwhan70,mcwhan73b},
see the phase diagram Fig.\ \ref{Fig:V2O3}. Held {\em et al.} \cite{Held01a,Keller04a} investigated  paramagnetic
V$_2$O$_3$ and Cr-doped  V$_2$O$_3$
 by LDA+DMFT(QMC), describing a Mott transition at a reasonable strength of the Coulomb interaction.
The authors 
reported reasonable
 agreement with photoemission spectroscopy  (PES) experiments by Schramme {\em et al.} \cite{Schramme00},
 see Fig.\  \ref{Fig:V2O3spectrum},
as well as with the experimentally expected
spin and orbital configuration \cite{park}.
Later,  Mo {\em et al.} \cite{Mo02,Mo06a} reinvestigated the 
PES spectrum of  V$_2$O$_3$ by 
new bulk-sensitive PES at the Spring-8 synchrotron.
Their results represent a big step forward on the experimental side since the V$_2$O$_3$ quasiparticle
peak could be resolved for the first time. The PES results
are also in better 
qualitative agreement with  LDA+DMFT(QMC),
but still show a  broader quasiparticle peak with
more spectral weight than theoretically expected.

The LDA+DMFT(QMC) results \cite{Held01a,Keller04a}
also include important differences to the Mott-Hubbard
transition in a one-band Hubbard model.
While the latter is characterised by the  divergence of the effective mass
(or  vanishing quasiparticle weight
$Z\rightarrow 0$), the effective mass for the
$a_{1g}$ band remains finite at the transition in 
 V$_2$O$_3$, only that of the $e_g$ band diverges. 
This reflects the more complicated nature of the Mott-Hubbard
transition in multi-orbital systems. 
As a detailed analysis by
Keller {\em et al.} \cite{Keller04a,Held05a} revealed, the
$a_{1g}$ orbital becomes insulating since the 
effective chemical potential $\mu-{\rm Re} \Sigma (\omega_0)$
moves out of the band edges of the non-interacting DOS,
see Fig.\ \ref{Fig:V2O3} (right panel).
Hence close to the Fermi energy, the
 $a_{1g}$ band behaves as if it becomes band-insulating at the transition. 
Shifts of the  $a_{1g}$ effective chemical potential
were also reported for BaVS$_3$ \cite{Lechermann05a},
albeit not leading to insulating  $a_{1g}$ bands in the
paramagnetic phase.
For BaVS$_3$, these shifts and corresponding changes of the orbital occupation explain the the nature of the charge density wave.

\begin{figure}[tb]
\vspace{-1cm}

{\includegraphics[clip=true,width=6.5cm]{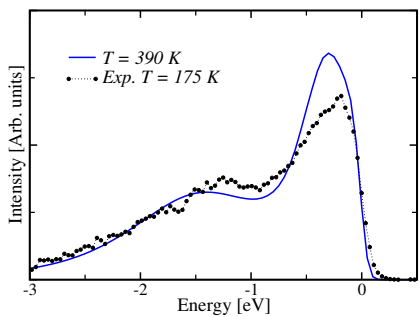}}
\vspace{-1.5cm}

\caption{\void \label{Fig:V2O3Ham}
Same comparison as on the left hand side of Fig.\ \ref{Fig:V2O3spectrum}, but
using the NMTO $t_{2g}$ Hamiltonian instead of the LDA DOS and slightly different
Coulomb interaction parameters (reproduced from \cite{Poteryaev07}).}
\end{figure}

Another important LDA+DMFT prediction was that  the Mott gap is
filled with spectral weight upon increasing temperature.
This characteristic feature 
 has been confirmed recently  in PES experiments by  Mo {\em et al.}
\cite{Mo04a}. It also reflects in the dc conductivity
which was reinvestigated by Limelette {\em et al.} \cite{Limelette03}
who found critical exponents which agree with those of
the DMFT Mott transition scenario \cite{DMFTMotte}. Laad {\em et al.}
\cite{Laad03b,Laad06a} also studied the Mott-Hubbard transition
in V$_2$O$_3$, using LDA+DMFT(IPT) and arriving at the conclusion that the Mott-Hubbard transition
is driven by changes of the trigonal distortion under pressure.
Using the NMTO-downfolded Hamiltonian instead the simplification
to  the LDA DOS employed in the earlier studies  \cite{Held01a,Keller04a,Laad03b,Laad06a}, Poteryaev {\em et al.} \cite{Poteryaev07}
recently recalculated the LDA+DMFT spectrum and reported somewhat 
better agreement with experiment \cite{Mo04a,Panaccione06}, see Fig.\ \ref{Fig:V2O3Ham}.

In model calculations 
corrections due to non-local correlations
have been discussed
\cite{Parcollet04a,Toschi06a}.
In particular close to the antiferromagnetic phase transition,
paramagnon-like excitations result in strong 
non-local correlations
\cite{Toschi06a}.
Besides, non-local correlations, also 
the electron-phonon coupling is expected
to be of relevance close to the
Mott-Hubbard transition \cite{Kotliar03a,Hassan04}.
The strong changes of the electronic degrees of freedom couple
to the lattice.
Hence, while the local
electronic correlations described by LDA+DMFT 
are the driving force
for the transition, we have to expect corrections to the present LDA+DMFT results in the immediate vicinity of the Mott-Hubbard transition and close to the onset of magnetic order.

\subsubsection{Peierls transition in VO$_2$}
\label{Sec:VO2}
There is  a metal-insulator
transition in another vanadate: VO$_2$
whose high-temperature rutile phase is metallic
while the low-temperature monoclinic phase is insulating.
One can make use of this effect in ``smart'' windows,
which become reflective (metallic) if  bright sunlight
heats them up. 

\begin{figure}[tb]
\vspace{-.5cm}

\centerline{\includegraphics[clip=true,width=8.5cm]{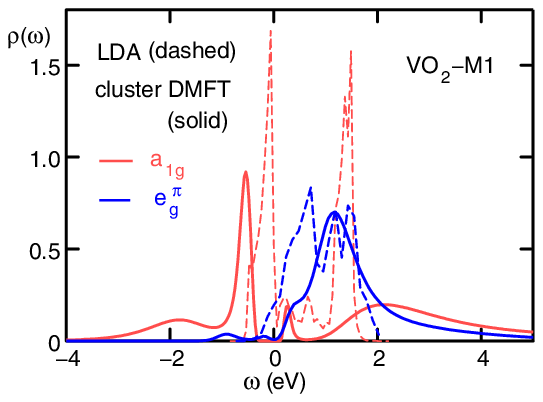}\includegraphics[clip=true,width=5.5cm]{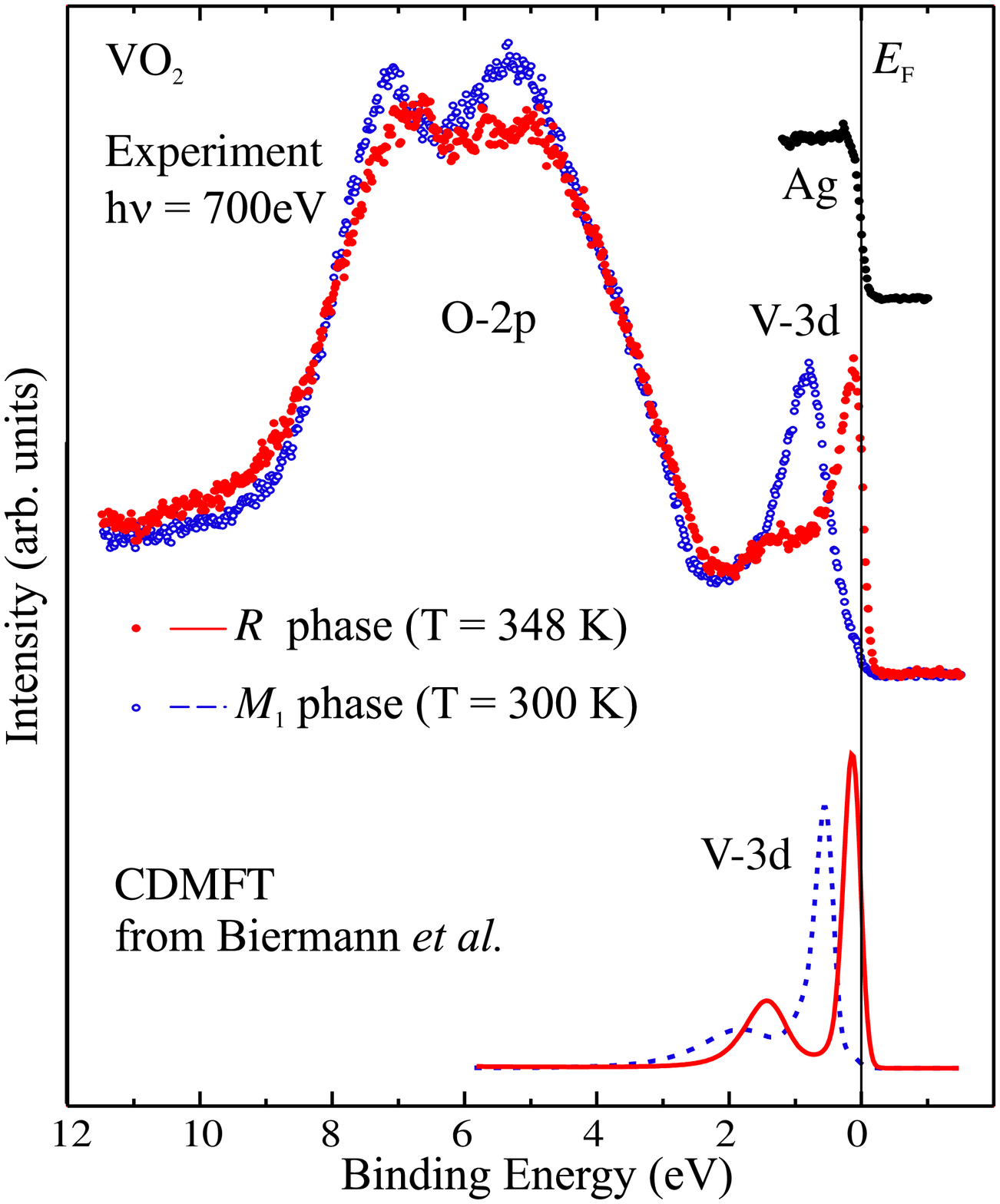}}
\vspace{-.5cm}

\caption{\void \label{Fig:VO2}
Left: Comparison of single-site and two-site  LDA+DMFT spectra 
for VO$_2$ for the (insulating) monoclinic crystal structure.
 (reproduced from \cite{Biermann05a}).
Right: Comparison of these LDA+cluster DMFT spectra with photoemission experiments for the rutile (R) and
 monoclinic (M$_1$) phase (reproduced from \cite{Koethe07a}).}
\end{figure}

 Already  within LDA \cite{Eyert02a},
the  monoclinic phase is almost gapped due to the dimerisation 
of V atoms.
Using single-site LDA+DMFT,  VO$_2$
and has been studied  by
Laad {\em et al.} \cite{Laad03d} and by Liebsch {\em et al.}
A more realistic scenario for the insulating nature of 
 VO$_2$ has been proposed however by Biermann {\em et al.} \cite{Biermann05a},
using a two site cluster DMFT. These two sites form a spin singlet,
triggering the insulating behaviour \cite{Biermann05a}. Hence, the gap in VO$_2$
shown in Fig.\ \ref{Fig:VO2}, is not
 a Mott-Hubbard gap
but a Peierls gap.

\subsubsection{Orbital selective Mott-Hubbard transition in  Ca$_{2-x}$Sr$_x$RuO$_4$}
\label{Sec:ruthenates}

\begin{figure}[tb]
\begin{minipage}{.5\textwidth}

\noindent{\bf Wide band -- SU(2) symmetry}
 \includegraphics[clip=true,width=6.5cm]{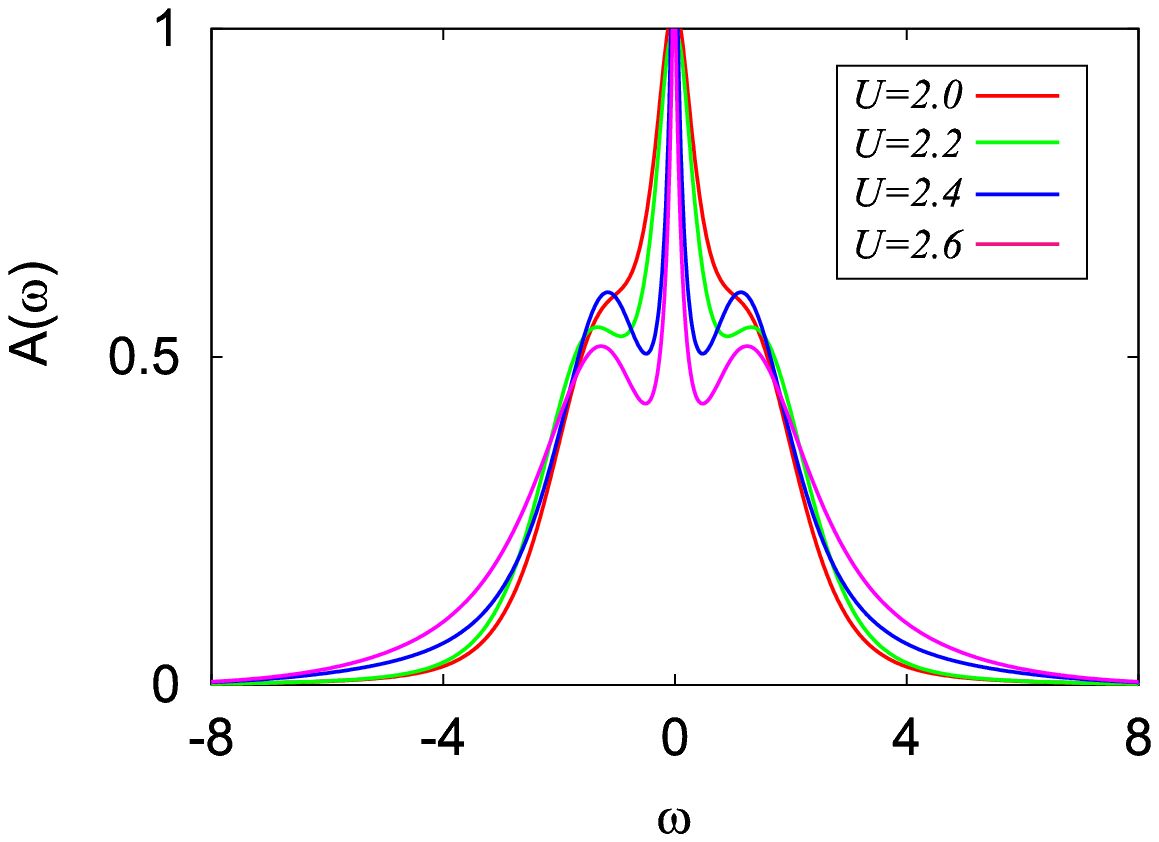}

\noindent{\bf Narrow band -- SU(2) symmetry}
\includegraphics[clip=true,width=6.5cm]{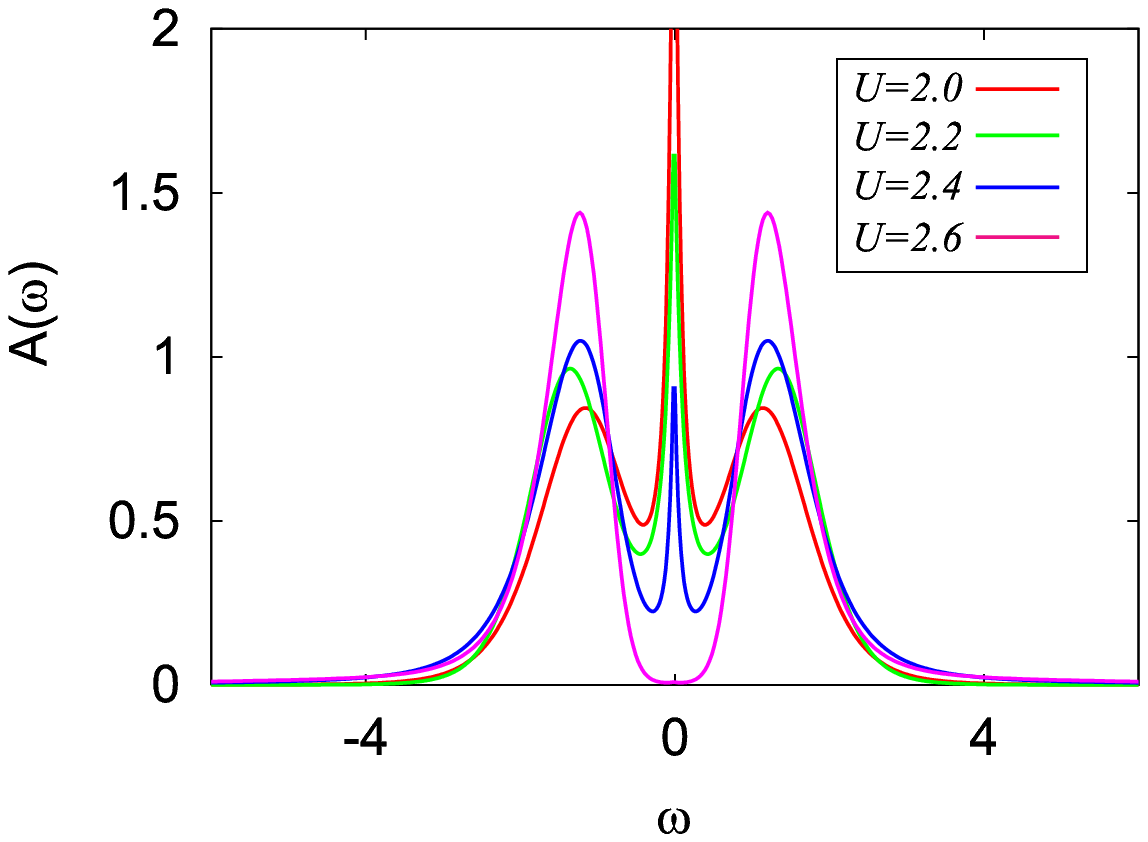}
\end{minipage}\hspace{0.5cm}
\begin{minipage}{.5\textwidth}
\vspace{1.2cm}

\noindent\hspace{.6cm} {\bf Z$_2$ symmetry}
\vspace{-1.9cm}

 \includegraphics[clip=true,width=6.5cm]{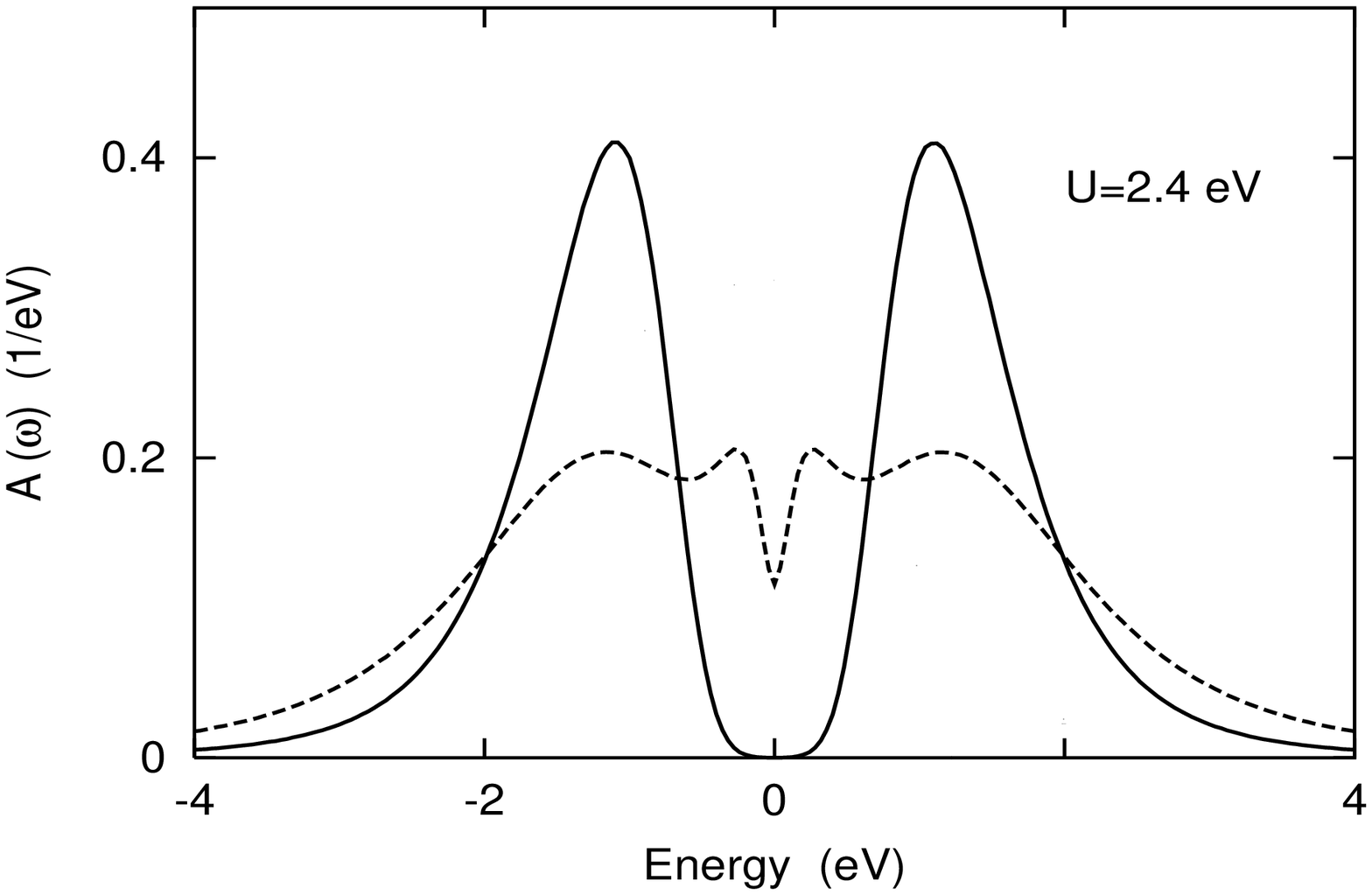}
\end{minipage}
\caption{\void \label{Fig:OSMT}
Orbital selective Mott-Hubbard transition for a two band model system with bandwidths 4 and 2 (a.u.) for the two bands respectively.
The left side shows the situation for SU(2) symmetric Hund's exchange. At U=2.6 (a.u.), the
wide band is still metallic (upper left panel) whereas the narrow band (lower left panel) is already insulating
(reproduced from \cite{Arita06b}).
this is in contrast to the right side, showing the Z$_2$ symmetric situation. Here, the intermediate phase
has a gap for the narrow band (full line) and a dip in the spectrum 
at the Fermi energy for the wide band (dashed line)
(reproduced from \cite{Liebsch04a}).}
\end{figure}

Among the first materials studied by LDA+DMFT was Ca$_{2-x}$Sr$_x$RuO$_4$
an unconventional superconductor. The LDA+DMFT
calculations by Liebsch and Lichtenstein
\cite{Liebsch00} and by Anisimov {\em et al.} \cite{Anisimov01,Pchelkina06a} are very different concerning the size of the Coulomb interaction of the 4d Ru orbitals, i.e. $\bar U = 0.8\,$eV, $J=0.2\,$eV in \cite{Liebsch00} and  $\bar U = 1.7\,$eV,$J=0.7\,$eV in \cite{Pchelkina06a}.
Both groups 
did not address the superconducting phase, and
restricted themselves to the normal phase. 
Actually, the study of superconducting transition metal oxides by
DMFT is very difficult, since low temperatures are required. Furthermore, 
non-s-wave superconductivity is only possible with more formidable
cluster DMFT calculations, which for the one-band Hubbard model
indeed show d-wave superconductivity, see \textcite{DCASC} and \textcite{LichtensteinDCA}.
These one-band calculations have some relevance for the cuprates but will not be discussed here as they are rather
model than material calculations.
Moreover these calculations have already been reviewed in 
Ref.\ \cite{Maier04}.
Coming back to  Sr$_2$RuO$_4$,
 Liebsch and Lichtenstein \cite{Liebsch00} reproduced the experimental ${\mathbf k}$-resolved
spectrum  very well, including the Hubbard side
structure and the Fermi surface. Anisimov {\em et al.}
\cite{Anisimov01} reported
that the Mott-Hubbard transition in Ca$_{2-x}$Sr$_x$RuO$_4$ occurs
subsequently for different orbitals so that there is a region where 
some orbitals are insulating 
and others are metallic, coining the name
orbital-selective Mott transition.
This orbital-selective Mott transition
arises due to different widths of the non-interacting 
bandwidth of the two-different types of $t_{2g}$ bands
and
has been a subject of intensive  model studies 
thereafter \cite{Liebsch03b,Liebsch03c,Koga04a,Liebsch04a,Koga05a,Koga05b,Inaba05a,Inaba05b,Biermann05b,deMedici05b,Arita06b,Knecht05a,Liebsch05a,vanDongen06a,Ferrero05a,Liebsch06a,Bluemer06a,Liebsch06b},
leading to, at first glance, puzzling discrepancies.
These discrepancies were settled when it was realised that
the Mott-Hubbard transition is very different
for a  system with Z$_2$ symmetric Hund's exchange
 \cite{Liebsch03b,Liebsch03c,Liebsch04a,Knecht05a,Liebsch05a,vanDongen06a,Bluemer06a} and SU(2) symmetric Hund's exchange
 \cite{Koga04a,Koga05a,Koga05b,Inaba05a,Inaba05b,Biermann05b,deMedici05b,Arita06b,deMedici05a,Pruschke06a,Dai06a}.
In the former case the Kondoesque quasiparticle peak 
cannot occur 
in the immediate vicinity of the Mott-Hubbard transition 
since the Z$_2$ symmetric  Hund's exchange forms
a $S_z=\pm 1$ spin and a spin-flip from
 $S_z=+1$ to $S_z=-1$ is not possible.
Consequently there is a dip in the spectral function 
for both the narrow and wide band,
see Fig.\ \ref{Fig:OSMT}.
Sakai {\em et al.} \cite{Sakai06a}
reinvestigated Sr$_2$RuO$_4$
using a SU(2) symmetric Hund's coupling.
But, 
for the rather small Coulomb interaction
of  \cite{Liebsch00}  (for which both bands have
pronounced quasiparticle peaks)
Sakai {\em et al.}  did not yet find pronounced differences
between SU(2) and Z$_2$ Hund's exchange.

\subsubsection{`Kinks' in SrVO$_3$} 
\label{SrVO3}

\begin{figure}
\includegraphics[clip=false,width=5cm]{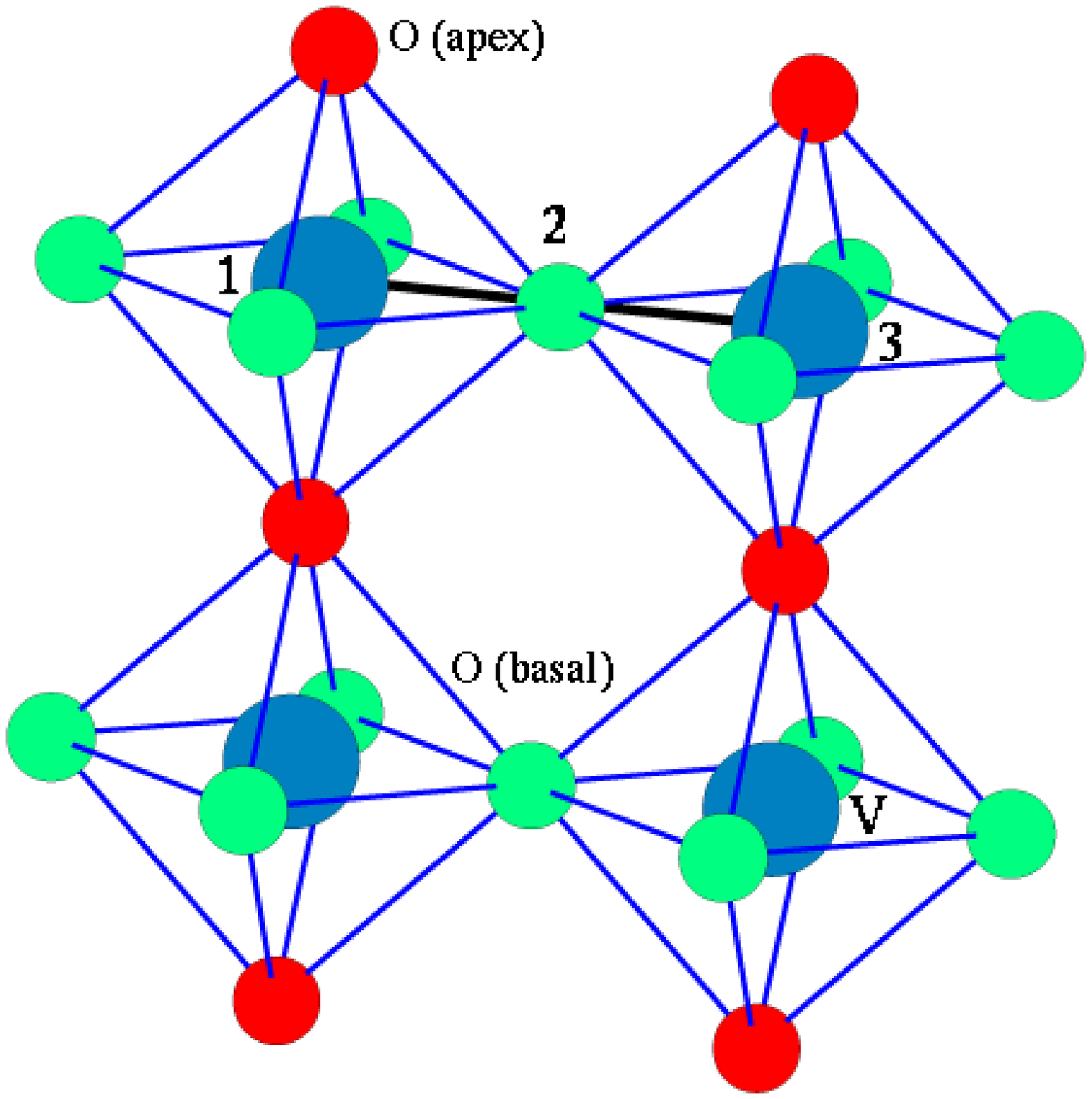}

\vspace{-5.3cm} 
\hfill \includegraphics[clip=false,width=5cm,angle=270]{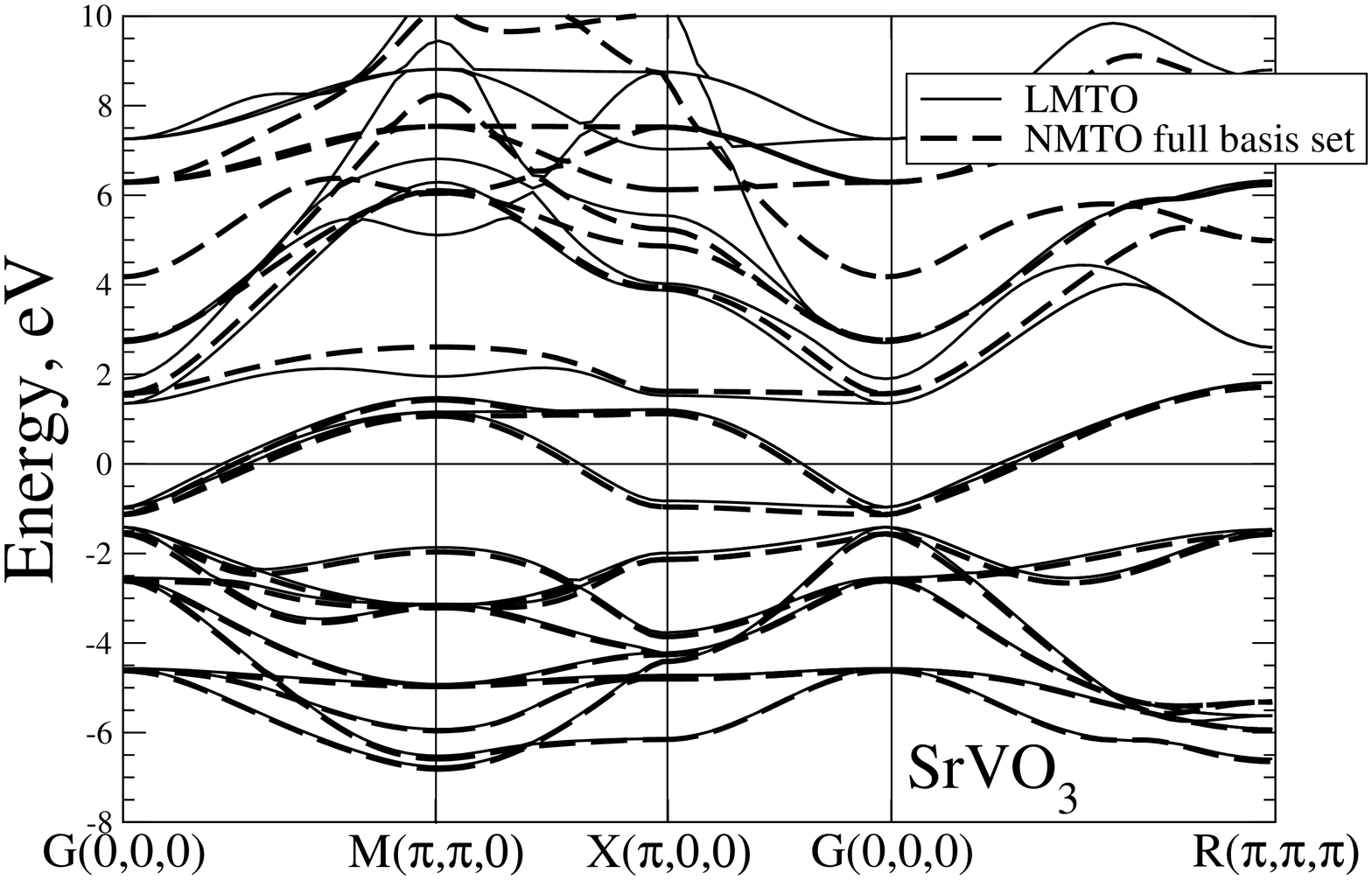}
\caption{\void Left: cubic perovskite crystal structure of SrVO$_3$; right: LDA bandstructure  calculated using LMTO and NMTO
(reproduced from \cite{Nekrasov05b,Nekrasov05a})
\label{Fig:SrVO3}}
\end{figure}

Among transition metal oxides, SrVO$_3$ is particularly simple 
because it has (i) a
3d$^1$ electronic configuration and (ii) 
a perfectly cubic perovskite lattice structure, see Fig.\ \ref{Fig:SrVO3}.
Due to (i), the effect of  Hund's exchange
interaction on the ground state properties is less crucial
since this exchange interaction only takes effect for two or more
electrons.
The  cubic symmetry (ii) on the other hand results 
in three degenerate $t_{2g}$ bands at the Fermi energy and allows for
the simplification described in Section \ref{SimpTMO}, i.e.
 using the DOS instead of the full LDA bandstructure.
Hence, SrVO$_3$ can be considered as a transition metal oxide
prototype.
Despite this simplicity, there 
has been some debate concerning the series Ca$_{x}$Sr$_{1-x}$VO$_3$
in which Ca doping $x$ leaves the  3$d^1$ configuration unchanged
but results in an 
orthorhombic distortion.
Interest in this  3$d^{1}$  series 
was initiated by Fujimori {\em et al.}
\cite{fujimori} who
reported
a pronounced lower Hubbard band in PES.
While
thermodynamic properties such as the Sommerfeld coefficient,
resistivity and paramagnetic susceptibility 
were reported to be 
essentially independent of $x$~\cite{Aiura93a,Inoue95a,Inoue98a},
 PES 
\cite{fujimori}
and Bremsstrahlungs isochromat spectra (BIS) 
\cite{Morikawa95a} suggested
 dramatic differences
between  CaVO$_{3}$ and SrVO$_{3}$, 
leading to the suggestion of a Mott-Hubbard transition
with increasing Ca doping $x$ \cite{Rozenberg96a}.

\begin{figure}[t]
\centerline{
\epsfig{file=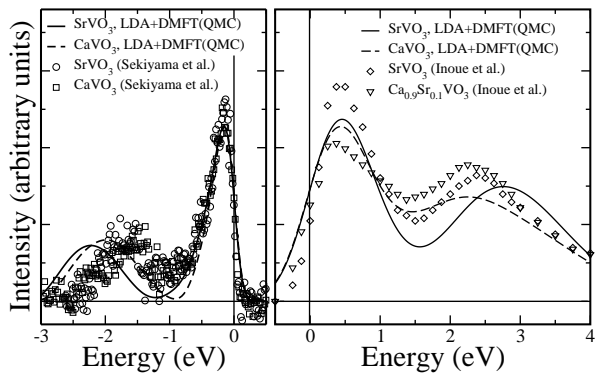,width=0.6\textwidth}
}
\caption{\void Parameter-free LDA+DMFT(QMC) spectra for SrVO$_3$ and CaVO$_3$
compared to PES and XAS experiments below and above the Fermi energy respectively
(reproduced from \cite{Nekrasov05b}).
\label{Fig:SrVO3spectra}}
\end{figure}

\begin{figure}[t]
\centering \epsfig{file=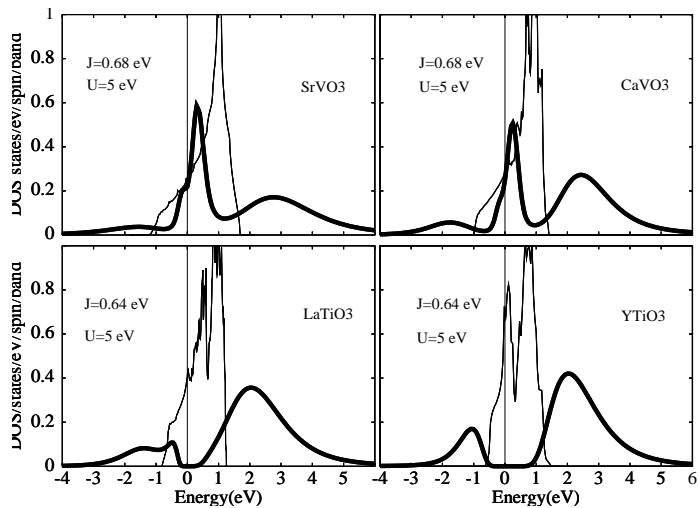,angle=270,width=0.7\textwidth,clip}
\caption{\void 
 LDA+DMFT(QMC) spectra (thick lines) obtained by \textcite{Pavarini03}
for the 3$d^1$ series SrVO$_3$, CaVO$_3$, LaTiO$_3$ and YTiO$_3$.
The thin lines represent the corresponding LDA DOSes  (reproduced from \textcite{Pavarini03}).
\protect\vspace{-0.3cm}
\label{Fig:d1}}
\end{figure}

This puzzling discrepancy was settled through
bulk-sensitive (high photon energy) PES by Maiti {\em et al.} \cite{Maiti01} and by Sekiyama {\em et al.}
\cite{Sekiyama02,Sekiyama03}
which, in mutual agreement with LDA+DMFT calculations
\cite{Nekrasov02a,Sekiyama03,Pavarini03,Nekrasov05b,Nekrasov05a},
showed similar spectra for CaVO$_{3}$ and SrVO$_{3}$.
Hence the earlier reported differences
were attributed to 
 the surface-sensitivity
of low photon energy PES. This is also supported by
LDA+DMFT calculations of 
Liebsch {\em et al.} \cite{Liebsch03a,Ishida06a} which show
pronounced differences between
  SrVO$_3$ surface and bulk spectra.
Fig.\ \ref{Fig:SrVO3} shows the LDA bandstructure,
and Fig.\ \ref{Fig:SrVO3spectra} the calculated LDA+DMFT(QMC) spectra
in comparison with experiment. Note that the LDA+DMFT spectra are parameter free since
the inter-orbital Coulomb interaction
$\bar{U}=3.55$~eV and  Hund'sxchange $J=1.0$~eV have been
obtained through constrained LDA calculations \cite{Nekrasov02a,Sekiyama03}.
Pavarini {\em et al.} \cite{Pavarini03} also systematically studied  the
similar $d^1$ systems  LaTiO$_3$ and YTiO$_3$ which have an increasingly smaller LDA
bandwidth so that these materials become indeed Mott-Hubbard insulators, see 
Fig.\ \ref{Fig:d1}.

Despite SrVO$_3$ being a simple material, the LDA+DMFT(QMC) results led nonetheless
to a surprise:
The dispersion of the quasiparticle peak shows `kinks' at $- 0.25\,$eV and (less pronounced) at  $0.25\,$eV. This was unexpected since such `kinks' are usually 
associated with phonon modes, particularly for high temperature superconductors \cite{Lanzara01a},
or the coupling to other bosonic degrees of freedom such as spin fluctuations.
The LDA+DMFT calculations showed that such `kinks' emerge naturally in strongly correlated systems with narrow quasiparticle peaks. Let us note that kinks have  also 
been observed experimentally for 
 SrVO$_3$ by Yoshida {\em et al.} 
in angular-resolved PES  \cite{Yoshida05a}.
Using the DMFT self-consistency equation,
Byczuk {\em et al.} \cite{Byczuk06a}
were able to show mathematically how a central peak in the spectrum necessarily results in a wiggle in the self energy
and, hence, a `kink' in the dispersion relation.
Landau's Fermi liquid theory is restricted to the excitations between Fermi energy and `kink'.
The rest of the central peak is strongly damped 
and follows another effective
mass renormalisation. That is,
\begin{eqnarray}
{E_{\mathbf k}}=& {Z_{\rm FL}} \epsilon_{\mathbf k} & \mbox{for} \;\;\;\;  |{E_{\mathbf k}}| \!<  \!\omega_* \\
{E_{\mathbf k}} =& {Z_{\rm CP}} \epsilon_{\mathbf k} \!\pm\! c \;\;\;\; & \mbox{for}|{E_{\mathbf k}}| \!>\! \omega_*,
\end{eqnarray}
where $\omega_*$ is the ``kink'' energy,
${E_{\mathbf k}}$ the dispersion relation of the interacting
system, $Z_{\rm FL}$  the Fermi liquid renormalisation factor, $Z_{\rm CP}$ the  renormalisation factor
for the dispersion beyond the Fermi-liquid regime,
$c$ is a constant.
From the bare (LDA) dispersion and one of the
renormalisation factors $Z_{\rm FL}$,  $Z_{\rm CP}$ and
$\omega_*$, the other two can be calculated in a 
simple way \cite{Byczuk06a}.
This allows for example to calculate from the linear
 specific heat coefficient ($\sim 1/Z_{\rm FL}$) the 
`position of the ``kink'' and the overall  bandwidth in 
angular resolved PES.

\begin{figure}[t]
\centering
\includegraphics[clip=true,width=0.55\textwidth]{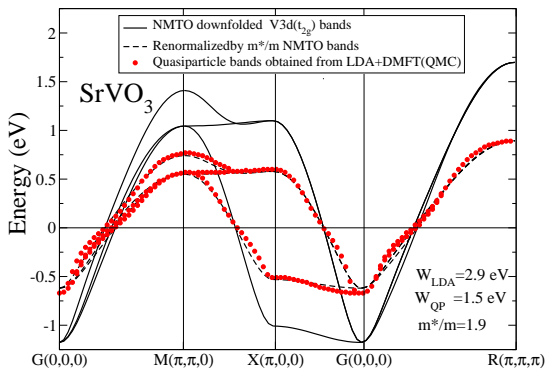}

\caption{\void Comparison of the LDA bandstructure for SrVO$_3$  (full lines)
with that from LDA+DMFT(QMC) (dots). 
Basically, the LDA+DMFT(QMC) describes renormalised 
quasiparticle bands with weight $Z^{-1}=1.9$.
However, there are deviations in the form of `kinks'
at $\pm 0.25\,$eV  (the Fermi energy is set to zero;
reproduced from \textcite{Nekrasov05a})}
\label{nmto_qmc_bands}
\end{figure}

\subsubsection{$e_{g}'$ hole pockets in  Na$_x$CO$_2$}
\label{Sec:cobaltates}

Unconventional superconductivity was also discovered in 
hydrated  Na$_x$CO$_2$ \cite{Takada03a}, resulting in
substantial interest in this material.
This material is difficult to characterise, particularly,
the  O-Co-O angle is affected by the hybridisation
and not known.
Presently, it is also unclear which orbitals are responsible for the
superconductivity, and even from which orbitals
the Fermi surface is composed of.
In LDA \cite{Koshiba03a,Singh03a,Marianetti04a,Johannes04a,Arita05a,Lechermann05b}
both $a_{1g}$ and $e_{g}'$ orbitals have similar
centres of gravity, leading to the possibility of
 so-called 
hole pockets in the  $e_{g}'$ bands.
Also the $a_{1g}$ bands seem to play an important role,
at least these orbitals change most dramatically in LDA
when going from
 normal-conducting mono-layer-hydrated cobaltate to  superconducting bi-layer-hydrated cobaltate \cite{Arita05a}.
Since these $t_{2g}$ bands are very narrow, Coulomb correlations
are expected to play a major role.
Within LDA+U, Zhang {\em et al.} \cite{Zhang04a,Lechermann05b}
reported the disappearance of the
 $e_{g}'$ hole pockets. In contrast, more accurate
LDA+DMFT
calculations  by Ishida, Perroni and Liebsch \cite{Ishida05a,Perroni06a}
predict the contrary, i.e.
an enhancement of these  pockets.
These results have been challenged recently by
Marianetti {\em et al.} \cite{Marianetti06b},
reporting the opposite behavior, i.e., the
suppression of the  $e_{g}'$ hole pockets in LDA+DMFT.
This would agree with experiments
where  these pockets seem to be absent.
Such conflicting results are possible since, in the
case of cobaltates, the results are very sensitive to
small changes of the LDA  crystal-field splitting,
see \textcite{Lechermann05b} for a discussion.
Finally, 
let us also mention two 
DMFT model studies, concentrating on
bad metallic behaviour \cite{Merino05a}
and the Na-induced potential
\cite{Marianetti06}, respectively.

\subsubsection{Heavy-Fermion behaviour in LiV$_2$O$_4$}
\label{Sec:LiVO}

Another interesting transition metal oxide is
LiV$_2$O$_4$, the first $d$ system signalling
heavy Fermion behaviour by, for example,
an unusually large  specific heat coefficient, one
order of magnitude larger than in other transition metal oxides. Nekrasov {\em et al.}
\cite{LiV2O4} did realistic  LDA+DMFT calculations for LiV$_2$O$_4$
and investigated whether the  scenario of Anisimov {\em et al.}
 \cite{Anisimov99a} for the 
heavy Fermion behaviour holds. The basic idea of \textcite{Anisimov99a}
is a separation of the partially filled t$_{2g}$ electrons
into localised ones forming local moments and delocalised ones producing
a partially filled metallic band. Then, the hybridisation between those two subsets
of electrons, as in $f$-electron materials, can give rise to heavy Fermion effects.
Nekrasov {\em et al.} \cite{LiV2O4} found a strong competition between {antiferromagnetic} direct (the exchange constant corresponds to an energy
$\approx$-450$\,k_B\,$K)
and ferromagnetic double exchange ($\approx1090$$\,k_B\,$K). 
With these estimates it appears to be reasonable that these two contributions almost cancel so that
the   Kondo exchange ($\approx$-630$\,k_B\,$K) prevails, resulting in heavy Fermion Kondo physics.
But, since the energy differences are rather small
and the calculations were performed at relatively high temperatures and with a numerical error,
the results of \cite{LiV2O4} do not allow for a final conclusion.
In their LDA+DMFT(IPT) calculations,
Laad {\em et al.} \cite{Laad03c} on the other hand stressed the importance of geometrical frustration
for the heavy Fermion behaviour which forms the basis of
another explanations of the unusual behaviour in LiV$_2$O$_4$ \cite{Lacroix,Shannon,Fulde,Burdin,Hopkinson,Fujimoto,Tsunetsugu,Yamashita}.

\begin{figure}

\includegraphics[clip=false,width=5.5cm]{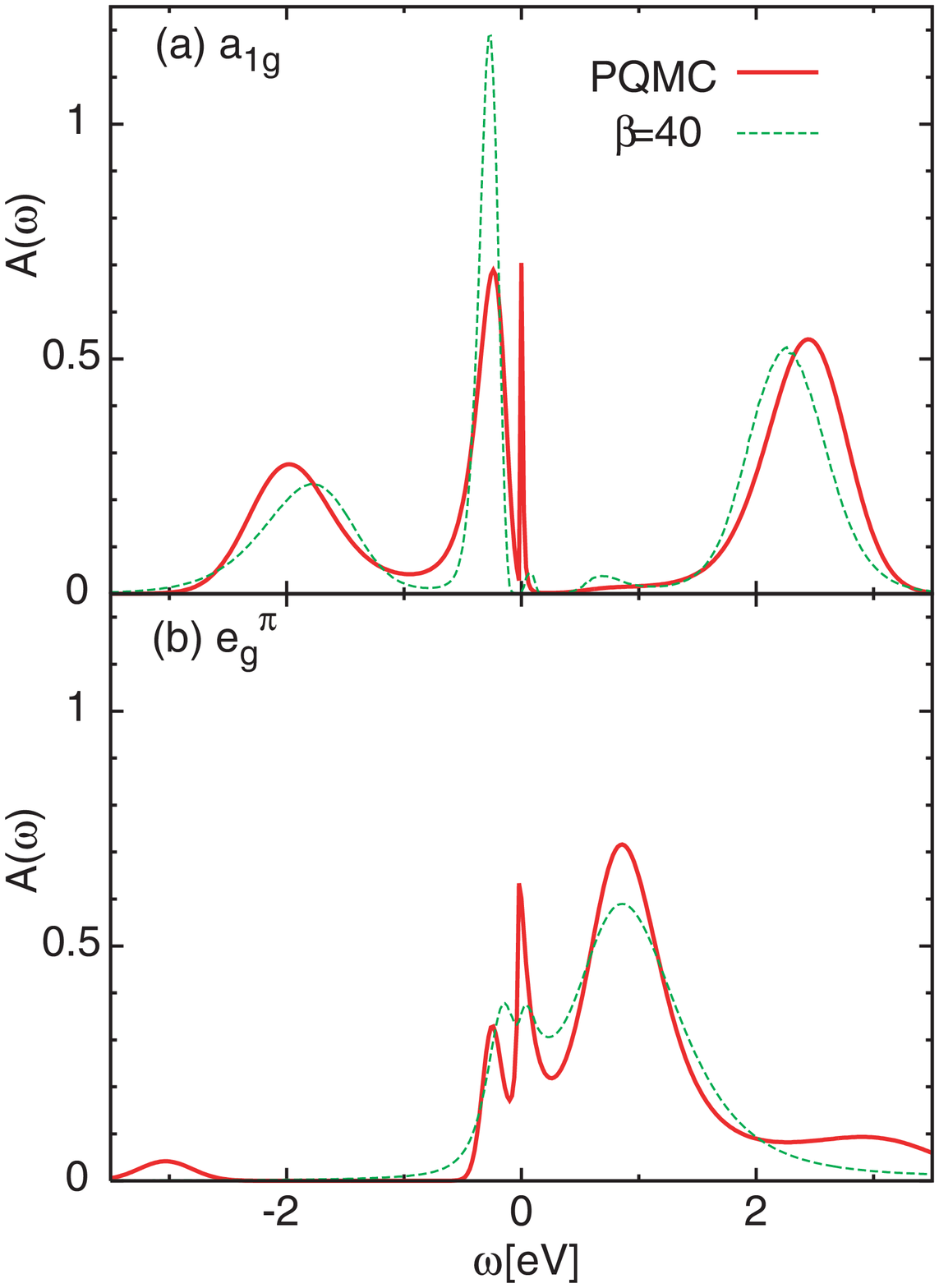}\hfill
\includegraphics[clip=false,width=6.5cm]{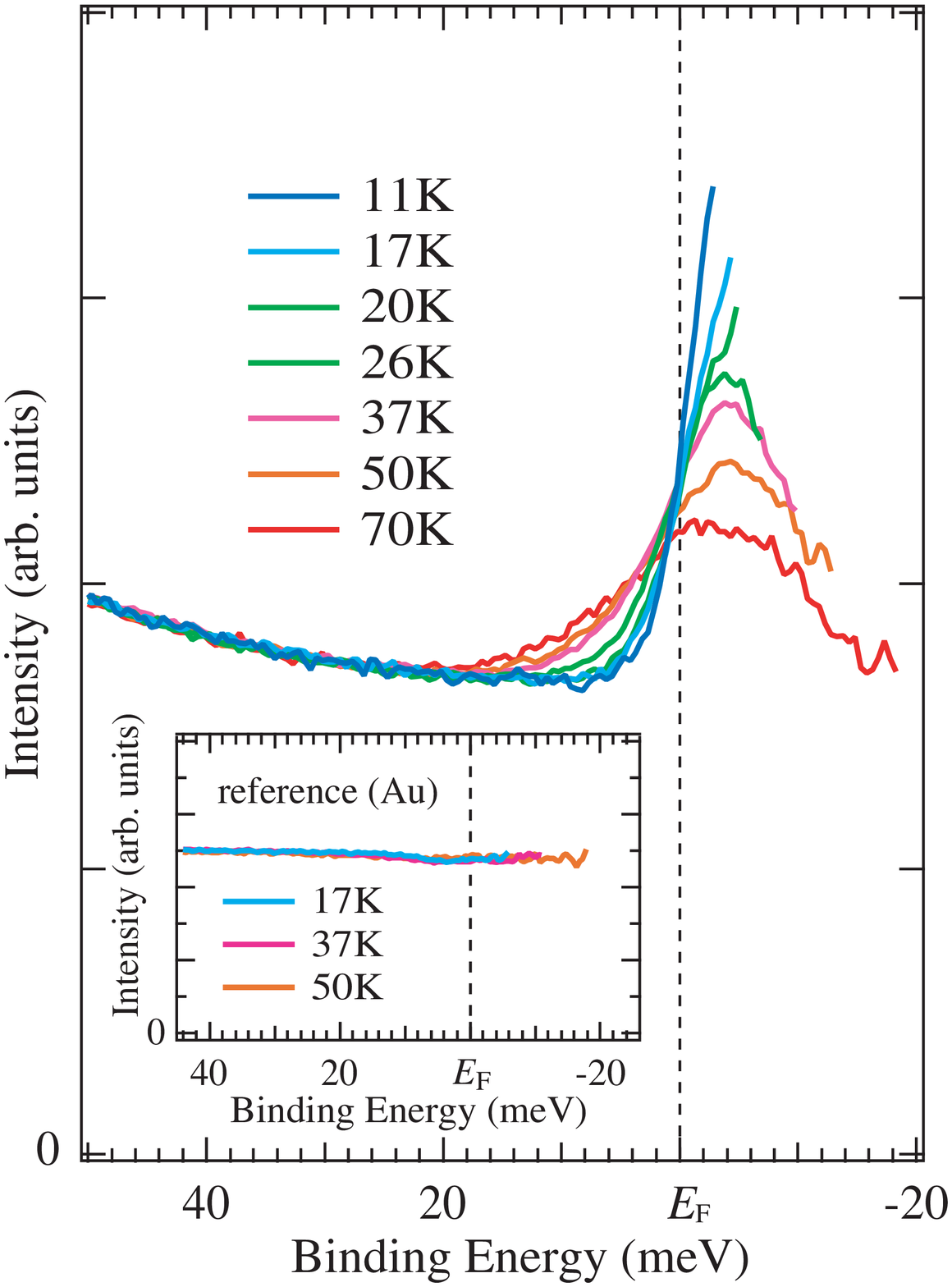}
\caption{\void Left:
Spectral function of   LiVO$_3$ as calculated
by LDA+DMFT for the $a_{1g}$ and the two $e_g$ orbitals
at zero temperature (PQMC) and
$T=1/\beta=1/40\,$eV (reproduced from  \cite{Arita06c}).
Right: Corresponding experimental PES spectrum in the vicinity of the
Fermi energy
(reproduced from  \cite{Shimoyamada06a}).
\label{Fig:LiVO}}
\end{figure}

More recently, Arita {\em et al.} \cite{Arita06c}
managed to investigate the low temperature behavior
of LiVO$_3$ using PQMC as an impurity solver.
These calculations indeed show a sharp peak above the Fermi energy, in agreement with experiments \cite{Shimoyamada06a}, see Fig.\ \ref{Fig:LiVO}.
The physical origin of this sharp peak
is however very different from the aforementioned scenarios.
The sharp peak emerges from the physics
of the  $a_{1g}$ orbital, which is a lightly doped Mott-Hubbard insulator,
i.e., a metal with a large mass renormalisation.

\subsubsection{Colossal magnetoresistance in manganites}
\label{Sec:CMR}

\begin{figure}

\noindent \includegraphics[clip=true,width=4.5cm,angle=270]{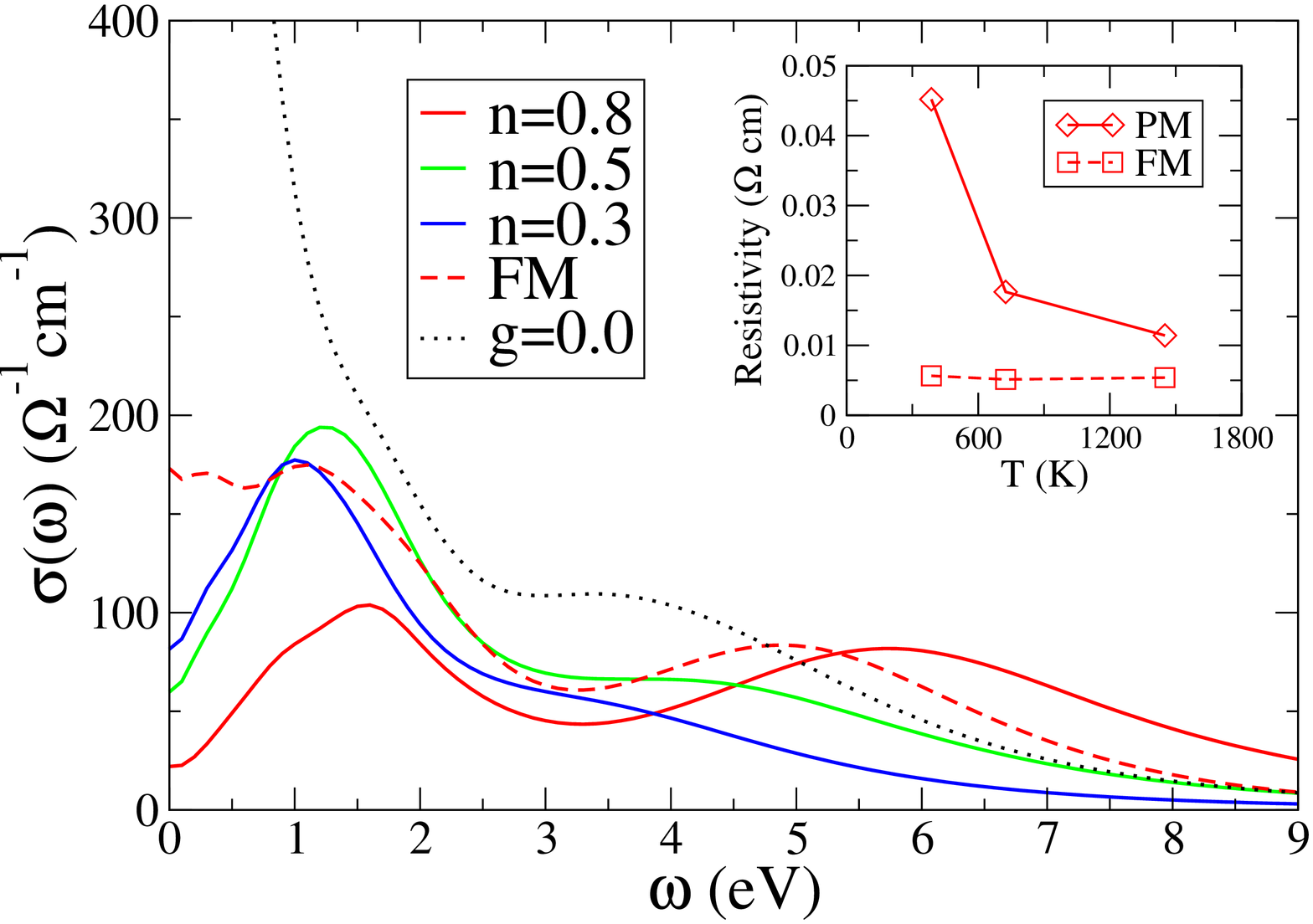}
\vspace{-5.4cm}

\hfill
\includegraphics[clip=false,width=7.cm]{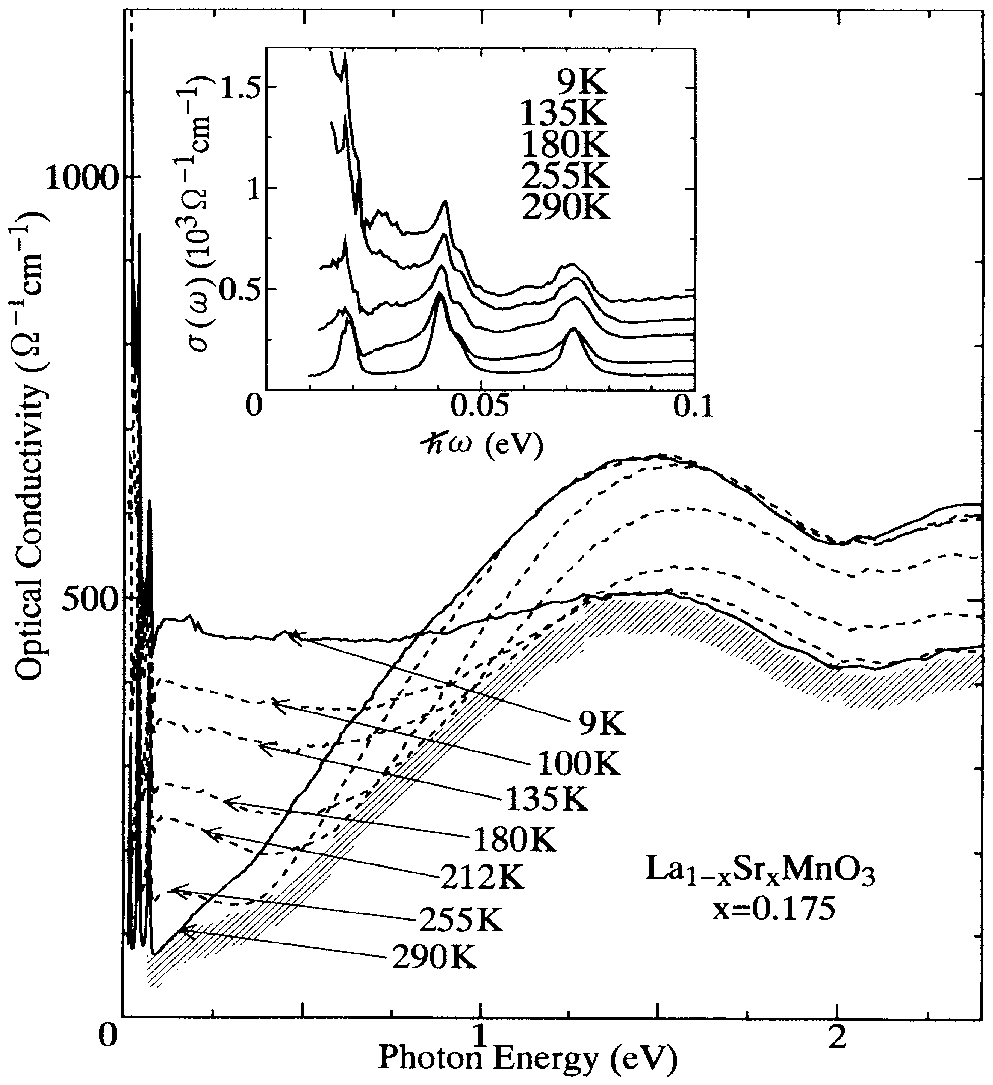}
\caption{\void 
Optical conductivity $\sigma(\omega)$. 
Left: theory (reproduced from \cite{Yang06a}), right: experiment  (reproduced from \cite{Okimoto95a})
for the paramagnetic (PM) phase of La$_{1-x}$Sr$_x$MnO$_3$, 
($n=1-x$ electrons/site in the $e_g$ orbitals).
The dotted line is without electron-phonon coupling,
 showing a metallic Drude peak; the
dashed line is the optical conductivity for
the ferromagnetic phase (FM) at  $x=0.2$.
Inset: The PM resistivity strongly increases with decreasing $T$,
resulting in a   `colossal' magnetoresistance
when going from the PM to the FM phase.
\label{Fig:CMR}\label{Fig:LaMnO3}}
\end{figure}

Because of the colossal magnetoresistance (CMR) \cite{Helmolt93a,Jin94a},
manganites such as La$_{1-x}$Ca$_x$MnO$_3$ 
have been at the focus of reasearch during the 
last years, including many DMFT
studies \cite{Furukawa94a,Millis95a,Millis96a,Millis96b,Rozenberg98a,Benedetti99a,Furukawa99a,Held00c,Pruschke01a,Michaelis03a,Laad04a,Ramakrishnan04a,Yamasaki06b,Yang06a}.
In the parent compount ($x=0$), Mn has a $d^4$ configuration
with cubic symmetry so that 3 electrons occupy the lower-lying
$t_{2g}$ orbitals, forming a spin $3/2$.
This spin and its coupling to the remaining itinerant
electron in the doubly degenerate $e_g$ bands
constitute the (ferromagnetic) Kondo lattice model, which
gives rise to ferromagnetism due to the so-called double exchange mechanism: A ferromagnetic allignment of the $t_{2g}$ spins is favorable since it maximises the kinetic energy of the $e_g$ electrons.
This model was solved exactly within DMFT by Furukawa  \cite{Furukawa94a,Furukawa99a}. Also employing DMFT,
Millis and coworkers \cite{Millis95a,Millis96a,Millis96b}
pointed out however that the double exchange of the Kondo lattice model is not enough for describing manganites and the 
CMR in particular. The importance of the electron-phonon coupling to Jahn-Teller phonons was stressed, and the Kondo lattice model plus
Jahn-Teller phonons was studied using DMFT \cite{Millis96a,Millis96b,Benedetti99a,Michaelis03a}.
This model gives large magnetoresistances, however only without Ca doping $x$ and also fails to describe other experimental aspects.
Held and Vollhardt \cite{Held00c} on the other hand stressed the importance of electronic correlations induced by the Coulomb interaction between the $e_g$ electrons.
The arising question,  whether the electron-phonon
coupling or the Coulomb interaction is responsible for the insulating nature of the parent compound,
was settled recently by LDA+DMFT calculations with a 
static Jahn-Teller distortion \cite{Yamasaki06b},
showing that both Jahn-Teller distortion and Coulomb interaction
are needed: The Jahn-Teller distortion gives rise to a crystal field splitting of the two $e_g$ orbitals which is largely enhanced by the Coulomb interaction so that a gap emerges. Since the $e_g$ electrons are also spin-alligned to the $t_{2g}$ spin, LaMnO$_3$ has
one spin-polarised $e_g$ electron per site and is hence  an insulator, albeit not a Mott-Hubbard insulator. With the crystal-field splitting being reduced under pressure,
 Yamasaki {\em et al.}  \cite{Yamasaki06b} were
also able to explain the pressure-induced insulator-to-metal transition found experimentally by Loa {\em at al.} \cite{Loa01a}.
Yang and Held \cite{Yang06a} subsequently studied doped manganites with dynamical Jahn-Teller phonons instead of a static distortion.
The authors reported the trapping of $e_g$ electrons as Jahn-Teller polarons, an effect already described
earlier without $e_g$-$e_g$ Coulomb interaction  \cite{Millis96a,Millis96b,Benedetti99a,Michaelis03a}. However, the electronic
correlations strongly enhance the tendencies towards polaron formation. 
The trapping of $e_g$ electrons as polarons 
results in an insulating-like paramagnetic phase with a pseudogap structure in the spectrum at the Fermi energy, irrespectively of doping $x$, in agreement with experiment
\cite{Bocquet92a,Chainani93a,Park96a,Saitoh97a}.
 Including the effects of electronic correlations,
also the optical conductivity \cite{Okimoto95a,Quijada98a,Jung98a,Takenaka99a,Kovaleva04a} could be described, see Fig.\ \ref{Fig:LaMnO3}, and a large CMR were reported
 for the doped system, see the inset of  Fig.\ \ref{Fig:LaMnO3}.

\subsection{Other Materials}
\label{OM}

\subsubsection{Half-metallic ferromagnetism in Heussler alloys}
\label{Sec:HMFM}

\begin{figure}[tb]

\noindent \includegraphics[clip=false,width=6cm]{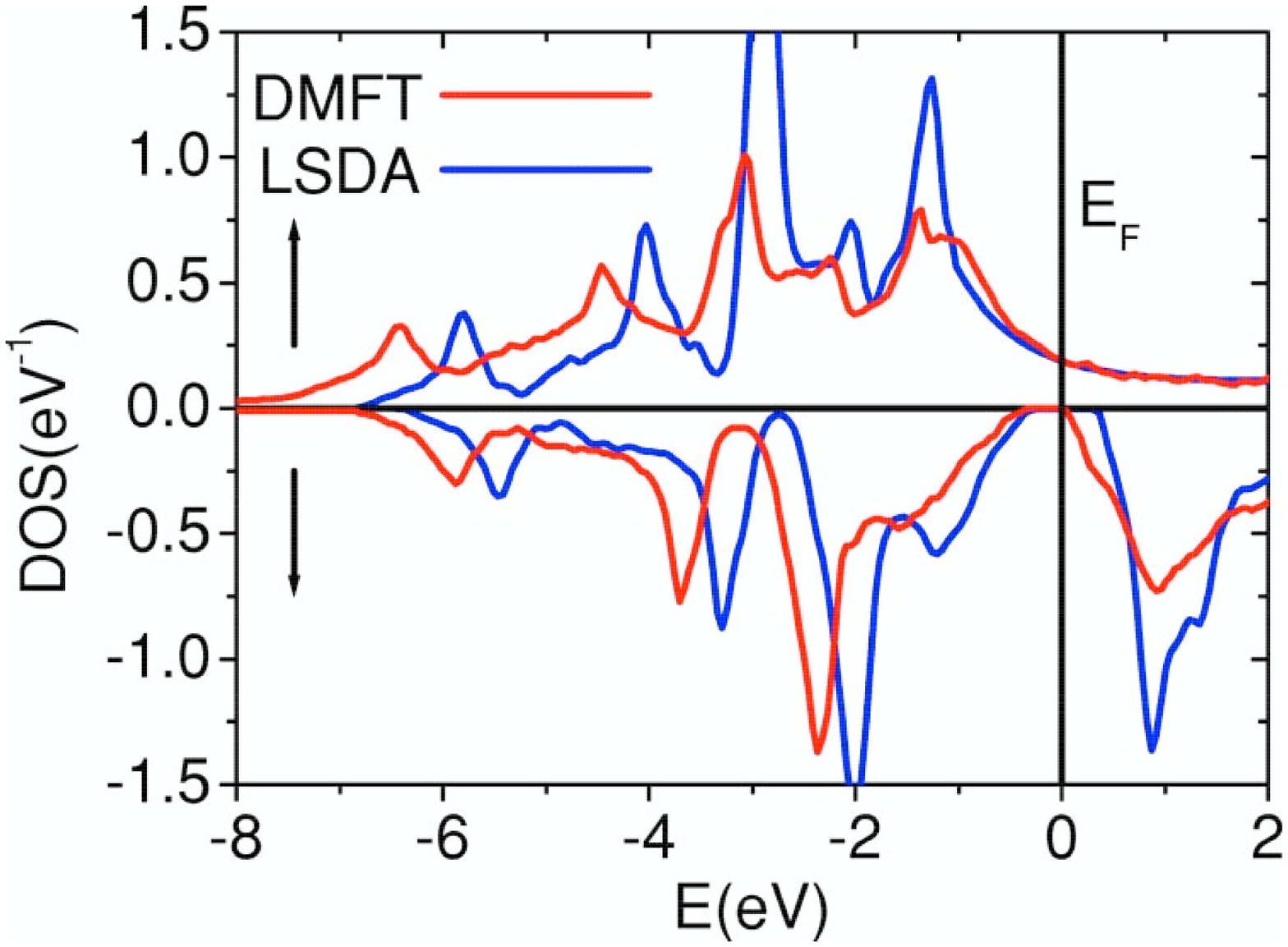}
\hfill
\includegraphics[clip=false,width=6cm]{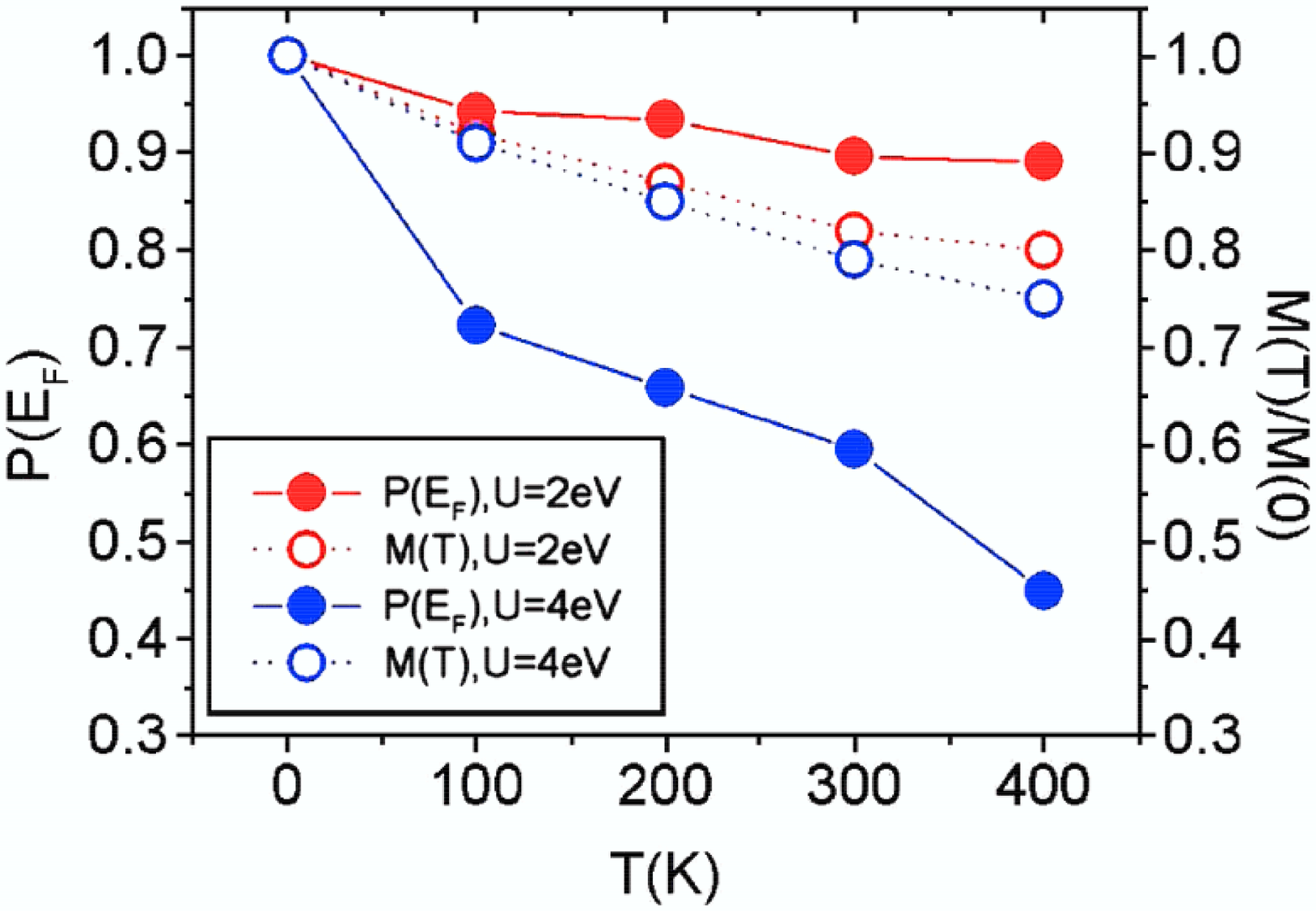}
\vspace{-1cm}

\caption{\void Left: Density of states for NiMnSb for spin-up (upper panel)
and down (lower pannel). Due to electronic correlations,
nonquasiparticle states appear within the LSDA gap
when using DMFT.
Right: Spin-polarisation $P$ and magnetization $M$ 
as a function of temperature for FeMnSb.
The nonquasiparticle states lead to a reduction of the
spin polarisation
(reproduced form \cite{Chioncel03b} 
and \cite{Chioncel06a}, respectively).
\label{Fig:Heussler}}
\end{figure}

For the technical realization of spintronic
\cite{Wolf01a,Zutic04a} devices,
materials with high spin-polarisation are necessary.
A promising candidate are half-metallic ferromagnets
which show metallic behavior for one-spin species and
insulating behavior for the other one,
allowing in principle for a fully spin-polarized current. 
Such materials have been intensively studied by 
LDA. However, many of the potential candidates have 
$d$ electrons so that electronic correlations
cannot be neglected.
An example of such correlation effects are
 nonquasiparticle states in the minority spin-band
above the Fermi energy
\cite{Edwards73a,Irkhin83a,Irkhin04a}.
This effect was also studied in realistic LDA+DMFT
calculations for the Heussler alloys  NiMnSb
\cite{Chioncel03b,Chadov06a} and FeMnSb \cite{Chioncel06a}, see Fig.\ \ref{Fig:Heussler}.
Due to these nonquasiparticle states, the polarisation is reduced at finite temperatures
if electronic 
correlations are strong, see left part of  Fig.\ \ref{Fig:Heussler}.
Since the emerging non-Fermi-liquid physics is related to magnons, which can only be described very rudimentary by 
the local DMFT,
the inclusion of non-local correlations is necessary for
a more accurate description.
The   nonquasiparticle states lead to  a reduced polarisationwhich is a severe disadvantage for spintronic sources which
optimally would be fully polarised.
Other candidates for half-metallic ferromagnetism which have been studied by LDA+DMFT are
Ga$_{1-x}$Mn$_x$As \cite{Craco03a,Aryanpour05b}, VAs
\cite{Chioncel06b},
 CrAs \cite{Chioncel05a} and
CrO$_2$
\cite{Yamasaki06a,Chioncel06c}.

\subsubsection{Superconductivity in A$_x$C$_{60}$}
\label{Sec:fullerenes}

\begin{figure}[tb]
\centerline{
\includegraphics[clip=false,width=7cm]{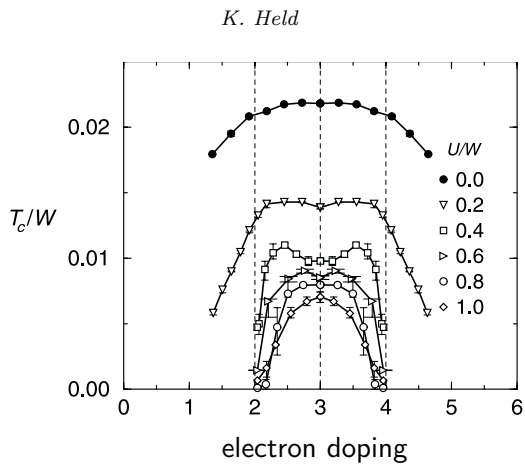}
}
\centerline{\phantom{aaaaaaa} \sf electron doping}
\vspace{.4cm}

\caption{\void $T_c$ as a function of electron doping
in C$_{60}$ fullerenes for different values of the ratio Coulomb interaction
over band-width $U/W$ (phonon frequency $\omega_{\rm ph}=0.24 W$; electron-phonon coupling strength $\lambda = 0.6$
(reproduced from \cite{Han03a}).
\label{Fig:C60}}
\end{figure}

Synthesized fullerenes display a varity of unusual properties, 
including superconductivity up to relatively high temperatures, e.g. 33K for Cs$_x$Rb$_y$C$_{60}$ \cite{Tanigaki91a}.
Such superconducting alkali-doped fullerenes A$_x$C$_{60}$  are molecular solids with a low bandwidth so that both the phonon frequencies and the Coulomb interaction are compareable to the bandwidth, and 
the dimensionless electron-phonon coupling  is
of order 1.
Hence, it is important to 
treat the electron-phonon coupling
and the Coulomb interaction on an equal footing which is possible through DMFT.
Such 
realistic DMFT calculations have been carried out by
Capone {\em et al.} 
\cite{Capone02a,Capone00a} and by Han {\em et al.}
\cite{Han03a}.
These authors found that the superconductivity mediated
by  electron-phonon coupling is surprisingly resistant
to the  Coulomb interaction.
Together with the strong electron-phonon coupling this hence 
explains the high superconducting transition temperatures,
without the need for a new electronic mechanism for 
superconducitivity.
Furthermore, Coulomb interaction and Jahn-Teller coupling
tend to localise the  electrons 
at the fillings $n=2,4$, which explains why  fillings
close to $n=3$ are most favorable for superconductivity,
see Fig. \ref{Fig:C60}. 
This correlation effect is beyond Eliashberg theory and
explains the   experimental oberserved \cite{Yildirim96a} change of $T_c$ with
doping.

\subsubsection{Mott-insulating zeolites}
\label{Sec:zeolites}

\begin{figure}[tb]
\vspace{-8cm}
\centerline{
\includegraphics[clip=false,width=12cm]{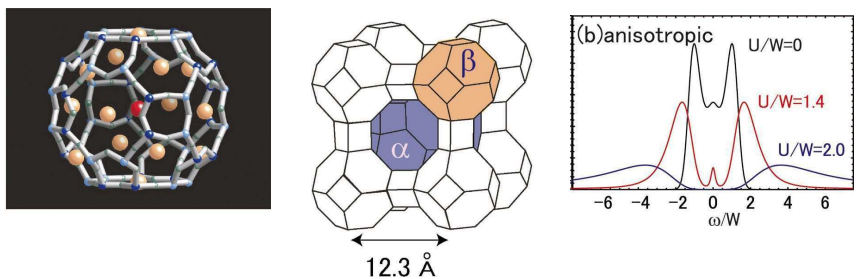}}
\vspace{-3cm}
\caption{\void Left: `Superatom' ($\alpha$ cage) of the undoped zeolite (dark blue: Si, light blue: Al, dark green: oxygen, orange and red: K); middle: crystallisation of the cages;
right: DMFT spectrum for realistic (unisotropic) hopping parameters showing Mott-insulating behavior for realistic $U$ values. (reproduced from \cite{Arita04b}) 
\label{Fig:zeolites}}
\end{figure}

 Zeolites 
loaded with alkali metals
such as K$_{n}$Al$_{12}$Si$_{12}$O$_{48}$ are at first glance an unlikely candidate for strong electronic correlations. These materials however form 
`superatoms', see Fig.\ \ref{Fig:zeolites},
which act like hydrogen atoms with well difined $s$ and $p$ orbitals. The superatoms in turn crystalise, see central panel of Fig.\ \ref{Fig:zeolites}, with a controllable tunneling from  superatom to  superatom. Since this tunneling is rather weak, these systems then indeed belong to the class of strongly correlated materials, despite the ingredients being K, Al, Si and O atoms. As Arita {\em et al.} 
\cite{Arita04b} showed by means of LDA and DMFT calculations,
 K$_{n}$Al$_{12}$Si$_{12}$O$_{48}$ is Mott-Hubbard insulating
even when the nominal doping level (averaged $n$) 
is fractional, which
explains the experimental fact that K-doped
zeolites are insulators \cite{Nakano99a,Nakano00a}.
Zeolites and fullerenes show that also 
materials with $s$ and $p$ valence electrons
can be strongly correlated because of a reduced hopping.
Tehn interesting many-body physics emerges.

\section{Summary and outlook}
\label{summary}

The physics of materials with strong electronic correlations
such as transition metal oxides and heavy Fermion systems
is characterised by renormalised quasiparticles
or Mott-insulating behaviour.
Conventional electronic structure calculations, for example, in the LDA or GW approximation,
cannot capture this kind of physics and the corresponding energy scales.
This became possible
by incorporating DMFT into realistic material calculations, merging
 the strength of  LDA (or  GW) as an {\em ab initio} approach
 with that of DMFT to deal with local electronic correlations between 
 $d$ or $f$ electrons.
LDA+DMFT proved to be a breakthrough and is
by now a standard approach, albeit standardised program packages are
still imperative and self-consistent calculations where the LDA bands are renormalised
due to the DMFT correlations are still rare.
Also the calculation of physical quantities other than
the electronic spectrum, the quantity inherently obtained as the Weiss field 
in  DMFT, is not yet standard.
Since in principle  all physical quantities can be 
obtained, there are many opportunities for future work.
One should be cautious however that non-local correlations,
which are neglected in DMFT, are not particularly important
 for the physical quantity at hand.
 
So far,
LDA+DMFT has been successfully applied to many  materials,
with the focus naturally  on those with strong electronic correlations.
Among others, the volume, the electronic  and the phonon spectrum of Pu,
the changes of physical properties in Ce through the volume collapse, and the ferromagnetism
of Fe and Ni, including the famous 6 eV in Ni, have been
calculated. Most work was devoted to transition metal oxides.
Here, the  Mott-Hubbard metal-insulator transition in
V$_2$O$_3$  was investigated and the filling of the Mott-Hubbard  gap with increasing temperature 
predicted.
The gap in VO$_2$ was identified as  a spin-Peirls gap, and the origin of the
colossal magnetoresistance in manganites understood: the paramagnetic phase
is insulating-like because of the localisation of electrons
as lattice polaron, 
assisted by strong electronic correlations. Even the, at first glance, simple
system SrVO$_3$ was found to show interesting physics, i.e. `kinks'.
In contrast to the calculations for 
transition metals and  $f$ electron systems, those for transition metal oxides were restricted to the
$d$ orbitals crossing the Fermi energy, except for
one recent calculation on NiO \cite{Kunes06a}.

A severe drawback of most calculations was 
that  the Coulomb interaction
has been used as a free parameter or was chosen ad hoc.
This is against the spirit of {\em ab-initio} calculations
which, 	as a matter of principle,  should be parameter-free.
Encouraging in this respect are the calculations for  the prototypical 
$4f$ and $3d$ materials Ce and SrVO$_3$ which show that parameter-free
LDA+DMFT calculations are possible if the constrained LDA method
is employed  for determining
 the Coulomb interaction.  In the case of GW+DMFT, the Coulomb interaction
is even inherently obtained from GW. 
Here, the challenge is to do DMFT calculations with a frequency-dependent (screened)   interaction.

One should always keep in mind that LDA+DMFT is an approximation
for solving the solid state Hamiltonian. 
The starting point, LDA, 
is already an approximation even for the extended $s$ and $p$ orbitals.
The second approximation involved is the selection of the `interacting' orbitals
which by principle is basic dependent. Because of this also the determination of
the Coulomb interaction  has a considerable uncertainty, less so for well localized $4f$
orbitals but more for $3d$ and particularly $4d$ orbitals. 
Similarly, the underlying  spectral density functional theory is basis dependent,
in very contrast to standard  density functional theory.
Thirdly,  DMFT  is a many-body approximation for dealing with strongly correlated electron
systems. It neglects non-local correlations.
Moreover, the solution of the DMFT impurity problem involves possibly
a fourth approximation
if for example the  non-crossing or iterated perturbation theory is employed.
While 
numerically exact quantum Monte Carlo (QMC) simulations are also possible, they
are restricted to  temperatures at or above (roughly) room temperature since they
become too expensive in terms of CPU time at lower temperatures.
In this respect density matrix renormalization group calculations, continuous-time 
and
projective quantum Monte Carlo simulations 
are promising alternatives for zero (or low) temperature.

A major challenge for the future will be to go beyond DMFT, i.e.
taking into account non-local correlations on top of the local
DMFT correlations in electronic structure calculations. In this respect,
cluster extensions  of DMFT have already been applied intensively to
the two-dimensional Hubbard model with a focus on unconventional superconductivity.
The numerical effort restricts these
cluster extensions to short range correlations within a (relatively small) cluster.
Realistic multi-band calculations were hitherto even restricted to 
only two sites. This puts particular importance onto the correlation within this pair of sites
and is hence naturally only appropriate for particular systems such as VO$_2$ where such pair correlations dominate.
Many important  phenomena in correlated electron systems on the other hand
stem from long range correlations, e.g.
magnons, the interplay between antiferromagnetic fluctuations and unconventional
superconductivity, quantum critical points, and critical behavior in general.
How these phenomena arise is quite well understood diagrammatically for weak coupling,
which as a matter of course is not the proper starting point for 
strongly correlated electrons. 
But if we replace the bare Coulomb interaction
of the weak coupling perturbation theory by the local (fully irreducible) vertex,
we generate a set of diagrams which includes both the local DMFT diagrams
and the dominant diagrams for long-range correlations.
This is the very idea of the dynamical vertex approximation which has been
 introduced 
most recently along with alternative approaches.
Another long term goal for electronic structure calculations with DMFT
is the calculation of the
the ionic positions hitherto taken from experiment,
the inclusion of lattice dynamics,
and eventually  molecular dynamics on the basis of LDA+DMFT.

Continuing the fruitful cooperation between
bandstructure and many-body physicists, we can optimistically
face these challenges in the future. Electronic structure calculations
employing DMFT or its diagrammatic and cluster extensions
will be a prospering area of research. In particular,
such approaches will
further
improve our understanding of systems with strongly correlated electrons.
The advances in electronic structure calculations through DMFT
put
 our ability to predict physical quantities of such strongly
correlated
materials onto a similar level as 
conventional electronic structure calculations
 for weakly correlated materials --- at last.

\section*{Acknowledgments}

This review has benefited from discussions with
too many people to list. But I would like to take the
opportunity to thank
my collaborators
 O.\ K.\ Andersen, V.\ I.\ Anisimov, R.\ Arita, N.\ Bl\"umer, M.\ Feldbacher, A.\ Katanin, G.\ Keller, M.\ Kollar,  A.\ K.\ McMahan,, W.\ Metzner, I.\ A.\ Nekrasov, Th.\ Pruschke, 
R.\ T.\ Scalettar,  A.\ Toschi,
 M.\ Ulmke, D.\ Vollhardt, A.\ Yamasaki and  Y.-F.\ Yang.
Moreover, I apologise to those whose work
could not be discussed in more detail due to
restrictions of time and length.
Financially, this work was supported by the Deutsche
Forschungsgemeinschaft through the Emmy Noether program.




\label{references}

\end{document}